\documentclass[twoside,12pt]{report}
\usepackage{tikz}
\usepackage{cite}
\usepackage{url}
\usepackage{amsmath}
\usepackage{amssymb}
\usepackage{amsthm}
\usepackage{latexsym}
\usepackage{graphicx}
\usepackage{feyn}
\usepackage{float}
\usepackage{epstopdf}
\usepackage{verbatim}
\usepackage[section]{placeins}
\usepackage{amsfonts}
\usepackage{color}
\usepackage{calligra}
\usepackage[T1]{fontenc}
\usepackage{lmodern}
\usepackage[utf8]{inputenc}
\usepackage[]{xcolor}
\usepackage{bm}
\usepackage{subcaption}
\usepackage{enumerate}
\usepackage[protrusion=true,expansion=true]{microtype}
\usepackage{fancyhdr}
\usepackage[hmarginratio=1:1]{geometry}
\usepackage[bottom]{footmisc}
\usepackage[pdfborder={0,0,0},pdfauthor={Rupert Small}, colorlinks=true, linkcolor=blue, citecolor=blue]{hyperref}
\usepackage{setspace}
\definecolor{apricot}{RGB}{253, 213, 177}
\newcommand{\bi}{\boldsymbol{i}}
\newcommand{\bj}{\boldsymbol{j}}

\newcommand{\RR}{\mathbb{R}}

\numberwithin{equation}{section} \numberwithin{lemma}{section}

\makeatletter
\newcommand{\contraction}[5][1ex]{%
  \mathchoice
    {\contraction@\displaystyle{#2}{#3}{#4}{#5}{#1}}%
    {\contraction@\textstyle{#2}{#3}{#4}{#5}{#1}}%
    {\contraction@\scriptstyle{#2}{#3}{#4}{#5}{#1}}%
    {\contraction@\scriptscriptstyle{#2}{#3}{#4}{#5}{#1}}}%
\newcommand{\contraction@}[6]{%
  \setbox0=\hbox{$#1#2$}%
  \setbox2=\hbox{$#1#3$}%
  \setbox4=\hbox{$#1#4$}%
  \setbox6=\hbox{$#1#5$}%
  \dimen0=\wd2%
  \advance\dimen0 by \wd6%
  \divide\dimen0 by 2%
  \advance\dimen0 by \wd4%
  \vbox{%
    \hbox to 0pt{%
      \kern \wd0%
      \kern 0.5\wd2%
      \contraction@@{\dimen0}{#6}%
      \hss}%
    \vskip 0.2ex%
    \vskip\ht2}}

\newcommand{\contraction@@}[3][0.06em]{%
  \hbox{%
    \vrule width #1 height 0pt depth #3%
    \vrule width #2 height 0pt depth #1%
    \vrule width #1 height 0pt depth #3%
    \relax}}
\newcommand{\Vk}{V}

\begin{document}
\fancyfoot{}
 \fancyhead[EL]{\fancyplain{}{\thepage}}
\fancyhead[ER]{\fancyplain{}\leftmark{}}
\fancyhead[OL]{\fancyplain{}\rightmark{}}
\fancyhead[OR]{\fancyplain{}{\thepage}}
\begin{titlepage}
\begin{center}
\rule{\textwidth}{.1cm}
{\huge \bfseries{On the Unification of \\[0.5cm]Random Matrix Theories}}
\rule{\textwidth}{.15cm}\\[0.5cm]
{By}\\[0.5cm]
{\Large {Rupert Small}}\\[1.5cm]
\includegraphics[width=4cm]{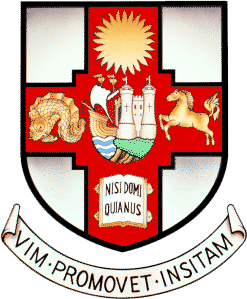}\\[0.5cm]
\textsc{University of Bristol}\\
\textsc{School of Mathematics}\\[3cm]
\small{\textit{A dissertation submitted to the University of Bristol in accordance with the requirements for the degree of Doctor of Philosophy in the School of Mathematics}}\\[0.2cm]
March 2015.
\end{center}
\end{titlepage}
\pagenumbering{roman}

\section*{Abstract}
Random Matrix Theory (RMT) is the study of matrices with random variables determining the entries, and various additional symmetry conditions imposed on the matrices. A comparitively young theory, it has its roots in Hungarian physicist and mathematician Eugene Paul Wigner's work in the early 1950's\cite{wigner1, wigner2}. In the following years and continuing unabated to the present day, it has permeated nearly every area of modern physics and even number theory\cite{mehta, guhr, akemann}. In recent decades attempts have been made to further refine what has become the \emph{canonical} random matrix theory with its associated symmetry conditions, by considering symmetries which allow the matrix representations of quantum potentials to impose $k$-body forces on the particles in a system containing $m$ particles ($k \le m$)\cite{weid, weidreview, srednicki, kota}. This is a generalisation of canonical RMT and is the topic of this thesis. It will be refered to as the \emph{unification of random matrix theories}, because every randomised $k$-body potential gives rise to a new ensemble. For $k=m$ these ensembles are exactly those known and studied already under the rubrik of canonical RMT, but for $k < m$ the resulting ensembles are different and little is known about them. The phase space of random matrix ensembles generated by randomised $k$-body potentials represents an opportunity to unify all random matrix theories into one, single mathematical theory of random matrices. The process involves \textit{embedding} the \textit{k}-body potential into the \textit{m}-particle state space creating what has become known as the \emph{embedded ensembles}. The embedded ensembles, first introduced by Mon and French\cite{mon} in 1975, gave physicists a framework for studying many-body interactions using random matrix theory, but also presented practitioners with the challenge of an incomplete theory, as it showed that the (canonical) form of random matrix theory under consideration at the time was a single instance of a much larger phase space of random matrix theories. This meant that only a small subsection of the possible statistics were being calculated. Canonical RMT was already providing scientists, engineers and mathematicians insights into a variety of different patterns and physical phenomena in number theory, engineering, physics and computer science. However, these connections only became visible when they were uncovered, so the potential that unification of random matrix theories offered presented both a big opportunity and new mathematical problems\cite{weid, weidreview,srednicki}. Alternative mathematical methods would be needed to manage the new complexities manifested by studying the unification of random matrix theories.

One of the greatest contributions to this area of study occured (in the author's opinion) at the turn of the last decade with the publication of the breakthrough paper by Benet, Rupp and Weidenm\"uller (BRW) which showed how a process of eigenvector expansions could be used to calculate certain statistical properties of $k$-body potentials\cite{weid}. These methods were a great advance but were also difficult to implement, and it remains unclear if they can practically be used to calculate moments of the level density of embedded ensembles which are higher than the fourth moment.

This thesis presents a completely different approach to the method proposed by (BRW). A new method involving \emph{particle diagrams} will be introduced and developed into a rigourous framework for carrying out embedded random matrix calculations. Using particle diagrams and the attendent methodology including \emph{loop counting} it becomes possible to calculate the fourth, sixth and eighth moments of embedded ensembles in a straightforward way. The method, which will be called the method of \textit{particle diagrams}, proves useful firstly by providing a means of classifying the components of moments into particle paths, or loops, and secondly by giving a simple algorithm for calculating the magnitude of combinatorial expressions prior to calculating them explicitly. By confining calculations to the limit case $m \ll l \to \infty$ this in many cases provides a sufficient excuse not to calculate certain terms at all, since it can be foretold using the method of particle diagrams that they will not survive in this asymptotic regime. Applying the method of particle diagrams washes out a great deal of the complexity intrinsic to the problem, with sufficient mathematical structure remaining to yield limiting statistics for the unified phase space of random matrix theories.

Finally, since the unified form of random matrix theory is essentially the set of all randomised $k$-body potentials, it should be no surprise that the early statistics calculated for the unified random matrix theories in some instances resemble the statistics currently being discovered for quantum spin hypergraphs and other randomised potentials on graphs\cite{hess, schroeder, huw}. This is just the beginning for studies into the field of unified random matrix theories, or embedded ensembles, and the applicability of the method of particle diagrams to a wide range of questions as well as to the more exotic symmetry classes such as the symplectic ensembles, is still an area of open-ended research.

\begin{center}
.....
\end{center}
\newpage

\section*{Acknowledgements}
I would like to give my gratitude and heartfelt thanks to Sara-Lea Small, Gale Pullen and Adam Zalcman for their role in my education, ultimately leading me to study Mathematics and Physics at the University of Bristol. I would also like to thank Prof.~John Hannay for his creative and inspirational teaching during my undergraduate studies. 
\begin{center}
..
\end{center}
Thanks to my postgraduate supervisor Dr.~Sebastian M\"uller of the School of Mathematics for pointing me towards such an interesting area of research and to Dr R\'emy Dubertrand for his flawless lectures on Random Matrix Theory.
\begin{center}
..
\end{center}
Finally, many thanks to Leo Brennan, Kildare County Council (Ireland) and the EPSRC (United Kingdom) for their generous grants which enabled me to pursue my studies to the level of postgraduate -- something which appeared to me as a child as being far beyond the realms of possibility.

\newpage
\section*{Author's Declaration}
I declare that the work in this thesis was carried out in accordance with the Regulations of the University of Bristol. The work is original except where indicated by special reference in the text. No part of the dissertation has been submitted for any other degree.
Any views expressed in the dissertation are those of the author and do not necessarily represent those of the University of Bristol. The thesis has not been presented to any other university for examination either in the United Kingdom or overseas.

\vspace*{15mm}
\begin{flushright}
\begin{tabular*}{0.7\textwidth}{cccr}
\empty &\empty &\empty &{\calligra{\LARGE Rupert Small}}\\
  \hline
  \empty & \empty & \empty & \empty \\
\end{tabular*}
\end{flushright}
\vspace*{-10mm}
\hspace*{45mm}
\vspace*{15mm}
\hspace*{65mm} March 2015
\newpage
\tableofcontents
\listoffigures%
\newpage
\thispagestyle{plain}
\section*{Notation}
 \vspace*{1cm}\hspace*{2cm}
\begin{table}[!th]
\begin{center}
\begin{tikzpicture}
\node (table) [inner sep=0pt] {
\begin{tabular}{c | l}
\empty & \empty \\
   {\bf{Symbol}} & {\bf{Definition}} \\
\empty & \empty \\
\hline
\hline
 \empty & \empty \\
   ${.}^{\dag}$ & Complex conjugate transpose \\
\empty & \empty \\
   $\overline{\mbox{\;.\;\raisebox{1.5mm}{}}}$ & Ensemble average \\
\empty & \empty \\
   $|.\rangle$ & Column vector \\
\empty & \empty \\
   $\otimes$ & Tensor product \\
\empty & \empty \\
   $\mathcal{O}$ & Growth rate\\
\empty & \empty \\
   ${\mathbf{C}}^{\infty}$ & Infinitely differentiable \\
\empty & \empty \\
\end{tabular}
};
\draw [rounded corners=.5em] (table.north west) rectangle (table.south east);
\end{tikzpicture}
\end{center}
\end{table}
\newpage
\pagenumbering{arabic}
\chapter[Introduction]{Introduction}\label{ch:intro}
\section{Physics and Unification}
Physics is the pursuit of natural laws, or models, which characterise and thereby predict the behaviour of events within our universe. This includes all measurable properties of the natural world, from how sub-atomic particles interact to the emergence of complicated molecules, proteins, plants and animals. Scientific knowledge can be seen as a vast collection of these models and measurable properties, collected over centuries from the efforts, insights and guesswork of all past scientists, both theoretical and experimental.
Each scientific discovery has its own unique history, interwoven into the fabric of previous discoveries. Sometimes a model discovered by a previous generation of scientists will be superseded by a new theory. Many physicists acknowledge that the cycle may never end\cite{feynman,bohr,weinberg, barrow}. That it is possible that the pursuit of knowledge is a never ending process of destruction and rediscovery, and that our understanding of nature can always be improved, albeit never completed. The examples which epitomise this process of continual reformation is the replacement of Newtonian laws of motion by Relativistic laws, and separately the discovery that Newtonian Mechanics is in fact a limiting case of the much more general theory of Quantum Mechanics\cite{feynman}. It also occurs that sometimes two models which previously were thought to be distinct are shown to be the result of a larger more general law. This is called \textit{unification}. Unification is one of the most important aspects of scientific labour, since it reveals the connections between phenomena in the natural world which were previously thought of as being different, but which are in fact a result of the same underlying patterns. A seminal example of unification is the family of mathematical expressions called Maxwell's equations\footnote{(i)~\unexpanded{$\nabla \cdot \mathbf{E} = \frac{\rho}{\epsilon_0}$}~~(ii)~\unexpanded{$\nabla\cdot\mathbf{B}=0$}~~(iii)~\unexpanded{$\nabla\times\mathbf{E} = -\frac{\partial \mathbf{B}}{\partial t}$}~~(iv)~\unexpanded{$\nabla\times\mathbf{B} = \mu_0 \left({\mathbf{J} +\epsilon_0\frac{\partial\mathbf{E}}{\partial t}}\right)$}.~ Together equations (iii) and (iv) describe the electric field as a function of the magnetic field and vice versa, showing that $\mathbf{E}$ and $\mathbf{B}$ are two aspects of a single unified physical process.}, which explain the relationship between electricity and magnetism, unifying two effects which were previously thought of as being distinct.

Since the process of discovery often leads to different approaches to similar problems\cite{feynman2} it is not always obvious that the underlying patterns governing different phenomena are the same. Not least of the problems is that seemingly disparate scientific questions are often studied within the confines of their own specific jargon. To break down the barriers between problems and to reveal their inherent similarities it is essential to look uncompromisingly at the underlying mathematics without being swayed by the layer of human language which has been added ``on top''. Another essential feature of unification theories is that they include mathematical parameters which connect things which were previously thought of as being unrelated. In this sense unification requires an expansion of the mathematical expressions to ``make room'' for effects which were not given a voice in previous theories.
\begin{figure}[!htb]\label{fig:unification}
\centering
\begin{tikzpicture}
\node[circle, draw=magenta!80,  inner sep=87pt] at (7,7) {}; 
\node[circle, draw=red!50,  inner sep=50pt] at (8,8) {}; 
\node[circle, draw=red!90,  inner sep=30pt] at (8.5,8.5) {}; 
\node[circle, fill=black,  inner sep=5pt] at (9.2,9.2) {}; 
\node[text=black,  inner sep=0pt] at (2.93,3.5) {\LARGE{\bf {\color{black}Physics}}}; 
\node[text=black,  inner sep=0pt] at (7.05,5.25) {\Large{\bf {\color{black}Many-Body Potentials}}}; 
\node[text=black,  inner sep=0pt] at (7.85,6.81) {\large\bf {\color{black}Unified RMT}}; 
\node[text=black,  inner sep=0pt] at (7.2,9.2) {\bf {Canonical RMT}}; 
\end{tikzpicture}
\caption[Unification Diagram]{The canonical form of random matrix theory is just a single point in a whole landscape of possible random matrix theories (unified RMT). Unified RMT in turn represents the set of randomised quantum many-body potentials, also known as the embedded ensembles.

\hrulefill}
\end{figure}
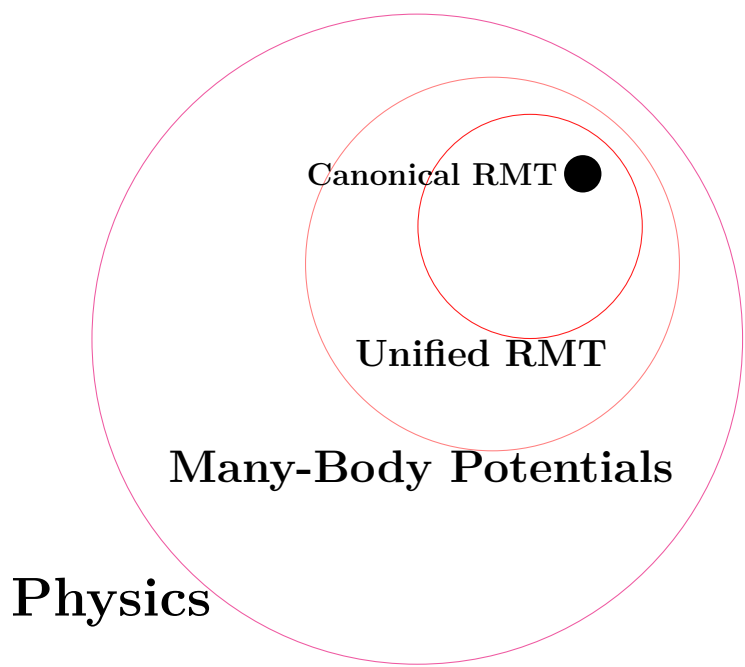
This thesis will look at a particular case of unfinished unification in physics and mathematics; the unification of random matrix theories. It will do this firstly by observing that the common, or canonical, form of random matrix theory is just a single point in a whole landscape of possible random matrix theories. This landscape of random matrix theories is known as the embedded ensembles (see Chapter \ref{ch:RMTs}) which are themselves a subclass of quantum $k$-body potentials. In Chapter \ref{sec:WSCL} it will be shown heuristically how Wigner's Semi-Circle law for canonical random matrices can be calculated using diagrammatic methods. In the same chapter an illustration of the method of supersymmetry will show the sometimes stark difference in mathematical approaches to the same problem, in this case Wigner's Semi-Circle Law. Finally in Chapters \ref{ch:MBRMTCH} and \ref{ch:THM} it will be shown that a method of particle diagrams, involving graphs of the relationships between quantum states, is sufficient for calculating the moments of the level density of the embedded (unified) ensembles. The majority of the results presented here have been published in the rapid communication \cite{small} and the article \cite{small2}.
\begin{center}.....\end{center}
\chapter[Random Matrix Theories]{Random Matrix Theories}\label{ch:RMTs}
\section[Canonical RMT]{Canonical RMT}\label{sec:CRMT}
\subsection{Statistical Landscapes}\label{sec:sl}
In the early 1960's theoretical physicist Freeman Dyson published a trio of papers with the heading \textit{Statistical Theory of the Energy Levels of Complex Systems}\cite{dyson1,dyson2,dyson3}. The results (which were also inspired by earlier works of Eugene Wigner and Madan Lal Mehta\cite{wigner1, wigner2, mehta}) and the related field of mathematics which grew around them, became known as random matrix theory (RMT). The basic tenet of random matrix theory as introduced by Dyson was the investigation of the statistics of ``\textit{all physical systems with equal probability}''. To paraphrase events, a challenge which researchers were facing at the time was to characterise the behaviour of large complex nuclei. Wigner, Dyson and Mehta envisioned approaching the problem by assuming nothing, or at least as close to nothing as possible. From this position they began to investigate the quantities that can be known about the behaviour of a particular physical system. Their starting point was the hamiltonian of the physical system with the single constraint
\begin{equation}\label{eq:sl1}
H=H^{\dag}
\end{equation}
for physicality. Assuming nothing else, one would not be able to say anything about a \textit{particular} hamiltonian $H$. However, by allowing each $H$ to occur with some probability $p(H)$ it becomes possible to study the whole space of probabilistically weighted hamiltonians. From this point onwards the investigation necessarily becomes statistical, and these ensembles of hamiltonians are studied by asking questions of the form ``\textit{What is the probability that...(etc.)}''.  The proposition which then needs to be verified with experiment is that a natural system is some unknown hamiltonian obeying (\ref{eq:sl1}) and therefore the typical behaviour of the system may obey the behaviour specified by the ensemble average of this. In other words, although the information extracted from these ensembles of hamiltonians takes the form ``\textit{What is the probability that...}'', it is possible to use the probabilistic results to say something about the properties of the ``typical'' (read average) of a hamiltonian, because
\begin{equation}\label{sl2}
\overline{f(X)} = \int f(X)\cdot (\mathrm{the~probability~that~X})~dX
\end{equation}
where the horizontal line above $f(x)$ indicates the average taken over the ensemble of $X$'s. Investigating random hamiltonians in this way is what gave fruit to the field of random matrix theory -- a family of sophisticated mathematical tools and equations for extracting as much information as possible about an ensemble of hamiltonians starting from only a handful of very basic assumptions\cite{mehta}. The additional assumptions would usually take the form
\begin{align}\label{eq:sl3}
\mathrm{1.~~}&H = H^{\dag}\notag\\
\mathrm{2.~~}&H = S^{-1}H S\\
\mathrm{3.~~}&\mathrm{Gaussian~probability~determining~the ~matrix~elements~of}~H\notag
\end{align}
where the matrix $S$ is some predefined matrix determined by the particular physical setup at hand. It imposes some symmetry on $H$ additional to the hermitian symmetry it already satisfies by virtue of (\ref{eq:sl1}). The three classical examples for $S$ as introduced by Dyson will be given in the next section (one of which simply takes $S=1$). These essentially divide the physicial landscape into four statistical landscapes; three groups defining a ``threefold way'' and a fourth being the complement of these (everything else not included in the three groups determing the threefold way). Although the initial development of a theory of random matrices focused on the first three symmetry groups defined by Dyson, it turns out that this classification defines just a single point in a much larger landscape of random matrix theories and a more nuanced approach to symmetry is required to unify them all.
\subsection[Symmetry Classes]{Symmetry Classes}\label{sec:SC}
Dyson's initial development of random matrix theory was placed within the context of a ``threefold way''. These are three symmetry classes which determine a set of the additional symmetries (see (\ref{eq:sl3})) obeyed by physical hamiltonians. Recreating some of Dyson's steps, this section will detail the formulation of the three symmetry classes determining the additional constraints on the matrix elements of $H$ when the hamiltonian refers to a quantum system from one of three specific groups. Later it will be shown why this analysis is incomplete, constituting just a single point in a phase space of additional symmetries. The three groups are distilled from the set of all possible hamiltonians by asking some fundamental questions. For example it is reasonable to ask the question ``What form does the potential take if the physical system contains only fermions, or only bosons?'' Similarly, it is natural to wonder ``What form does the potential take if the system is time-reversal invariant compared to the case when it is not?'' These are questions of symmetry which are integral to the original purpose of RMT, which is to assess the statistical properties of ensembles of hamiltonians. Although random matrices do not have to be hamiltonians per se -- they can also be scattering matrices\cite{akemann}, or any other matrix for that matter -- the classical categorisation of RMT into three symmetry classes as given by Dyson takes its starting point as the hamiltonian of a physically permitted quantum system. In other words, the starting point is the hamiltonian satisfying the Schr\"odinger equation of the system
\begin{equation}\label{SC1}{H}\psi = i\hbar\frac{\partial {\psi}}{\partial t} = E\psi.\end{equation}
Even with a modest amount of complexity Schr\"odinger's equation becomes too difficult to solve analytically, which is why RMT was developed in the first place. In RMT it is assumed that the Hamiltonian ${H}$ is highly complex and only some rudimentary symmetry properties of the physical system are known. After placing these restrictions on the hamiltonian \textit{all remaining free parameters are randomised}. Thereafter an attempt is made to calculate whatever statistical properties can be yielded as a result of the randomisation and symmetrisation process. Using this \textit{probabalistic} view the problem becomes one of studying \textit{ensembles} of matrices, each grouped into families depending on the set of minimum conditions imposed upon them and each weighted by some probability of occurring within the ensemble. Instead of studying the eigenvalues of a single instance of $H$ one calculates \textit{distributions} of eigenvalues defined by how likely they are to occur across the entire ensemble.
\subsubsection{Particle Wave Functions}
The starting point for calculating the three symmetry classes proposed by Dyson is a remarkable law of nature which through repeated experimentation has established that any quantum state (\emph{a.k.a wave function} or \emph{p.d.f}) describing a physical experiment containing elementary particles is either even-symmetric or odd-symmetric under particle exchange. Particles with even wave functions (bosons) and particles with odd wave functions (fermions) can be represented mathematically with the respective expressions
\begin{align}\label{SC2}{\psi}(r_{P_1}, \ldots ,r_{P_m}) &= \psi(r_{1}, \ldots ,r_{m}) \mbox{   (bosons)}\notag\\
{\psi}(r_{P_1}, \ldots ,r_{P_m}) &= (-1)^{P}\psi(r_{1}, \ldots ,r_{m}) \mbox{   (fermions)}\end{align}
where $\{P_1,\ldots,P_m\}$ is a permutation of $\{1,\ldots,m\}$ and $P$ is the number of pairwise permutations between the single particle states required to bring $\{P_1,\ldots,P_m\}$ back to the initial configuration. More concisely, for the wave function of both bosons and fermions the following symmetry condition holds 
\begin{equation}\label{SC3}{\psi}(r_{P_1}, \ldots ,r_{P_m}) = \xi^{\,P}\psi(r_{1}, \ldots ,r_{m})\end{equation}
\begin{table}[!t]
\begin{center}
\begin{tikzpicture}
\node (table) [inner sep=0pt] {
\begin{tabular}{l|l}
\textbf{Bosons} $(\xi = 1 )$ & \textbf{Fermions} $(\xi = -1)$\\
\hline
photons & protons \\
\hline
pions & neutrons \\
\hline
mesons & electrons \\
\hline
gluons & muons \\
\hline
${}^{4}$He & neutrinos \\
\hline
{} & quarks \\
\hline
{} & ${}^{3}$He \\
\end{tabular}
};
\draw [rounded corners=.5em] (table.north west) rectangle (table.south east);
\end{tikzpicture}
\caption[Table of common bosons and fermions]{Table of common bosons and fermions, the two classes of quantum particles in nature. Bosons can always be represented by an even probability density function while fermions, obeying the Pauli Exclusion Principle, can always be represented by an odd probability density function.

\hrulefill}
\label{ex:table}
\end{center}
\end{table}
where as before $\xi=1$ for bosons and $\xi=-1$ for fermions. A composite particle composed of any number of bosons and an even(odd) number of fermions will have an even(odd) state representation. That is, the composite particle itself be a boson(fermion). Hence it is possible to divide up the vast physical landscape of possibilities into just four groups;
\begin{enumerate}
  \item \textit{Time Reversal Invariant Bosonic systems.}
  \item \textit{Non Time Reversal Invariant Bosonic and Fermionic systems.}
  \item \textit{Time Reversal Invariant Fermionic systems.}
  \item \textit{None of these.}\\
\end{enumerate}
\noindent The first trio of systems define the three symmetry classes of canonical RMT. One may protest that all possible realities are taken into account by the first three groups. However, each group is defined in terms of the \textit{minimum} set of restrictions on each member so systems requiring additional restrictions must go into the fourth group. What are the attendant symmetries that must be satisfied by a potential ${H}$ for a system in each group?
\subsubsection{Invariance under Time Reversal}\label{sec:IUTR}
To form a meaningful mathematical picture of how time reversal invariance relates to a physical measurement it is first necessary to look at unitary operators, or more precisely in the present context, \textit{unitary matrices}. A good introduction to the following ideas can also be found in \cite{mehta, haake}. A unitary matrix $U$, is a matrix which conserves the overlap between states
\begin{equation}\label{eq:unitary1}\langle U\psi|U\phi \rangle  = \langle \psi|\phi \rangle. \end{equation}
Rearrangment gives
\begin{equation}\label{eq:unitary2}UU^{\dag} = 1.\end{equation}
Expressed in terms of a summation over the individual elements of the matrix this is equivalent to
\begin{equation}\label{eq:unitary3}\sum_k U_{ik}U_{jk}^{*} = \delta_{ij}.\end{equation}
In other words the rows (columns) of a unitary matrix are orthonormal. Replacing $\phi$ with $\psi$ in (\ref{eq:unitary1}) gives ||$\langle U\psi|U\psi \rangle||  = ||\langle \psi|\psi \rangle||$ so that Unitary transforms also conserve the length of a vector. From the above it follows that $U \in \mathbb{C}^{n\times n}$ rigidly rotates a vector in $\mathbb{C}^{n}$ while conserving its length. Notably, a state transformed as $\psi \to U\psi$ is indistinguishable from the original state when the measurement operator is also transformed as
\begin{equation}\label{eq:measurement.operator.transform}
  A \to UAU^{\dagger}.
\end{equation}
Additionally assuming a system of particles whose states are of the form $\psi = \psi_x\,e^{i\omega_1 t}$, it follows immediately that the overlap between two arbitrary quantum states is
\begin{equation}
\langle{\psi|\phi}\rangle = \langle{\psi_x|\phi_x}\rangle \cdot e^{i(\omega_2 - \omega_1)t}.
\end{equation}
If time is flowing in the forward direction the coordinates of a particle will be given by its position, momentum and $t$-variable $\{{\bm x}, {\bm p}, t\}$ whereas if time runs backwards the particle's momentum and time variable will reverse at every position so that the coordinates become $\{{\bm x}, -{\bm p}, -t\}$ and the overlap between two states will be
\begin{equation}\label{eq:trev1}
\langle{\psi^R|\phi^R}\rangle = \langle{\psi_x|\phi_x}\rangle \cdot e^{- i(\omega_2 - \omega_1)t} = \langle{\psi|\phi}\rangle^{*}
\end{equation}
where $|\psi^R\rangle$ is the wave function $|\psi\rangle$ under time reversal. Representing the time reversal of a state as an operator gives the tautological expression
\begin{equation}\label{eq:trev2}{T}\psi = \psi^R\end{equation}
where by (\ref{eq:trev1}) ${T}$ must obey
\begin{equation}\label{eq:trev3}\langle{{T}\psi|{T}\phi}\rangle = \langle{\psi|\phi}\rangle^{*}.\end{equation}
That is, the operator ${T}$ is anti-unitary. Additionally, under time reversal the Hamiltonian of the system becomes
\begin{equation}
  H^{R} = {T}H{T}^{-1}.
\end{equation}
Comparing (\ref{eq:trev3}) with (\ref{eq:unitary1}) it can be seen that the only difference is complex conjugation, so ${T}$ can be written as
\begin{equation}{T} = KC\end{equation}
where $K$ is unitary and the operator $C$ is defined by its action of taking the complex conjugate of the state on which it operates. Since double application of time reversal to a state should leave it in a state indistinguishable from itself \textit{i.e.} differing only in phase, ${T}$ must without exception obey the relation
\begin{equation}\label{eq:trev4}{T}^{ 2} = \alpha\end{equation}
with $|\alpha| = 1$, so
\begin{equation}\label{eq:trev5}KK^{*} = \alpha.\end{equation}
Additionally, by (\ref{eq:measurement.operator.transform}) $T$ transforms as
\begin{equation}\label{eq:trev5.2}{T} \to UTU^{-1}\end{equation}
under unitary transform of the wave-function by $U$, and the matrix $K$ defined previously transforms as
\begin{equation}\label{eq:trev5.3}K \to UKCU^{-1} = UKC\left({U^{T}}\right)^{*} = UKU^{T}.\end{equation}
\noindent A system is called invariant under time reversal if
\begin{equation}\label{eq:trev5.4}H^R = H\end{equation}
where $H$ is the hamiltonian. A matrix satisfying (\ref{eq:trev5.4}) is called self-dual. A time reversal invariant system therefore satisfies the following sequence of equalities
\begin{align}\label{eq:trev5.5}H &= THT^{-1} = KCH(KC)^{-1}\notag\\
&= KCHCK^{-1} = KH^{T}K^{-1}\end{align}
where the fact that $H$ is hermitian has been used, so that $H^{*} = H^{T}$. Recalling that $K$ is unitary, implying $K^{*}K^{T} = 1$, and motivated by a stroke of insight to multiply (\ref{eq:trev5}) by $K^{T}$ it is seen that
\begin{equation}\label{eq:trev6}K = \alpha K^{T} = \alpha^2 K\end{equation}
which means that the value of $\alpha$ is restricted to satisfying
\begin{equation}\label{eq:trev7}\alpha = \pm 1.\end{equation}
\subsubsection{Twofold Way}
The insight that $\alpha$ can take two possible values reveals mathematically that there are two distinct types of systems, both satisfying time-reversal invariance. Equation (\ref{eq:trev7}) implies two possible outcomes
\begin{align}\label{eq:trev8}KK^{*} &= 1 ~~ \textrm{or}\\
\label{eq:trev9}KK^{*} &= -1.\end{align}
In addition to being unitary, if (\ref{eq:trev8}) holds $K$ is symmetric and (\ref{eq:trev9}) implies $K$ is anti-symmetric. Following, but not proven here, is that in the symmetric case there is a unitary transform which by (\ref{eq:trev5.3}) gives
\begin{equation}\label{eq:trev10}K = 1\end{equation}
which implies that the system is bosonic (even spin\footnote{\label{note1}This thesis will use bosonic and fermionic interchangeably with, respectively, even spin and odd spin throughout.}) and in the anti-symmetric case there is a unitary transform such that $K$ becomes
\begin{equation}\label{eq:trev11}
 K =
 \begin{pmatrix}
  0 & +1 &  & \\
  -1 & 0 &  &  \textrm{\Huge{o}} &  \\
   & & \ddots &  &  \\
   & \textrm{\Huge{o}} & & 0 & +1 \\
   & & & -1 & 0\\
 \end{pmatrix}
\end{equation}
which implies that the system is fermionic (odd spin${}^{\ref{note1}}$)\cite{mehta, haake}. That is, a basis can be chosen $\psi \to U\psi$ for each of the two possible time reversal invariant systems such that $K$ has the above forms for each case respectively. After choosing such a basis subsequent transforms on $K$ can only be performed if the relevant relation (\ref{eq:trev10}) or (\ref{eq:trev11}) remains true. Hence for bosonic systems further transforms are restricted by (\ref{eq:trev5.3}) and (\ref{eq:trev10}) to being of the form
\begin{equation}\label{eq:trev12}K \to OKO^{T}\end{equation}
where $O$ \textit{must} be orthogonal (real unitary) and hence by (\ref{eq:trev5.5}) $H$ is real hermitian symmetric. To summarise, the above developments gives justification for the following definition
\newtheorem{definition}{Definition}
\begin{definition}
\label{trev.even1} The ensemble of Hamiltonians of Bosonic time reversal invariant systems are hermitian symmetric matrices statistically invariant under
\begin{equation}\label{eq:trev13}H \to OHO^{T}\end{equation}
where $O$ is orthogonal.
\end{definition}
\noindent Here statistical invariance means that the probability of a given $H$ occuring within the ensemble is the same as the probability of any other $H$ found after applying the transform, which in this case is (\ref{eq:trev13}). The ensemble defined by Definition \ref{trev.even1} along with the condition that the \textit{p.d.f.} of the elements of a member of the ensemble are statistically independent defines the Gaussian Orthogonal Ensemble (GUE) which is denoted by $\beta = 1$.

Turning now to the fermionic case note that by (\ref{eq:trev11}) the number of rows (columns) of $K$ is always even. Once a basis is chosen such that (\ref{eq:trev11}) holds, further transforms on $K$ are restricted to satisfying
\begin{equation}\label{eq:trev14}K = ZKZ^{T}\end{equation}
where $Z$ is unitary. Matrices $Z$ satisfying (\ref{eq:trev14}) form what in the literature is called the symplectic group. Notice that by substituting (\ref{eq:trev11}) into (\ref{eq:trev5.5}) it can be seen that the hamiltonian of a time reversal invariant Fermionic system satisfies
\begin{equation}\label{eq:ud9} (THT^{-1})_{ij} = (-1)^{i+j}H^{*}_{i+(-1)^{i+1},~j+(-1)^{j+1}} = H_{ij} = H^{\dag}_{ij}\end{equation}
which can be written as
\begin{equation}\label{eq:trev15}
H =
 \begin{pmatrix}
  e_1 & 0 & a_{11} & b_{11} & a_{12} & b_{12} & \cdots & a_{1m} & b_{1m} \\
   & e_1 & -b_{11}^{*} & a_{11}^{*} & -b_{12}^{*} & a_{12}^{*} & \cdots & -b_{1m}^{*} & a_{1m}^{*}  \\
   & & e_2 & 0 & a_{21} & b_{21} & \cdots & a_{2m} & b_{2m} \\
   & & & e_2 & -b_{21}^{*} & a_{21}^{*} & \cdots & -b_{2m}^{*} & a_{2m}^{*} \\
   & & & & \ddots & & & \vdots & \vdots \\
   & & & & & \ddots & & \vdots & \vdots \\
   & & \textrm{\Huge{*}} & & & & & e_m & 0 \\
   & & & & & & & & e_m \\
 \end{pmatrix}
\end{equation}
where the * in the lower diagonal denotes that the matrix is hermitian so that the lower diagonal is defined by the upper diagonal elements. The diagonal of $H$ is made up of $2 \times 2$ blocks of the form
\begin{equation}\label{eq:trev16}
\begin{pmatrix}
e & 0 \\
0 & e \\
\end{pmatrix}
\end{equation}
where by hermitian symmetry $e$ must be real. The off diagonal components consist of $2 \times 2$ blocks of the form
\begin{equation}\label{eq:trev17}
\begin{pmatrix}
a & b \\
-b^{*} & a^{*} \\
\end{pmatrix}
\end{equation}
where $a, b \in \mathbb{C}$. A useful shortcut to the same conclusion is to calculate the matrix elements of $H$ relative to the basis
\begin{equation}\label{eq:trev171}\psi_1,~ T\psi_1, ~\psi_2,~ T\psi_2, ~\ldots ,~\psi_{2n},~ T\psi_{2n}. 
\end{equation}
Since $T=KC$ and for this (symplectic) case $T^2=-1$ it follows straightforwardly that
\begin{align}
\langle \psi|H|T\psi \rangle &= -\langle T\psi |H|\psi \rangle^{*}\label{eq:trev172}\\
\langle \psi|H|\psi \rangle &=\langle T\psi|H|T\psi \rangle^{*}\label{eq:trev173}
\end{align}
which is the equivalent of (\ref{eq:trev17}). Taking summary once again, the above discussion gives motivation for the following definition.
\begin{definition}
\label{trev.odd1} The ensemble of Hamiltonians of Fermionic time reversal invariant systems are hermitian self-dual ($H^{R} = H$) matrices statistically invariant under
\begin{equation}\label{eq:trev18}H \to ZHZ^{-1} = ZH(KZ^{T}K^{-1}) = ZHZ^{R}\end{equation}
where $Z$ is symplectic.
\end{definition}
\noindent Matrices which additionally have the property that entries not related by symmetry are statistically independent form the Gaussian Symplectic Ensemble (GSE), which is denoted by $\beta =4$.
\subsubsection{Non Time-Reversal Invariance}
The Hamiltonian of non time-reversal invariant Bosonic and Fermionic systems is unrestricted other than by statistical invariance under a unitary transformation. Although this has been a guiding restriction for the cases $\beta=1$ and $\beta=4$, the additional demands of time reversal invariance confined these to having the symmetries defined above. Relaxing these restrictions, which implies $S=1$  in (\ref{eq:sl3}), gives the following definition of non time-reversal invariant systems.
\begin{definition}
\label{trev.gue1} The ensemble of Hamiltonians of non time-reversal invariant Bosonic or Fermionic systems are hermitian matrices statistically invariant under
\begin{equation}\label{eq:trev19}H \to UHU^{-1}\end{equation}
where $U$ is unitary.
\end{definition}
\noindent If in addition the entries of the matrices not related by symmetry are statistically independent Definition \ref{trev.gue1} defines the Gaussian Unitary Ensemble (GUE), which is denoted by $\beta = 2$.
Definitions \ref{trev.even1}, \ref{trev.odd1} and \ref{trev.gue1} form Dyson's threefold way. They give the restrictions on the matrix representation of the potential $H$ depending on which class of quantum system it describes. By classifying the physical landscape into these classes the matrix $H$ is found to obey additional symmetry (on top of $H=H^{\dag}$), even if everything else remains, for the present, unknown. The symmetrising matrices $O$, $Z$ and $U$ define the matrix $S$ of (\ref{eq:sl3}).
It should be emphasized that the additional symmetries satisfied by $H$ in each of these three groups are the \emph{minimal} constraints satisfied by any $H$ belonging to the set. Hence the set of non time-reversal invariant Hamiltonians defined by the GUE \emph{do not exclude} the time-reversal invariant systems defined by the GOE and GSE ensembles. Those matrices are present in the ensemble of GUE matrices as well, but occur with a lower probability. The set of non time-reversal invariant hamiltonians is then more strictly the set of ``not necessarily but possibly time-reversal invariant'' hamiltonians.
\subsubsection{``None of These'' -- Embedded RMT}
While the canonical approach to random matrix theory places a single random variable in each cell of the matrix, it will be shown in subsequent sections that there are in fact additional ways to define the potential $H$, these being defined in terms of the \textit{order} of the potential, $k$, which will be defined later under the framework of many-body potentials. In this way the canonical form of RMT will be extended and a new model will be proposed which has the canonical form as a special case. Under the new model what is normally referred to as RMT will be shown to be just one of the possible points in a \textit{phase space of random matrix theories}, each with its own set of statistical properties. This unified phase space is sometimes referred to as \textit{Embedded Random Matrix Theory}.

Instead of looking at classical, or canonical RMT hamiltonians, the purpose of this thesis will be to investigate the statistics of Embedded RMT hamiltonians. These define a unified form of RMT which is still in its infancy; each embedded hamiltonian represents a particular instance of a random matrix theory, and a unified theory of random matrices must classify the statistical properties of each one. This thesis will show how the symmetry restrictions imposed in any particular instance of a random matrix theory can be represented in terms of \textit{particle diagrams} and a new mathematical methodology will be introduced to make calculations of the statistics of quantum many-body systems from a range of random matrix theories in the phase space.
\subsection[Many-Body Potentials]{Many-Body Potentials}\label{sec:second.quantization}
It will be seen that the hermitian matrices of canonical RMT with GOE, GSE or GUE symmetry imposed on the matrix elements of the hamiltonian $H$ represent only one possible flavour of a random matrix theory. To address this problem and in an attempt to unify the field of random matrix theories, K.~K.~Mon and J.~B.~French introduced the embedded RMT ensembles\cite{mon}, which are hamiltonians written in second-quantised form and determined by a trio of parameters $k$, $m$ and $l$ (defined later). For each set of values $\{k, m, l\}$ one attains a distinct random matrix theory. These second-quantised hamiltonians were already studied by physicists before the advent of RMT in the context of Many-Body Quantum Mechanics where they appeared under the rubric of \textit{Many-Body Potentials}. In this section the intention is to describe in some detail what a quantum many-body potential is, which will involve an introduction to the basic notational norms of second-quantisation. A good induction into the following formalism can also be found in \cite{negele, dickhoff}. The preliminary aim will be to express an arbitrary many-body operator in terms of creation and annihilation operators. The hamiltonian operator $H$ expressed in this way will form the extension to canonical RMT under consideration. It will then be seen that there is a particular set of values for the parameters $\{k, m, l\}$ determining this hamiltonian where it returns to the form used in classical RMT which, as noted, is a specific case of the more general class.

In the forthcoming model the number of quantum particles in the system of interest is a variable denoted by the letter $m$. The tensor product of $m$ single-particle states
\begin{equation}\label{eq:sq1} | {\alpha}_{1} \ldots {\alpha}_{m} ) \equiv |  {\alpha}_{1} \rangle \otimes \ldots \otimes |  {\alpha}_{m} \rangle \end{equation}
gives a natural way of describing an $m$-particle state. A simple list of the single-particle states is written in a single packet ($ket$) and defined as the quantum state. In position representation this $m$-particle state can be rewritten as
\begin{align}
{\psi}_{{\alpha}_{1} \ldots {\alpha}_{m}}(r_{1}, \ldots ,r_{m}) &= (r_{1}, \ldots ,r_{m}|{\alpha}_{1} \ldots {\alpha}_{m}) = \langle{r_{1}|{\alpha}_{1}}\rangle \langle{r_{2}|{\alpha}_{2}}\rangle \ldots \langle{{\alpha}_{m}|r_{m}}\rangle \notag\\ & = {\phi}_{{\alpha}_{1}}(r_{1}){\phi}_{{\alpha}_{2}}(r_2)\ldots{\phi}_{{\alpha}_{m}}(r_m).
\end{align}
Given an $m$-particle state $\psi_m$ the properly symmetrized and normalised bosonic (\textit{B}) and fermionic (\textit{F}) state is therefore given by
\begin{equation}P_{\{ B,F\}}\psi(r_{1}, \ldots ,r_{m})  = \frac{1}{m!}{\sum}_{P}^{}{\xi }^{P}\psi(r_{P_1}, \ldots ,r_{P_m}) \end{equation}
where the symbol $\xi = 1$ for states describing bosons and $\xi = -1$ for states describing fermions. $P$ is the set of permutations on $\{1, 2, \ldots, m\}$ and the same letter is also used to signify the parity of the permutation. To simplify notation later, define the properly symmetrized (but not necessarily normalised) $m$-particle state by
\begin{equation}
|{\alpha}_{1}\ldots{\alpha}_{m}\} \equiv \sqrt{m!}P_{\{B,F\}}|{\alpha}_{1}\ldots{\alpha}_{m}) = \frac{1}{\sqrt{m!}} \sum_{P}^{} {\xi}^{P}|{\alpha}_{P_1}\ldots{\alpha}_{P_m}).
\end{equation}
The \textit{closure relation} for this space is then given as
\begin{align}
\sum_{{\alpha}_{1}\ldots  {\alpha}_{m}}^{}[P_{\{B,F\}}|{\alpha}_{1}\ldots{\alpha}_{m})][P_{\{B,F\}}|{\alpha}_{1} \ldots{\alpha}_{m})]^{*} &= \frac{1}{m!}\sum_{{\alpha}_{1}\ldots{\alpha}_{m}}^{}|{\alpha}_{1}\ldots{\alpha}_{m}\}\{{\alpha}_{1}\ldots{\alpha}_{m}| \notag\\&= 1.
\end{align}
Since $m$-particle states are regarded as describing a probability density function they should be normalized to unity. The odd symmetry of fermionic states prohibits them from containing any two particles with the same state, so the overlap of two $m$-particle Fermionic states when non-zero is
\begin{equation}\label{eq:normalization1}\{{\alpha}_{1}'\ldots{\alpha}_{m}'|{\alpha}_{1}\ldots{\alpha}_{m}\} = (-1)^{P}.\end{equation}
If the two states do not contain the same number of particles there is certainly a zero overlap. Bosonic $m$-particle states on the other hand -- having even symmetry -- are allowed to have single particle states equal to other single particle states. This means that when the overlap is taken between two bosonic states containing $k$ unique states with $n_i$ single particle states in state $i, 1\le i\le k$ the result is
\begin{equation}\label{eq:normalization2}\{{\alpha}_{1}'\ldots{\alpha}_{m}'|{\alpha}_{1}\ldots{\alpha}_{m}\} = n_{1}!n_{2}!\ldots n_{k}!\end{equation}
where the number of particles is $m = \sum_{\alpha}^{}n_{\alpha}$. One can fortunately condense this notation further, expressing (\ref{eq:normalization1}) and (\ref{eq:normalization2}) as the single expression
\begin{equation}\label{eq:normalization3}\{{\alpha}_{1}'\ldots{\alpha}_{m}'|{\alpha}_{1}\ldots{\alpha}_{m}\} = {\xi}^{P} \prod_{\alpha}^{}n_{\alpha}!\end{equation}
Hence for functions representing a collection of bosons the orthonormal states are given by
\begin{equation}\label{eq:normalization4}|{\alpha}_{1}\ldots {\alpha}_{m}\rangle = \frac{1}{\sqrt{\prod_{\alpha}^{}n_{\alpha}!}}|{\alpha}_{1} \ldots {\alpha}_{m}\}\end{equation}
whereas for states representing a collection of fermions this simplifies to
\begin{equation}\label{eq:norm.state.def}|{\alpha}_{1}\ldots {\alpha}_{m}\rangle = |{\alpha}_{1} \ldots {\alpha}_{m}\}.\end{equation}
To conclude, the orthonormal closure relation for bosons and fermions is
\begin{equation} \frac{1}{m!}\sum_{{\alpha}_{1}\ldots{\alpha}_{m}}|{\alpha}_{1} \ldots {\alpha}_{m}\}\{{\alpha}_{1} \ldots {\alpha}_{m}| = \sum_{{\alpha}_{1}\ldots{\alpha}_{m}}\frac{\prod_{\alpha}n_{\alpha}!}{m!}|{\alpha}_{1} \ldots {\alpha}_{m} \rangle \langle {\alpha}_{1} \ldots {\alpha}_{m}| = 1.\end{equation}
\subsubsection{$k$-Body Operators}
The above formulates a way to describe states which represent collections of bosons and fermions. How to mathematically describe the Hamiltonians which exert forces on these states? The hamiltonians will be expressed as $k-$body operators, where $k$ is a variable determined by the particular system under consideration. An operator is called a $k$-body operator if its effect on an $m$-particle state is the sum of its effect on each of the ${m}\choose{k}$ single particle \textit{k}-tuples contained in the $m$-particle state. For example a 1-body operator is an operator satisfying
\begin{equation}H|{\alpha}_{1}\ldots{\alpha}_{m}) = \sum_{i=1}^{m} {H_{i}}|{\alpha}_{1}\ldots{\alpha}_{m})\end{equation}
where ${H_{i}}$ is the effect of $H$ hitting only the $i$'th single-particle state. It can therefore be seen that
\begin{equation}({\alpha}_{1}\ldots{\alpha}_{m}|H|{\beta}_{1}\ldots{\beta}_{m}) = \sum_{i}({\alpha}_{1}\ldots {\alpha}_{m}|{H_{i}}|{\beta}_{1}\ldots{\beta}_{m}) = \sum_{i}\prod_{k \neq i}\langle{\alpha_{k}|\beta_{k}}\rangle\langle{\alpha_i|{H_{i}}|\beta_i}\rangle\end{equation}
and for  non-orthogonal $|{\alpha}_{1}\ldots{\alpha}_{m})$ and $|{\beta}_{1}\ldots{\beta}_{m})$ one can conclude that
\begin{equation}\frac{({\alpha}_{1}\ldots{\alpha}_{m}|H|{\beta}_{1}\ldots{\beta}_{m})}{({\alpha}_{1}\ldots{\alpha}_{m}|{\beta}_{1}\ldots{\beta}_{m})} = \sum_{i}\frac{\langle{\alpha_i|H|\beta_i}\rangle}{\langle{\alpha_{i}|\beta_{i}}\rangle}.\end{equation}
The definition for 2-body operators is the natural extention of this idea. The effect of $H_2$ on an $m$-particle state is the sum of its effect on each of the ${m}\choose{2}$ single-particle pairs contained within the composite $m$-particle state so that
\begin{equation}H|{\alpha}_{1}\ldots {\alpha}_{m}) = \sum_{i<j} {H}_{ij}|{\alpha}_{1}\ldots{\alpha}_{m})= \frac{1}{2}\sum_{i \neq j} {H}_{ij}|{\alpha}_{1}\ldots{\alpha}_{m}).\end{equation}
As with the 1-body case one can write the following relation
\begin{align}({\alpha}_{1}\ldots{\alpha}_{m}|H|{\beta}_{1}\ldots{\beta}_{m}) &= \frac{1}{2}\,\sum_{i \ne j}({\alpha}_{1}\ldots {\alpha}_{m}|{H}_{ij}|{\beta}_{1}\ldots{\beta}_{m})\notag\\ &= \frac{1}{2}\,\sum_{i \neq j}\prod_{k \neq i,j}\langle{\alpha_{k}|\beta_{k}}\rangle\langle{\alpha_i\alpha_j|{H}_{ij}|\beta_i\beta_j}\rangle\end{align}
and diving across by non-orthogonal $|{\alpha}_{1}\ldots{\alpha}_{m}),|{\beta}_{1}\ldots{\beta}_{m})$ gives the expression
\begin{equation}\frac{({\alpha}_{1}\ldots{\alpha}_{m}|H|{\beta}_{1}\ldots{\beta}_{m})}{({\alpha}_{1}\ldots{\alpha}_{m}|{\beta}_{1}\ldots {\beta}_{m})}  = \frac{1}{2} \sum_{i \neq j}\frac{\langle{\alpha_i\alpha_j|{H}_{ij}|\beta_i\beta_j}\rangle}{\langle{\alpha_{i}|\beta_{i}}\rangle\langle{\alpha_{j}|\beta_{j}}\rangle}.\end{equation}
Following in direct analogy to the 1- and 2-body potentials above, the definition for the general $k$-body case is given by
\begin{equation}\label{eq:k.body.pot}H|{\alpha}_{1}\ldots {\alpha}_{m}) = \frac{1}{k!}\sum_{i_1 \neq i_2 \neq \ldots \neq i_k} {H}_{i_1 \ldots i_k}|{\alpha}_{1}\ldots{\alpha}_{m})\end{equation}
where the result of ${H}_{i_1 \ldots i_k}$ hitting the entire $m$-particle state means the same thing as $H$ hitting the $k$-body state $|\alpha_{i_1} \ldots \alpha_{i_k})$. The normalized matrix elements of the operator are therefore given by
\begin{equation}\frac{({\alpha}_{1}\ldots{\alpha}_{m}|H|{\beta}_{1}\ldots{\beta}_{m})}{({\alpha}_{1}\ldots{\alpha}_{m}|{\beta}_{1}\ldots {\beta}_{m})}  = \frac{1}{k!} \sum_{i_1 \neq i_2 \ldots \neq i_k}\frac{\langle{\alpha_{i_1}\ldots\alpha_{i_k}|{H}_{i_1 \ldots i_k}|\beta_{i_1}\ldots\beta_{i_k}}\rangle}{\langle{\alpha_{i_1}|\beta_{i_1}}\rangle\ldots\langle{\alpha_{i_k}|\beta_{i_k}}\rangle}.\end{equation}
The above explains how the many-body potential $H$ \textit{acts} on many-body quantum states, but what is $H$ \textit{actually}, mathematically speaking?
\subsubsection{Second Quantisation}
To describe $H$ mathematically it is necessary to reformulate its properties as described already (which is to say, how it acts) in terms of creation and annihilation operators. These form an efficient way of talking about $m$-particle states because by using creation and annihilation operators it subsequently becomes possible to talk about the existence or non-existence of the single-particle states comprising each $m$-particle state. This binary way of thinking matters. It opens up a variety of shortcuts, for example when taking overlaps between $m$-body states; the answer can be calculated simply by looking at whether certain single-particle states exist in the $m$-body states. This ``superpower'' will prove extremely useful later on.

To begin, the $m$-particle creation operator $a^{\dag}_{\lambda}$ adds a particle in the state $|\lambda\rangle$ to the $m$-particle state on which it acts
\begin{equation}a^{\dag}_{\lambda}|{\alpha}_{1}\ldots{\alpha}_{m}\} \equiv |\lambda{\alpha}_{1}\ldots{\alpha}_{m}\}.\end{equation}
This is consistent even if $\lambda$ is already present in the $m$-particle state. By (\ref{eq:normalization4}) it follows that
\begin{align}\left(\sqrt{\prod_{\alpha}n_{\alpha}!}\right) & a^{\dag}_{\lambda}|{\alpha}_{1}\ldots{\alpha}_{m}\rangle = \sqrt{n_{\lambda} + 1}\sqrt{\prod_{\alpha}n_{\alpha}!}|\lambda{\alpha}_{1}\ldots{\alpha}_{m}\rangle
\end{align}
giving
\begin{align}
a^{\dag}_{\lambda}|{\alpha}_{1}\ldots{\alpha}_{m}\rangle = \sqrt{n_{\lambda} + 1}|\lambda{\alpha}_{1}\ldots{\alpha}_{m}\rangle\end{align}
where $n_{\lambda}$ is the number of single-particle states equal to $|\lambda\rangle$ in the original $m$-body state $|{\alpha}_{1}\ldots{\alpha}_{m}\rangle$. Defining $|0\rangle$ as the vacuum state, any other state can then be expressed in the form
\begin{equation}|{\lambda}_{1}\ldots{\lambda}_{m}\rangle = \frac{1}{\sqrt{\prod_{\lambda} n_{\lambda}!}} a_{\lambda_1}^{\dag}\ldots a_{\lambda_m}^{\dag}|0\rangle.\end{equation}
To express an arbitrary $k$-body operator in terms of creation and annihilation operators in the standard form the \textit{commutation relations} between them are needed. First it can be noticed that since
\begin{align} a_{\lambda}^{\dag}a_{\mu}^{\dag}|\lambda_1 \ldots \lambda_m\} = |\lambda \mu \lambda_1 \ldots \lambda_m\} &= \xi|\mu \lambda \lambda_1 \ldots \lambda_m\}= \xi a_{\mu}^{\dag} a_{\lambda}^{\dag}|\lambda_1 \ldots \lambda_m\} \end{align}
one has $a_{\lambda}^{\dag}a_{\mu}^{\dag} \equiv \xi a_{\mu}^{\dag} a_{\lambda}^{\dag}$ which yields
\begin{equation}\label{eq:comrel1}[a_{\lambda}^{\dag}, a_{\mu}^{\dag}]_{-\xi} := a_{\lambda}^{\dag}a_{\mu}^{\dag} - \xi a_{\mu}^{\dag} a_{\lambda}^{\dag} = 0.\end{equation}
Taking the complex conjugate gives
\begin{equation}\label{eq:comrel2}[a_{\lambda}, a_{\mu}]_{-\xi} = 0.\end{equation}
Equations (\ref{eq:comrel2}) and (\ref{eq:comrel2}) give the first two commutation relations. To find the third requires an investigation into the operator $a_{\lambda} := (a_{\lambda}^{\dag})^{\dag}$. Taking the overlap
\begin{equation}\{{\alpha}_{1}\ldots{\alpha}_{m}|a_{\lambda}|{\beta}_{1}\ldots{\beta}_{n}\} = \{{\lambda\alpha}_{1}\ldots{\alpha}_{m}|{\beta}_{1}\ldots{\beta}_{n}\}.\end{equation}
one sees this can be non-zero only if $m+1=n$. Representing the action of $a_{\lambda}$ in terms of the identity over all possible quantum states \textit{i.e.} including states containing different numbers of particles, and assuming that $|\lambda\rangle$ is in $|{\beta}_{1}\ldots{\beta}_{n}\}$, gives
\begin{align}\label{eq:annhilation.general}a_{\lambda}|{\beta}_{1}\ldots{\beta}_{n}\} &= \sum_{m=0}^{\infty} \frac{1}{m!} \sum_{{\alpha}_{1}\ldots{\alpha}_{m}} \{{\alpha}_{1}\ldots{\alpha}_{m}|a_{\lambda}|{\beta}_{1}\ldots{\beta}_{n}\}|{\alpha}_{1}\ldots{\alpha}_{m}\}\notag\\
&= \sum_{m=0}^{\infty} \frac{1}{m!} \sum_{{\alpha}_{1}\ldots{\alpha}_{m}} \{\lambda\alpha_{1}\ldots{\alpha}_{m}|{\beta}_{1}\ldots{\beta}_{n}\}|{\alpha}_{1}\ldots{\alpha}_{m}\}\notag\\
&= \sum_{i}^{n} {\xi}^{i-1} {\delta}_{\lambda\beta_i}|{\beta}_{1}\ldots\beta^{i-1}\beta^{i+1}\ldots{\beta}_{n}\}\notag\\
&= \sum_{i}^{n} {\xi}^{i-1} {\delta}_{\lambda\beta_i}|{\beta}_{1}\ldots\hat{\beta_i}\ldots{\beta}_{n}\}\end{align}
where $\hat{\beta_i}$ denotes that the $i$'th particle is removed from the set. Hence the effect of $a_{\lambda}$ on the state is to remove a particle in the state $\lambda$. The operator $a_{\lambda}$ is therefore referred to as an \textit{annihilation operator}, its normalized formulation being
\begin{align}a_{\lambda}|{\beta}_{1}\ldots{\beta}_{n}\rangle &= \frac{1}{\sqrt{\prod_j n_j !}}\sum_{i}^{n} \left(\sqrt{\prod_i n_i !}\right){\xi}^{i-1} {\delta}_{\lambda\beta_i}|{\beta}_{1}\ldots\hat{\beta_i}\ldots{\beta}_{n}\rangle\notag \\
&= \frac{1}{\sqrt{n_{\lambda}}} \sum_{i}^{n} {\xi}^{i-1} {\delta}_{\lambda\beta_i}|{\beta}_{1}\ldots\hat{\beta_i}\ldots{\beta}_{n}\rangle.\end{align}
As an aside note that for bosons ($\xi=1$) this becomes just
\begin{equation}a_{\lambda}|n_{{\beta}_{1}}\ldots n_{\lambda} \ldots n_{{\beta}_{q}}\rangle = \sqrt{n_{\lambda}}|n_{{\beta}_{1}}\ldots (n_{\lambda}-1) \ldots n_{{\beta}_{q}}\rangle\end{equation}
where $|n_{{\beta}_{1}}\ldots n_{{\beta}_{q}}\rangle$ denotes the state with $n_{\beta_i}$ particles in the state $\beta_i$. It should be noted that the final commutation relation $[a_{\lambda},a_{\mu}^{\dag}]_{- \xi}$ is still unknown. It can be found firstly by noticing that
\begin{align}a_{\lambda}a_{\mu}^{\dag}|\alpha_1 \ldots \alpha_n\} &= a_{\lambda}|\mu\alpha_1 \ldots \alpha_n\}\notag\\
&= {\delta}_{\lambda\mu}|\alpha_1 \ldots \alpha_n\} + \sum_{i=1}^{n}{\xi}^i \delta_{\lambda\alpha_i}|\mu\alpha_1 \ldots \hat{\alpha_i} \ldots \alpha_n\}\end{align}
and likewise that
\begin{align}a_{\mu}^{\dag}a_{\lambda}|\alpha_1 \ldots \alpha_n\} &= a_{\mu}^{\dag} \sum_{i=1}^{n}{\xi}^{i-1} \delta_{\lambda\alpha_i}|\alpha_1 \ldots \hat{\alpha_i} \ldots \alpha_n\}\notag\\
&= \sum_{i=1}^{n}{\xi}^{i-1} \delta_{\lambda\alpha_i} |\mu\alpha_1 \ldots \hat{\alpha_i} \ldots \alpha_n\}.\end{align}
Combining these two equations gives
\begin{align}a_{\lambda}a_{\mu}^{\dag} &= {\delta}_{\lambda\mu}|\alpha_1 \ldots \alpha_n\} + a_{\mu}^{\dag}a_{\lambda}|\alpha_1 \ldots \alpha_n\}\notag\\
&= [{\delta}_{\lambda\mu} + \xi a_{\mu}^{\dag}a_{\lambda}]|\alpha_1 \ldots \alpha_n\}\end{align}
which is the final commutation relation sought. Namely
\begin{equation}\label{eq:comrel3}[a_{\lambda},a_{\mu}^{\dag}]_{- \xi} = {\delta}_{\lambda\mu}.\end{equation}
Given (\ref{eq:comrel1}), (\ref{eq:comrel2}) and (\ref{eq:comrel3}) an expression can now be found for an arbitrary $k$-body operator in terms of creation and annihilation operators, initially by working in a basis where the operator is in its diagonal form, and finally by expressing the result in any basis. To begin, following (\ref{eq:k.body.pot}) and assuming that ${H_k}$ is diagonal
\begin{equation}{H}_{i_1 \ldots i_k} = (\alpha_{i_1} \ldots \alpha_{i_k} | {H_k}| \alpha_{i_1} \ldots \alpha_{i_k})\end{equation}
so that for arbitrary $m$-body states $|\alpha_{i_1} \ldots \alpha_{i_m})$ and $|\beta_{i_1} \ldots \beta_{i_m})$ the following sequence of equalities are attained
\allowdisplaybreaks[1]
\begin{align}(\alpha_{i_1} \ldots &\alpha_{i_m}|{H}_k|\beta_{i_1} \ldots \beta_{i_m}) = (\alpha_{i_1} \ldots \alpha_{i_m}|\frac{1}{k!}\sum_{i_1 \neq i_2 \neq \ldots \neq i_k}^{m} {H}_{i_1 \ldots i_k}|\beta_{i_1} \ldots \beta_{i_m})\notag\\
&= \frac{1}{k!}\sum_{i_1 \neq \ldots \neq i_k}\prod_{j \neq i_1,\ldots,i_k}\langle \alpha_j|\beta_j\rangle (\alpha_{i_1} \ldots \alpha_{i_k}|{H}_{i_1 \ldots i_k}|\beta_{i_1} \ldots \beta_{i_k})\notag\\
&= \frac{1}{k!}\sum_{i_1 \neq \ldots \neq i_k}\prod_{j \neq i_1,\ldots,i_k}\langle \alpha_j|\beta_j\rangle(\beta_{i_1} \ldots \beta_{i_k}|{H}_k|\beta_{i_1} \ldots \beta_{i_k})(\alpha_{i_1} \ldots \alpha_{i_k}|\beta_{i_1} \ldots \beta_{i_k})\notag\\
&= \frac{1}{k!}(\alpha_{i_1} \ldots \alpha_{i_m}|\beta_{i_1} \ldots \beta_{i_m})\sum_{i_1 \neq \ldots \neq i_k}(\beta_{i_1} \ldots \beta_{i_k}|{H}_k|\beta_{i_1} \ldots \beta_{i_k})
\label{eq:sqx}\end{align}
the sum being over all $k$-tuples in the state $\beta := |\beta_{i_1} \ldots \beta_{i_m})$.
\subsubsection{Tuple Counting}
Expressing (\ref{eq:sqx}) in terms of the matrix elements of the operator gives
\begin{equation}(\alpha_{i_1} \ldots \alpha_{i_m}|{H}_k|\beta_{i_1} \ldots \beta_{i_m}) = \frac{1}{k!} (\alpha_{i_1} \ldots \alpha_{i_n}|\sum_{{\bm k}_u - tuples}{H}_{i_1 \ldots i_k}{T}_{\bm k}|\beta_{i_1} \ldots \beta_{i_n})\end{equation}
the sum now running over \textit{unique} $k$-tuples ${\bm k}_u$ (any permutation of a given \textit{k}-tuple is considered non-unique) and the tuple-counting operator ${T}_{\bm k}$ giving the total number of $k$-tuples denoted by the labels ${\bm k}=\{i_1 \ldots i_k\}$ in the state upon which it acts. For a 1-body operator the number of particles in the state $\alpha$ is in fact given by the \textit{number operator} $n_{\alpha} = a_{\alpha}^{\dag}a_{\alpha}$ so that
\begin{equation}{T}_1 = a_{\alpha}^{\dag}a_{\alpha}\end{equation}
and similarly the number of pairs consisting of the single-particle states $|\alpha\rangle$ and $|\beta\rangle$ is $n_{\alpha}n_{\beta}$ if $\alpha \neq \beta$ and $n_{\alpha}(n_{\alpha}-1)$ otherwise. These can be combined into the single condition
\begin{equation}{T}_2 = n_{\alpha}(n_{\alpha}-\delta_{\alpha\beta}) = a_{\alpha}^{\dag}a_{\beta}^{\dag}a_{\beta}a_{\alpha}\end{equation}
Generalizing immediately for a $k$-body operator the number of $k$-tuples are given by the operation
\begin{equation}\label{eq:k.body.number}{T}_k = \sum_{i_1 \ldots i_k} n_{i_1}(n_{i_2} - \delta_{i_1 i_2})(n_{i_3} - \delta_{i_1 i_3} - \delta_{i_2 i_3}) \ldots (n_{i_k} - \delta_{i_1 i_k} - \delta_{i_2 i_k} - \ldots - \delta_{i_{k-1} i_k}).\end{equation}
Using the identity
\begin{equation}a_p \delta_{pq} = a_q \delta_{pq}\end{equation}
equation (\ref{eq:k.body.number}) reduces neatly to
\begin{equation}{T}_k = a_{i_1}^{\dag}\ldots a_{i_k}^{\dag}a_{i_k} \ldots a_{i_1}\end{equation}
so that the second-quantized formulation of a diagonal $k$-body operator is
\begin{equation}\label{eq:diagonalized.canonical}H = \frac{1}{k!}\sum_{i_1 \ldots i_k} {H}_{i_1 \ldots i_k} a_{i_1}^{\dag} \ldots a_{i_k}^{\dag}a_{i_k} \ldots a_{i_1}.\end{equation}
To express this in terms of an arbitrary basis observe that in terms of some basis $|\lambda\rangle$ one can write
\begin{equation}a_{\lambda_i}^{\dag} = \sum_{a} \langle a | a_{\lambda_i} \rangle a^{\dag}_a.\end{equation}
Taking the complex conjugate gives
\begin{equation}a_{\lambda_i} = \sum_{a} \langle a_{\lambda_i}|a \rangle a_a\end{equation}
and substituting these expressions into (\ref{eq:diagonalized.canonical}) leads to the following
\begin{align}\label{eq:canonical.potential1}H &= \frac{1}{k!}\sum_{i_1^{'} \ldots i_k^{'}} {H}_{i_1^{'} \ldots i_k^{'}}\sum_{j_1}\langle j_1 | i_1^{'}\rangle a_{j_1}^{\dag} \ldots \sum_{j_k}\langle j_k | i_k^{'}\rangle a_{j_k}^{\dag} \sum_{i_k}\langle i_k^{'} | i_k\rangle a_{i_k} \ldots \sum_{i_1}\langle i_1^{'} | i_1\rangle a_{i_1}\notag\\
&= \frac{1}{k!}\sum_{i_1^{'} \ldots i_k^{'}} \sum_{{j_1 \ldots j_k}\atop{i_1 \ldots i_k}} {H}_{i_1^{'} \ldots i_k^{'}}\langle j_1 \ldots j_k|i_1^{'} \ldots i_k^{'} \rangle \langle i_1^{'} \ldots i_k^{'}|i_1 \ldots i_k \rangle a_{j_1}^{\dag} \ldots a_{j_k}^{\dag} a_{i_k} \ldots a_{i_1} \notag\\
&= \frac{1}{k!}\sum_{{j_1 \ldots j_k}\atop{i_1 \ldots i_k}} {H}_{j_1 \ldots j_k;i_1 \ldots i_k} a_{j_1}^{\dag} \ldots a_{j_k}^{\dag} a_{i_k} \ldots a_{i_1}\end{align}
which is the second-quantized form of a $k$-body operator in an arbitrary basis.
This formula expresses the hamiltonian matrix $H$ of (\ref{eq:sl1}) as a second-quantised operator i.e.~an operator defined by a sequence of creation and annihilation operators. Implicit in the model are three important parameters. Firstly there is $m$, the variable determining the number of single particles in the system. In other words for any state $|\mu\rangle$ one has
\begin{equation}
|\mu\rangle = |\alpha_1\ldots\alpha_m\rangle
\end{equation}
for some set of single-particle states $\{\alpha_1,\ldots,\alpha_m\}$ as seen in (\ref{eq:norm.state.def}).
Secondly there is $k \le m$, the order of the interaction or in other words the ``number of bodies'' involved in each interaction under the force of the potential. Finally, there is the implicit parameter $l$ which determines the number of energy levels available to each of the single-particle states in the compound $m$-body state. Hence $l$ is the size of the set from which the single-particle states $\alpha_1,\ldots,\alpha_m$ can take their values. As mentioned in section \ref{sec:second.quantization} the trio of values $\{k,m,l\}$ together with the symmetry conditions imposed on $H$ form a single instance of a random matrix theory. As an unrestricted phase space the hamiltonian of (\ref{eq:canonical.potential1}) represents the unification of these random matrix theories. Next it will be shown that canonical RMT coincides with the case $k=m$ with $l\to \infty$.
\subsection{Canonical RMT as a Single Point}\label{sec:rmtsp}
To paraphrase section \ref{sec:sl} Canonical RMT is the study of random matrices with some predefined symmetry conditions and a single random variable determining the quantity in each cell of the matrix. The unified form of random matrix theory determined by the hamiltonian of (\ref{eq:canonical.potential1}) however, allows for the possibility that more than one $p.d.f.$ determines any given cell of the matrix. For the special case where $k=m$ this becomes
\begin{equation}H = \frac{1}{m!}\sum_{{j_1 \ldots j_m}\atop{i_1 \ldots i_m}} {H}_{j_1 \ldots j_m;i_1 \ldots i_m} a_{j_1}^{\dag} \ldots a_{j_m}^{\dag} a_{i_m} \ldots a_{i_1}\end{equation}
so that for any two $m$-body states
\begin{equation}|\mu\rangle = |\alpha_1, \ldots,\alpha_m\rangle\end{equation}
\begin{equation}|\nu\rangle = |\beta_1,\ldots,\beta_m\rangle\end{equation} the matrix elements of $H$ become
\begin{align}H_{\mu\nu} &= \frac{1}{m!}\sum_{{j_1 \ldots j_m}\atop{i_1 \ldots i_m}} {H}_{j_1 \ldots j_m;i_1 \ldots i_m} \langle\mu|a_{j_1}^{\dag} \ldots a_{j_m}^{\dag} a_{i_m} \ldots a_{i_1}|\nu\rangle\notag\\
&={H}_{\alpha_1 \ldots\alpha_m;\beta_1 \ldots \beta_m}\label{eq:single_point_H}\end{align}
Hence the cells of the matrix $H$ for the special case $k=m$ contain only one element -- the final line of (\ref{eq:single_point_H}). There is no summation as there would be for the case $k < m$. This single element, being defined in terms of a probability density function and symmetrised by some condition on $H$ will of course give back the corresponding canonical RMT ensemble symmetrised under the same condition. This tells us that $k=m$ is the point in the phase space $\{k, m, l\}$ of the unified theory which coincides with canonical random matrix theory. Moreover, since the random matrices of the canonical theory are usually assumed to be infinite one takes $l\to\infty$, giving canonical RMT as the theory coinciding with the point $\{m, m, \infty\}$ in the unified phase space.
\begin{center}.....\end{center}
\chapter[Wigner's Semi-Circle Law]{Wigner's Semi-Circle Law}\label{sec:WSCL}
\section[Wigner's Law for $k=m$]{Wigner's Law for $k=m$}
Wigner's Semi-Circle Law is one of the more iconic and widely known results to come out of the field of random matrix theory. This is the rule, proven mathematically, which states that the average level density of Hamiltonians from the GUE, GOE and GSE ensembles take the form of a semi-circle\cite{mehta}. In other words, it says that on average the $p.d.f.$ of the energy values of these systems is a semi-circle.\\
\begin{figure}\label{fig:wscl}
\begin{subfigure}{.5\textwidth}
  \centering
  \includegraphics[width=\linewidth]{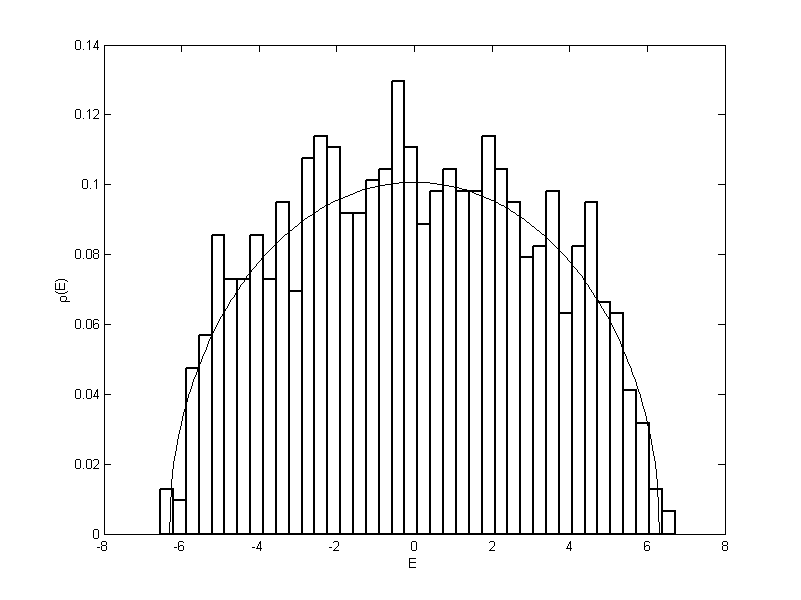}
  \caption{N=10}
  \label{fig:sfig1}
\end{subfigure}%
\begin{subfigure}{.5\textwidth}
  \centering
  \includegraphics[width=\linewidth]{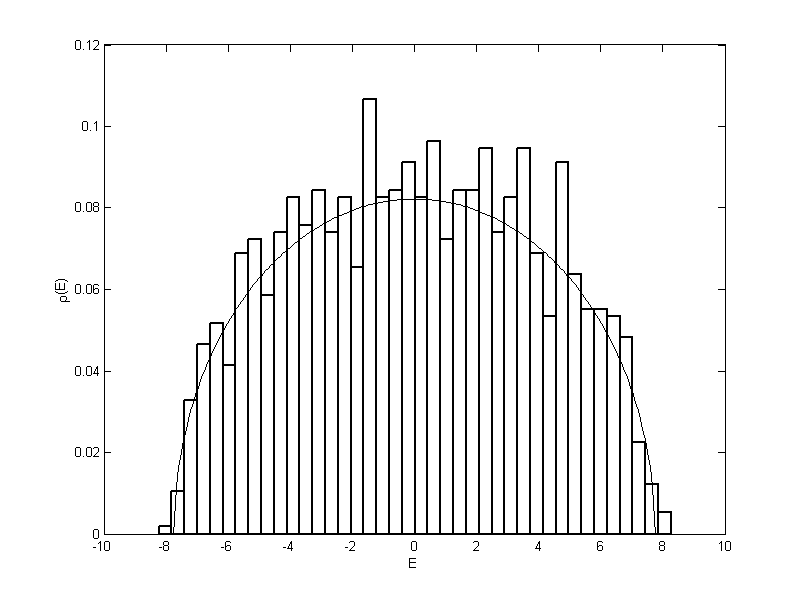}
  \caption{N=15}
  \label{fig:sfig2}
\end{subfigure}\\
\begin{subfigure}{.5\textwidth}
  \centering
  \includegraphics[width=\linewidth]{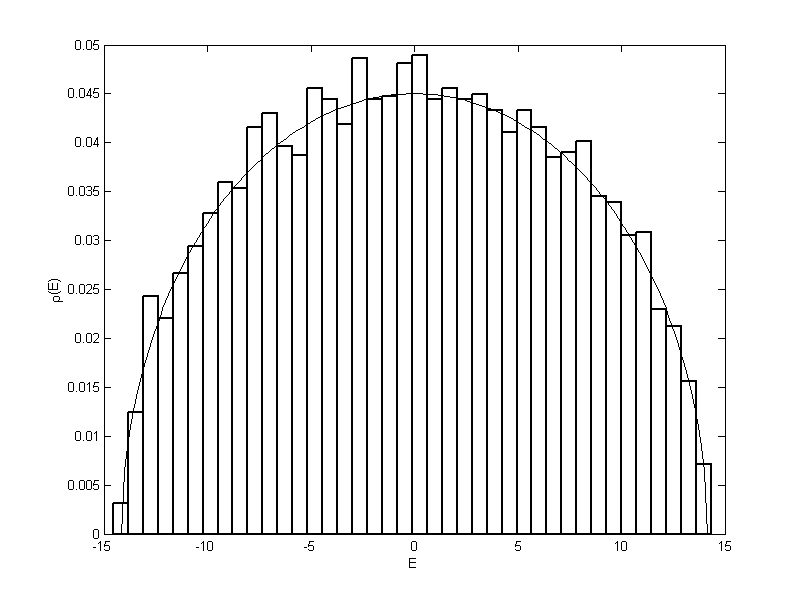}
  \caption{N=50}
  \label{fig:sfig1}
\end{subfigure}%
\begin{subfigure}{.5\textwidth}
  \centering
  \includegraphics[width=\linewidth]{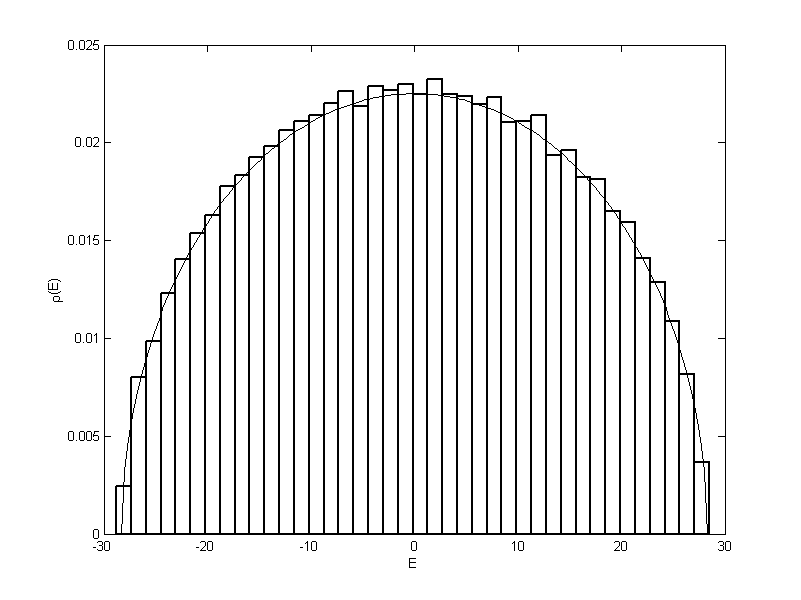}
  \caption{N=200}
  \label{fig:sfig2}
\end{subfigure}
\caption[Eigenvalue distributions for the GUE]{Plots of the eigenvalue distributions for $N\times N$ matrices sampled from the GUE ensemble with $N$ increasing in each simulation. The entries of the matrix are taken to be complex gaussian random variables with variance $1$. The symmetry condition placed on the matrices is $H=H^{\dag}$.}
\end{figure}
As shown in the previous section Wigner's Semi-circle law is in fact a statement about many-body hamiltonians, specifically those where the order of the interaction $k$ is taken to be equal to the number of particles in the system, $m$. Assuming that all particles interact \emph{simultaneously} under the force of a potential was an assumption made implicitly in the canonical form of random matrix theory, because it led to important simplifications in a challenging and very technical new field of mathematical endeavor. There was a second rationale as well. Measurements of particular nuclear energy levels in the 1950s and 1960s \cite{rosen1, rosen2, rosen3, rosen4, rosen5} involved nuclear interactions with a high degree of random mixing. This was caused by high excitation energies due to the particles being accelerated within a synchrocyclotron. The interactions appeared to be so complex that a ``black box'' approach to studying the nuclei became a practical and reasonable way forward as it involved assuming the bare minimum of facts about the specific details of the potential $H$, beyond its symmetry. The most natural assumption was therefore to represent the physical hamiltonian as a hermitian matrix with some distribution (say a gaussian) determining the matrix elements; this was the black box. So apart from the mathematical simplification the initial approach to studying nuclear resonances, which resulted in the foundation of canonical random matrix theory, was motivated by experiment. The fact that it implicitly implied $k=m$ did not detract from the possibility that nature did indeed appear to behave this way, statistically speaking, in some complex systems.
\newline However, it soon became apparent that many-body potentials taking values of $k < m$ would need to be investigated as well. Indeed since the entire range $0 \le k \le m$ could plausibly model some experiment it became an imperative to find the statistics of the energy values for all of these possibilities, as this would be the only way to check whether or not the resultant models predicted behaviour which is also measured in actual experiments. This unified approach to random matrix theory was proposed by K.~K.~Mon and J.~B.~French \cite{mon} and the unified ensembles determined by the modified hamiltonian (\ref{eq:canonical.potential1}) were called the \textit{Embedded Ensembles}. Hence, theoretical physicists and mathematicians first solved the case $k=m$ and only later turned to the general case (still unsolved in its entirety) determined by allowing $k$ to take any value within the permitted range $0 \le k \le m$. The next sections will explain how to calculate Wigners semi-circle law in two ways. One method will use a complex technique called \emph{supersymmetry} involving the liberal application of anti-commuting variables and their properties, which will be discussed beforehand. The second way will use a simple method involving some basic diagrams and combinatorics. Both methods will be used to derive the semi-circle law of canonical random matrix theory, i.e. for the special case $k=m$. An attempt to calculate Wigner's law for the general case $k \le m$ will be presented in later sections and will rely on new diagrammatic methods which are the primary topic of this thesis. There it will be seen that the analogue of Wigner's Semi-Circle Law in the unified phase space of embedded RMT is, for many values of $k$, not even a semi-circle.

\section[Diagrams]{Diagrams}\label{sec:diagrams}
In this section Wigner's semi-circle law will be shown only for the case $k=m$ with $l\to \infty$ and just one restriction on the symmetry of the potential, namely $H= H^{\dag}$. This is the canonical class of RMT known as the GUE ensemble (see section \ref{sec:SC}). For the purpose of the proof the quantum states containing $m$ particles are assumed to consist entirely of fermions and these $m$-body states will be denoted by the greek symbols $|\mu\rangle, |\nu\rangle, |\rho\rangle, |\sigma\rangle$ et cetera. Each of the $m$ particles in these states will be one of $l$ single-particle occupation levels, with $m \ll l$ and the caveat that no two single-particle states in the same $m$-body state can be the same because the states are fermionic. Hence there is a total of $N = {l\choose m}$ orthogonal $m$-body states in the basis of the system. \subsubsection{Notation}
The single-particle creation and annihilation operators are as before written as $a_j^{\dag}$ and $a_j$ respectively with $j = 1, \ldots, l$. With the intention of simplifying the notation define the shorthand expression
\begin{equation}\label{eq:diags1}\bj = (j_1,\ldots,j_k)\end{equation}
 and similarly for $\bi$. This furnishes a suitable abbreviation for the creation and annihilation operators which will now be written as
 \begin{equation}\label{eq:diags2}a_{\bj} = a_{j_k}\ldots a_{j_1}.\end{equation}
A useful corollary of this is the equality
\begin{equation}\label{eq:ee30}a_{\bj}^{\dag} = a_{j_1}^{\dag} \ldots a_{j_k}^{\dag}.\end{equation}
Each of the $l\choose m$ states in the basis can be written in the form
\begin{equation}\label{eq:ee301}a_{{j_m}}^{\dag}\ldots a_{{j_1}}^{\dag}|0\rangle\end{equation}
with $|0\rangle$ denoting the vacuum state and the restriction $1\leq j_1 < j_2 < \ldots < j_m\leq l$ since these are $m$-body states containing only fermions. From this it can be seen that for the special case $k=m$ each $m$-body state can in fact be rewritten as
\begin{equation}\label{eq:ee302}a_{\bj}^{\dag}|0\rangle\end{equation}
for some particular set $\bj$. The $k$-body potential as derived in section {\ref{sec:second.quantization} is written as
\begin{equation}\label{eq:ee27}{H}_k = \sum_{{{1 \le j_1 < \ldots < j_k \le l} \atop {1 \le i_1 < \ldots < i_k \le l}}} {v}_{j_1 \ldots j_k;i_1 \ldots i_k} a_{j_1}^{\dag} \ldots a_{j_k}^{\dag} a_{i_k} \ldots a_{i_1}\end{equation}
with the stricter sum now running over $k$-tuples with non-repeating states. With the notational abbreviations defined above this can now be rewritten in the following way
\begin{equation}\label{eq:egue01}{H}_k = \sum_{\bj \bi}v_{\bj\bi} a_{\bj}^{\dag}  a_{\bi}.\end{equation}
Because present attention will be focused on the case where the potential is a hermitian matrix from the GUE ensemble, the only symmetry condition which $H_k$ needs to satisfy is given by
\begin{align}\label{eq:ee33}   \langle \mu|H_k|\nu \rangle = \langle \nu|H_k|\mu \rangle ^{*} \end{align}
for all $\mu, \nu$. This implies that
\begin{align}\label{eq:ee331}
\sum_{\bj \bi} {v}_{\bj\bi} \langle\mu|a_{\bj}^{\dag} a_{\bi}|\nu\rangle &= \sum_{\bj \bi} {v}_{\bi\bj}^{*} \langle\nu|a_{\bj}^{\dag} a_{\bi}|\mu\rangle.
\end{align}
Hence for the specific case $k=m$ the expression (\ref{eq:ee302}) can be used to rewrite (\ref{eq:ee331}) as
\begin{align}\label{eq:ee34}\sum_{\bj \bi} {v}_{\bj\bi} \langle0|a_{\mu}a_{\bj}^{\dag} a_{\bi}a_{\nu}^{\dag}|0\rangle = \sum_{\bj \bi} {v}_{\bi\bj}^{*} \langle 0|a_{\nu}a_{\bj}^{\dag} a_{\bi}a_{\mu}^{\dag}|0\rangle.
\end{align}
This grants a further simplification, since
\begin{equation}\label{eq:diags3}a_{\boldsymbol{p}}^{~}a_{\boldsymbol{q}}^{\dag}|0\rangle = \delta_{\boldsymbol{pq}}
\end{equation}
implying that (\ref{eq:ee331}) is equivalent to
\begin{equation}\label{eq:ee34}\sum_{\bj, \bi} {v}_{\bj\bi} \delta_{{\bi}\nu}\delta_{{\bj}\mu}= \sum_{\bj, \bi} {v}_{\bi\bj}^{*}\delta_{{\bi}\mu}\delta_{{\bj}\nu}
\end{equation}
so that ultimately
\begin{align}\label{eq:ee35} \langle \mu|H|\nu\rangle^{*} = {v}_{\mu\nu}^{*} = {v}_{\nu\mu}=  \langle \nu|H|\mu\rangle.\end{align}
It will become apparent shortly why this equality is going to be useful in subsequent calculations.
\subsection{Moments and Ensemble Averages}\label{sec:MAEA}
The exact form of the average $p.d.f.$ for the eigenvalues of the GUE ensemble will be calculated by finding the moments of the density function, the $2n$-th moment being given by the expression
\begin{equation}\label{eq:moments}
\beta_{2n} = \frac{\frac{1}{N}\mathrm{tr}({\overline{H^{2n}_k}})}{\left(\frac{1}{N}\mathrm{tr}({\overline{H^2_k}})\right)^n}.
\end{equation}
Fortunately (the moments would otherwise be infinite), in the denominator there is always the normalisation expression
\begin{equation}\label{eq:diag1}
\frac{1}{N}\mathrm{tr}({\overline{H^2_k}}) = \frac{1}{N}\overline{v_{\mu\nu}v_{\nu\mu}} = \frac{1}{N}\overline{|v_{\mu\nu}|^2}
\end{equation}
with implicit summation over the repeated indices $\mu, \nu$ and the overbar denoting the ensemble average (which is the value weighted by the probability that it occurs within the GUE ensemble).
Yet to be determined is the probability distribution for those elements of the matrix which are not determined by symmetry. Following the norms of canonical RMT the gaussian distribution will be used, though practitioners in the field do also study ensembles where this is not the case (e.g. \cite{beenakker,hackenbroich}). The convergence of Random Matrix results for differing distributions of the matrix elements is referred to in the literature as \emph{universality}\cite{mehta}. Assuming that all $v_{\bj\bi}$ are i.i.d. complex random variables sampled from a gaussian with mean zero and standard deviation 1 yields
\begin{equation}\label{eq:diag2}
\frac{1}{N}\mathrm{tr}({\overline{H^2_k}}) = N.
\end{equation}
This permits another neat simplifcation; for the current case with $k=m$ the moments are in fact given by
\begin{equation}\label{eq:moments2}
\beta_{2n} = \frac{1}{N^{n+1}}\mathrm{tr}({\overline{H^{2n}_k}}).
\end{equation}
To discover what the value of this trace is, re-write it explicitly in the form
\begin{equation}\label{eq:diag3}
\mathrm{tr}({\overline{H^{2n}_k}}) = \overline{v_{\bj^{(1)}\bi^{(1)}} v_{\bj^{(2)}\bi^{(2)}} \ldots v_{\bj^{(2n)}\bi^{(2n)}}} \langle\mu|a_{\bj^{(1)}}^{\dag}  a_{\bi^{(1)}} a_{\bj^{(2)}}^{\dag}  a_{\bi^{(2)}} \ldots a_{\bj^{(2n)}}^{\dag}  a_{\bi^{(2n)}}|\mu\rangle
\end{equation}
with summation over the repeated indices again implicit in the notation. Since these are complex gaussian random variables with mean zero the ensemble average denoted by the overbar will only be non-zero when it is made over even powers of the $v$'s. The ensemble average $\overline{v^{2}}$ is the standard deviation which, to simplify notation, has been predetermined to be of unit value
\begin{equation}\label{eq:diag4}
\overline{v^{2}} = \sqrt{\frac{1}{2\pi}}\int v^{2} e^{- {v^2}/{2}}dv = 1.
\end{equation}
The general case can be shown, given unit standard deviation, to give the ensemble average of every even power of $v$ as
\begin{equation}\label{eq:diag5}
\overline{v^{2n}} = \sqrt{\frac{1}{2\pi}}\int v^{2n} e^{- {v^2}/{2}}dv = (2n-1)!!
\end{equation}
This, coincidentally, is the number of ways of uniquely pairing a set of $2n$ objects, which is also the sequence determining the moments of a gaussian distribution. It will be shown that the moments in this case are not in fact the moments of a gaussian distribution.
\subsubsection{Products of Pairs of Random Variables}
Using just (\ref{eq:diag5}) equation (\ref{eq:diag3}) can be written as an expression determined by summing over only the product of the average of all possible pairs of the $v$. Specifically, let $\{\sigma\}$ denote the set of $(2n-1)!!$ unique pairings of the sequence $1, 2, \ldots, 2n$ with $\sigma(x)$ taking the value of the integer paired with $x$ for a given permutation $\sigma$ from the set $\{\sigma\}$. Note that there is no $x$ for which $\sigma(x)=x$. Then the average of the product of $v$'s from (\ref{eq:diag3}) can be rewritten in terms of averages of pairs of $v$
\begin{equation}\label{eq:diag6}
\overline{v_{\bj^{(1)}\bi^{(1)}} v_{\bj^{(2)}\bi^{(2)}} \ldots v_{\bj^{(2n)}\bi^{(2n)}}} = \sum_{\sigma} \prod_{x=1}^{2n}\overline{v_{\bj^{(x)}\bi^{(x)}}v_{\bj^{\sigma(x)}\bi^{\sigma(x)}}}.
\end{equation}
This result is used frequently in statistical mechanics and is sometimes referred to as \emph{Wick's Theorem} (even when it doesn't involve creation and annihilation operators). Hence (\ref{eq:diag3}) becomes
\begin{equation}\label{eq:diag7}
\mathrm{tr}({\overline{H^{2n}_k}}) =\left[\sum_{\sigma} \prod_{x=1}^{2n}\overline{v_{\bj^{(x)}\bi^{(x)}}v_{\bj^{\sigma(x)}\bi^{\sigma(x)}}}\right]\langle\mu|a_{\bj^{(1)}}^{\dag}  a_{\bi^{(1)}} a_{\bj^{(2)}}^{\dag}  a_{\bi^{(2)}} \ldots a_{\bj^{(2n)}}^{\dag}  a_{\bi^{(2n)}}|\mu\rangle.
\end{equation}
Of note is that since $k=m$, operating on a state $|\mu\rangle$ with annihilation operator $a_{\bi}$ will give
\begin{equation}\label{eq:diag8}
a_{\bi}|\mu\rangle = \delta_{\bi\mu}|0\rangle
\end{equation}
This, and the fact that the state $|\mu\rangle$ occurs on both the right and left hand side of the sequence of creation and annihilation operators of (\ref{eq:diag7}) means that this sequence must follow some path in state-space, first removing $|\mu\rangle$ to give the vacuum state $|0\rangle$ then replacing it with some other state $|t_{2n}\rangle$, removing this to give $|0\rangle$, replacing it with some other state, $|t_{2n-1}\rangle$ and so on, with the final step being a replacement of a vacuum state with the initial state $|\mu\rangle$. This process of annihilation and creation occurs $4n$ times, with $2n$ annihilation operations and $2n$ creation operations. The result is the formation a path in state-space beginning and ending with the state $|\mu\rangle$ which never removes a state which is not already there. Any terms in the sum of (\ref{eq:diag7}) which do not contain such a sequence will have a value of zero. Explicitly, the rightmost component of the product given in (\ref{eq:diag7}) either takes the value zero or can be written in the form
\begin{align}\label{eq:diag10}
\langle\mu|a_{\bj^{(1)}}^{\dag}  a_{\bi^{(1)}} a_{\bj^{(2)}}^{\dag}  a_{\bi^{(2)}} \ldots a_{\bj^{(2n)}}^{\dag}  a_{\bi^{(2n)}}|\mu\rangle = \langle\mu|a_{\mu}^{\dag}  a_{t_2} a_{t_2}^{\dag} \ldots a_{t_{2n}} a_{t_{2n}}^{\dag}  a_{\mu}|\mu\rangle = 1.
\end{align}
Hence
\begin{align}\label{eq:diag102}
\langle\mu|a_{\bj^{(1)}}^{\dag}  a_{\bi^{(1)}} a_{\bj^{(2)}}^{\dag}  a_{\bi^{(2)}} \ldots a_{\bj^{(2n)}}^{\dag}  a_{\bi^{(2n)}}|\mu\rangle =  \prod_a\delta_{\bj^{(a)}t_a}\prod_b\delta_{\bi^{(b)}t_{b+1}}
\end{align}
where $t_1$ is identified with $\mu$ and $t_{2n+1}$ is identified with $t_1$. The above allows (\ref{eq:diag7}) to attain the final form
\begin{equation}\label{eq:diag101}
\mathrm{tr}({\overline{H^{2n}_k}}) =\sum_{\sigma} \prod_{a=1}^{2n}\overline{v_{t_a t_{a+1}}v_{t_{\sigma(a)}t_{\sigma(a)+1}}}.
\end{equation}
Each component of the product can be expressed as
\begin{equation}\label{eq:diag11}
\overline{v_{t_{m}t_{m+1}} v_{t_{n}t_{n+1}}}
\end{equation}
for some $m,n$. For uncorrelated $v_{t_{m}t_{m+1}}$ and $ v_{t_{n}t_{n+1}}$ the ensemble average will yield null, since the gaussian determining the p.d.f. of these elements is an even function. However, for correlated $v_{t_{m}t_{m+1}}$ and $v_{t_{n}t_{n+1}}$ (\ref{eq:diag11}) will yield unity if and only if
\begin{equation}\label{eq:diag12}
v_{t_{m}t_{m+1}} =   v_{t_{n}t_{n+1}}^{*}
\end{equation}
because the ensemble average will then yield the second moment of $v$ which has been fixed to take the value 1. By (\ref{eq:ee35}) this is equivalent to saying
\begin{equation}\label{eq:diag13}
\overline{v_{t_{m}t_{m+1}} v_{t_{n}t_{n+1}}} = \delta_{t_{m}t_{n+1}}\delta_{ t_{m+1}t_{n}}
\end{equation}
so that (\ref{eq:diag101}) becomes
\begin{equation}\label{eq:diag14}
\mathrm{tr}({\overline{H^{2n}_k}}) =\sum_{\sigma} \prod_{a=1}^{2n} \delta_{t_{a}t_{\sigma(a)+1}}\delta_{ t_{a+1}t_{\sigma(a)}}.
\end{equation}
It will be shown in subsequent sections how this can be represented by diagrams. These will reveal that the product of delta functions giving (\ref{eq:diag14}) leads to a sequence of paths on the labels $a=1, 2, \ldots, 2n$ and each path will be identified with a partition of the set $1, 2, \ldots, 2n$. Each partition will in turn be identified with a diagram, and these diagrams will be counted to give (\ref{eq:diag14}) which can then be inserted back into (\ref{eq:moments2}) to give the moments of the level density for the GUE.

\subsection{Partitions}\label{sec:partitions}
Equation (\ref{eq:diag14}), and through it the moments of the level density (\ref{eq:moments2}), are related to certain combinatorial diagrams. This will be shown by observing that each instance of the sum of (\ref{eq:diag14}) can be mapped to a partition on the set of integers $\{1, 2, \ldots, 2n\}$. Using diagrammatic methods for the $k=m$ case of canonical random matrix theory is not new \cite{kreweras}. It will be seen that the moments of the level density are in fact given by the Catalan Numbers\footnote{the sequence defined by $C_n = \frac{1}{n+1} {{2n}\choose{n}}$}, which occur in numerous other combinatorial problems outside of the field of random matrix theory. For these -- some of which are much older than random matrix theory itself and have their own distinct literature -- diagrams are frequently used. For $2n=4$ (\ref{eq:diag14}) gives
\begin{equation}\label{eq:partitions1}
\mathrm{tr}({\overline{H^{4}_k}}) =\overline{v_{t_1 t_2}v_{t_2 t_3}v_{t_3 t_4}v_{t_4 t_1}} = \sum_{\sigma} \prod_{a=1}^{4} \delta_{t_{a}t_{\sigma(a)+1}}\delta_{ t_{a+1}t_{\sigma(a)}}.
\end{equation}
the right hand side of which can be read as the ``product of the average of all unique pairings of $v$''. Hence
\begin{align}\label{eq:partitions2}
\mathrm{tr}({\overline{H^{4}_k}}) =\overline{v_{t_1 t_2}v_{t_2 t_3}v_{t_3 t_4}v_{t_4 t_1}} &= \overline{v_{t_1 t_2}v_{t_2 t_3}} ~\overline{v_{t_3 t_4}v_{t_4 t_1}} \notag\\
&+ \overline{v_{t_1 t_2}v_{t_3 t_4}} ~\overline{v_{t_2 t_3}v_{t_4 t_1}} \notag\\
&+ \overline{v_{t_1 t_2}v_{t_4 t_1}} ~\overline{v_{t_2 t_3}v_{t_3 t_4}}\notag\\
&= \delta_{t_1 t_3} + \delta_{t_2 t_4} + \delta_{t_1 t_2}\delta_{t_2 t_3}\delta_{t_3 t_4} \delta_{t_4 t_1}
\end{align}
\subsubsection{Partitions as Cycles}
Identifying $t_1$ with $1$, $t_2$ with $2$ and so on, the components of (\ref{eq:partitions2}) can be represented as cycles in the following way
\begin{align}
\delta_{t_1 t_3} &\equiv (13)(2)(4)\label{eq:partitions31}\\
\delta_{t_2 t_4} &\equiv (24)(1)(3)\label{eq:partitions32}\\
\delta_{t_1 t_2}\delta_{t_2 t_3}\delta_{t_3 t_4} \delta_{t_4 t_1} &\equiv (1234).\label{eq:partitions33}
\end{align}
Note that where $c_i$ denotes the number of orbits in a given cycle with length $i$ that for each cycle of (\ref{eq:partitions31} -- \ref{eq:partitions33}) it holds that
\begin{equation}\label{eq:partitions4}
\sum_i i c_i = 4 = 2n.
\end{equation}
Each orbit represents a degree of freedom in the variables $t_1, t_2, \ldots$ (etc) of (\ref{eq:partitions1}).  The value of tr$(\overline{H_k^4})$ is given by implicitly summing over all values that each of these free variables can take, the result being that each orbit contributes a factor $l\choose m$ to the final expression. Hence the cycle $(13)(2)(4)$ of (\ref{eq:partitions31}) is to be identified with ${l\choose m}^3$ and likewise for the cycle $(24)(1)(3)$. The cycle $(1234)$ however, consists only of a single orbit, so that it is identified with the value $l\choose m$. Hence
\begin{equation}\label{eq:partitions5}
\mathrm{tr}({\overline{H^{4}_k}}) = 2{l\choose m}^3 + {l\choose m}
\end{equation}
so that the fourth moment for the GUE with $k=m$ given by (\ref{eq:moments2}) is
\begin{equation}\label{eq:partitions6}
\beta_4 = \frac{\mathrm{tr}({\overline{H^{4}_m}})}{N^{2+1}} = \frac{2{l\choose m}^3 + {l\choose m}}{{l\choose m}^3} = 2 + \mathcal{O}\left(\frac{1}{N}\right).
\end{equation}
For $2n=6$ (\ref{eq:diag14}) reads
\allowdisplaybreaks[1]
\begin{align}\label{eq:partitions7}
\mathrm{tr}({\overline{H^{6}_k}}) =\overline{v_{t_1 t_2}v_{t_2 t_3}v_{t_3 t_4}v_{t_4 t_1}} &= 
\overline{v_{t_1 t_2}v_{t_2 t_3}} ~\overline{v_{t_3 t_4}v_{t_4 t_5}} ~\overline{v_{t_5 t_6}v_{t_6 t_1}}\notag\\
&+ \overline{v_{t_1 t_2}v_{t_2 t_3}} ~\overline{v_{t_3 t_4}v_{t_5 t_6}}~\overline{v_{t_4 t_5}v_{t_6 t_1}}\notag\\
&+ \overline{v_{t_1 t_2}v_{t_2 t_3}} ~\overline{v_{t_3 t_4}v_{t_6 t_1}}~\overline{v_{t_4 t_5}v_{t_5 t_6}}\notag\\
&+ \overline{v_{t_1 t_2}v_{t_3 t_4}} ~\overline{v_{t_2 t_3}v_{t_4 t_5}}~\overline{v_{t_5 t_6}v_{t_6 t_1}}\notag\\
&+ \overline{v_{t_1 t_2}v_{t_3 t_4}} ~\overline{v_{t_2 t_3}v_{t_5 t_6}}~\overline{v_{t_4 t_5}v_{t_6 t_1}}\notag\\
&+ \overline{v_{t_1 t_2}v_{t_3 t_4}} ~\overline{v_{t_2 t_3}v_{t_6 t_1}}~\overline{v_{t_4 t_5}v_{t_5 t_6}}\notag\\
&+ \overline{v_{t_1 t_2}v_{t_4 t_5}} ~\overline{v_{t_2 t_3}v_{t_3 t_4}}~\overline{v_{t_5 t_6}v_{t_6 t_1}}\notag\\
&+ \overline{v_{t_1 t_2}v_{t_4 t_5}} ~\overline{v_{t_2 t_3}v_{t_5 t_6}}~\overline{v_{t_3 t_4}v_{t_6 t_1}}\notag\\
&+ \overline{v_{t_1 t_2}v_{t_4 t_5}} ~\overline{v_{t_2 t_3}v_{t_6 t_1}}~\overline{v_{t_3 t_4}v_{t_5 t_6}}\notag\\
&+ \overline{v_{t_1 t_2}v_{t_5 t_6}} ~\overline{v_{t_2 t_3}v_{t_3 t_4}}~\overline{v_{t_4 t_5}v_{t_6 t_1}}\notag\\
&+ \overline{v_{t_1 t_2}v_{t_5 t_6}} ~\overline{v_{t_2 t_3}v_{t_4 t_5}}~\overline{v_{t_3 t_4}v_{t_6 t_1}}\notag\\
&+ \overline{v_{t_1 t_2}v_{t_5 t_6}} ~\overline{v_{t_2 t_3}v_{t_6 t_1}}~\overline{v_{t_3 t_4}v_{t_4 t_5}}\notag\\
&+ \overline{v_{t_1 t_2}v_{t_6 t_1}} ~\overline{v_{t_2 t_3}v_{t_3 t_4}}~\overline{v_{t_4 t_5}v_{t_5 t_6}}\notag\\
&+ \overline{v_{t_1 t_2}v_{t_6 t_1}} ~\overline{v_{t_2 t_3}v_{t_4 t_5}}~\overline{v_{t_3 t_4}v_{t_5 t_6}}\notag\\
&+ \overline{v_{t_1 t_2}v_{t_6t_1}} ~\overline{v_{t_2 t_3}v_{t_5 t_6}}~\overline{v_{t_3 t_4}v_{t_4 t_5}}.
\end{align}
Again identifying $t_1$ with $1$, $t_2$ with $2$ (etc), the components of (\ref{eq:partitions7}) can be represented as the cycles
\begin{align}\label{eq:partitions8}
\delta_{t_1 t_3}\delta_{t_3 t_5}\delta_{t_5 t_1}  &\equiv (135)(2)(4)(6)\notag\\
\delta_{t_1 t_3}\delta_{t_3 t_6}\delta_{t_4 t_5} \delta_{t_5 t_6} \delta_{t_4 t_1} &\equiv (13654)(2)\notag\\
\delta_{t_1 t_3}\delta_{t_3 t_1}\delta_{t_4 t_6} &\equiv (13)(46)(2)(5)\notag\\
\delta_{t_1 t_4}\delta_{t_2 t_3}\delta_{t_2 t_5} \delta_{t_3 t_4}\delta_{t_5 t_1} &\equiv (14325)(6)\notag\\
\delta_{t_1 t_4}\delta_{t_2 t_3}\delta_{t_2 t_6} \delta_{t_3 t_5} \delta_{t_5 t_6} &\equiv (14)(2356)\notag\\
\delta_{t_1 t_4}\delta_{t_2 t_3}\delta_{t_1 t_2} \delta_{t_3 t_6}\delta_{t_4 t_6} &\equiv (14632)(5)\notag\\
\delta_{t_1 t_5}\delta_{t_2 t_4} &\equiv (15)(24)(3)(6)\notag\\
\delta_{t_1 t_5}\delta_{t_2 t_4}\delta_{t_2 t_6} \delta_{t_3 t_5}\delta_{t_1 t_3} \delta_{t_4 t_6} &\equiv (153)(246)\notag\\
\delta_{t_1 t_5}\delta_{t_2 t_4}\delta_{t_1 t_2} \delta_{t_3 t_6}\delta_{t_4 t_5} &\equiv (1542)(36)\notag\\
\delta_{t_1 t_6}\delta_{t_2 t_5}\delta_{t_2 t_4} \delta_{t_4 t_1}\delta_{t_5 t_6} &\equiv (16524)(3)\notag\\
\delta_{t_1 t_6}\delta_{t_2 t_5}\delta_{t_3 t_4} \delta_{t_1 t_3}\delta_{t_4 t_6} &\equiv (1643)(25)\notag\\
\delta_{t_1 t_6}\delta_{t_2 t_5}\delta_{t_1 t_2} \delta_{t_3 t_6}\delta_{t_3 t_5} &\equiv (16352)(4)\notag\\
\delta_{t_2 t_6}\delta_{t_2 t_4}\delta_{t_4 t_6} &\equiv (246)(1)(3)(5)\notag\\
\delta_{t_2 t_6}\delta_{t_2 t_5}\delta_{t_3 t_4} \delta_{t_3 t_6}\delta_{t_4 t_5} &\equiv (26345)(1)\notag\\
\delta_{t_2 t_6}\delta_{t_3 t_5} &\equiv (26)(35)(1)(4).
\end{align}
This time the cycles obey
\begin{equation}\label{eq:partitions9}
\sum_i i c_i = 6 = 2n
\end{equation}
and since the value of each term is the number of orbits in the cycle times $l\choose m$ one gains
\begin{equation}\label{eq:partitions10}
\mathrm{tr}({\overline{H^{6}_m}}) = 5{l\choose m}^4 + 10{l\choose m}^2
\end{equation}
so
\begin{equation}\label{eq:partitions11}
\beta_6 = \frac{\mathrm{tr}({\overline{H^{6}_m}})}{N^{3+1}} = 5 + \mathcal{O}\left(\frac{1}{N}\right).
\end{equation}
For $2n=8$ (\ref{eq:diag14}) reads
\begin{align}\label{eq:partitions12}
\mathrm{tr}({\overline{H^{8}_k}}) =\overline{v_{t_1 t_2}v_{t_2 t_3}v_{t_3 t_4}v_{t_4 t_5}v_{t_5 t_6}v_{t_6 t_7}v_{t_7 t_8} v_{t_8 t_1}}.
\end{align}
Now there are $(2n-1)!! = 105$ ways to pair the $v$'s, each giving rise to a unique partition on the integers $1, 2, \ldots, 8$. This time there are just $14$ cycles determining the leading behaviour of the $8$-th moment as $N\to\infty$, given by
\begin{align}
&(2468)(1)(3)(5)(7)\notag\\
&(1357)(2)(4)(6)(8)\label{eq:partitions13}\\
\notag\\
&(15)(24)(68)(3)(7)\notag\\
&(17)(26)(35)(4)(8)\notag\\
&(28)(37)(46)(1)(5)\notag\\
&(13)(48)(57)(2)(6)\label{eq:partitions14}\\
\notag\\
&(135)(68)(2)(4)(7)\notag\\
&(248)(57)(1)(3)(6)\notag\\
&(137)(46)(2)(5)(8)\notag\\
&(246)(17)(3)(5)(8)\notag\\
&(357)(28)(1)(4)(6)\notag\\
&(268)(35)(1)(4)(7)\notag\\
&(468)(13)(2)(5)(7)\notag\\
&(157)(24)(3)(6)(8).\label{eq:partitions15}
\end{align}
Note that the cycles $(1357)(2)(4)(6)(8)$ and $(2468)(1)(3)(5)(7)$ of (\ref{eq:partitions13}) are equivalent up to a constant mod $9$. In other words the orbit $(1357)$ with each element incremented by $1$ mod $9$ is identical to the orbit $(2468)$ and similarly the orbits $(2)(4)(6)(8)$ incremented by $1$ mod $9$ are equal to $(3)(5)(7)(1)$. The same applies to the four cycles of (\ref{eq:partitions14}), each being equivalent to eachother up to an increment of $1$ mod $9$, $2$ mod $9$ or $3$ mod $9$. Finally, the eight cycles given by (\ref{eq:partitions15}) are equivalent up to an increment of $i$ mod $9$ where $1 \le i \le 7$.
Assigning each orbit of each cycle the value $l\choose m$ as before and dividing by the normalisation constant $N^{n+1}$ gives the eighth moment for the GUE as
\begin{equation}\label{eq:partitions16}
\beta_8 = \frac{\mathrm{tr}({\overline{H^{8}_m}})}{N^{4+1}} = 14 + \mathcal{O}\left(\frac{1}{N}\right).
\end{equation}
The process for calculating the moments could continue indefinitely in this way. But what is the underlying pattern? And in what form can this pattern be codified in order to find a general expression for all moments of the $k=m$ scenario? In the next section the cycles introduced above will be represented as diagrams. This will illustrate the possibility of transferring the problem (of calculating the moments) into a ``diagram space'' and using the new space to make calculations. The method will act as a precurser to a much more complex diagrammatic technique which will be used in subsequent chapters to calculate moments for random matrix ensembles where $k\le m$.
\subsection{Polygons}\label{sec:polygons}
The first method for representing (\ref{eq:diag14}) will be to simply draw the cycles from the previous section as segments of $2n$-sided polygons. It will be seen that the leading order behaviour of the moments, being the cycles with $n+1$ orbits as shown in section \ref{sec:partitions}, are given by those partitioned polygons where (i) no partitions intersect, (ii) no two isolated vertices neighbour eachother and (iii) any two vertices which can be connected by a non-intersecting edge are connected as such. For the $4$-th moment there are just two cycles with three orbits, namely $\delta_{t_1 t_3} \equiv (13)(2)(4)$ and $\delta_{t_2 t_4} \equiv (24)(1)(3)$ which are illustrated in Fig \ref{fig:poly4}.
\begin{figure}[htp]
\hrulefill
\vspace{.1cm}

\begin{subfigure}{.5\textwidth}
  \centering
\begin{tikzpicture}
\def \n {1}
\def \R {1.5cm}
\def \RR {1.8cm}
\draw (-45:\R) \foreach \x in {-135,-225,-315} {
            -- (\x:\R)
        } -- cycle;
\foreach \x in {-45,-135,-225,-315} {
        \draw [fill] (\x:\R)  circle [radius=0.1];
        }
\foreach \x/\i in {-45/1,-135/2,-225/3,-315/4} {
        \node at (\x:\RR) {\i};
\draw (-225:\R) -- (-45:\R);
        }
\end{tikzpicture}
  \label{fig:p41}
\end{subfigure}%
\begin{subfigure}{.33\textwidth}
  \centering
 \begin{tikzpicture}
\def \n {1}
\def \R {1.5cm}
\def \RR {1.8cm}
\draw (-45:\R) \foreach \x in {-135,-225,-315} {
            -- (\x:\R)
        } -- cycle;
\foreach \x in {-45,-135,-225,-315} {
        \draw [fill] (\x:\R)  circle [radius=0.1];
        }
\foreach \x/\i in {-45/1,-135/2,-225/3,-315/4} {
        \node at (\x:\RR) {\i};
\draw (-135:\R) -- (-315:\R);
        }
\end{tikzpicture} 
  \label{fig:p42}
\end{subfigure}%
\caption[Polygon partitions for fourth moment]{Plot of polygons for fourth moments with the left hand side giving an illustration of the cycle $(13)(2)(4)$ on a $2n=4$ sided polygon and the right hand side illustrating the only unique rotation of this cycle, which is $(24)(1)(3)$.}
\label{fig:poly4}
\end{figure}
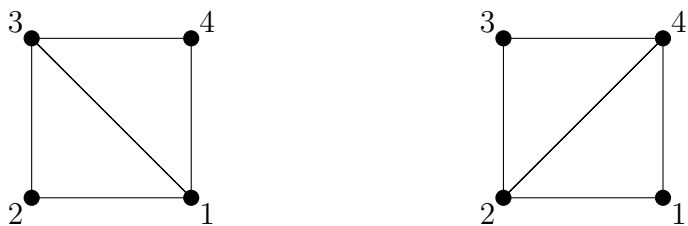
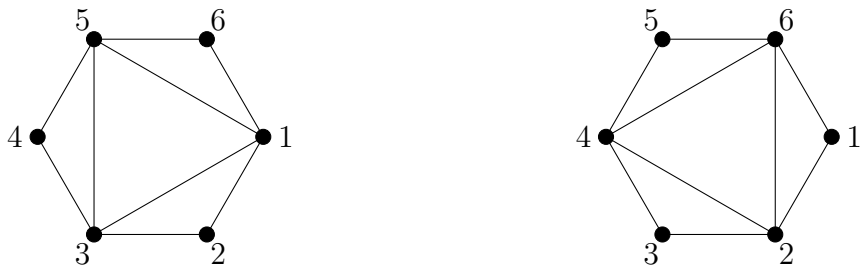
\begin{figure}[htp]
\begin{subfigure}{.5\textwidth}
  \centering
\begin{tikzpicture}
\def \n {1}
\def \R {1.5cm}
\def \RR {1.8cm}
\def \one {0}
\def \two {-60}
\def \three {-120}
\def \four {-180}
\def \five {-240}
\def \six {-300}
\draw (0:\R) \foreach \x in {-60,-120,-180,-240,-300} {
            -- (\x:\R)
        } -- cycle;
\foreach \x in {0,-60,-120,-180,-240,-300} {
        \draw [fill] (\x:\R)  circle [radius=0.1];
        }
\foreach \x/\i in {0/1,-60/2,-120/3,-180/4,-240/5,-300/6} {
        \node at (\x:\RR) {\i};
        }
\draw (\one:\R) -- (\three:\R) -- (\five:\R) --cycle;
\end{tikzpicture}
  \label{fig:p61}
\end{subfigure}%
\begin{subfigure}{.5\textwidth}
  \centering
 \begin{tikzpicture}
\def \n {1}
\def \R {1.5cm}
\def \RR {1.8cm}
\def \one {0}
\def \two {-60}
\def \three {-120}
\def \four {-180}
\def \five {-240}
\def \six {-300}
\draw (0:\R) \foreach \x in {-60,-120,-180,-240,-300} {
            -- (\x:\R)
        } -- cycle;
\foreach \x in {0,-60,-120,-180,-240,-300} {
        \draw [fill] (\x:\R)  circle [radius=0.1];
        }
\foreach \x/\i in {0/1,-60/2,-120/3,-180/4,-240/5,-300/6} {
        \node at (\x:\RR) {\i};
        }
\draw (\two:\R) -- (\four:\R) -- (\six:\R) --cycle;
\end{tikzpicture} 
  \label{fig:p62}
\end{subfigure}%
\caption[The cycles $(135)(2)(4)(6)$ and $(246)(1)(3)(5)$]{Illustration of the cycle $(135)(2)(4)(6)$ and it's unique rotation $(246)(1)(3)(5)$ given by incrementing each element by $1$ mod $7$.}
\label{fig:poly6}
\end{figure}
\begin{figure}[htp]
\begin{subfigure}{.33\textwidth}
  \centering
 \begin{tikzpicture}
\def \one {0}
\def \two {-60}
\def \three {-120}
\def \four {-180}
\def \five {-240}
\def \six {-300}
\def \n {1}
\def \R {1.5cm}
\def \RR {1.8cm}
\draw (0:\R) \foreach \x in {-60,-120,-180,-240,-300} {
            -- (\x:\R)
        } -- cycle;
\foreach \x in {0,-60,-120,-180,-240,-300} {
        \draw [fill] (\x:\R)  circle [radius=0.1];
        }
\foreach \x/\i in {0/1,-60/2,-120/3,-180/4,-240/5,-300/6} {
        \node at (\x:\RR) {\i};
        }
\draw (\one:\R) -- (\three:\R);
\draw (\four:\R) -- (\six:\R);
\end{tikzpicture}
  \label{fig:p63}
\end{subfigure}%
\begin{subfigure}{.33\textwidth}
  \centering
\begin{tikzpicture}
\def \one {0}
\def \two {-60}
\def \three {-120}
\def \four {-180}
\def \five {-240}
\def \six {-300}
\def \n {1}
\def \R {1.5cm}
\def \RR {1.8cm}
\draw (0:\R) \foreach \x in {-60,-120,-180,-240,-300} {
            -- (\x:\R)
        } -- cycle;
\foreach \x in {0,-60,-120,-180,-240,-300} {
        \draw [fill] (\x:\R)  circle [radius=0.1];
        }
\foreach \x/\i in {0/1,-60/2,-120/3,-180/4,-240/5,-300/6} {
        \node at (\x:\RR) {\i};
        }
\draw (\one:\R) -- (\five:\R);
\draw (\four:\R) -- (\two:\R);
\end{tikzpicture}
  \label{fig:p64}
\end{subfigure}%
\begin{subfigure}{.33\textwidth}
  \centering
 \begin{tikzpicture}
\def \one {0}
\def \two {-60}
\def \three {-120}
\def \four {-180}
\def \five {-240}
\def \six {-300}
\def \n {1}
\def \R {1.5cm}
\def \RR {1.8cm}
\draw (0:\R) \foreach \x in {-60,-120,-180,-240,-300} {
            -- (\x:\R)
        } -- cycle;
\foreach \x in {0,-60,-120,-180,-240,-300} {
        \draw [fill] (\x:\R)  circle [radius=0.1];
        }
\foreach \x/\i in {0/1,-60/2,-120/3,-180/4,-240/5,-300/6} {
        \node at (\x:\RR) {\i};
        }
\draw (\two:\R) -- (\six:\R);
\draw (\three:\R) -- (\five:\R);
\end{tikzpicture}
  \label{fig:p65}
\end{subfigure}%
\caption[Three equivalent cycles]{Illustration of the cycles $(13)(46)(2)(5)$~(left), $(15)(24)(3)(6)$~(center) and $(26)(35)(1)(4)$~(right) of (\ref{eq:partitions8}).}
\label{fig:poly62}
\vspace{2cm}
\begin{subfigure}{.5\textwidth}
  \centering
\begin{tikzpicture}
\def \n {1}
\def \R {1.5cm}
\def \RR {1.8cm}
\def \one {-22.5}
\def \two {-67.5}
\def \three {-112.5}
\def \four {-157.5}
\def \five {-202.5}
\def \six {-247.5}
\def \seven {-292.5}
\def \eight {-337.5}
\draw (-22.5:\R) \foreach \x in {-67.5,-112.5,-157.5,-202.5,-247.5,-292.5,-337.5} {
            -- (\x:\R)
        } -- cycle;
\foreach \x in {-22.5,-67.5,-112.5,-157.5,-202.5,-247.5,-292.5,-337.5} {
        \draw [fill] (\x:\R)  circle [radius=0.1];
        }
\foreach \x/\i in {-22.5/1,-67.5/2,-112.5/3,-157.5/4,-202.5/5,-247.5/6,-292.5/7,-337.5/8} {
        \node at (\x:\RR) {\i};
        }
\draw (\two:\R) -- (\four:\R) -- (\six:\R) -- (\eight:\R) --cycle;
\end{tikzpicture}
  \label{fig:p81}
\end{subfigure}%
\begin{subfigure}{.5\textwidth}
  \centering
 \begin{tikzpicture}
\def \n {1}
\def \R {1.5cm}
\def \RR {1.8cm}
\def \one {-22.5}
\def \two {-67.5}
\def \three {-112.5}
\def \four {-157.5}
\def \five {-202.5}
\def \six {-247.5}
\def \seven {-292.5}
\def \eight {-337.5}
\draw (-22.5:\R) \foreach \x in {-67.5,-112.5,-157.5,-202.5,-247.5,-292.5,-337.5} {
            -- (\x:\R)
        } -- cycle;
\foreach \x in {-22.5,-67.5,-112.5,-157.5,-202.5,-247.5,-292.5,-337.5} {
        \draw [fill] (\x:\R)  circle [radius=0.1];
        }
\foreach \x/\i in {-22.5/1,-67.5/2,-112.5/3,-157.5/4,-202.5/5,-247.5/6,-292.5/7,-337.5/8} {
        \node at (\x:\RR) {\i};
        }
\draw (\one:\R) -- (\three:\R) -- (\five:\R) -- (\seven:\R) --cycle;
\end{tikzpicture} 
  \label{fig:p82}
\end{subfigure}%
\caption[The cycles $(2468)(1)(3)(5)(7)$ and $(1357)(2)(4)(5)(8)$]{The cycles $(2468)(1)(3)(5)(7)$~(left) and $(1357)(2)(4)(5)(8)$.}
\label{fig:poly8}
\end{figure}
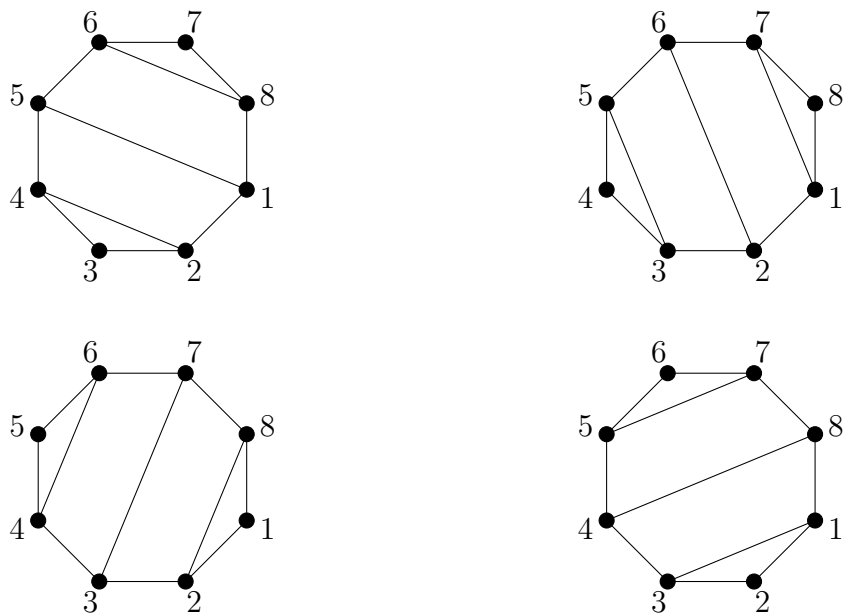
\begin{figure}[htp]
\begin{subfigure}{.5\textwidth}
  \centering
 \begin{tikzpicture}
\def \n {1}
\def \R {1.5cm}
\def \RR {1.8cm}
\def \one {-22.5}
\def \two {-67.5}
\def \three {-112.5}
\def \four {-157.5}
\def \five {-202.5}
\def \six {-247.5}
\def \seven {-292.5}
\def \eight {-337.5}
\draw (-22.5:\R) \foreach \x in {-67.5,-112.5,-157.5,-202.5,-247.5,-292.5,-337.5} {
            -- (\x:\R)
        } -- cycle;
\foreach \x in {-22.5,-67.5,-112.5,-157.5,-202.5,-247.5,-292.5,-337.5} {
        \draw [fill] (\x:\R)  circle [radius=0.1];
        }
\foreach \x/\i in {-22.5/1,-67.5/2,-112.5/3,-157.5/4,-202.5/5,-247.5/6,-292.5/7,-337.5/8} {
        \node at (\x:\RR) {\i};
        }
\draw (\two:\R) -- (\four:\R);
\draw (\one:\R) -- (\five:\R);
\draw (\eight:\R) -- (\six:\R);
\end{tikzpicture}
  \label{fig:p83}
\end{subfigure}%
\begin{subfigure}{.5\textwidth}
  \centering
\begin{tikzpicture}
\def \n {1}
\def \R {1.5cm}
\def \RR {1.8cm}
\def \one {-22.5}
\def \two {-67.5}
\def \three {-112.5}
\def \four {-157.5}
\def \five {-202.5}
\def \six {-247.5}
\def \seven {-292.5}
\def \eight {-337.5}
\draw (-22.5:\R) \foreach \x in {-67.5,-112.5,-157.5,-202.5,-247.5,-292.5,-337.5} {
            -- (\x:\R)
        } -- cycle;
\foreach \x in {-22.5,-67.5,-112.5,-157.5,-202.5,-247.5,-292.5,-337.5} {
        \draw [fill] (\x:\R)  circle [radius=0.1];
        }
\foreach \x/\i in {-22.5/1,-67.5/2,-112.5/3,-157.5/4,-202.5/5,-247.5/6,-292.5/7,-337.5/8} {
        \node at (\x:\RR) {\i};
        }
\draw (\three:\R) -- (\five:\R);
\draw (\two:\R) -- (\six:\R);
\draw (\one:\R) -- (\seven:\R);
\end{tikzpicture}
  \label{fig:p84}
\end{subfigure}\\[5mm]
\begin{subfigure}{.5\textwidth}
  \centering
 \begin{tikzpicture}
\def \n {1}
\def \R {1.5cm}
\def \RR {1.8cm}
\def \one {-22.5}
\def \two {-67.5}
\def \three {-112.5}
\def \four {-157.5}
\def \five {-202.5}
\def \six {-247.5}
\def \seven {-292.5}
\def \eight {-337.5}
\draw (-22.5:\R) \foreach \x in {-67.5,-112.5,-157.5,-202.5,-247.5,-292.5,-337.5} {
            -- (\x:\R)
        } -- cycle;
\foreach \x in {-22.5,-67.5,-112.5,-157.5,-202.5,-247.5,-292.5,-337.5} {
        \draw [fill] (\x:\R)  circle [radius=0.1];
        }
\foreach \x/\i in {-22.5/1,-67.5/2,-112.5/3,-157.5/4,-202.5/5,-247.5/6,-292.5/7,-337.5/8} {
        \node at (\x:\RR) {\i};
        }
\draw (\two:\R) -- (\eight:\R);
\draw (\three:\R) -- (\seven:\R);
\draw (\four:\R) -- (\six:\R);
\end{tikzpicture}
  \label{fig:p85}
\end{subfigure}%
\begin{subfigure}{.5\textwidth}
  \centering
\begin{tikzpicture}
\def \n {1}
\def \R {1.5cm}
\def \RR {1.8cm}
\def \one {-22.5}
\def \two {-67.5}
\def \three {-112.5}
\def \four {-157.5}
\def \five {-202.5}
\def \six {-247.5}
\def \seven {-292.5}
\def \eight {-337.5}
\draw (-22.5:\R) \foreach \x in {-67.5,-112.5,-157.5,-202.5,-247.5,-292.5,-337.5} {
            -- (\x:\R)
        } -- cycle;
\foreach \x in {-22.5,-67.5,-112.5,-157.5,-202.5,-247.5,-292.5,-337.5} {
        \draw [fill] (\x:\R)  circle [radius=0.1];
        }
\foreach \x/\i in {-22.5/1,-67.5/2,-112.5/3,-157.5/4,-202.5/5,-247.5/6,-292.5/7,-337.5/8} {
        \node at (\x:\RR) {\i};
        }
\draw (\eight:\R) -- (\four:\R);
\draw (\one:\R) -- (\three:\R);
\draw (\five:\R) -- (\seven:\R);
\end{tikzpicture}
  \label{fig:p86}
\end{subfigure}%
\caption[Four equivalent cycles]{The four equivalent cycles (up to addition modulo $9$) given by $(24)(15)(68)(3)(7)$~(top left), $(17)(26)(35)(4)(8)$~(top right), $(28)(37)(46)(1)(5)$~(bottom left) and $(13)(48)(57)(2)(6)$~(bottom right).}
\label{fig:poly82}
\end{figure}
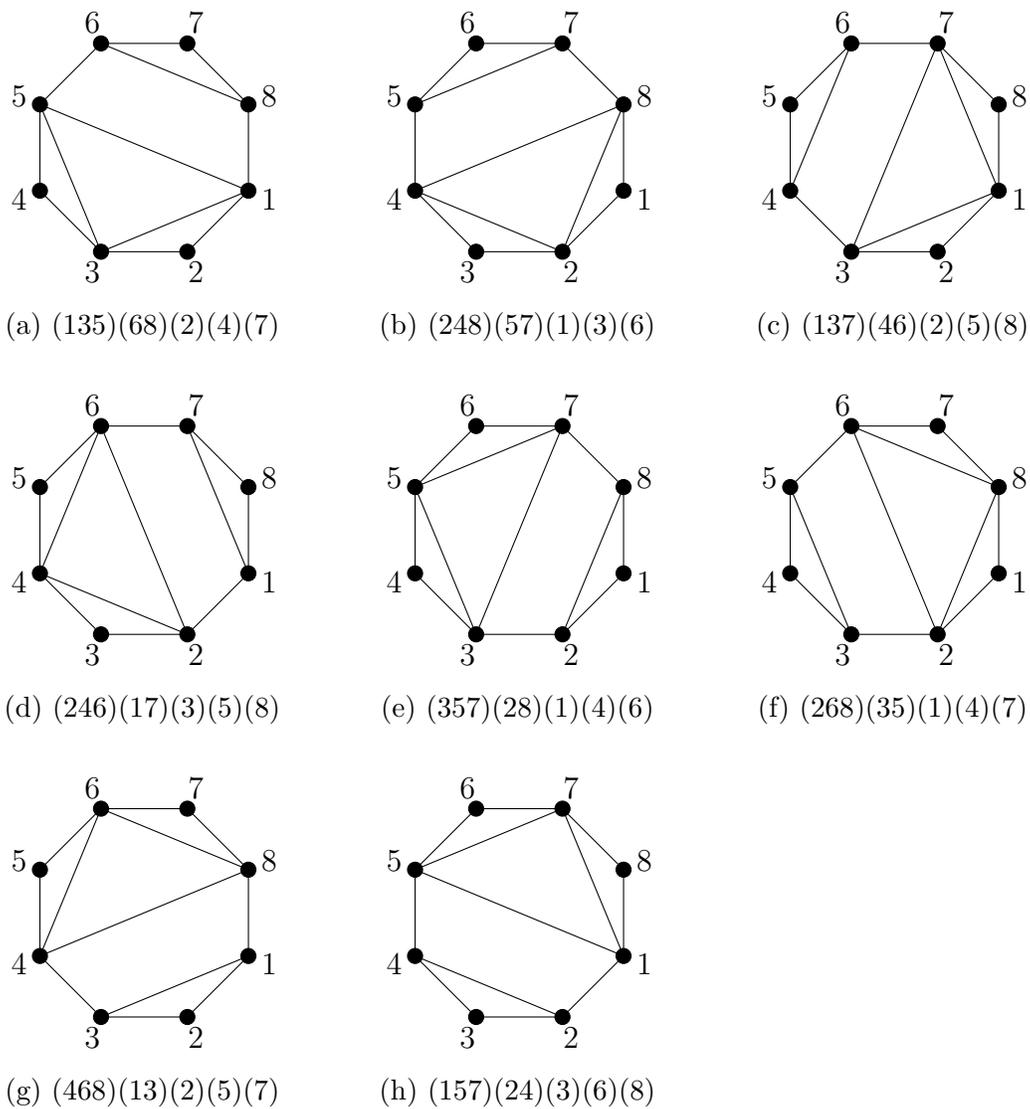
\begin{figure}[htp]
\begin{subfigure}{.33\textwidth}
  \centering
 \begin{tikzpicture}
\def \n {1}
\def \R {1.5cm}
\def \RR {1.8cm}
\def \one {-22.5}
\def \two {-67.5}
\def \three {-112.5}
\def \four {-157.5}
\def \five {-202.5}
\def \six {-247.5}
\def \seven {-292.5}
\def \eight {-337.5}
\draw (-22.5:\R) \foreach \x in {-67.5,-112.5,-157.5,-202.5,-247.5,-292.5,-337.5} {
            -- (\x:\R)
        } -- cycle;
\foreach \x in {-22.5,-67.5,-112.5,-157.5,-202.5,-247.5,-292.5,-337.5} {
        \draw [fill] (\x:\R)  circle [radius=0.1];
        }
\foreach \x/\i in {-22.5/1,-67.5/2,-112.5/3,-157.5/4,-202.5/5,-247.5/6,-292.5/7,-337.5/8} {
        \node at (\x:\RR) {\i};
        }
\draw (\one:\R) -- (\three:\R) -- (\five:\R) -- cycle;
\draw (\eight:\R) -- (\six:\R);
\end{tikzpicture} 
\caption{(135)(68)(2)(4)(7)}
  \label{fig:p87}
\end{subfigure}%
\begin{subfigure}{.33\textwidth}
  \centering
 \begin{tikzpicture}
\def \n {1}
\def \R {1.5cm}
\def \RR {1.8cm}
\def \one {-22.5}
\def \two {-67.5}
\def \three {-112.5}
\def \four {-157.5}
\def \five {-202.5}
\def \six {-247.5}
\def \seven {-292.5}
\def \eight {-337.5}
\draw (-22.5:\R) \foreach \x in {-67.5,-112.5,-157.5,-202.5,-247.5,-292.5,-337.5} {
            -- (\x:\R)
        } -- cycle;
\foreach \x in {-22.5,-67.5,-112.5,-157.5,-202.5,-247.5,-292.5,-337.5} {
        \draw [fill] (\x:\R)  circle [radius=0.1];
        }
\foreach \x/\i in {-22.5/1,-67.5/2,-112.5/3,-157.5/4,-202.5/5,-247.5/6,-292.5/7,-337.5/8} {
        \node at (\x:\RR) {\i};
        }
\draw (\two:\R) -- (\four:\R) -- (\eight:\R) -- cycle;
\draw (\five:\R) -- (\seven:\R);
\end{tikzpicture}
\caption{(248)(57)(1)(3)(6)}
  \label{fig:p88}
\end{subfigure}%
\begin{subfigure}{.33\textwidth}
  \centering
\begin{tikzpicture}
\def \n {1}
\def \R {1.5cm}
\def \RR {1.8cm}
\def \one {-22.5}
\def \two {-67.5}
\def \three {-112.5}
\def \four {-157.5}
\def \five {-202.5}
\def \six {-247.5}
\def \seven {-292.5}
\def \eight {-337.5}
\draw (-22.5:\R) \foreach \x in {-67.5,-112.5,-157.5,-202.5,-247.5,-292.5,-337.5} {
            -- (\x:\R)
        } -- cycle;
\foreach \x in {-22.5,-67.5,-112.5,-157.5,-202.5,-247.5,-292.5,-337.5} {
        \draw [fill] (\x:\R)  circle [radius=0.1];
        }
\foreach \x/\i in {-22.5/1,-67.5/2,-112.5/3,-157.5/4,-202.5/5,-247.5/6,-292.5/7,-337.5/8} {
        \node at (\x:\RR) {\i};
        }
\draw (\one:\R) -- (\three:\R) -- (\seven:\R) -- cycle;
\draw (\four:\R) -- (\six:\R);
\end{tikzpicture}
\caption{(137)(46)(2)(5)(8)}
  \label{fig:p89}
\end{subfigure}\\[5mm]
\begin{subfigure}{.33\textwidth}
  \centering
 \begin{tikzpicture}
\def \n {1}
\def \R {1.5cm}
\def \RR {1.8cm}
\def \one {-22.5}
\def \two {-67.5}
\def \three {-112.5}
\def \four {-157.5}
\def \five {-202.5}
\def \six {-247.5}
\def \seven {-292.5}
\def \eight {-337.5}
\draw (-22.5:\R) \foreach \x in {-67.5,-112.5,-157.5,-202.5,-247.5,-292.5,-337.5} {
            -- (\x:\R)
        } -- cycle;
\foreach \x in {-22.5,-67.5,-112.5,-157.5,-202.5,-247.5,-292.5,-337.5} {
        \draw [fill] (\x:\R)  circle [radius=0.1];
        }
\foreach \x/\i in {-22.5/1,-67.5/2,-112.5/3,-157.5/4,-202.5/5,-247.5/6,-292.5/7,-337.5/8} {
        \node at (\x:\RR) {\i};
        }
\draw (\two:\R) -- (\four:\R) -- (\six:\R) -- cycle;
\draw (\one:\R) -- (\seven:\R);
\end{tikzpicture}
\caption{(246)(17)(3)(5)(8)}
  \label{fig:p810}
\end{subfigure}%
\begin{subfigure}{.33\textwidth}
  \centering
 \begin{tikzpicture}
\def \n {1}
\def \R {1.5cm}
\def \RR {1.8cm}
\def \one {-22.5}
\def \two {-67.5}
\def \three {-112.5}
\def \four {-157.5}
\def \five {-202.5}
\def \six {-247.5}
\def \seven {-292.5}
\def \eight {-337.5}
\draw (-22.5:\R) \foreach \x in {-67.5,-112.5,-157.5,-202.5,-247.5,-292.5,-337.5} {
            -- (\x:\R)
        } -- cycle;
\foreach \x in {-22.5,-67.5,-112.5,-157.5,-202.5,-247.5,-292.5,-337.5} {
        \draw [fill] (\x:\R)  circle [radius=0.1];
        }
\foreach \x/\i in {-22.5/1,-67.5/2,-112.5/3,-157.5/4,-202.5/5,-247.5/6,-292.5/7,-337.5/8} {
        \node at (\x:\RR) {\i};
        }
\draw (\seven:\R) -- (\three:\R) -- (\five:\R) -- cycle;
\draw (\eight:\R) -- (\two:\R);
\end{tikzpicture}
\caption{(357)(28)(1)(4)(6)}
  \label{fig:p811}
\end{subfigure}%
\begin{subfigure}{.33\textwidth}
  \centering
 \begin{tikzpicture}
\def \n {1}
\def \R {1.5cm}
\def \RR {1.8cm}
\def \one {-22.5}
\def \two {-67.5}
\def \three {-112.5}
\def \four {-157.5}
\def \five {-202.5}
\def \six {-247.5}
\def \seven {-292.5}
\def \eight {-337.5}
\draw (-22.5:\R) \foreach \x in {-67.5,-112.5,-157.5,-202.5,-247.5,-292.5,-337.5} {
            -- (\x:\R)
        } -- cycle;
\foreach \x in {-22.5,-67.5,-112.5,-157.5,-202.5,-247.5,-292.5,-337.5} {
        \draw [fill] (\x:\R)  circle [radius=0.1];
        }
\foreach \x/\i in {-22.5/1,-67.5/2,-112.5/3,-157.5/4,-202.5/5,-247.5/6,-292.5/7,-337.5/8} {
        \node at (\x:\RR) {\i};
        }
\draw (\two:\R) -- (\six:\R) -- (\eight:\R) -- cycle;
\draw (\three:\R) -- (\five:\R);
\end{tikzpicture}
\caption{(268)(35)(1)(4)(7)}
  \label{fig:p812}
\end{subfigure}\\[5mm]
\begin{subfigure}{.33\textwidth}
  \centering
 \begin{tikzpicture}
\def \n {1}
\def \R {1.5cm}
\def \RR {1.8cm}
\def \one {-22.5}
\def \two {-67.5}
\def \three {-112.5}
\def \four {-157.5}
\def \five {-202.5}
\def \six {-247.5}
\def \seven {-292.5}
\def \eight {-337.5}
\draw (-22.5:\R) \foreach \x in {-67.5,-112.5,-157.5,-202.5,-247.5,-292.5,-337.5} {
            -- (\x:\R)
        } -- cycle;
\foreach \x in {-22.5,-67.5,-112.5,-157.5,-202.5,-247.5,-292.5,-337.5} {
        \draw [fill] (\x:\R)  circle [radius=0.1];
        }
\foreach \x/\i in {-22.5/1,-67.5/2,-112.5/3,-157.5/4,-202.5/5,-247.5/6,-292.5/7,-337.5/8} {
        \node at (\x:\RR) {\i};
        }
\draw (\four:\R) -- (\six:\R) -- (\eight:\R) -- cycle;
\draw (\one:\R) -- (\three:\R);
\end{tikzpicture}
\caption{(468)(13)(2)(5)(7)}
  \label{fig:p813}
\end{subfigure}%
\begin{subfigure}{.33\textwidth}
  \centering
 \begin{tikzpicture}
\def \n {1}
\def \R {1.5cm}
\def \RR {1.8cm}
\def \one {-22.5}
\def \two {-67.5}
\def \three {-112.5}
\def \four {-157.5}
\def \five {-202.5}
\def \six {-247.5}
\def \seven {-292.5}
\def \eight {-337.5}
\draw (-22.5:\R) \foreach \x in {-67.5,-112.5,-157.5,-202.5,-247.5,-292.5,-337.5} {
            -- (\x:\R)
        } -- cycle;
\foreach \x in {-22.5,-67.5,-112.5,-157.5,-202.5,-247.5,-292.5,-337.5} {
        \draw [fill] (\x:\R)  circle [radius=0.1];
        }
\foreach \x/\i in {-22.5/1,-67.5/2,-112.5/3,-157.5/4,-202.5/5,-247.5/6,-292.5/7,-337.5/8} {
        \node at (\x:\RR) {\i};
        }
\draw (\one:\R) -- (\five:\R) -- (\seven:\R) -- cycle;
\draw (\two:\R) -- (\four:\R);
\end{tikzpicture}
\caption{(157)(24)(3)(6)(8)}
  \label{fig:p814}
\end{subfigure}%
\caption[Polygons rigidly rotated about the origin]{Illustration of the eight cycles of equation (\ref{eq:partitions15}) each with the same shape, but rigidly rotated about the origin.}
\label{fig:poly83}
\end{figure}
\newline For the sixth moment the contributing cycles from (\ref{eq:partitions8}) are illustrated in Figs \ref{fig:poly6} and \ref{fig:poly62}. Finally the 14 contributing cycles for the $8$-th moment are illustrated in Figs \ref{fig:poly8} -- \ref{fig:poly83}. These illustrations prompt the following conclusion; the $2n$-th moment of the level density of the GUE for the canonical case $k=m$ is equal to the number of was of correctly partitioning a $2n$-sided poygon, where the conditions on the partitions are (i) no partitions intersect, (ii) no two isolated vertices neighbour eachother and (iii) any two vertices which can be connected by a non-intersecting edge are connected as such. It turns out this is an old problem and has been studied before by the mathematician Germain Kreweras in the early 1970's\cite{kreweras}. In the next section the problem of counting the number of correct partitians of a $2n$-sided polygon will be translated into another old problem; Dyck Words. In this way it will be shown that the $2n$-th moment of the GUE with $k=m$ is given by the sequence of Catalan Numbers $1, 2, 5, 14, 42, 132, 429, 1430, 4862, 16796, \ldots$ defined by
\begin{align}
\beta_{2n} = \frac{1}{n+1}{{2n}\choose n}.
\end{align}
\subsection{Dyck Words}
Each instance of the sum (\ref{eq:diag101}) gives a cycle and each cycle can be associated with the partitioning of a $2n$-sided polygon. These in turn can be translated into Dyck Words. Equations (\ref{eq:partitions31}, \ref{eq:partitions32}) give the two cycles of the fourth moment as
\begin{align}\label{eq:dw1}
(13)(2)(4)\notag\\
(24)(1)(3)
\end{align}
These are translated as follows
\begin{enumerate}[(i)]
\item Write the sequence $123..2n$ in order without brackets
\item A number with a left bracket in the partition notation is also given a left bracket
\item A number with a right bracket in the partition notation is also given a right bracket
\item Remove all numbers, leaving only sequences of opening and closing brackets
\item Replace each opening bracket with a $X$ and each closing bracket with a $Y$.
\end{enumerate}
The result is a Dyck Word. As an example take the cycle $(13)(2)(4)$. For this $2n=4$ cycle the ordered sequence $1234$ firstly becomes $(1(2)3)(4)$ because the partition $(13)(2)(4)$ has a left bracket for the integers $1, 2$ and $4$ and a right bracket for the integers $2, 3$ and $4$, as does the ordered sequence of numbers and brackets $(1(2)3)(4)$. This in turn becomes $(())()$ by removing all numbers, and finally by (iv) the result is a Dyck word of length $2n+2=6$
\begin{align}\label{eq:dw2}
(13)(2)(4)\equiv XXYYXY
\end{align}
by replacing each $($ with $X$ and each $)$ with $Y$. Similarly the cycle $(24)(1)(3)$ becomes $(1)(2(3)4)$ becomes $()(())$ giving the 6 letter Dyck word
\begin{align}\label{eq:dw3}
(24)(1)(3)\equiv XYXXYY.
\end{align}
It should be clear that this translation can be performed for every cycle (\ref{eq:diag101}) determining the moments of the level density. However the converse is not true. Not every Dyck word of length $2n$ refers to a permitted cycle of length $2n+2$. The Dyck words of length $6$ for example are given by
\begin{align}\label{eq:dw4}
XXXYYY,\;\;\;\;
XXYXYY,\;\;\;\;
XYXYXY ,\;\;\;\;
XXYYXY,\;\;\;\;
XYXXYY
\end{align}
whereas the fourth moment of the level density results in the cycles (13)(2)(4) and (24)(1)(3) (see \ref{eq:partitions31}, \ref{eq:partitions32}) corresponding only to the last two Dyck words of (\ref{eq:dw4}).
Hence the amount of information contained in a permitted partition of $2n$ elements is not equal to the information represented by a $2n+2$ Dyck word. This can be appreciated by observing that the cycles (\ref{eq:partitions31}-- \ref{eq:partitions32}), (\ref{eq:partitions8}) and (\ref{eq:partitions13}--\ref{eq:partitions15}) with polygon representations given in their respective diagrams Figs \ref{fig:poly4} -- \ref{fig:poly83} are all completely determined by the cycles containing all the odd (or equivalently even) integers of the sequence. In other words, given the restrictions of section (\ref{sec:polygons}) each diagram can be reproduced knowing only the cycles involving odd integers, or conversely knowing only the cycles with even integers. For the fourth moment the relevant cycles (\ref{eq:partitions31}-- \ref{eq:partitions32}) become
\begin{align}\label{eq:dw5}
(13)(2)(4)\equiv (13)\notag\\
(24)(1)(3)\equiv (1)(3).
\end{align}
In this way each polygon for the $2n$-th moment is associated to a partition of $n$ odd numbers as illustrated in (\ref{eq:dw5}). Because these sequences consist of odd numbers it helps to make the identification $1\to1, 3\to 2, 5\to 3, \ldots, 2n-1 \to n$ so that (\ref{eq:dw5}) ultimately is mapped to
\begin{align}\label{eq:dw6}
(13)(2)(4)\equiv (13)&\equiv(12)\notag\\
(24)(1)(3)\equiv (1)(3)&\equiv(1)(2).
\end{align}
where the information given by the partitions on the l.h.s of the relation is equivalent to the information contained on the r.h.s and both can be used to reproduce a contributing partition for a $2n$-sided polygon. The cycles for the sixth moment can similarly be rewritten as
\begin{align}\label{eq:dw7}
 (135)(2)(4)(6)\equiv(135)&\equiv(123)\notag\\
(13)(46)(2)(5)\equiv(13)(5)&\equiv(12)(3)\notag\\
 (15)(24)(3)(6)\equiv(15)(3)&\equiv(13)(2)\notag\\
 (246)(1)(3)(5)\equiv(1)(3)(5)&\equiv(1)(2)(3)\notag\\
(26)(35)(1)(4)\equiv(35)(1)&\equiv(23)(1).
\end{align}
It can be seen that each of the permitted cycles contributing to the leading order term of the $2n$-th moment is equivalently represented as a non-crossing partition of $n$ integers, so that the number of cycles which give the value of the $2n$-th moment is equal to the number of non-crossing partitions of the set $\{1, 2, \ldots, n\}$. This implies that summation is identical to the summation of \emph{Dyck words} of length $2n$, not $2n+2$. This is a very famous, beautiful and old result which has long been solved. As illustrated above, it has numerous forms including the enumeration of non-crossing partitions of $\{1, 2, \ldots , n\}$, as well as the partitioning of polygons in various ways.
\subsubsection{Catalan Numbers}
Every Dyck word of length $2n$ can be represented as monotonic paths drawn in a $n \times n$ grid.
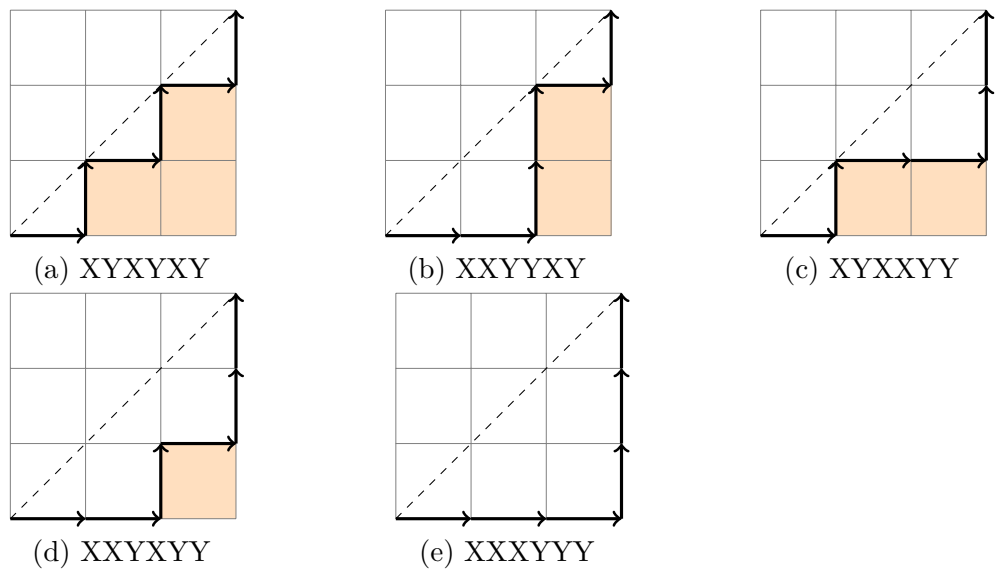
\begin{figure}[htp]\label{fig:dyck.paths}
\begin{subfigure}{.33\textwidth}
  \centering
 \begin{tikzpicture}
\fill[orange!25] (1,0) rectangle (2,1);
\fill[orange!25] (2,0) rectangle (3,1);
\fill[orange!25] (2,1) rectangle (3,2);
\draw[step=1cm,gray,very thin] (0,0) grid (3,3);
\draw[dashed] (0,0) -- +(3,3);
\draw[very thick,->] (0,0) -- (1,0);
\draw[very thick,->] (1,0) -- (1,1);
\draw[very thick,->] (1,1) -- (2,1);
\draw[very thick,->] (2,1) -- (2,2);
\draw[very thick,->] (2,2) -- (3,2);
\draw[very thick,->] (3,2) -- (3,3);
\end{tikzpicture}
  \caption{XYXYXY}
  \label{fig:dp1}
\end{subfigure}%
\begin{subfigure}{.33\textwidth}
  \centering
  \begin{tikzpicture}
\fill[orange!25] (2,0) rectangle (3,1);
\fill[orange!25] (2,1) rectangle (3,2);
\draw[step=1cm,gray,very thin] (0,0) grid (3,3);
\draw[dashed] (0,0) -- +(3,3);
\draw[very thick,->] (0,0) -- (1,0);
\draw[very thick,->] (1,0) -- (2,0);
\draw[very thick,->] (2,0) -- (2,1);
\draw[very thick,->] (2,1) -- (2,2);
\draw[very thick,->] (2,2) -- (3,2);
\draw[very thick,->] (3,2) -- (3,3);
\end{tikzpicture} 
  \caption{XXYYXY}
  \label{fig:dp2}
\end{subfigure}%
\begin{subfigure}{.33\textwidth}
  \centering
  \begin{tikzpicture}
\fill[orange!25] (1,0) rectangle (2,1);
\fill[orange!25] (2,0) rectangle (3,1);
\draw[step=1cm,gray,very thin] (0,0) grid (3,3);
\draw[dashed] (0,0) -- +(3,3);
\draw[very thick,->] (0,0) -- (1,0);
\draw[very thick,->] (1,0) -- (1,1);
\draw[very thick,->] (1,1) -- (2,1);
\draw[very thick,->] (2,1) -- (3,1);
\draw[very thick,->] (3,1) -- (3,2);
\draw[very thick,->] (3,2) -- (3,3);
\end{tikzpicture}
  \caption{XYXXYY}
  \label{fig:dp3}
\end{subfigure}\\
\begin{subfigure}{.33\textwidth}
  \centering
 \begin{tikzpicture}
\fill[orange!25] (2,0) rectangle (3,1);
\draw[step=1cm,gray,very thin] (0,0) grid (3,3);
\draw[dashed] (0,0) -- +(3,3);
\draw[very thick,->] (0,0) -- (1,0);
\draw[very thick,->] (1,0) -- (2,0);
\draw[very thick,->] (2,0) -- (2,1);
\draw[very thick,->] (2,1) -- (3,1);
\draw[very thick,->] (3,1) -- (3,2);
\draw[very thick,->] (3,2) -- (3,3);
\end{tikzpicture}
  \caption{XXYXYY}
  \label{fig:dp4}
\end{subfigure}
\begin{subfigure}{.33\textwidth}
  \centering
  \begin{tikzpicture}
\draw[step=1cm,gray,very thin] (0,0) grid (3,3);
\draw[dashed] (0,0) -- +(3,3);
\draw[very thick,->] (0,0) -- (1,0);
\draw[very thick,->] (1,0) -- (2,0);
\draw[very thick,->] (2,0) -- (3,0);
\draw[very thick,->] (3,0) -- (3,1);
\draw[very thick,->] (3,1) -- (3,2);
\draw[very thick,->] (3,2) -- (3,3);
\end{tikzpicture}
  \caption{XXXYYY}
  \label{fig:dp5}
\end{subfigure}%
\caption{Plot of the Dyck Paths.}
\end{figure}
The number of Dyck words of length $2n$ is given by the $n$'th Catalan number
\begin{equation}\label{eq:dw8}
C_n = {\frac{1}{n+1}} {{2n}\choose{n}}
\end{equation}
which is equivalent to saying that the level density of (\ref{eq:egue01}) is a semi-circle, since these are the moments of a semi-circular probability density function. In other words
\begin{equation}\label{eq:dw9}
\frac{2}{\pi r^2} \int_{-r}^{r} x^{2n} \sqrt{r^2 - x^2}~dx = C_n.
\end{equation}
We will briefly look at one way of showing that the number of Dyck words of length $2n$ is indeed given by \ref{eq:dw8}. There are many cute proofs to this and the one presented next is to illustrate the fact that a diagrammatic method can swiftly yield the semi-circle law.\\[5mm]
\noindent{\bf{Proof}}\\[0.5mm]
The proof begins with some monotonic sequence involving $n$ vertical steps and $n$ horizontal steps which {\emph{do}} indeed cross the horizontal line, and therefore  do not constitute an acceptable Dyck word. Identifying the first node lying above the diagonal, and drawing a line through this point which runs parrallel to the diagonal gives the picture illustrated in fig {\ref{fig:dpp1}} where all steps taken after this point are coloured in red. Defining the number of horizontal steps taken up to this point as $k$, it follows that the number of vertical steps taken is $k+1$, since there is one arrow peaking up above the diagonal.
\begin{figure}[htp]\label{fig:dyck.paths.proof}
\begin{subfigure}{.5\textwidth}
  \centering
\begin{tikzpicture}
\draw[step=1cm,gray,very thin] (0,0) grid (6,6);
\draw[dashed] (0,0) -- +(6,6);
\draw[red!80, very thick,->] (2,3) -- (2,4);
\draw[red!80, very thick,->] (2,4) -- (3,4);
\draw[red!80, very thick,->] (3,4) -- (3,5);
\draw[red!80, very thick,->] (3,5) -- (4,5);
\draw[red!80, very thick,->] (4,5) -- (5,5);
\draw[red!80, very thick,->] (5,5) -- (5,6);
\draw[red!80, very thick,->] (5,6) -- (6,6);
\fill[gray!50](2,3) circle (.1cm) ;
\draw[gray, dashed] (0,1) -- +(5,5);
\draw[very thick,->] (0,0) -- (1,0);
\draw[very thick,->] (1,0) -- (1,1);
\draw[very thick,->] (1,1) -- (2,1);
\draw[very thick,->] (2,1) -- (2,2);
\draw[very thick,->] (2,2) -- (2,3);
\end{tikzpicture} 
  \caption{A non-Dyck word on a \normalsize{$n\times n$} grid.}
  \label{fig:dpp1}
\end{subfigure}%
\begin{subfigure}{.5\textwidth}
  \centering
\begin{tikzpicture}
\draw[step=1cm,gray,very thin] (0,0) grid (5,7);
\draw[dashed] (0,0) -- +(5,5);
\draw[red!80, very thick,->] (2,3) -- (3,3);
\draw[red!80, very thick,->] (3,3) -- (3,4);
\draw[red!80, very thick,->] (3,4) -- (4,4);
\draw[red!80, very thick,->] (4,4) -- (4,5);
\draw[red!80, very thick,->] (4,5) -- (4,6);
\draw[red!80, very thick,->] (4,6) -- (5,6);
\draw[red!80, very thick,->] (5,6) -- (5,7);
\fill[gray!50](2,3) circle (.1cm) ;
\draw[gray, dashed] (0,1) -- +(5,5);
\draw[very thick,->] (0,0) -- (1,0);
\draw[very thick,->] (1,0) -- (1,1);
\draw[very thick,->] (1,1) -- (2,1);
\draw[very thick,->] (2,1) -- (2,2);
\draw[very thick,->] (2,2) -- (2,3);
\end{tikzpicture}
  \caption{Monotonic path on \normalsize{$(n-1)\times(n+1)$} grid.}
  \label{fig:dpp2}
\end{subfigure}%
\caption[Dyck words represented on a grid]{Illustration of the transformation showing that every non-Dyck word on a $n\times n$ grid can be mapped to a monotonic path on a $(n-1)\times (n+1)$ grid.}

\hrulefill
\end{figure}
Now reflect the remaining $n-k$ horizontal steps, and the remaining $n-k-1$ vertical steps in the line drawn parrallel to the diagonal passing through the first node above it. This yields the diagram shown in fig \ref{fig:dpp2}. The flipping step has turned $n-k$ horizontal steps into vertical steps and $n-k-1$ vertical steps into horizontal steps. Hence the resulting picture is a monotonic path on a $(n-k-1 + k) \times (n-k+k+1)$ grid. In other words, any monotonic path from the initial $n\times n$ grid which is not a Dyck word is a monotonic path on a $(n-1)\times(n+1)$ grid. The number of Dyck words of length $2n$ is then given by
\begin{equation}\label{eq:dw10}
{{2n}\choose {n}} -  {{2n}\choose{n+1}} = \frac{1}{n+1} {{2n}\choose{n}}
\end{equation} 
since we first find all ways of making a monotonic path from $n$ vertical steps and $n$ horizontal steps, and then remove all monotonic paths making $(n-1)$ horizontal steps and $(n+1)$ vertical steps. \qed

With relative simplicity it has been shown how to calculate a random matrix theory result for the $k=m$ case. But what about when $k<m$? In later chapters it will be shown how diagrams can once again be utilised in calculations of the moments. Whereas for $k<m$ these diagrams will appear and function differently to what has been shown above, for $k=m$ they become equivalent to the representation of cycles on polygons as shown in section \ref{sec:polygons}.
\section[Supersymmetry]{Supersymmetry}
In this section it will be seen how an existing technique involving grassmannian variables and an innovation by \cite{weid} can be used to show that the level density of the embedded (a.k.a unified) GUE ensemble is a semi-circle, as may be expected. However, this result will only cover the domain $m<2k$ with $l\to\infty$. In the context of unified RMT the result covers the volume of the phase space $\{k>{\frac{m}{2}}, m, \infty\}$. In the next chapter a new method will be introduced to calculate statistics of the unified RMT phase space for values of $2k < m$ as well. It will be seen that in this portion of the volume of the phase space the analogue of Wigner's semi-circle law is \emph{not} a semi-circle!
\subsection{Anti-commutative Grassmann variables}
The proof of Wigner's semi-circle law for $k$-body potentials begins using the same syperanalysis as used for the $m$-body case. A good account of this can be found in \cite{haake, fyodorov}. One begins with definitions which describe the form and behaviour of anti-commutative variables, which are non-numerical ``\textit{things}'' with the following algebraic rules attached
\begin{equation}\label{eq:gr1}\xi_p \xi_q = -\xi_q \xi_p,~p, q = 1 \ldots d \end{equation}
which implies, among other things, that $\xi_p^2 = 0$. Defining $\xi_p^{*} := (\xi_p)^{*}$ to be the complex conjugate of $\xi_p$, where $\xi_p^{*}$ is another grassmannian variable which therefore must also obey (\ref{eq:gr1}), and additionally enforcing the rules
\begin{equation}\label{eq:gr2} (\xi_p \xi_q \xi_r)^*  = \xi_p^{*} \xi_q^{*} \xi_r^{*}\end{equation}
and
\begin{equation}\label{eq:gr3} \xi_p^{**} = -\xi_p\end{equation}
it follows that $(\xi_p^{*} \xi_p)^{*} = \xi_p^{**} \xi_p^{*} = \xi_p^{*} \xi_p$. In other words, what has been called ``complex conjugation'' leaves the grassmannian $\xi_p^{*} \xi_p$ unchanged. In this sense it can be thought of similarly to the length of a standard complex variable. Finding a complex analogy for every grassmanian property, however, is not the goal. They are ultimately very different, distinct objects. Another property of grassmannians which illustrates this well is that they do \textit{not} have an inverse. Namely, assuming that $\xi_p^{-1}$ exists gives $\xi_p^{-1}\xi_p\xi_p^{-1} = - \xi_p^{-1}(\xi_p^{-1}\xi_p) = -\xi_p^{-1}$ and likewise $\xi_p^{-1}\xi_p\xi_p^{-1} = (\xi_p^{-1}\xi_p)\xi_p^{-1} = \xi_p^{-1}$ so that $\xi_p^{-1} = -\xi_p^{-1}$. Namely, it is not possible to abide by the original assumptions, (\ref{eq:gr1}), (\ref{eq:gr2}) and (\ref{eq:gr3}) without also contradicting the existence of an inverse. Analogously to the way in which a real or complex $\mathbf{C^{\infty}}$ function can be expressed in the form of a Taylor series $f(x) = a + b x + c x^2 + \ldots$ any function of purely grassmannian variables can be expressed in the form
\begin{equation}\label{eq:gr4}F(\xi_1, \ldots, \xi_N) =  \sum_{m_i = 0,1} f(m_1, \ldots, m_N) \xi_1^{m_1} \ldots \xi_N^{m_N}\end{equation}
since $\xi_p^n = 0$ for any $n \geq 2$. Having defined the form of grassmanian functions it is now possible to describe a formal meaning for differentiation and integration. That is, since grassmannian objects are different to numerical objects it is necessary to redefine what it means to differentiate and integrate. Moreover, there is no constraint to make these definitions in any way analogous to the numerical case, but rather the definitions are only required to be commensurate with (\ref{eq:gr1}), (\ref{eq:gr2}) and (\ref{eq:gr3}). The definitions are
\begin{equation}\label{eq:gr5} \int d\xi_i \xi = 1\end{equation}
and
\begin{equation}\label{eq:gr6} \int d \xi_i = 0 \end{equation}
where the differential $d \xi_i$ obeys the same commutation relations as a normal grassmannian variable. Grassmannian differentiation is defined analogously to real differentiation, namely ${\partial}/{\partial \xi_i} (a \xi_i)  = a$ for any commuting variable $a$. Moreover, the operator ${\partial}/{\partial \xi_i}$ obeys the same commutation relations as grassmannian variables so that
\begin{equation}\label{eq:gr7} \frac{\partial}{\partial \xi_i}\frac{\partial}{\partial \xi_j} = - \frac{\partial}{\partial \xi_j}\frac{\partial}{\partial \xi_i}\end{equation}
which yields ${{\partial}^2}/{\partial \xi_i^2} = 0$ and ${\partial}/{\partial \xi_i} (\xi_j) = - (\xi_j) {\partial}/{\partial \xi_i}$ for $i \ne j$. It follows immediately that
\begin{equation}\label{eq:gr8}\frac{\partial}{\partial \xi_N} \ldots \frac{\partial}{\partial \xi_1} F = f(1,1, \ldots, 1) \end{equation}
and likewise
\begin{equation}\label{eq:gr9} \int d\xi_n \ldots d\xi_1 F = f(1,1, \ldots, 1).\end{equation}
That is, added to the already remarkable properties of anti-commuting variables is the fact that the operation of differentiation is identical to that of integration.
\subsection{Determinants}
From this comes
\begin{align}\label{eq:gr10}D &= \int \left ( {\prod_j^{1 \ldots N} d\xi_j^{*} d\xi_j} \right ) exp( - \sum_{ik}^{1 \ldots N} \xi_i^{*} A_{ik} \xi_k) \notag \\
&=  \int \left ( {\prod_j^{1 \ldots N} d\xi_j^{*} d\xi_j} \right ) exp( - \xi^{*} A \xi)\notag\\ 
&= \int \left ( {\prod_j^{1 \ldots N} d\xi_j^{*} d\xi_j} \right ) \frac{(-1)^N}{N!}(\xi^{*} A \xi)^N \notag\\
&= \int \left ( {\prod_j^{1 \ldots N} d\xi_j d\xi_j^{*}} \right ) \frac{1}{N!}(\xi^{*} A \xi)^N
\end{align}
since by (\ref{eq:gr6}) the $N$'th term is the only one contributing a non-zero term to the integral. There are $(N!)^2$ terms in the product of the sum. $N!$ for the number of ways a non-zero term can be chosen, times $N!$ for the number of different orderings for such a term. Hence 
\begin{align}\label{eq:gr11}D = \int \left ( {\prod_j^{1 \ldots N} d\xi_j^{*} d\xi_j} \right ) \frac{1}{N!} &\sum_{\bi\bj} \xi_{i_1}^{*}\xi_{j_1} \xi_{i_2}^{*}\xi_{j_2} \ldots  \xi_{i_N}^{*}\xi_{j_N} ~ A_{i_1 j_1}A_{i_2 j_2} \ldots A_{i_N j_N}\end{align}
where ${\bi}$ and ${\bj}$ are permutations on the indices. Since each permutation on ${\bi}$ gives the same contribution it follows that
\begin{align}\label{eq:gr12}D &= \int \left ( {\prod_j^{1 \ldots N} d\xi_j^{*} d\xi_j} \right ) \sum_{P} \xi_{i_1}^{*}\xi_{P_1} \xi_{i_2}^{*}\xi_{P_2} \ldots  \xi_{i_n}^{*}\xi_{P_n} ~ A_{i_1 P_1}A_{i_2 P_2} \ldots A_{i_N P_N}\notag \\
& = \sum_{P} (-1)^P A_{i_1 P_1}A_{i_2 P_2} \ldots A_{i_N P_N} \notag \\ 
& = \mathrm{det}~A.\end{align}
So using grassmannian variables the determinant of a matrix can be expressed as the integral of an exponential function of anti-commuting variables. This will prove important later on. Another important is the observation that
\begin{align}\label{eq:gr13}\int \frac{1}{\pi^N} \left ( {\prod_{k=1}^{N}d^2 z_k} \right ) exp \left ( {- \sum_{ij} z_i^{*}A_{ij} z_j} \right ) = \int \frac{1}{\pi^N} \left ( {\prod_{k=1}^{N}d^2 Q_k} \right ) exp \left ( {- \sum_{k} \lambda_k Q_k^{*} Q_k } \right ) \end{align}
where $d^2 z = dRe\{z\}dIm\{z\}$ and it has been assumed that $A$ is hermitian, and making the variable transformation $Q_k = (Uz)_k$ for $A = U^{\dag}\Lambda U$. The determinant of the Jacobian relating the differentials $d^2 [Z]$ and $d^2 [Q]$ is subsequently unity. Hence
\begin{equation}\label{eq:gr14}\int \frac{1}{\pi^N} \left ( {\prod_{k=1}^{N}d^2 z_k} \right )  exp \left ( {- \sum_{ij} z_i^{*}A_{ij} z_j}\right )  = \frac{1}{\prod_k \lambda_k} = \frac{1}{\mathrm{det}~A}.\end{equation}
\subsection{The Generating Function}
It will now be shown how the Green's function can be expressed in terms of a generating function. The grassmannian variables and the resulting expressions for determinants found in the preceding section will then be used to express the generating functions in terms of gaussian integrals. This process of changing the form of the determinant into a gaussian integral using functions of grassmannian and complex variables is called \textit{the method of supersymmetry}. The level density of the system is $\rho(E) = \frac{1}{\pi} Im ~G(E^{-})$ where the Green's function $G(E) := \frac{1}{N} Tr \left ( {\frac{1}{E - H}} \right )$. Introducing the generating function
\begin{equation}\label{eq:gr15}Z(E, j) := \frac{\mathrm{det}~(E - H)}{\mathrm{det}~(E - H - j)} = \frac{\mathrm{det}~i(E - H)}{\mathrm{det}~i(E - H - j)} \end{equation}
and observing that
\begin{align}\label{eq:gr16}\frac{\partial}{\partial j} & \left [ {\frac{1}{\mathrm{det}~(E - H - j)}} \right ]_{j=0} = \frac{\partial}{\partial j} exp \left [ - Tr (ln(E - H - j)) \right ]_{j=0} \notag \\
&  = \frac{\partial}{\partial j} exp \left [ - Tr (ln(E - H)) + j  Tr (E - H)^{-1} + \mathcal{O}(j^2)\right ]_{j=0} \notag \\
& = \frac{1}{\mathrm{det}~(E - H)} ~ Tr \left [ {\frac{1}{E - H}} \right ]\end{align}
gives the Green's function as
\begin{equation}\label{eq:gr17}G(E) = \left. {\frac{1}{N} \frac{\partial}{\partial j} Z(E, j)} \right |_{j=0}.\end{equation}
Given that $\rho(E_1, E_2) = \rho(E_1)\rho(E_2)$ it is now possible to write the two point generating function as
\begin{align}\label{eq:gr18}Z(E_1^-, E_2^+, E_3^-, E_4^+) &= \frac{\mathrm{det}(E_3^- - H)\mathrm{det}(E_4^+ - H)}{\mathrm{det}(E_1^- - H)\mathrm{det}(E_2^+ - H)} \notag\\
&=  (-1)^N \frac{\mathrm{det}[i(E_3^- - H)]\mathrm{det}[i(E_4^+ - H)]}{\mathrm{det}[i(E_1^- - H)]\mathrm{det}[-i(E_2^+ - H)]} \end{align}
where using the notation of (\ref{eq:gr15}) gives $E_1^- = (E_3 - j) - i\epsilon_1$, $E_2^+ = (E_4 - k)+ i\epsilon_2$,  $E_3^- = E_3 - i\epsilon_1$ and $E_4^+ = E_4 + i\epsilon_2$, implicitly with $\epsilon_{1,2} \rightarrow 0$ so that
\begin{equation}\label{eq:gr19}G(E_1^-)G(E_2^+) = \frac{1}{N^2} \left. {\frac{\partial^2 Z(E_1^-, E_2^+, E_3^-, E_4^+)}{\partial E_1\partial E_2}}\right|_{E_1=E_3, E_2=E_4}\end{equation}
Plugging (\ref{eq:gr12}) and (\ref{eq:gr14}) into the above immediately yields
\begin{align}\label{eq:gr20}Z &=  {\left ( \frac{-1}{\pi^2} \right )}^N\int d[z_1, \eta_1^*, \eta_1] d[z_2, \eta_2^*, \eta_2]\notag\\
&~~\times \mathrm{exp}[i(z_1^{\dag}(H - E_1^-)z_1 + \eta_1^{\dag}(H - E_3^-)\eta_1)]\notag\\
&~~~~~\times \mathrm{exp}[i( - z_2^{\dag}(H - E_2^+)z_2 + \eta_2^{\dag}(H - E_4^+)\eta_2)]\end{align}
where $z_1, z_2$ are $N$-component complex variables and $\eta_1, \eta_2$ are $N$-component grassmannians. Now for brevity define
\begin{equation}\label{eq:gr21}\Phi = \left( \begin{array}{c}
z_1 \\
z_2 \\
\eta_1 \\
\eta_2\end{array} \right) ~~~\mathrm{and}~~~ L = \left( \begin{array}{cccc}
1 & 0 & 0 & 0\\
0 & -1 & 0 & 0 \\
0 & 0 & 1 & 0\\
0 & 0 & 0 & 1 \end{array} \right)\end{equation}
so that $d\Phi^* d\Phi = {\left ( \frac{1}{\pi} \right ) }^{2N}d[z_1, \eta_1^*, \eta_1] d[z_2, \eta_2^*, \eta_2]$, the matrix $L$ is $4N\times4N$ with $L_{\alpha} := L_{\alpha \alpha}$ and the generating function becomes
\begin{equation}\label{eq:gr22}Z = (-1)^N \int d\Phi^* d\Phi~\mathrm{exp}\left ( {i \sum_{\alpha}^4 \Phi_{\alpha}^{\dag} L_{\alpha}[H - E_{\alpha}]\Phi_{\alpha}} \right ).\end{equation}
Hence, for the ensemble average the aim is to calculate
\begin{align}\label{eq:gr23}W &:= \overline{\mathrm{exp}(i\sum\Phi_{\alpha}L_{\alpha}H\Phi_{\alpha})}\notag\\
& = \int d[v]~ \mathrm{exp}\left ( i\sum\Phi_{\alpha}L_{\alpha}H\Phi_{\alpha} - \sum_{{\bj \bi}} \frac{v_{{\bj \bi}}v_{{\bi \bj}}}{2v_o^2} \right )\notag \\
& = \int d[v]~ \mathrm{exp} \left ( -\frac{1}{2v_o^2}\sum_{{\bj \bi}}\left[ |v_{{\bj \bi}}|^2 \right.\right.  \left. \left.- 2iv_o^2\sum \Phi_{\alpha\mu}^{\dag}L_{\alpha}\langle\mu|a_{{\bj}}^{\dag}a_{{\bi}}|\nu\rangle\Phi_{\alpha\nu}v_{{\bj \bi}}\right ]\right).\end{align}
Making the change of variables $v_{{\bj \bi}} \rightarrow v_{{\bj \bi}} - iv_o^2\sum  \Phi_{\alpha\mu}^{\dag}L_{\alpha}\langle\mu|a_{{\bi}}^{\dag}a_{{\bj}}|\nu\rangle\Phi_{\alpha\nu}$ (notice the change in order of the creation and annihilation operators in the second term) and integrating over $v$ gives $W$ as
\begin{equation}\label{eq:gr24}\mathrm{exp}\left(-\frac{v_o^4}{2}\sum  \Phi_{\alpha\mu}^{\dag} \Phi_{\alpha\nu} \Phi_{\beta\rho}^{\dag} \Phi_{\beta\sigma}L_{\alpha}L_{\beta}\langle\mu|a_{{\bi}}^{\dag}a_{{\bj}}|\nu\rangle  \langle\rho|a_{{\bj}}^{\dag}a_{{\bi}}|\sigma\rangle  \right).\end{equation}
Using the method of expansion in the eigenvalues detailed in \cite{weid} one has $A_{\mu\nu\rho\sigma} := \overline{\langle\mu|V|\sigma\rangle\langle\rho|V|\nu\rangle} = \frac{v_o^2}{N}\sum_{sa}\Lambda^s C_{\mu\sigma}^{sa} C_{\rho\nu}^{sa}$. Substituting this into (\ref{eq:gr24}) and taking caution with the anti-commutative properties of the grassmannian components of $\Phi$ grants the following equality
\begin{equation}\label{eq:gr25}W = \mathrm{exp}\left ( -\frac{1}{2N} \sum_{\alpha\beta s a} L_{\alpha}L_{\beta}' Q_{\alpha\beta}Q_{\beta\alpha}\right )\end{equation}
where
\begin{equation}\label{eq:gr26} L' = \left( \begin{array}{cccc}
1 & 0 & 0 & 0\\
0 & -1 & 0 & 0 \\
0 & 0 & -1 & 0\\
0 & 0 & 0 & -1 \end{array} \right)\end{equation}
and $Q_{\alpha\beta} = \sum_{\mu\sigma}\Phi_{\alpha\mu}^{\dag} (v_o^2 \lambda^s C_{\mu\sigma}^{sa}) \Phi_{\beta\sigma}$ with $\lambda^s := \sqrt{\Lambda^s}$.
\subsection{Supertrace and Superdeterminant}
By writing out the 16 components of (\ref{eq:gr25}) explicitly it can be rewritten as
\begin{equation}\label{eq:gr27}W = \mathrm{exp} \left ( -\frac{1}{2N} \sum_{sa} \mathrm{\bf Str}~\tilde{Q}^2\right)\end{equation}
with $\tilde{Q} = L_{\alpha} Q_{\alpha\beta}$ and the supertrace {\bf Str} defined as
\begin{equation}
\mathrm{\bf Str}F = \mathrm{Tr}A - \mathrm{Tr}D
\end{equation}
where $F = \left( \begin{array}{cc}
A & B \\
C & D \end{array} \right)$ is a supermatrix, the matrices $A$ and $D$ consisting only of commuting variables (such as complex numbers), and the matrices $B$ and $C$ consisting only of grassmanian variables.
It is now possible to introduce an additional dummy variable by noting that up to a normalization factor
\begin{equation}\label{eq:gr28}\mathrm{exp}\left ( - \frac{1}{2N} \mathrm{\bf Str}~\tilde {Q}^2 \right ) = \int \mathrm{exp} \left (- \frac{N}{2}\mathrm{\bf Str}~\sigma^2 +i~\mathrm{\bf Str}~\sigma \tilde{Q} \right ) d[\sigma].\end{equation}
Then writing out the expression component-wise and rearranging it can be seen that
\begin{equation}\label{eq:gr29}\mathrm{\bf Str}~\sigma \tilde{Q} = \Phi^{\dag}L(s\sigma)^{T}(v_o^2\lambda^s C^{sa}I_4)\Phi\end{equation}
where $s=diag(I_2, -I_2)=s^T$, so that up to a normalisation factor the ensemble average of the generating function $\overline{Z}$ is given by
\begin{equation}\label{eq:gr30}\int \mathrm{exp} \left (- \frac{N}{2}\mathrm{\bf Str}~\sigma^2  + \Phi^{\dag} [ i L(s\sigma)^{T}(v_o^2\lambda^s C^{sa}I_4) - i LE]\Phi \right ) \end{equation}
to which the identity $\int d\Phi^{\dag}d\Phi ~\mathrm{exp} (- \Phi^{\dag}F\Phi) = (\mathrm{\bf{Sdet}}F)^{-1}$ can now be applied, with the superdeterminant {\bf Sdet} being defined by
\begin{equation}
\mathrm{\bf Sdet} = \frac{\mathrm{det}(D - CA^{-1} B)}{\mathrm{det}A}.
\end{equation}
Noting that this will be a $4N \times 4N$ superdeterminant (denoted in bold to distinguish it from the $4 \times 4$ superdeterminant) and again ignoring normalisation constants (\ref{eq:gr30}) can now be written as
\begin{equation}\label{eq:gr31}\int(\mathrm{\bf{Sdet}}[L(s\sigma)^{T}(v_o^2\lambda^s C^{sa}) - LE])^{-1} \mathrm{exp} \left (- \frac{N}{2}\mathrm{Str}~\sigma^2  \right ).\end{equation}
Using $\mathrm{ln\bf{Sdet}}~F = \mathrm{{\bf Str}{ln}}~F$ and factoring out $L$ in addition to noticing that
\begin{equation}\label{eq:gr32}\mathrm{exp}( \mathrm{{\bf Str}ln} L) =  \mathrm{exp}~\mathrm{\bf Str} \left( \begin{array}{cccc}
0 & 0 & 0 & 0\\
0 & i \pi & 0 & 0 \\
0 & 0 & 0 & 0\\
0 & 0 & 0 & 0 \end{array} \right) = \mathrm{exp}(iN\pi)\end{equation}
and ignoring the $\pm 1$ implied by (\ref{eq:gr32}) it can be concluded that $\overline{Z}$ without normalization can be written as
\begin{equation}\label{eq:gr33}\int \mathrm{exp} \left ( - \frac{N}{2}\mathrm{\bf Str}~\sigma^2  + \mathrm{{\bf Str}ln}\left (E - (s\sigma)^{T}(v_o^2\lambda^s C^{sa}) \right) \right ).\end{equation}
Finally, dropping the transpose due to the diagonal nature of the term $v_o^2\lambda^s C^{sa}$ in the $4 \times 4$ sense and reconciling the dimensions of the two supertraces via $\mathrm{{\bf Str}ln} = \mathrm{tr{\bf{Str}}ln}$ yields
\begin{equation}\label{eq:gr34}\overline{Z} \propto \int \mathrm{exp} \left ( - N [ \frac{1}{2}\mathrm{\bf Str}~\sigma^2  + \frac{1}{N}\mathrm{ tr {\bf{Str}}~ln} (\sigma s v_o^2\lambda^s C^{sa} - E )] \right ) .\end{equation}
\subsection{Saddle-point Approximation}
A saddle-point approximation will now be applied over the $\sigma$ variables. Minimising the argument of the exponential yields
\begin{equation}\label{eq:gr35}\sigma^{sa} = - \frac{1}{N}\mathrm{tr} \left ( \frac{v_o^2\lambda^s C^{sa} - E }{\chi - E} \right )\end{equation}
where $\chi = \sum_{sa}v_o^2 \sigma^{sa} \lambda^s C^{sa}$. Multiplying across by $v_o^2 \lambda^s C_{\mu\nu}^{sa}$ and summing over ${sa}$ as well as expressing the trace in component form gives
\begin{align}\label{eq:gr36}\chi_{\mu\nu} = \frac{v_o^4}{N}~\sum_{sa\rho\sigma}(E - \chi)_{\rho\sigma}^{-1}\Lambda^s C_{\rho\sigma}^{sa} C_{\mu\nu}^{sa}
= v_o^2 ~\sum_{\rho\sigma} (E - \chi)_{\rho\sigma}^{-1}\overline{\langle\sigma|V|\nu\rangle\langle\mu|V|\rho\rangle}.\end{align}
Assuming that $\chi$ is of the form $r\hat{I}$ for some constant $r$ and given that all the eigenvectors $C^{sa}$ are traceless save for $C^{0} = \delta_{\mu\nu}$ gives, self-consistent with the prior assumptions on $\chi$, that
\begin{equation}\label{eq:gr37}\chi = \frac{v_o^4 \Lambda^0}{E - \chi}\end{equation}
where it is shown in \cite{weid} that
\begin{equation}
\Lambda^{0}(k) = {{m}\choose{k}}{{l-m+k}\choose{k}}.
\end{equation}
Finally, solving this for $\chi$ yields
\begin{equation}\label{eq:gr38}\chi = v_o^2 \lambda^0 \tau_{\pm}^0\end{equation}
with $\tau_{\pm}(E) = \frac{E}{2v_o^2 \lambda^0} \pm i~\sqrt{1 - \left (\frac{E}{2v_o^2 \lambda^0} \right )^2}$. Plugging this into (\ref{eq:gr35}) implies $\sigma^0 = \tau$ and $\sigma^{sa} = 0$ for all $s \geq 1$. Construct the supermatrix $\sigma$ as
\begin{equation}
\sigma^0 = \mathrm{diag}(\tau_{+}(E_1) , \tau_{-}(E_2) , \tau_{+}(E_3), \tau_{-}(E_4))
\end{equation}
so that the supertrace for $\sigma^2$ doesn't yield zero and the advanced and retarded components of the generating function are matched. Then by setting $E_2 = E_4$ in (\ref{eq:gr34}) the two-point generating function reduces to a level density generating function $\overline{Z(E_1, E_3)}$ proportional to
\begin{align}\label{eq:gr39}\hspace{-.1cm}\int \mathrm{exp}\left ( -N\left [ \frac{\tau_{+}^2 (E_1)}{2} - \frac{\tau_{+}^2 (E_3)}{2}\right]\right)\mathrm{exp}\left(-N\left[\mathrm{ln}(\lambda^0 v_o^2 \tau_{+}(E_1) - E_1)  - \mathrm{ln}(\lambda^0 v_o^2 \tau_{+}(E_3) - E_3)\right ] \right ).\end{align}
Taking the derivative with respect to $E_1$ as per (\ref{eq:gr19}) and selecting the imaginary component of the resulting Green's function gives the embedded Wigner's semi-circle law for the $k$-body EGUE as
\begin{equation}\label{eq:gr40}\lim_{N \to \infty}\overline{\rho (E)} \propto \sqrt{\left(v_o^2 \lambda^0\right)^2 - \left(\frac{E}{2}\right)^2}.\end{equation}
Due to a technicality involving the identity $A_{\mu\nu\rho\sigma}  = \frac{v_o^2}{N}\sum_{sa}\Lambda^s C_{\mu\sigma}^{sa} C_{\rho\nu}^{sa}$ the saddle-point approximation converges to this result only for $2k > m$. In the next chapter a simpler method will be introduced, which avoids the grassmannian gymnastics illustrated above and can be used to find the moments of embedded RMT systems for all $k$.
\begin{center}.....\end{center}
\chapter[Many-Body RMT]{Many-Body RMT}\label{ch:MBRMTCH}
\section[RMT vs. \emph{Embedded} RMT]{RMT vs. \emph{Embedded} RMT}\label{sec:embedded.rmt}
Many-body random matrix theory is the application of random matrix concepts to the study of the $k$-body potentials (see section \ref{sec:second.quantization}). This is a \emph{superset} of random matrix theory with hamiltonian
\begin{equation}\label{eq:mb1}{H}_k = \sum_{{{1 \le j_1 \le \ldots \le j_k \le l} \atop {1 \le i_1 \le \ldots \le i_k \le l}}} {v}_{j_1 \ldots j_k;i_1 \ldots i_k} a_{j_1}^{\dag} \ldots a_{j_k}^{\dag} a_{i_k} \ldots a_{i_1}\end{equation}
and coefficients ${v}_{j_1 \ldots j_k;i_1 \ldots i_k}$ taking the value of a random variable. By determining the coefficients from a probability density function the static potential (\ref{eq:mb1}) generates an \emph{ensemble} of matrices which can be studied statistically, just as in the case of canonical RMT. As shown in \ref{sec:rmtsp} the canonical form of RMT coincides with just a single point $k=m$ in the phase space of the unified random matrix theory, also known as \emph{embedded RMT}. One of the purposes of unification is to study all of these random matrix theories as a group, rather than creating new methods and theorems for each particular case. In this sense embedded RMT treats all randomised $k$-body systems equally, allowing $k$ to take any value in the range $1\le k\le m$. While there is a vast literature for the statistics and theorems of canonical random matrix ensembles ($k=m$) including Wigner's semi-circle law described in chapter \ref{sec:WSCL}, little was known about the properties of unified RMT ensembles ($1\le k\le m$) including the analogue of Wigner's Law for these embedded systems until relatively recently. In fact it turns out that for some values of $k < m$ the analogue of Wigner's semi-circle law is not even semi-circular.

While some progress has previously been made towards calculating various statistics for embedded ensembles\cite{mon, weid}, the moments for the level density for these systems when $k \le \frac{m}{2}$ were unknown beyond the fourth moment until the method invented by the author was introduced. The current chapter will introduce this new method, called the method of \emph{particle diagrams}, and illustrate how it can be used to calculate the fourth, sixth and eighth moments of the level density for the eGUE ensemble. Recall from see section \ref{sec:second.quantization} that the eGUE ensemble represents the hamiltonian of non time-reversal invariant quantum systems of $m$ particles interacting under the force of a $k$-body potential. Interestingly, the moments calculated in this way point to a convergence in the statistical behaviours across a wide range of many-body hamiltonians of a similar form, albeit used independently to study the statistics of quantum spin chains, graphs and hypergraphs \cite{hess,schroeder,huw}.

In the context of embedded RMT the three symmetry groups introduced by Dyson for canonical RMT become the embedded GUE (eGUE), embedded GOE (eGOE) and embedded GSE (eGSE). For each of these classes the hamiltonian (\ref{eq:mb1}) obeys the same symmetry rules as in the canonical case even though the resulting statistics can be completely distinct due to the changing structure of $H$ itself. The only non-deterministic components of the potential are the random variables ${v}_{j_1 \ldots j_k;i_1 \ldots i_k}$. With the foresight that their symmetry properties will be required in calculations later, the next sections will investigate the second moments of $v_{\bj \bi}$ for the eGUE, eGOE and eGSE. This will also illustrate how the restrictions of symmetry from each of Dyson's three groups affect the symmetry properties of an embedded $k$-body hamiltonian.

\subsection{Second Moments of the Random Variables}
It was shown in section \ref{sec:second.quantization} that the hermitian potential describing the energy of a system of $m$ fermions interacting under the influence of a $k$-body potential is ${H}_k = \sum_{{{1 \le j_1 \le \ldots \le j_k \le l} \atop {1 \le i_1 \le \ldots \le i_k \le l}}} {v}_{j_1 \ldots j_k;i_1 \ldots i_k} a_{j_1}^{\dag} \ldots a_{j_k}^{\dag} a_{i_k} \ldots a_{i_1}$. The \emph{second moments} of the random variables ${v}_{j_1 \ldots j_k;i_1 \ldots i_k}=v_{\bj \bi}$ are defined as $\overline{v_{\bj \bi}v_{\bj^{'} \bi^{'}}}$. For the method proposed in this thesis for calculating the moments of the level density, the socalled  second moments form the first of several essential ingredients. The normalised $2n$-th moments of the level density are given by 
\begin{equation}\label{eq:smrv1}
\beta_{2n} = \frac{\frac{1}{N}\mathrm{tr}({\overline{H^{2n}_k}})}{\left(\frac{1}{N}\mathrm{tr}({\overline{H^2_k}})\right)^n}
\end{equation}
for which the numerator can be re-written as
\begin{equation}\label{eq:smrv2}
\mathrm{tr}({\overline{H^{2n}_k}}) = \overline{v_{\bj^{(1)}\bi^{(1)}} v_{\bj^{(2)}\bi^{(2)}} \ldots v_{\bj^{(2n)}\bi^{(2n)}}} \langle\mu|a_{\bj^{(1)}}^{\dag}  a_{\bi^{(1)}} a_{\bj^{(2)}}^{\dag}  a_{\bi^{(2)}} \ldots a_{\bj^{(2n)}}^{\dag}  a_{\bi^{(2n)}}|\mu\rangle.
\end{equation}
Even before seeing the forthcoming calculations for the moments of the eGUE it can be appreciated that in order to calculate (\ref{eq:smrv2}) it is necessary to also calculate the ensemble average
\begin{equation}\label{eq:smrv22}
\overline{v_{\bj^{(1)}\bi^{(1)}} v_{\bj^{(2)}\bi^{(2)}} \ldots v_{\bj^{(2n)}\bi^{(2n)}}}.
\end{equation}
It has been noted before in section \ref{sec:MAEA} that the second moment $\overline{v_{\bj \bi}v_{\bj^{'} \bi^{'}}}=0$ for uncorrelated $v_{\bj \bi}, v_{\bj^{'} \bi^{'}}$ whereas for correlated $v_{\bj \bi}, v_{\bj^{'} \bi^{'}}$ with $v_{\bj \bi} = v_{\bj^{'} \bi^{'}}^{*}$ the second moment is unity (by virtue of normalisation). This and the additional knowledge that in general for a gaussian random variable
\begin{equation}\label{eq:smrv3}
\overline{v^{2n}} = \sqrt{\frac{1}{2\pi}}\int v^{2n} e^{- {v^2}/{2}}dv = (2n-1)!!
\end{equation}
gives the ensemble average of the product of random variables (\ref{eq:smrv22}) in terms of the product of the average of all possible pairings of the random variables
\begin{equation}\label{eq:smrv4}
\overline{v_{\bj^{(1)}\bi^{(1)}} v_{\bj^{(2)}\bi^{(2)}} \ldots v_{\bj^{(2n)}\bi^{(2n)}}} = \sum_{\sigma} \prod_{x=1}^{2n}\overline{v_{\bj^{(x)}\bi^{(x)}}v_{\bj^{\sigma(x)}\bi^{\sigma(x)}}}.
\end{equation}
This result is used frequently in statistical mechanics and is sometimes referred to as \emph{Wick's Theorem} (even when it doesn't involve creation and annihilation operators). Since the first step for calculating the higher moments is to calculate these \emph{second moments} the next sections will explain how to calculate the quantity $\overline{v_{\bj \bi}v_{\bj^{'} \bi^{'}}}$ for the eGUE, eGOE and eGSE.

\subsection{Unitary Symmetry ($\beta=2$)}\label{sec:usym}
It was shown in section \ref{sec:SC} that the GUE ensemble refers to the set of random matrices $H_k$ obeying hermitian symmetry
\begin{equation}\label{eq:us1}
H_k = H_k^{\dag}
\end{equation}
which are the class of hamiltonians referring to time-reversal invariant fermionic quantum systems. Since by the Pauli exclusion principle many-body fermionic quantum states cannot contain repeated single-particle states (\ref{eq:mb1}) can be rewritten as the restricted sum
\begin{equation}\label{eq:us2}{H}_k = \sum_{{{1 \le j_1 < \ldots < j_k \le l} \atop {1 \le i_1 < \ldots < i_k \le l}}} {v}_{j_1 \ldots j_k;i_1 \ldots i_k} a_{j_1}^{\dag} \ldots a_{j_k}^{\dag} a_{i_k} \ldots a_{i_1}.\end{equation}
As seen before in section \ref{sec:diagrams} there is a useful vectorised abbreviation of this given by
\begin{equation}\label{eq:us3}
{H}_k = \sum_{\bj \bi}v_{\bj\bi} a_{\bj}^{\dag} a_{\bi}
\end{equation}
where $\bj = (j_1,\ldots,j_k)$ (similarly for $\bi$) and the corrolary $a_{\bj}^{\dag} = a_{j_1}^{\dag} \ldots a_{j_k}^{\dag}$. In a fermionic state space components of the sum where the vectors $\bj$ or $\bi$ contain repeated indices will anyway be zero, so restricting the summation is not actually necessary. To generate an ensemble of matrices from this potential each of the variables $v_{\bj\bi}$ in the sum is taken to be a complex valued random variable with a gaussian p.d.f. of mean zero and unit variance. The system described by the hamiltonian is assumed to have some number $l>m$ non-degenerate single particle levels so that the dimension of the space is $N = {l \choose m}$. A matrix symmetrised with the condition (\ref{eq:us1}) obeys
\begin{align}\label{eq:us4} \langle \mu|H_k|\nu \rangle = \langle \nu|H_k|\mu \rangle ^{*} \end{align}
for all quantum states $|\mu\rangle$ and $|\nu\rangle$. This implies that
\begin{align}\label{eq:us5}\sum_{\bj \bi} {v}_{\bj\bi} \langle\mu|a_{\bj}^{\dag} a_{\bi}|\nu\rangle &= \sum_{\bj \bi} {v}_{\bj\bi}^{*} \langle\nu|a_{\bj}^{\dag} a_{\bi}|\mu\rangle
\end{align}
and matching coefficients for each term in the sum gives
\begin{align}\label{eq:us6}v_{\bj \bi}^{*} &= v_{\bi \bj}.\end{align}
The coefficients $v_{\bj\bi}$ not related by (\ref{eq:us6}) are then \emph{uncorrelated} i.i.d. gaussian random variables with mean zero and variance $1$. This means that for uncorrelated ${v_{{\bj \bi}}}$ and ${v_{{\bj^{'} \bi^{'}}}}$
\begin{align}\label{eq:us7}\overline {v_{{\bj \bi}}v_{{\bj^{'} \bi^{'}}}} = \frac{1}{{2\pi}} \int v_{{\bj \bi}}v_{{\bj^{'} \bi^{'}}} \mathrm{exp}\left({-\frac{1}{2}v_{{\bj \bi}}^2}\right)\mathrm{exp}\left({-\frac{1}{2}v_{{\bj^{'} \bi^{'}}}^2}\right)dv_{{\bj \bi}}dv_{{\bj^{'}\bi^{'}}} = 0
\end{align}
since the gaussians are even functions, making the total argument of the integral odd with respect to the integration variables. Likewise if $v_{{\bj \bi}} = v_{{\bj^{'} \bi^{'}}}$ then $v_{{\bj \bi}}$ and $v_{{\bj^{'} \bi^{'}}}$ are complex and their product yields two even functions of opposite sign, again giving a zero average. However in the final scenario with
\begin{equation}\label{eq:us8}v_{{\bj \bi}} = v_{{\bj^{'} \bi^{'}}}^{*}\end{equation}
the second moment becomes
\begin{equation}\label{eq:us9}\overline {v_{{\bj \bi}}^{2}} = \frac{1}{\sqrt{2\pi}} \int v_{{\bj \bi}}^2 \mathrm{exp}\left({-\frac{1}{2}v_{{\bj \bi}}^2}\right)dv_{{\bj \bi}} = \end{equation}
from which it can be directly concluded that
\begin{align}\label{eq:us10}
\overline {v_{{\bj \bi}}v_{{\bj^{'} \bi^{'}}}} =  \delta_{{\bj \bi^{'}}}\delta_{{\bi \bj^{'}}}.
\end{align}
This is the same identity calculated in section \ref{sec:diagrams} for the case $k=m$. It will be an important identity in subsequent sections for calculations made in the space of unified random matrix theory. The reason it will play such a fundamental role in calculations is based on the idea behind Wick's Theorem; each term in certain summations of products will be expressed in terms of summations of products of pairs, for which (\ref{eq:us10}) will become an important aid. Analogous identities for the eGOE and eGSE will be derived next.

\subsection{Orthogonal Symmetry ($\beta=1$)}\label{sec:osym}
The embedded gaussian orthogonal ensemble (eGOE) refers to the set of potentials of bosonic time-reversal invariant quantum systems, as discussed in section \ref{sec:SC}. The hamiltonian
\begin{equation}\label{eq:os1}{H}_k = \sum_{{{1 \le \bj_1 \le \ldots \le \bj_k \le l} \atop {1 \le \bi_1 \le \ldots \le \bi_k \le l}}} {v}_{\bj_1 \ldots \bj_k \bi_1 \ldots \bi_k} a_{\bj_1}^{\dag} \ldots a_{\bj_k}^{\dag} a_{\bi_k} \ldots a_{\bi_1}\end{equation}
sums over $m$-body bosonic states which are permitted to have repeated single-particle indices. The number of bosonic $m$-particle states containing repeats of $z\le m$ unique single-particle states is ${l\choose z}{{m-1}\choose{z-1}}$. The dimension of the state space is then just the sum of all these
\begin{align}\label{eq:os2}N = \sum_{z=0}^{m}{l\choose z}{{m-1}\choose{z-1}} = {{l+m-1}\choose{m}}.\end{align}
As explained in section \ref{sec:SC} the symmetry properties imposed on $H_k$ imply that it consists only of real values, while also obeying the hermitian symmetry satisfied by matrices in the eGUE (see previous section). Hence for $\beta = 1$
\begin{align}\label{eq:os3}\sum v_{{\bj}{\bi}} \langle\mu|a_{{\bj}}^{\dag} a_{{\bi}}|\nu\rangle&= \sum \left( v_{{\bj}{\bi}} \langle\nu|a_{{\bj}}^{\dag} a_{{\bi}}|\mu\rangle\right)^{*}\notag\\
&= \sum v_{{\bj}{\bi}}^{*} \langle\mu|a_{{\bi}}^{\dag} a_{{\bj}}|\nu\rangle
= \sum v_{{\bj}{\bi}} \langle\nu|a_{{\bj}}^{\dag}a_{{\bi}}|\mu\rangle.
\end{align}
Matching coefficients gives
\begin{equation}\label{eq:os4}
v_{\bj \bi} = v_{\bj \bi}^{*} = v_{\bi\bj}
\end{equation}
so that with $v_{{\bj \bi}} = v_{{\bj^{'} \bi^{'}}}$ or $v_{{\bj \bi}} = v_{{\bj^{'} \bi^{'}}}^{*}$ the value of $\overline {v_{{\bj \bi}}v_{{\bj^{'} \bi^{'}}}}$ will be $1$ while for uncorrelated ${v_{{\bj \bi}},~v_{{\bj^{'} \bi^{'}}}}$ the value of $\overline {v_{{\bj \bi}}v_{{\bj^{'} \bi^{'}}}}$ will be zero. There will be no cancellations in the integral since these are real random variables, so the product of $v_{{\bj \bi}}$ with itself yields only a single term. There are no cross terms as there would be if $v_{{\bj \bi}}$ were complex valued as in the case of the eGUE. Hence for the eGOE
\begin{align}\label{eq:os5}
\overline {v_{{\bj \bi}}v_{{\bj^{'} \bi^{'}}}} =  \delta_{{\bj \bi^{'}}}\delta_{{\bi \bj^{'}}} + \delta_{{\bj \bj^{'}}}\delta_{{\bi \bi^{'}}}.
\end{align}
Analogously to the eGUE, this identity is important for calculations of the moments of the level density for $H_k$ taken from the eGOE because the resulting expressions involve summations of products of $v_{{\bj \bi}}$ which can be expressed as summations of products of \emph{pairs} of $v_{{\bj \bi}}$, for which (\ref{eq:os5}) then plays a central role.

\subsection{Symplectic Symmetry $(\beta=4)$}\label{sec:symsym}
For $\beta=4$ the hamiltonian must satisfy both the hermitian symmetry of (\ref{eq:us1}) and the symmetry implied by the fact that the potential takes the form of (\ref{eq:trev15}). To begin, define the pairs of entries in the upper(lower) diagonal plane of (\ref{eq:trev17}) related by complex conjugation as \textit{blue} pairs, and those related by complex conjugation times $-1$ as \textit{red} pairs. The additional symmetry enforced by (\ref{eq:trev15}) implies that for every blue matrix element there is a map to the element with row number and column number a distance $+1$ away, and related to it by complex conjugation, while for every red element there is a map to the element with row number a distance $+1$ away and column number a distance $-1$ away, and related to it by complex conjugation times $-1$. Assuming that it is possible to use the same map $\sigma$ to increase the column number for each block of the matrix it follows, in addition to (\ref{eq:us1}), that
\begin{align}\label{eq:sysy1}   \langle \mu|H_k|\nu \rangle = \langle \sigma_{\mu}|H_k|\sigma_{\nu} \rangle ^{*} \end{align}
for a blue matrix cell $(\mu, \nu)$ and
\begin{align}\label{eq:sysy2}   \langle \mu'|H_k|\sigma_{\nu'} \rangle = - \langle \sigma_{\mu^{'}}|H_k|\nu^{'} \rangle ^{*} \end{align}
for a red matrix cell $(\mu', \nu')$ where $\sigma_{\mu}$ is the action of the map $\sigma$ on the $m$-body state $|\mu\rangle$. Hence a blue cell corresponds to those given by (\ref{eq:trev172}) with
\begin{equation}\label{eq:sysy3}
\langle \mu|H_k|\nu \rangle = \langle T{\mu}|H_k|T{\nu} \rangle ^{*}
\end{equation}
for some $\mu, \nu$ while a red cell corresponding to (\ref{eq:trev173}) can be written as
\begin{equation}\label{eq:sysy4}
 \langle \mu' |H_k| T\nu' \rangle = - \langle T\mu'|H_k|{\nu'} \rangle ^{*}
\end{equation}
for some $\rho', \nu'$. Setting $\mu' = T\rho'$ and comparing equations (\ref{eq:sysy1}), (\ref{eq:sysy2}) with (\ref{eq:sysy3}) and (\ref{eq:sysy4}) gives 
\begin{equation}\label{eq:sysy5}
T|\mu\rangle = \sigma_{\mu}.
\end{equation}
Equations (\ref{eq:sysy1}) and (\ref{eq:sysy2}) imply that the random variables of the potential are subject to the condition
\begin{equation}\label{eq:sysy6}
{v}_{{\bj_k} {\bi_k}} = {v}_{\sigma_{\bj_k} \sigma_{\bi_k}}^{*}
\end{equation}
when the condition is yielded by applying (\ref{eq:sysy1}) and are subject to the condition
\begin{equation}\label{eq:sysy7}
{v}_{{\bj_k} {\sigma_{\bi_k}}} = - {v}_{\sigma_{\bj_k} {\bi_k}}^{*}
\end{equation}
when the condition is yielded by applying (\ref{eq:sysy2}). Hence for the eGSE the random variable ${v}_{{\bj_k} {\bi_k}}$ does not satisfy the same condition for all ${{\bj_k}, {\bi_k}}$. This distinctive feature of the eGSE indicates that it is necessary to count the number of sets of $\{{{\bj_k}, {\bi_k}}\}$ for which the two conditions  (\ref{eq:sysy6}) and  (\ref{eq:sysy7}) overlap in order to calculate the moments of the level density of the eGSE.  The overlapping of these conditions gives the ``\textit{symplectic zeros}'' which are the values of the random variable $\{{{\bj_k}, {\bi_k}}\}$ for which ${v}_{{\bj_k}{\bi_k}} = - {v}_{{\bj_k} {\bi_k}}$ implying ${v}_{{\bj_k}{\bi_k}} = 0$. In other words, it may be that the sets of $k$ single-particle states $\bj, \bi$ are present in $m$-body states defining a blue cell as well as other distinct $m$-body states defining a red cell. This situation only arises for $k < m$. For $k=m$ there can be no overlap between the conditions (\ref{eq:sysy6}) and (\ref{eq:sysy7}). When making the transition from studying the eGUE to studying the eGSE the first difference, and the most significant hurdle, is this overlap in the conditions on the random variables ${v}_{{\bj_k} {\bi_k}}$.

\subsubsection{Permutations}
As shown in section \ref{sec:SC} the assumptions underlying the symplectic symmetry of $H$ coincides with a fermionic time-reversal invariant quantum system. If there are $m$ fermions in the system the state space will consist of $m$-body fermionic states. Defining the system by a $k$-body potential with $l$ non-degenerate single-particle states there must be $l \choose m$ non-degenerate states in the basis describing the system and the matrix elements of the symplectic potential are given by $H_{\mu\nu} := \langle \mu |H|\nu \rangle$. Since $K$ is even (see section \ref{sec:SC} ) it follows that $H$ is even as well. If $l = 2n$ is even and $m$ odd, Glaisher's Rule gives
\begin{equation}\label{eq:perms1}
{l \choose m} \equiv \left\{ \begin{array}{ll}
         0 ~\mathrm{mod}~ 2 & \mbox{if $l$ is even and $m$ is odd};\\
        {{\lfloor l/2 \rfloor} \choose {\lfloor z/2 \rfloor}} ~\mathrm{mod}~ 2 & \mbox{otherwise}.\end{array} \right \}
\end{equation}
which ensures that there are an even number $l \choose m$ of $m$-body states and therefore an even number of rows and columns in the matrix representation of the potential $H$. This is just an example of the interesting dependency of $m, l$ and $l\choose m$ that occurs for potentials of the eGSE. It is known that a basis can be chosen in which the time reversal operator $K$ takes the form of (\ref{eq:trev11}). With $K$ taking this block diagonal form, the operator $T=KC$ defining time-reversal invariance, in addition to taking the complex conjugate of a state, maps one basis vector to another and multiplies it by -1. The operator $T^2$ being that which applies time reversal twice to a state, maps a basis vector to itself times a phase of -1 \emph{sans} complex conjugation (see section \ref{sec:IUTR}). By letting the state labels be the set $\{1,2, \ldots, l\}$ for the basis in which $H$ takes the canonical symplectic form (\ref{eq:trev15}) it must be the case that the time reversal operator $T$ maps any given $m$-body state to another, times at most a phase. Hence the action $T|\mu\rangle = \sigma_{\mu}$ is limited to permuting the $m$ single-particle labels in the set $\mu$ and multiplying the result by some phase $e^{i\phi}$. Assuming that this phase is the same for all states it follows that
\begin{equation}\label{eq:perms2}
T\mu = \sigma_{\mu} = e^{i\phi}\sigma(\mu)
\end{equation}
where $\sigma(\cdot)$ is a permutation on the set $\{1,2, \ldots, l\}$. Applying $T^2=-1$ gives
\begin{equation}
e^{2i\phi}\sigma^2 = -1
\end{equation}
so that
\begin{equation}\phi = \frac{2p-1}{2}\pi
\end{equation}
with $p=1, 2, 3, etc.$, and also
\begin{equation}
\sigma^2=1
\end{equation}
which implies that the permutation map $\sigma$ is a \emph{pairwise} permutation map. These permutations can be thought of as \textit{single-particle} state maps which exchange one single-particle state with another and vice versa as a function only of the initial state. For example, the pairwise permutation map
\begin{equation}\label{eq:perms3}
\sigma = \begin{pmatrix}
1 & 2 & 3 & 4 & \ldots & 2n-1 & 2n\\
2 & 1 & 4 & 3 & \ldots & 2n & 2n - 1\\
\end{pmatrix}
\end{equation}
exchanges the single-particle state $1$ with $2$, $2$ with $1$, $3$ with $4$, $4$ with $3$ and so on.

\subsubsection{The Zeros of the Random Variable}
The considerations so far for the second moments of the symplectic ensemble suggest that the eGSE involves far more subtlety than the eGUE and eGOE. A summary of the required approach to problems relying on the second moments, including all calculations of the higher moments of the eGSE, can be expressed as follows:

\vspace{1cm}\hspace{-1.1cm}\setlength{\fboxsep}{5pt}\fcolorbox{black}{red!15}{Fix $T$ to canonical form} $\rightarrow$ \fcolorbox{black}{red!15}{Define a map $\sigma$}  $\rightarrow$ \fcolorbox{black}{red!15}{Count Zeros} $\rightarrow$ \fcolorbox{black}{red!15}{Calculate Moments}\vspace{1cm}

\noindent The form of $T=KC$ is fixed in order to predetermine the form of the symmetry relations on $H$, namely (\ref{eq:sysy3}) and (\ref{eq:sysy4}) which are the analogue of the symmetry conditions of (\ref{eq:trev15}) written in terms of $T$. Next a map $\sigma$ is chosen to determine the action of $T$. Any resulting zeros from an overlap of (\ref{eq:sysy6}) and (\ref{eq:sysy7}) need to be identified to completely characterise the second moments $\overline{v_{\bj\bi}v_{\bj'\bi'}}$ for all $\bj, \bi, \bj', \bi'$. Hence a ``zeros theorem'' needs to be stated for each permutation map $\sigma$ in order to identify the $k$-sets $\{\bj, \bi\}$ for which the random variable is \emph{not} in fact random but instead equal to zero. Given these additional steps for the eGSE however, the form of the equations do not change dramatically. As for the eGUE, when ${v_{{\bj \bi}}}$ and ${v_{{\bj^{'} \bi^{'}}}}$ are not related by (\ref{eq:sysy6}) or (\ref{eq:sysy7})
\begin{align}\label{eq:zrv1}
\overline {v_{{\bj \bi}}v_{{\bj^{'}\bi^{'}}}} = \frac{1}{{2\pi}} \int v_{{\bj \bi}}v_{{\bj^{'} \bi^{'}}} \mathrm{exp}\left({-\frac{1}{2}v_{{\bj \bi}}^2}\right)~\mathrm{exp}\left({-\frac{1}{2}v_{{\bj^{'} \bi^{'}}}^2}\right)dv_{{\bj \bi}}dv_{{\bj^{'} \bi^{'}}} = 0
\end{align}
since the gaussians are even functions, making the total argument of the integral odd with respect to the integration variables. Likewise if $v_{{\bj \bi}} = v_{{\bj^{'} \bi^{'}}}$ then $v_{{\bj \bi}}$ and $v_{{\bj^{'} \bi^{'}}}$ are complex and their product yields two even functions of opposite sign, again giving a zero average. However, for
\begin{equation}\label{eq:zrv2	}
v_{{\bj \bi}} = \pm v_{{\bj^{'} \bi^{'}}}^{*}
\end{equation}
the average becomes
\begin{equation}\label{eq:zrv3}
\overline {v_{{\bj \bi}}^{2}} = \frac{\pm 1}{\sqrt{2\pi}} \int v_{{\bj \bi}}^2 \mathrm{exp}\left({-\frac{1}{2}v_{{\bj \bi}}^2}\right)dv_{{\bj \bi}} = \pm 1
\end{equation}
The sign of (\ref{eq:zrv2	}) is determined by the permutation map $\sigma$. By substituting $T|\mu\rangle = e^{i\phi}\sigma(\mu)$ from (\ref{eq:perms2}) the socalled blue matrix elements with symmetry condition given by (\ref{eq:sysy3}) obey
\begin{equation}\label{eq:zrv4}
\langle \mu|H|\nu \rangle = \langle e^{i\phi}\sigma(\mu)|H| e^{i\phi}\sigma(\nu)\rangle^{*} = \langle \sigma(\mu)|H| \sigma(\nu)\rangle^{*}
\end{equation}
while red matrix elements (\ref{eq:sysy4}) obey
\begin{equation}\label{eq:zrv5}
\langle e^{i\phi}\sigma(\mu)|H| \nu\rangle = -\langle\mu |H|e^{i\phi}\sigma(\nu) \rangle^{*}
\end{equation}
implying
\begin{equation}\label{eq:zrv6}
\langle \sigma(\mu)|H| \nu\rangle = -\langle\mu |H|\sigma(\nu) \rangle^{*}.
\end{equation}
From these it follows that
\begin{equation}\label{eq:zrv7}
v_{\bj\bi} = v_{\sigma(\bj)\sigma(\bi)}^{*}
\end{equation}
for $k$-tuples $\bj, \bi$ which are elements of the $m$-particle states determining blue matrix elements and
\begin{equation}\label{eq:zrv8}
v_{\bj\bi} = -v_{\sigma(\bj)\sigma(\bi)}^{*}
\end{equation}
for $k$-tuples $\bj, \bi$ which are elements of the $m$-particle states determining red matrix elements. This finally yields the non-zero second moments as
\begin{equation}\label{eq:zrv9}
\overline{v_{\bj\bi}v_{\bj'\bi'}} = \mathrm{sgn}({\bj \bi}) \cdot \delta_{{\bj \sigma(\bj^{'})}}\delta_{{\bi \sigma(\bi^{'})}}
\end{equation}
Hence by the above argument and by (\ref{eq:sysy6}) and  (\ref{eq:sysy7}) it can be concluded that
\begin{align}\label{eq:zrv10}
\overline {v_{{\bj \bi}}v_{{\bj^{'} \bi^{'}}}} = \delta_{{\bj \bi^{'}}}\delta_{{\bi \bj^{'}}} + \delta_{\beta4} \cdot \mathrm{sgn}({\bj \bi}) \cdot \delta_{{\bj \sigma(\bj^{'})}}\delta_{{\bi \sigma(\bi^{'})}}
\end{align}
where the amplitude of $\delta_{\beta 4}$ gives the purely Symplectic component of the average. The $sgn$ function determines the sign of the $k$-tuple, where the sign is positive for tuples resulting from (\ref{eq:zrv7}), negative for those resulting from (\ref{eq:zrv8}) and zero for those tuples satisfying both.

\section[The Embedded GUE]{The Embedded GUE}\label{sec:egue}
\subsection{Introduction}
\noindent Given the results from sections \ref{sec:usym}, \ref{sec:osym} and \ref{sec:symsym} it follows that the second moment for the embedded gaussian unitary, orthogonal and symplectic ensembles can be written as the compound expression
\begin{equation}
{v_{{\bj \bi}}v_{{\bj^{'} \bi^{'}}}} = \delta_{{\bj \bi^{'}}}\delta_{{\bi \bj^{'}}} + \delta_{\beta 1}\delta_{{\bj \bj^{'}}}\delta_{{\bi \bi^{'}}}  + \delta_{\beta4} \cdot \mathrm{sgn}({\bj \bi}) \cdot \delta_{{\bj \sigma(\bj^{'})}}\delta_{{\bi \sigma(\bi^{'})}}.
\end{equation}
This is just the linear combination of equations (\ref{eq:us10}), (\ref{eq:os5}) and (\ref{eq:zrv10}). The second moment is therefore greatly simplified in the case of the eGUE, reducing to read just
\begin{equation}
{v_{{\bj \bi}}v_{{\bj^{'} \bi^{'}}}} = \delta_{{\bj \bi^{'}}}\delta_{{\bi \bj^{'}}}.
\end{equation}
It is for this reason that calculating the moments of the embedded gaussian unitary ensemble is the least difficult of the three cases. This thesis will therefore focus primarily on the eGUE case, even though the methods introduced can be applied to other symmetry classes as well. To begin with, a known result for the normalised fourth moment, or kurtosis, of the level density for the eGUE will be derived using a simpler alternate calculation to that found in \cite{weid}. This will serve as a platform for introducing both the basic definitions and the new methodology involving particle diagrams which this thesis aims to explain. In subsequent sections the same methods will be used to calculate the sixth and eighth moments as well, albeit with the addition of some extra methodological features. For the case of the eGUE we will consider $m$ spinless fermions in a system with $m \ll l$ single-particle states, all interacting under the action of a \textit{k}-body potential ($k \le m$) with an identical gaussian \textit{p.d.f} determining its independent entries. The single-particle creation and annihilation operators are $a_j^{\dag}$ and $a_j$ respectively with $j = 1, \ldots, l$. As introduced in prior chapters, a useful shorthand notation for products of these is the abbreviation $\bj=(j_1,\ldots,j_k)$, $a_{\bj}=a_{j_k}\ldots a_{j_1}$ (similarly  for $\bi)$. A corollary of this is
\begin{equation}\label{eq:ee30}a_{\bj}^{\dag} = a_{j_1}^{\dag} \ldots a_{j_k}^{\dag}\end{equation}
For the sake of continuity and comparison with extant literature, notational traditions will be preserved for the calculation of the kurtosis by writing the orthonormal $m$-particle states as $|\mu\rangle, |\nu\rangle, |\rho\rangle, etc.$ where each state takes the form $a_{{j_m}}^{\dag}\ldots a_{{j_1}}^{\dag}|0\rangle$ with $|0\rangle$ denoting the vacuum state and the restriction $1\leq j_1 < j_2 < \ldots < j_m\leq l$. The $k$-body potential is given by
\begin{equation}\label{eq:ee27}{H}_k = \sum_{{{1 \le j_1 < \ldots < j_k \le l} \atop {1 \le i_1 < \ldots < i_k \le l}}} {v}_{j_1 \ldots j_k;i_1 \ldots i_k} a_{j_1}^{\dag} \ldots a_{j_k}^{\dag} a_{i_k} \ldots a_{i_1}\end{equation}
which will be abbreviated to
\begin{equation}\label{eq:egue01}{H}_k = \sum_{\bj,\bi}v_{\bj\bi} a_{\bj}^{\dag}  a_{\bi}\end{equation}
as before. In this case all the $m$-particle states are assumed to contain only non-repeating single-particle states and since the number of allowed single-particle states is $l$, the dimension of the $m$-body state space is $N={l\choose m}$. For the embedded GUE ensemble presently under consideration the only symmetry condition on the potential is that it be hermitian $\langle \mu|H_k|\nu \rangle = \langle \nu|H_k|\mu \rangle ^{*}$ for all $\mu, \nu$ giving the simplified second moment ${v}_{\bj\bi}^{*} = {v}_{\bi\bj}$ derived in section \ref{sec:usym}. As in canonical RMT (fixed $k=m$) the matrix elements not related by hermitian symmetry are uncorrelated \textit{i.i.d} complex gaussian random variables with mean zero and variance $v_o^2 = 1$ (wlog).
\subsection{Particle Diagrams}\label{sec:particlediagrams}
Of particular use in calculations is the abbreviation
\begin{align}\label{eq:ee46}A_{\mu\nu\rho\sigma} &:= \overline{\langle \mu|H_k|\sigma \rangle\langle \rho|H_k|\nu \rangle} =  \langle \mu|a_{\bj}^{\dag}  a_{\bi} |\sigma \rangle \langle \rho|a_{\bi}^{\dag}a_{\bj} |\nu \rangle.\end{align}
Here summation over the repeated indices $\bi, \bj$ is implied. For $A_{\mu\nu\rho\sigma}$ to be non-vanishing $|\sigma\rangle$ and $|\rho\rangle$ must both contain the $k$ single-particle states included in $\bi$, and  $|\mu\rangle$ and $|\nu\rangle$ must both contain the $k$ single-particle states included in $\bj$. In addition $a_{\bi}|\mu\rangle$ and $a_{\bj}|\sigma\rangle$ have to contain the same single-particle states implying that $|\mu\rangle$ and $|\sigma\rangle$ coincide in the $m-k$ single-particle states not included in $\bi$ or $\bj$. The same applies to $|\rho\rangle$ and $|\nu\rangle$. These relations are illustrated in Fig. \ref{fig:qterm_1} where solid bonds $~\feyn{f}~$ connect states sharing $m-k$ single-particle states and dashed bonds $~\feyn{h}~$ connect many-particle states sharing $k$ single-particle states. Note that in this figure the overlaps indicated by neighboring bonds are disjoint, e.g., the single-particle states of $\bi$ form the overlap $\sigma\feyn{h}\rho $ but are excluded from the overlap $\mu\feyn{f}\sigma$ because the state $|\sigma\rangle$ contains $m$ non-repeated single-particle states. The ``particle diagrams'' drawn in this way form an essential ingredient for evaluating the moments of the level density. With the odd moments being zero the normalised $2n$'th moments of the level density are given by
\begin{equation}\label{eq:moments}
\beta_{2n} = \frac{\frac{1}{N}\mathrm{tr}({\overline{H^{2n}_k}})}{\left(\frac{1}{N}\mathrm{tr}({\overline{H^2_k}})\right)^n}
\end{equation}
so the fourth moment, or kurtosis, which will be calculated next is
\begin{equation}\label{eq:ee48}\kappa = \frac{\frac{1}{N}\mathrm{tr}(\overline{H_{k}^4})}{\left(\frac{1}{N}\mathrm{tr}(\overline{H_{k}^2})\right)^2}\;.\end{equation}
The terms in the numerator and in the denominator of $\kappa$ can be evaluated using Wick's theorem which expresses the Gaussian average as a sum over all ways to draw contraction lines between the factors $V_k$. This gives
\begin{align}
\mathrm{tr}(\overline{V_k^2})&=\langle\mathrm{tr} \contraction[2ex]{}{V_k}{}{V_k}{}
V_k V_k\rangle\\
\label{trV4}
\mathrm{tr}(\overline{V_k^4})&=
\langle\mathrm{tr}
\contraction[2ex]{}{V_k}{}{V_k}
\contraction[2ex]{V_kV_k}{V_k}{}{V_k}
V_k V_k V_k V_k\rangle
+ \langle
\mathrm{tr}
\contraction[4ex]{}{V_k}{V_kV_k}{V_k}
\contraction[2ex]{V_k}{V_k}{}{V_k}
V_k V_k V_k V_k\rangle
+\langle
\mathrm{tr}
\contraction[2ex]{}{V_k}{V_k}{V_k}
\contraction[4ex]{V_k}{V_k}{V_k}{V_k}
V_k V_k V_k V_k\rangle.
\end{align}
Here the brackets $\langle.\rangle$ are used as an alternative notation for the ensemble average.
Note that the first two contributions to (\ref{trV4}) coincide due to cyclic invariance of the trace. Wick's theorem then allows us to compute the averages of contracted matrix elements as if the remaining elements were absent. The result can be expressed in terms of (\ref{eq:ee46}), leading to
\begin{equation}\label{eq:ee49}
\mathrm{tr}(\overline{H_{k}^2}) = \sum_{\mu}\overline{\langle\mu|H_{k}^2|\mu\rangle} = A_{\mu\mu\rho\rho}\\
\end{equation}
in the denominator, while in the numerator
\begin{equation}\label{eq:ee52}
\mathrm{tr}(\overline{H_{k}^4}) =  2 A_{\sigma\sigma\rho\rho}A_{\sigma\sigma\mu\mu} + A_{\mu\nu\rho\sigma}A_{\sigma\mu\nu\rho}
\end{equation}
with the summations over repeated indices $\mu, \nu, \rho, \sigma$ implicit. One can alternatively derive (\ref{eq:ee52}) from first principles by observing that the random variables ${v}_{\bj \bi}$ are gaussian so $\sqrt{\frac{\alpha}{\pi}}\int x^2 e^{-\alpha x^2}dx=\frac{1}{2\alpha}$ and $\sqrt{\frac{\alpha}{\pi}}\int x^4 e^{-\alpha x^2}dx=\frac{3}{4\alpha^2}$.
Given the restrictions from Fig. \ref{fig:qterm_1} it can be seen that it is necessary to sum over all $|\mu\rangle$ and $|\rho\rangle$ sharing $m-k$ single-particle states in order to calculate (\ref{eq:ee49}).
There are $N = {l\choose m}$ states in the sum over all possible $|\mu\rangle$, ${m\choose m-k}$ ways to choose the overlap with $|\rho\rangle$, and ${l-(m-k)\choose k}$ ways to choose the rest of $|\rho\rangle$. Hence the result is
\begin{equation}\label{eq:arg3} \mathrm{tr}(\overline{H_{k}^2}) = {l \choose m}{m \choose k}{{l-m+k}\choose k}.\end{equation}
The trace in the numerator of $\kappa$ is given by (\ref{eq:ee52}). The calculation for the first term is almost identical to the calculation yielding (\ref{eq:arg3}) giving
\begin{equation}\label{eq:arg8} 2  A_{\sigma\sigma\rho\rho}A_{\sigma\sigma\mu\mu} = 2 {l \choose m}{m \choose k}^2{{l-m+k} \choose k}^2\end{equation}
(summation implicit) so that the only remaining term needed to complete the calculation for $\kappa$ is $A_{\mu\nu\rho\sigma}A_{\sigma\mu\nu\rho}$. This term,  as well as subsequent quotients defining the sixth and eighth moments, requires the summation of a series of binomial expressions not all of which are simple enough to write down, as has been done with (\ref{eq:arg3}) and (\ref{eq:arg8}).
\subsection{Binomial Arguments}
To determine which of these binomial expressions survive in the limit of large $l$ a new quantity called the {\it argument} will now be defined as the power of a binomial expression in $l$ in the limit $l\to\infty$. For a quotient to give a non-vanishing result in that limit the argument of the numerator must be at least as large as the argument of the denominator. The argument can be obtained using Stirling's formula. Taking the dimension of the state space $N={l\choose m}$ as an example and applying Stirling's formula $lim_{n\to\infty}n! = \sqrt{2\pi n}{\left (\frac{n}{e}\right )}^n$ the value of the argument of $N$ is taken as the power of $N(m!) \sim l^{l - (l - m)}$ which is $m$, so $\arg(N) = m$. More generally
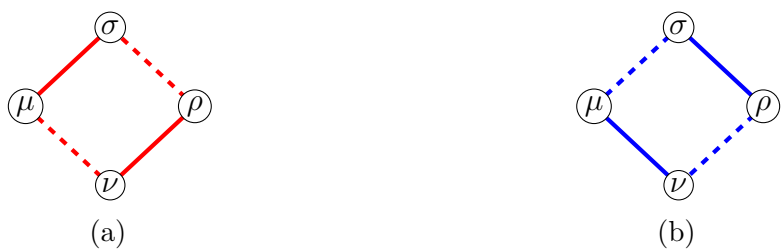
\begin{figure}[t]
\begin{subfigure}{.5\textwidth}
\centering
\begin{tikzpicture}[scale=1.5]\draw[ultra thick, red] (0,0) -- (0.75,.7);\draw[ultra thick, dashed, red] (0.75,.7) -- (1.5,0);\draw[ultra thick, dashed, red] (0,0) -- (0.75,-.7);\draw[ultra thick, red] (0.75,-.7) -- (1.5,0);\node[circle, fill=white, draw=black, text=black, inner sep=1pt] at (0,0) {$\mu$}; \node[circle, fill=white, draw=black, text=black,  inner sep=1pt] at (1.5,0) {$\rho$}; \node[circle, fill=white, draw=black, text=black,  inner sep=1pt] at (0.75,-.7) {$\nu$}; \node[circle, fill=white, draw=black, text=black,  inner sep=1pt] at (0.75,.7) {$\sigma$};
\end{tikzpicture}
  \caption{}
  \label{fig:qterm_1}
\end{subfigure}%
\begin{subfigure}{.5\textwidth}
\centering
\begin{tikzpicture}[scale=1.5]
\draw[ultra thick, dashed, blue] (0,0) -- (0.75,.7);\draw[ultra thick, blue] (0.75,.7) -- (1.5,0);\draw[ultra thick, blue] (0,0) -- (0.75,-.7);\draw[ultra thick, dashed, blue] (0.75,-.7) -- (1.5,0);\node[circle, fill=white, draw=black, text=black, inner sep=1pt] at (0,0) {$\mu$}; \node[circle, fill=white, draw=black, text=black, inner sep=1pt] at (1.5,0) {$\rho$}; \node[circle, fill=white, draw=black, text=black, inner sep=1pt] at (0.75,-.7) {$\nu$}; \node[circle, fill=white, draw=black, text=black, inner sep=1pt] at (0.75,.7) {$\sigma$};\end{tikzpicture}
  \caption{}
  \label{fig:qterm_2}
\end{subfigure}%
\caption[Particle diagrams for the 4'th moment]{(a) illustrates the particle diagram implied by  $A_{\mu\nu\rho\sigma}=\langle \mu|a_{\bj}^{\dag} a_{\bi}  |\sigma \rangle \langle \rho|a_{\bi}^{\dag}  a_{\bj} |\nu \rangle$ of (\ref{eq:ee46}) and (\ref{eq:ee52}), and (b) illustrates the particle diagram implied by the factor $A_{\sigma\mu\nu\rho}$ in (\ref{eq:ee52}). Each bond between compound $m$-body states represents a set of single-particles shared by both states.

\hrulefill}
\label{fig:qtermfig}
\end{figure}
\begin{equation}\label{eq:arg1}\arg\left [\prod_n{{l - a_n} \choose b_n}\right ] = \sum_n  b_n \end{equation}
or alternatively in terms of multinomials
\begin{equation}
\label{eq:tr4A}
\arg{l-a\choose b_1 \; b_2 \; b_3\; \ldots}=\sum_nb_n.
\end{equation}
Note that any additional factors independent of $l$ do not have any impact on the argument -- they contribute only finite coefficients to the product and therefore do not affect the power of $l$. From (\ref{eq:arg3}) it then follows that $\arg [ ( \frac{1}{N} \mathrm{tr}(\overline{V_{k}^2}) )^2 ] = 2k$. Since it is expected that the value of the fourth moment will converge to some finite value the term $\frac{1}{N}\mathrm{tr}(\overline{V_{k}^4})$ should have terms with arguments equal to $2k$  and possibly some terms with an argument less than $2k$, which indeed turns out to be the case. Those terms with an argument less than $2k$ will be ignored, contributing values of order no higher than $l^{-1}$ to $\kappa$ as $l\to\infty$, whereas those with an argument equal to $2k$ must be calculated. Using (\ref{eq:arg8}) one easily sees that after division by $N$ the term $2  A_{\sigma\sigma\rho\rho}A_{\sigma\sigma\mu\mu}$ has the argument $2k$ as expected.

For the second summand $A_{\mu\nu\rho\sigma}A_{\sigma\mu\nu\rho}$ in (\ref{eq:ee52}) non-zero contributions arise if the indices of $A_{\mu\nu\rho\sigma}$ obey the same restrictions as introduced earlier with reference to (\ref{eq:ee46}) and depicted in Fig. \ref{fig:qterm_1}. The analogous restrictions for non-zero $A_{\sigma\mu\nu\rho}$ are given by Fig. \ref{fig:qterm_2}. Together Figs. \ref{fig:qterm_1} and \ref{fig:qterm_2} form the particle diagram for $A_{\mu\nu\rho\sigma}A_{\sigma\mu\nu\rho}$, formally defined as the collection of bonds incorporating all restrictions for the indices. Since there will be occasion to use the combined diagram in later calculations for the higher order moments as well, it will be referred to subsequently as the \textit{standard diagram}. Of interest is the number of unique $m$-body states satisfying this diagram (giving the value of the trace) but only the term for which the argument of this number reaches its maximal value $2k$ is needed. This is the term for which the maximal number of single-particle states can be chosen which satisfy all of the bonds of the particle diagram, \textit{ie}. the term for which there are maximal degrees of freedom in the choice of single-particle states which satisfy the bonds of the diagram. From Fig. \ref{fig:qterm_1} it can be observed that the state $|\mu\rangle$ determines the $k$ single-particle states in the state $|\nu\rangle$ which are not determined by the bond $\nu\feyn{f}\rho$ so that $|\mu\rangle$ and $|\rho\rangle$ together fully determine the single-particle states contained in $|\nu\rangle$, and similarly for the state $|\sigma\rangle$. Hence all the single-particle states satisfying an instance of the diagram are completely determined by $|\mu\rangle$ and $|\rho\rangle$ and the argument of the diagram is maximal when  $|\mu\rangle$ determines the smallest possible number of single-particle states in $|\rho\rangle$ and vice versa. This means that e.g. the bond $\mu\feyn{h}\nu$ of Fig. \ref{fig:qterm_1} must share the minimal number of states possible with the bond $\nu\feyn{h}\rho$ of Fig. \ref{fig:qterm_2}, and likewise the bond $\mu\feyn{f}\sigma$ of Fig. \ref{fig:qterm_1} must share the minimal number of states possible with the bond $\sigma\feyn{f}\rho$ of Fig. \ref{fig:qterm_2}. The minimum overlap possible between the smaller pair is zero, while the minimum overlap between the larger is $s=|k-(m-k)|=m-2r$ where $r:=\min(k,m-k)$. This overlap being minimal additionally implies that the larger of the two bonds between two nodes (states) contains all single-particle states included in the smaller one, leaving $r$ states participating in both bonds and $s$ states participating only in the larger of the two bonds. Analogous reasoning applies to each of the corresponding pairs of bonds of Fig. \ref{fig:qterm_1} and Fig. \ref{fig:qterm_2}. The sets of $s$ states for each pair must be disjoint relative to the sets of $r$ overlapping states, as a consequence of neighboring bonds in the same diagram \ref{fig:qterm_1} or \ref{fig:qterm_2} being disjoint. Hence the possible choices for states are given by partitioning the $l$ available single-particle states into one set of $s$ states shared by both $|\mu\rangle$ and $|\rho\rangle$, and four sets of $r$ states. Thus to leading order the expression can be written as the multinomial
\begin{align}\label{eq:arg11}&A_{\mu\nu\rho\sigma}A_{\sigma\mu\nu\rho} \sim{l\choose {s\; r\; r\; r\; r}}\nonumber\\
~~~~&= {l \choose m}{{l - m} \choose r}{{l - m - r}\choose r}{m \choose r}{{m - r}\choose r}\;.\end{align}
Recalling $N={l\choose m}$ the argument of the corresponding term in the numerator of $\kappa$ is then $\arg \left [ \frac{1}{N}\sum A_{\mu\nu\rho\sigma}A_{\sigma\mu\nu\rho}\right ] = 2r = 2\cdot min(k, m-k)$ so it is only for $k \le m-k$ that the argument of this term is equal to $2k$, while for $k > m - k$ it is always less. Fitting all the surviving terms into the expression for $\kappa$ gives the limit form of the fourth moment as $l \to \infty$
\begin{align}\label{eq:arg16}\lim_{N\to\infty}\kappa = 2 + \lim_{N\to\infty}\frac{\frac{1}{N}A_{\mu\nu\rho\sigma}A_{\sigma\mu\nu\rho}}{\left [ \frac{1}{N}\sum A_{\mu\mu\rho\rho} \right ] ^2}
= 2 + \frac{{{m - k} \choose k}}{{m \choose k}}\end{align}
which corroborates the result found by Benet \textit{et. al.} \cite{weid} using the eigenvector expansion method and for $2k > m$ agrees with what is expected using the method of supersymmetry, namely $\kappa=2$.
\begin{figure}[H]
\centering
\includegraphics[scale=.7]{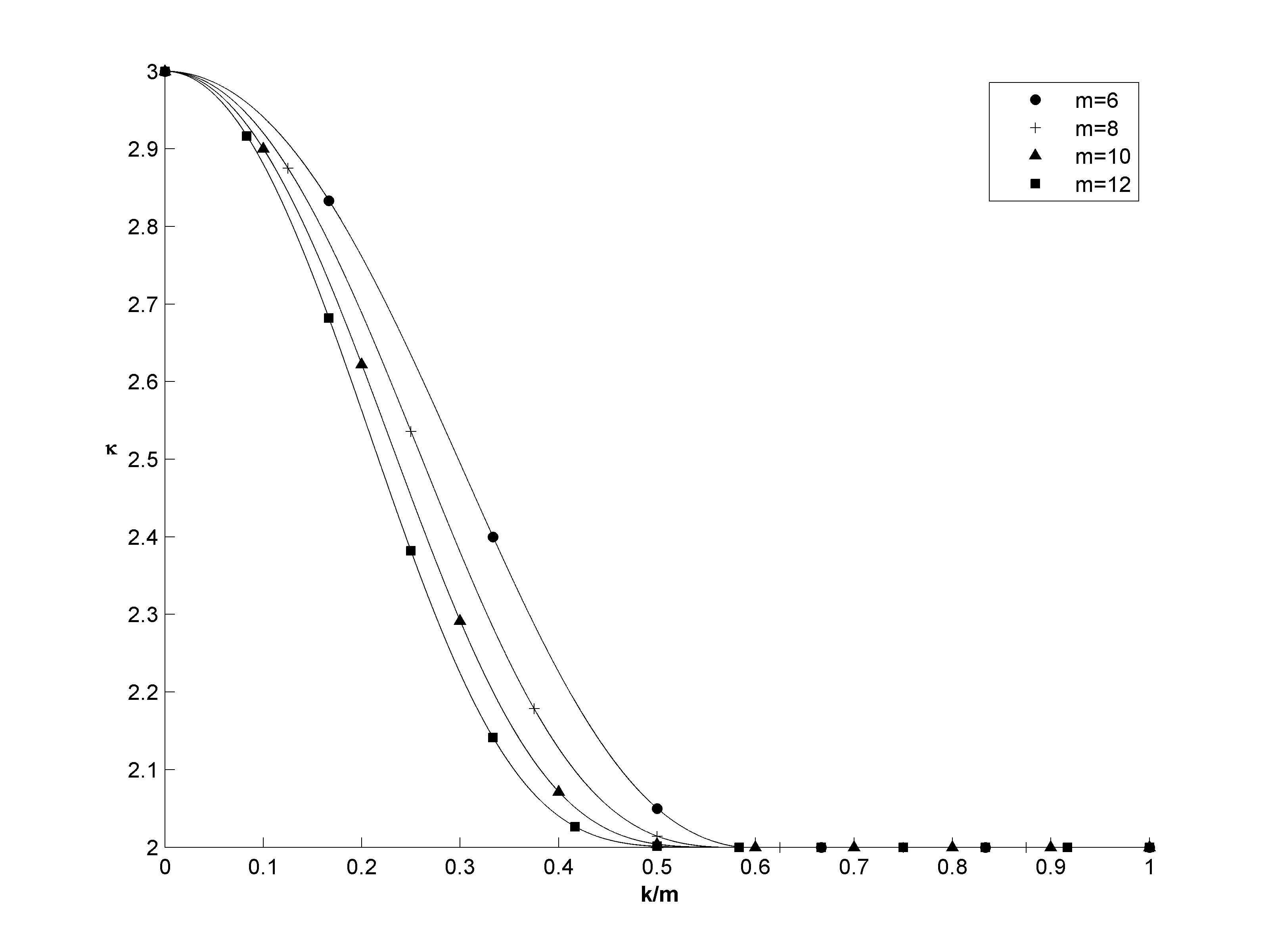}
\caption[The fourth moment $\kappa$ against $k/m$]{$\kappa$ against $k/m$ for $m=6, 8, 10, 12$ showing how the kurtosis has a semi-circular value, $\kappa=2$ for $k/m > \frac{1}{2}$ after which it transitions to a gaussian kurtosis, $\kappa = 3$ at $k=0$. Higher values of $m$ give faster convergence to the semi-circular moment.}
\label{fig:fourth_moment}
\end{figure}
\subsection{Paths and Loops}
Particle diagrams represent diagramatically the factors of $A_{\mu\nu\rho\sigma}$ (\ref{eq:ee46}) which form the constituent parts of $\overline{H^{2n}}$ (see (\ref{eq:ee52}) for instance). What will be needed next in order to calculate higher moments for embedded ensembles is an extension of the notation of particle diagrams to combine \emph{arguments} with particle diagrams in a mathematically robust way. In so doing, it becomes possible to calculate any diagram, no matter how complex. This will be done by using arguments (powers of $l$) and diagrams (representations of $A_{\mu\nu\rho\sigma}$) as before, with the additional concept of \emph{closed loops}. It is a necessary to introduce closed loops on diagrams because as we look to calculate higher and higher moments the resulting diagrams become increasingly complex and less responsive to such a simple analysis as that which yielded the fourth moment. Closed loops form the unbreakable link between a diagram and it's maximal argument.

As explained in section \ref{sec:particlediagrams} a particle diagram is a set of bonds drawn between $m$-body states indicating the minimal overlap between them. Each $m$-body state in the diagram can be thought of as a node. In calculations of the moments of the level density each node is connected by four bonds, two of size $k$ and two of size $m-k$. See Fig \ref{fig:loop_illustration} for a representation. This is a property that emerges from the application of Wick's theorem to tr$\overline{H^{2n}}$ which can always be expressed in terms of sums of factors of $A_{\mu\nu\rho\sigma} = \langle\mu|H|\sigma\rangle\langle\rho|H|\nu\rangle$. Since from first principles the trace is simply
\begin{equation}\label{eq:PALS}
\mathrm{tr}\overline{H^{2n}} = \overline{H_{a_1 a_2}H_{a_2 a_3}H_{a_3 a_4} \ldots H_{a_{2n} a_1}}
\end{equation}
every index appears twice, which directly implies that each index will appear twice in the resulting expansion of (\ref{eq:PALS}) in terms of $A_{\mu\nu\rho\sigma}$. Hence apart from a single exception (discussed next) every index will appear once in one factor of $A$ and a second time in another distinct factor of $A$. This in turn  implies that every node in every diagram will take the form of Fig \ref{fig:loop_illustration} wherein the node (a set of $m$ state labels) is completely determined by a $k$ set $\feyn{h}$ disjoint from the set $\feyn{f}$ of $m-k$ labels. That these must be disjoint is given by the fact that the nodes represent fermionic states which are sets of $m$ non-repeated labels. Morever, since there are two such disjoint sets connected to each node, one for each appearance of a state as an index of $A$, these must overlap entirely in order to conserve the number of states in the node as $m$. In other words, with reference to Fig \ref{fig:loop_illustration}, the state $|\mu\rangle $ is the union of the two disjoint black bonds
\begin{equation}
\mu = \mathbf{\feyn{h}} \cup \mathbf{\feyn{f}}
\end{equation}
from the first factor of $A$ containing $\mu$ as an index, as well as the union of the two disjoint blue bonds
\begin{equation}
\mu = {\color{blue}\mathbf{\feyn{h}}} \cup {\color{blue}\mathbf{\feyn{f}}}
\end{equation}
from the second factor of $A$ containing the state $\mu$ as an index. Hence, for example, the $m-k$ states determined by the blue solid bond ${\color{blue}\mathbf{\feyn{f}}} \mu$ must themselves form a disjoint set; those which are shared with the black solid bond $\mathbf{\feyn{f}} \mu$ and those shared with the black dashed bond $\mathbf{\feyn{h}} \mu$. In this way, the single-particle states which form the set denoted by the blue solid bond will ``travel'' (read: be a member of) through either the black dashed bond or the black solid bond, and onwards to the next node (see Fig \ref{fig:loop_illustration2}). At the next node this process will be repeated and the single-particle states will seperate again into disjoint sets and travel further, continuing along a path around the diagram until forming a loop. 
\begin{figure}[b!]
\hrulefill
\vspace{1cm}

\centering
\begin{tikzpicture}[scale=1.7]
\draw[ultra thick, dashed, blue] (0.78,.85) -- (1.7,0);
\draw[ultra thick, blue] (0.78,-.85) -- (1.7,0);
\begin{scope}[shift={(1.76,0)}]
\draw[ultra thick, black] (-0.2,0) -- (0.72,.85);
\draw[ultra thick, dashed, black] (-.2,0) -- (0.72,-.85);
\node[circle, fill=white, draw=black, text=black, inner sep=3pt] at (-.1,0) {$\mu$}; 
\end{scope}
\end{tikzpicture}
\caption[Loops Illustration]{Each node in a particle diagram takes the basic form shown. Since the blue bonds are disjoint and together completely determine the $m$ single-particle labels in $|\mu\rangle$ they must overlap completely with the set defined by the black bonds, which are also disjoint and completely determine the single-particle states in $|\mu\rangle$.}
\label{fig:loop_illustration}
\end{figure}
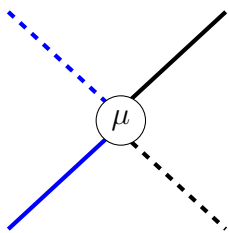

For the singular case where a single factor of $A_{\mu\mu\rho\rho}$ includes both instances of a label the two bonds $\feyn{f}$ connecting $\mu$ and $\rho$ are equal and are represented by a so called ``tail'' $\mu\feyn{f}\rho$, and the $k$-bonds $\feyn{h}$ are implicit in that they connect each state with itself, as opposed to the more common case where for example $\mu$ shares a $k$-bond with two distinct states, neither of which is equal to $\mu$.

These loops on the particle diagram indicate the single-particle states shared by all bonds and nodes (states) which participate in the loop. There are only ever a finite number of possible loops in any diagram. This relates directly to the argument of the diagram, since the argument counts the degrees of freedom in a diagam, i.e. the number of labels which together determine, through bonds, all states in the diagram. Hence the number of elements in a loop is the number which that loop contributes to the total argument. In this way loops will form a fundamental tool for (a) expressing the argument as a sum of loops and (b) maximising the arguments by optimising the number of elements in each loop.
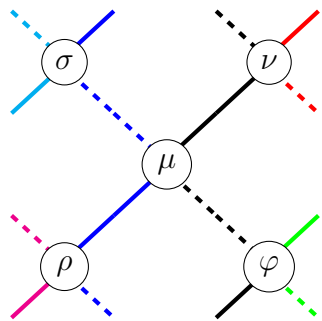
\begin{figure}[b!]
\hrulefill
\vspace{1cm}

\centering
\begin{tikzpicture}[scale=.8]
\draw[ultra thick, dashed, blue] (0.78,.85) -- (1.7,0); 
\draw[ultra thick, blue] (0.78,-.85) -- (1.7,0); 
\begin{scope}[shift={(1.76,0)}]
\draw[ultra thick, black] (-0.2,0) -- (0.72,.85); 
\draw[ultra thick, dashed, black] (-.2,0) -- (0.72,-.85); 
\node[circle, fill=white, draw=black, text=black, inner sep=3pt] at (-.1,0) {$\mu$}; 
\end{scope}
\begin{scope}[shift={(1.7,1.7)}]
\draw[ultra thick, dashed, black] (0.78,.85) -- (1.7,0); 
\draw[ultra thick, black] (0.78,-.85) -- (1.7,0); 
\begin{scope}[shift={(1.76,0)}]
\draw[ultra thick, red] (-0.2,0) -- (0.72,.85); 
\draw[ultra thick, dashed, red] (-.2,0) -- (0.72,-.85); 
\node[circle, fill=white, draw=black, text=black, inner sep=3pt] at (-.1,0) {$\nu$}; 
\end{scope}
\end{scope}
\begin{scope}[shift={(-1.7,-1.7)}]
\draw[ultra thick, dashed, magenta] (0.78,.85) -- (1.7,0); 
\draw[ultra thick, magenta] (0.78,-.85) -- (1.7,0); 
\begin{scope}[shift={(1.76,0)}]
\draw[ultra thick, blue] (-0.2,0) -- (0.72,.85); 
\draw[ultra thick, dashed, blue] (-.2,0) -- (0.72,-.85); 
\node[circle, fill=white, draw=black, text=black, inner sep=3pt] at (-.1,0) {$\rho$}; 
\end{scope}
\end{scope}
\begin{scope}[shift={(-1.7,1.7)}]
\draw[ultra thick, dashed, cyan] (0.78,.85) -- (1.7,0); 
\draw[ultra thick, cyan] (0.78,-.85) -- (1.7,0); 
\begin{scope}[shift={(1.76,0)}]
\draw[ultra thick, blue] (-0.2,0) -- (0.72,.85); 
\draw[ultra thick, dashed, blue] (-.2,0) -- (0.72,-.85); 
\node[circle, fill=white, draw=black, text=black, inner sep=3pt] at (-.1,0) {$\sigma$}; 
\end{scope}
\end{scope}
\begin{scope}[shift={(1.7,-1.7)}]
\draw[ultra thick, dashed, black] (0.78,.85) -- (1.7,0); 
\draw[ultra thick, black] (0.78,-.85) -- (1.7,0); 
\begin{scope}[shift={(1.76,0)}]
\draw[ultra thick, green] (-0.2,0) -- (0.72,.85); 
\draw[ultra thick, dashed, green] (-.2,0) -- (0.72,-.85); 
\node[circle, fill=white, draw=black, text=black, inner sep=3pt] at (-.1,0) {$\varphi$}; 
\end{scope}
\end{scope}
\end{tikzpicture}
\caption[Loops Illustration]{The fact that each node in a diagram takes the form of Fig \ref{fig:loop_illustration} implies that the combined diagram can always be ``flattened'' into the form of a cyclic lattice. This implies that every single-particle state label in the particle diagram is the member of a loop which travels around the lattice along those paths not excluded by the fermionic nature of the nodes (states).}
\label{fig:loop_illustration2}
\end{figure}

\subsection{Example: The Fourth Moment with Loops}\label{sec:TFMWL}
To illustrate the process of calculating arbitrarily high moments using arguments, particle diagrams and loops, an example will now be made with of fourth moment, which has already been calculated from first principles as (\ref{eq:arg16}). It has been seen that the fourth moment, or kurtosis, $\kappa = {\frac{1}{N}\mathrm{tr}(\overline{H_{k}^4})}/{\left(\frac{1}{N}\mathrm{tr}(\overline{H_{k}^2})\right)^2}$ where in the denominator one has
\begin{equation}\label{eq:tfmwl1}
\mathrm{tr}(\overline{H_{k}^2}) = \sum_{\mu}\overline{\langle\mu|H_{k}^2|\mu\rangle} = A_{\mu\mu\rho\rho}
\end{equation}
and in the numerator
\begin{equation}\label{eq:tfmwl2}
\mathrm{tr}(\overline{H_{k}^4}) =  2 A_{\sigma\sigma\rho\rho}A_{\sigma\sigma\mu\mu} + A_{\mu\nu\rho\sigma}A_{\sigma\mu\nu\rho}.
\end{equation}
As forecast in the previous section (i) both components of the fourth moment can be expressed entirely in terms of the quantity $A_{\mu\nu\rho\sigma}$ and moreover (ii) every index appears twice, either as an instance of a tail with repeated indices of the form $A_{aabb}$ or once each in two distinct factors of $A$ such as the second term $A_{\mu\nu\rho\sigma}A_{\sigma\mu\nu\rho}$ of (\ref{eq:tfmwl2}). For terms of the form $A_{aabb}$ no further sophistication is necessary because these simply add a single factor of ${m\choose k}{{l-m+k}\choose{k}}$ in cases where only $b$ is a free choice, or alternatively ${l\choose m}{m\choose k}{{l-m+k}\choose{k}}$ in cases such as (\ref{eq:arg3}) where both $a$ and $b$ have not yet been determined by other bonds in the diagram. These considerations give
\begin{figure}[t]
\centering
\begin{tikzpicture}[scale=1.5]
\draw[ultra thick, red] (-.2,0) -- (0.75,.9);
\draw[ultra thick, dashed, red] (0.75,.9) -- (1.7,0);
\draw[ultra thick, dashed, red] (-.2,0) -- (0.75,-.9);
\draw[ultra thick, red] (0.75,-.9) -- (1.7,0);
\draw[ultra thick, dashed, blue] (0,0) -- (0.75,.7);
\draw[ultra thick, blue] (0.75,.7) -- (1.5,0);
\draw[ultra thick, blue] (0,0) -- (0.75,-.7);
\draw[ultra thick, dashed, blue] (0.75,-.7) -- (1.5,0);
\node[circle, fill=white, draw=black, text=black, inner sep=1.2pt] at (-.05,0) {$\mu$}; 
\node[circle, fill=white, draw=black, text=black,  inner sep=1.2pt] at (1.55,0) {$\rho$}; 
\node[circle, fill=white, draw=black, text=black,  inner sep=1.2pt] at (0.75,-.78) {$\nu$};
 \node[circle, fill=white, draw=black, text=black,  inner sep=1.2pt] at (0.75,.78) {$\sigma$};
\end{tikzpicture}
\caption[The Standard Diagram]{The particle diagram of the term $A_{\mu\nu\rho\sigma}A_{\sigma\mu\nu\rho}$ is the standard diagram, as seen before in Fig \ref{fig:qterm_1} and \ref{fig:qterm_2} but now with both factors illustrated as a single combined particle diagram.

\hrulefill}
\label{fig:combinedqterm}
\end{figure}
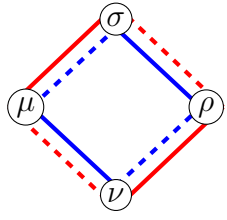
\begin{equation}\label{eq:tfmwl3}
\mathrm{tr}(\overline{H_{k}^2}) = {l\choose m}{m\choose k}{{l-m+k}\choose{k}}
\end{equation}
and
\begin{equation}\label{eq:tfmwl4}
\mathrm{tr}(\overline{H_{k}^4}) =  2 {l\choose m}\left[{m\choose k}{{l-m+k}\choose{k}}\right]^2 + A_{\mu\nu\rho\sigma}A_{\sigma\mu\nu\rho}.
\end{equation}
The term $A_{\mu\nu\rho\sigma}A_{\sigma\mu\nu\rho}$ can be expressed as a particle diagram as in Fig \ref{fig:qterm_1} and \ref{fig:qterm_2}, however this time the diagram will be drawn as a single combined picture, Fig \ref{fig:combinedqterm}. This is the form in which it is possible to calculate the moments by identifying all loops within the diagram. Because of the fermionic nature of the states, loops cannot contain two bonds from the same factor of $A$, but apart from this caveat all loops are permitted. The loops for Fig \ref{fig:combinedqterm} are illustrated in Fig \ref{fig:paths_fourth}.
\begin{figure}
\begin{subfigure}{.3\textwidth}
\centering
\begin{tikzpicture}[node distance=1.6cm]
\node (sigma) [draw, circle, inner sep=1pt, fill=blue!20, font=\bfseries] {\footnotesize $\sigma$};
\node (rho) [draw, circle, inner sep=1pt, fill=blue!20, font=\bfseries, below right of=sigma] {\footnotesize $\rho$};
\node (mu) [draw, circle, inner sep=1pt, fill=blue!20, font=\bfseries, below left of=sigma] {\footnotesize $\mu$};
\node (nu) [draw, circle, inner sep=1pt, fill=blue!20, font=\bfseries, below right of =mu] {\footnotesize $\nu$};
\draw (mu) to [out=90, in=180] (sigma);
\draw (sigma) to [out=-90, in=0] (mu);
\end{tikzpicture}
\vspace{.5cm}\caption[q_term paths]{The loop {\normalsize{$\overrightarrow{\mu\sigma\mu}$}} defining the set of single-particle states shared by the bonds {\normalsize{$\mu$}}\protect\tikz\draw[thick, red] (0,0) -- (1,0);{\normalsize{$\sigma$}} and {\normalsize{$\mu$}}\tikz\draw[thick, dashed, blue] (0,0) -- (1,0);{\normalsize{$\sigma$}} of Fig \ref{fig:combinedqterm}.}
  \label{fig:p4}
\end{subfigure}\hfill%
\begin{subfigure}{.3\textwidth}
\centering
\begin{tikzpicture}[node distance=1.6cm]
\node (sigma) [draw, circle, inner sep=1pt, fill=apricot, font=\bfseries] {\footnotesize $\sigma$};
\node (rho) [draw, circle, inner sep=1pt, fill=apricot, font=\bfseries, below right of=sigma] {\footnotesize $\rho$};
\node (mu) [draw, circle, inner sep=1pt, fill=apricot, font=\bfseries, below left of=sigma] {\footnotesize $\mu$};
\node (nu) [draw, circle, inner sep=1pt, fill=apricot, font=\bfseries, below right of =mu] {\footnotesize $\nu$};
\draw (sigma) to [out=0, in=90] (rho);
\draw (rho) to [out=180, in=-90] (sigma);
\end{tikzpicture}
 \vspace{.5cm} \caption{The loop {\normalsize{$\overrightarrow{\sigma\rho\sigma}$}} defining the set of single-particle states shared by the bonds {\normalsize{$\sigma$}}\protect\tikz\draw[thick, dashed, red] (0,0) -- (1,0);{\normalsize{$\rho$}} and {\normalsize{$\sigma$}}\tikz\draw[thick, blue] (0,0) -- (1,0);{\normalsize{$\rho$}} of Fig \ref{fig:combinedqterm}.}
  \label{fig:p4}
\end{subfigure}\hfill%
\begin{subfigure}{.3\textwidth}
\centering
\begin{tikzpicture}[node distance=1.6cm]
\node (sigma) [draw, circle, inner sep=1pt, fill=blue!20, font=\bfseries] {\footnotesize $\sigma$};
\node (rho) [draw, circle, inner sep=1pt, fill=blue!20, font=\bfseries, below right of=sigma] {\footnotesize $\rho$};
\node (mu) [draw, circle, inner sep=1pt, fill=blue!20, font=\bfseries, below left of=sigma] {\footnotesize $\mu$};
\node (nu) [draw, circle, inner sep=1pt, fill=blue!20, font=\bfseries, below right of =mu] {\footnotesize $\nu$};
\draw (mu) to [out=0, in=90] (nu);
\draw (nu) to [out=180, in=-90] (mu);
\end{tikzpicture}
  \vspace{.5cm}\caption{The loop {\normalsize{$\overrightarrow{\mu\nu\mu}$}} defining the set of single-particle states shared by the bonds {\normalsize{$\mu$}}\protect\tikz\draw[thick, dashed, red] (0,0) -- (1,0);{\normalsize{$\nu$}} and {\normalsize{$\mu$}}\tikz\draw[thick, blue] (0,0) -- (1,0);{\normalsize{$\nu$}} of Fig \ref{fig:combinedqterm}.}
  \label{fig:p4}
\end{subfigure}\hfill\vspace{2cm}
\begin{subfigure}{.3\textwidth}
\centering
\begin{tikzpicture}[node distance=1.6cm]
\node (sigma) [draw, circle, inner sep=1pt, fill=apricot, font=\bfseries] {\footnotesize $\sigma$};
\node (rho) [draw, circle, inner sep=1pt, fill=apricot, font=\bfseries, below right of=sigma] {\footnotesize $\rho$};
\node (mu) [draw, circle, inner sep=1pt, fill=apricot, font=\bfseries, below left of=sigma] {\footnotesize $\mu$};
\node (nu) [draw, circle, inner sep=1pt, fill=apricot, font=\bfseries, below right of =mu] {\footnotesize $\nu$};
\draw (rho) to [out=-90, in=0] (nu);
\draw (nu) to [out=90, in=180] (rho);
\end{tikzpicture}
 \vspace{.4cm} \caption{The loop {\normalsize{$\overrightarrow{\nu\rho\nu}$}} defining the set of single-particle states shared by the bonds {\normalsize{$\nu$}}\protect\tikz\draw[thick, red] (0,0) -- (1,0);{\normalsize{$\rho$}} and {\normalsize{$\nu$}}\tikz\draw[thick, dashed, blue] (0,0) -- (1,0);{\normalsize{$\rho$}} of Fig \ref{fig:combinedqterm}.}
  \label{fig:p4}
\end{subfigure}\hfill%
\begin{subfigure}{.3\textwidth}
\centering
\begin{tikzpicture}[node distance=1.6cm]
\node (sigma) [draw, circle, inner sep=1pt, fill=blue!20, font=\bfseries] {\footnotesize $\sigma$};
\node (rho) [draw, circle, inner sep=1pt, fill=blue!20, font=\bfseries, below right of=sigma] {\footnotesize $\rho$};
\node (mu) [draw, circle, inner sep=1pt, fill=blue!20, font=\bfseries, below left of=sigma] {\footnotesize $\mu$};
\node (nu) [draw, circle, inner sep=1pt, fill=blue!20, font=\bfseries, below right of =mu] {\footnotesize $\nu$};
\draw (mu) to [out=90, in=180] (sigma);
\draw (sigma) to [out=-90, in=180] (rho);
\draw (rho) to [out=-90, in=0] (nu);
\draw (nu) to [out=90, in=0] (mu);
\end{tikzpicture}
 \vspace{.5cm} \caption{The first loop {\normalsize{$\overrightarrow{\mu\sigma\rho\nu\mu}$}} defining the set of single-particle states shared by the bonds {\normalsize{$\mu$}}\protect\tikz\draw[thick, red] (0,0) -- (1,0);{\normalsize{$\sigma$}},~ {\normalsize{$\sigma$}}\tikz\draw[thick, blue] (0,0) -- (1,0);{\normalsize{$\rho$}},~\\ {\normalsize{$\rho$}}\tikz\draw[thick, red] (0,0) -- (1,0);{\normalsize{$\nu$}} and {\normalsize{$\nu$}}\tikz\draw[thick, blue] (0,0) -- (1,0);{\normalsize{$\mu$}} of Fig \ref{fig:combinedqterm}.}
  \label{fig:p4}
\end{subfigure}\hfill%
\begin{subfigure}{.3\textwidth}
\centering
\begin{tikzpicture}[node distance=1.6cm]
\node (sigma) [draw, circle, inner sep=1pt, fill=apricot, font=\bfseries] {\footnotesize $\sigma$};
\node (rho) [draw, circle, inner sep=1pt, fill=apricot, font=\bfseries, below right of=sigma] {\footnotesize $\rho$};
\node (mu) [draw, circle, inner sep=1pt, fill=apricot, font=\bfseries, below left of=sigma] {\footnotesize $\mu$};
\node (nu) [draw, circle, inner sep=1pt, fill=apricot, font=\bfseries, below right of =mu] {\footnotesize $\nu$};
\draw (mu) to [out=0, in=-90] (sigma);
\draw (sigma) to [out=0, in=90] (rho);
\draw (rho) to [out=180, in=90] (nu);
\draw (nu) to [out=180, in=-90] (mu);
\end{tikzpicture}
 \vspace{.5cm} \caption{The second loop {\normalsize{$\overrightarrow{\mu\sigma\rho\nu\mu}^{'}$}} defining the set of single-particle states shared by the bonds {\normalsize{$\mu$}}\protect\tikz\draw[thick, dashed, red] (0,0) -- (1,0);{\normalsize{$\sigma$}},~ {\normalsize{$\sigma$}}\tikz\draw[thick, dashed, blue] (0,0) -- (1,0);{\normalsize{$\rho$}},~\\ {\normalsize{$\rho$}}\tikz\draw[thick, dashed, red] (0,0) -- (1,0);{\normalsize{$\nu$}} and {\normalsize{$\nu$}}\tikz\draw[thick, dashed, blue] (0,0) -- (1,0);{\normalsize{$\mu$}} of Fig \ref{fig:combinedqterm}.}
  \label{fig:p4}
\end{subfigure}\hfill\vspace{1cm}
\caption[Loops Illustration]{Illustration of the loops of the factor  $A_{\mu\nu\rho\sigma}A_{\sigma\mu\nu\rho}$ seen in (\ref{eq:ee52}) with the factor $A_{\mu\nu\rho\sigma}$ being represented by the particle diagram of Fig \ref{fig:qterm_1} and the factor $A_{\sigma\mu\nu\rho}$ being represented by the particle diagram of Fig \ref{fig:qterm_2}. Each bond between compound states represents a set of single-particle states shared by both of the compound states, and each loop represents the set of single-particle states shared by all of the compound quantum states which form the nodes of the loop.}
\label{fig:paths_fourth}
\end{figure}
Lines curving outwards indicate a path through a bond of the red component $A_{\sigma\mu\nu\rho}$ while curves bending inwards indicate that the path is crossing a bond in the blue component $A_{\mu\nu\rho\sigma}$.
The sum of the single-particle labels in every loop in the diagram is necessarily the sum of all dinstinct single-particle labels contained in the diagram and hence is equal to the argument. Labelling each loop gives
\begin{align}\label{eq:tfmwl5}
\alpha&=\overrightarrow{\mu\sigma\mu}\notag\\
\beta&=\overrightarrow{\sigma\rho\sigma}\notag\\
\gamma&=\overrightarrow{\mu\nu\mu}\notag\\
\delta&=\overrightarrow{\nu\rho\nu}\notag\\
\epsilon&=\overrightarrow{\mu\sigma\rho\mu\nu}\notag\\
\kappa&=\overrightarrow{\mu\sigma\rho\mu\nu}^{'}
\end{align}
so that 
\begin{equation}\label{eq:tfmwl6}
arg (A_{\mu\nu\rho\sigma}A_{\sigma\mu\nu\rho}) = \alpha+\beta+\gamma+\delta+\epsilon+\kappa.
\end{equation}
Since the total number of single-particle states in each bond $\feyn{h}$ is $k$ and the sum of all single-particle labels in each of the solid bonds $\feyn{f}$ must be $m-k$ it becomes possible to write the following ``conservation equations'' wherein the sum of the elements (read single-particle labels) in all loops passing through a given bond must equal the total number of elements in the bond (either $k$ or $m-k$). Now each loop is taken to denote the number of elements in that loop. The conservation equations for the bonds of the red component of the diagram read
\begin{align}
\alpha+\epsilon&=m-k\label{eq:tfmwl7}\\
\beta+\kappa&=k\label{eq:tfmwl8}\\
\gamma+\kappa&=k\label{eq:tfmwl9}\\
\delta+\epsilon&=m-k\label{eq:tfmwl10}
\end{align}
while the conservation equations for bonds belonging to the blue component of the diagram are
\begin{align}
\alpha+\kappa&=k\label{eq:tfmwl11}\\
\beta+\epsilon&=m-k\label{eq:tfmwl12}\\
\gamma+\epsilon&=m-k\label{eq:tfmwl13}\\
\delta+\kappa&=k.\label{eq:tfmwl14}
\end{align}
Plugging equations (\ref{eq:tfmwl7}) and (\ref{eq:tfmwl8}) into (\ref{eq:tfmwl6}) gives
\begin{equation}\label{eq:tfmwl15}
arg (A_{\mu\nu\rho\sigma}A_{\sigma\mu\nu\rho}) = m+\gamma+\delta.
\end{equation}
Now additionally substituting (\ref{eq:tfmwl9}) and (\ref{eq:tfmwl14}) gives
\begin{equation}\label{eq:tfmwl16}
arg (A_{\mu\nu\rho\sigma}A_{\sigma\mu\nu\rho}) = m+2k-2\kappa
\end{equation}
which is maximised when $\kappa=0$, implying that to attain the maximum argument
\begin{align}\label{eq:tfmwl17}
\alpha&=k-\kappa=k\\
\beta&= k-\kappa=k\\
\gamma&=k-\kappa=k\\
\delta&=k-\kappa=k\\
\epsilon&=m-k-\gamma=m-2k.
\end{align}
The final step is simply selecting the single-particle states for each loop from the total number of possible single-particle energy levels $l$ to give the leading order term
\begin{equation}\label{eq:tfmwl18}
A_{\mu\nu\rho\sigma}A_{\sigma\mu\nu\rho} \sim {{l}\choose{m-2k\; k\; k \; k\; k}}.
\end{equation}
This shows, as found already by direct arguments, that the fourth moment is
\begin{align}\label{eq:tfmwl19}
\lim_{N\to\infty}\kappa &= \frac{2 {\frac{1}{N}}A_{\sigma\sigma\rho\rho}A_{\sigma\sigma\mu\mu} + {\frac{1}{N}} A_{\mu\nu\rho\sigma}A_{\sigma\mu\nu\rho}}{{\left ( {\frac{1}{N}}A_{\mu\mu\rho\rho} \right )^2 }}\notag\\
&=\frac{2 {\frac{1}{N}}{l\choose m}\left[{m\choose k} {{l-m+k}\choose{k}} \right]^2 + {\frac{1}{N}} {{l}\choose{m-2k\; k\; k \; k\; k}}}{\left( \frac{1}{N} {l\choose m}{m\choose k}{{l-m+k}\choose{k}} \right)^2}\notag\\
&=2+ \frac{{{m-k}\choose{k}}}{{m\choose k}}.
\end{align}
The above explanation illustrates the complete application of the method of particle diagrams. This involves firstly the notion of an argument which identifies the order of magnitude of terms in powers of $l$. Numerators in a quotient must necessarily attain the same argument as the denominator or they will not survive in the limit $l\to\infty$. Those terms which will not survive can be ignored. Secondly, the method of particle diagrams maps every term into the form of a diagram of nodes and bonds.

Finally, the method of particle diagrams is completed by identifying all loops within a diagram and maximising the argument by maximising the sum of the (number of single-particle labels in) loops as seen in the example of (\ref{eq:tfmwl6}). After this the practitioner must simply select the components of each loop from the set of possible single-particle state labels $\{1, 2, \ldots, l\}$ as seen in the example of (\ref{eq:tfmwl18}).

\begin{center}.....\end{center}

\chapter[The 6'th and 8'th moments]{The 6'th and 8'th Moments}\label{ch:THM}

\section[Sixth Moment]{Sixth Moment}\label{sec:SM}
\noindent If the fourth moment confirmed that the method of particle diagrams can greatly simplify calculations, the sixth moment illustrates the flexibility of the method as well as its ability to scale. For the sixth moment a transition is made from flat particle diagrams to three-dimensional graph-like diagrams. Indeed, the terminology used will be increasingly that of graphs and sets. The $m$-body states will be nodes on the graph. Graphs will contain paths -- sequences of neighbouring nodes. And the concept of loops on the particle diagrams, where the first state in a path equals the last state, will become an increasingly important tool for maximising the argument of attendant binomial expressions as the complexity of the particle diagrams become more complex. The sixth moment of the level density is given by
\begin{equation}\label{eq:s1}h = \frac{\frac{1}{N}\mathrm{tr}(\overline{V_{k}^6})}{\left(\frac{1}{N}\mathrm{tr}(\overline{V_{k}^2})\right)^3}.\end{equation}
Using Wick's theorem as before furnishes
\begin{align}
\mathrm{tr}(\overline{V_k^6})
&=2 \langle\mathrm{tr}
\contraction[2ex]{}{V_k}{}{V_k}
\contraction[2ex]{V_kV_k}{V_k}{}{V_k}
\contraction[2ex]{V_kV_kV_kV_k}{V_k}{}{V_k}
V_k V_k V_k V_kV_k V_k\rangle
+
3\langle\mathrm{tr}
\contraction[2ex]{}{V_k}{}{V_k}
\contraction[4ex]{V_kV_k}{V_k}{V_kV_k}{V_k}
\contraction[2ex]{V_kV_kV_k}{V_k}{}{V_k}
V_k V_k V_k V_kV_k V_k\rangle
+
6\langle\mathrm{tr}
\contraction[2ex]{}{V_k}{}{V_k}
\contraction[2ex]{V_kV_k}{V_k}{V_k}{V_k}
\contraction[4ex]{V_kV_kV_k}{V_k}{V_k}{V_k}
V_k V_k V_k V_kV_k V_k\rangle\nonumber\\
&+
3\langle\mathrm{tr}
\contraction[2ex]{}{V_k}{V_k}{V_k}
\contraction[2ex]{V_k V_k V_k}{V_k}{V_k}{V_k}
\contraction[4ex]{V_k}{V_k}{V_k V_k}{V_k}
V_k V_k V_k V_kV_k V_k\rangle
+\langle
\mathrm{tr}
\contraction[2ex]{}{V_k}{V_kV_k}{V_k}
\contraction[4ex]{V_k}{V_k}{V_kV_k}{V_k}
\contraction[6ex]{V_kV_k}{V_k}{V_kV_k}{V_k}
V_k V_k V_k V_kV_k V_k\rangle.
\end{align}
Here the prefactors indicate the number of equivalent diagrams that can be obtained by cyclic permutation of the trace. Written in terms of $A_{\mu\nu\rho\sigma}$ the summands are
\begin{align}\label{eq:s3} \frac{1}{N}\mathrm{tr}(\overline{V_{k}^6}) & = \frac{1}{N}\big[2~A_{ptqq}A_{tvuu}A_{vpww} + 3A_{ptqq}A_{uwvv}\left ( A_{tpwu} \right ) \notag\\
&+ 6A_{ptqq}\left (A_{twvu}A_{upwv} \right ) + 3A_{putq}A_{qwvt}A_{upwv} \notag\\
& + A_{pvuq}A_{qwvt}A_{tpwu}\big].\end{align}
Terms involving $A$'s with identical first and second (or third and fourth) indices simplify as they give a contribution only if the two other indices coincide as well. For instance for $A_{ptqq}=\langle p|a_{\bj}^{\dag}a_{\bi}|q\rangle\langle q|a_{\bi}^{\dag} a_{\bj}|t\rangle$ to be nonzero the states $a_{\bj}|p\rangle$
and $a_{\bj}|t\rangle$ both have to coincide with $a_{\bi}|q\rangle$. Adding the single-particle states with indices in $\bj$ then gives coinciding $|p\rangle$ and $|t\rangle$. Using this idea as well as the reasoning leading to (\ref{eq:arg3}) the first two terms in (\ref{eq:s3}) can be evaluated to give
\begin{equation}\label{eq:s5} 2~A_{ptqq}A_{tvuu}A_{vpww} + 3~A_{ptqq}A_{uwvv}\left ( A_{tpwu} \right ) = 5{l\choose m}\left[{m\choose k}{{l-m+k}\choose{k}} \right]^3.\end{equation}
\subsubsection{The term $\mathbf{6A_{ptqq}\left (A_{twvu}A_{upwv} \right )}$}
For the third component $6A_{ptqq}\left (A_{twvu}A_{upwv} \right )$ it is similarly required that  $|p\rangle=|t\rangle$ for a non-zero contribution. The particle diagram for this term is illustrated in Fig. \ref{fig:fish}. Note that it is nearly identical to the particle diagram of Fig \ref{fig:combinedqterm} except for the addition of a tail ${p}\feyn{f}{q}$ which adds a factor ${m\choose k}{{l-m+k}\choose k}$ obtained as for Eq. (\ref{eq:arg3}). This additional factor added to the expression we know already gives the leading order expression when $l\to\infty$ as
\begin{equation}\label{eq:s6}A_{ptqq}\left (A_{twvu}A_{upwv} \right )\sim{m\choose k}{{l-m+k}\choose{k}} {{l}\choose{s~r~r~r~r}}.\end{equation}
\begin{figure}[ht!]
\begin{subfigure}{.5\linewidth}
\centering
\begin{tikzpicture}[scale=1.7]
\draw[ultra thick, black] (1.5,0) -- (2.8,0);\node[circle, fill=white, draw=black, text=black,  inner sep=1pt] at (2.8,0) {$p$};
\draw[ultra thick, red] (0,0) -- (0.75,.7);\draw[ultra thick, dashed, red] (0.75,.7) -- (1.5,0);\draw[ultra thick, dashed, red] (0,0) -- (0.75,-.7);\draw[ultra thick, red] (0.75,-.7) -- (1.5,0);
\draw[ultra thick, dashed, blue] (-0.2,0) -- (0.72,.85);\draw[ultra thick, blue] (0.78,.85) -- (1.7,0);\draw[ultra thick, blue] (-.2,0) -- (0.72,-.85);\draw[ultra thick, dashed, blue] (0.78,-.85) -- (1.7,0);
\node[circle, fill=white, draw=black, text=black, inner sep=1.5pt] at (-.1,0) {$\mu$}; \node[circle, fill=white, draw=black, text=black,  inner sep=1.5pt] at (1.6,0) {$\rho$}; \node[circle, fill=white, draw=black, text=black,  inner sep=1.5pt] at (0.75,-.8) {$\nu$}; \node[circle, fill=white, draw=black, text=black,  inner sep=1.5pt] at (0.75,.8) {$\sigma$};
\end{tikzpicture}
\caption{}
\label{fig:fish}
\end{subfigure}
\begin{subfigure}{.5\linewidth}
\centering
\begin{tikzpicture}[scale=1.7]
 \draw[ultra thick, dashed, red] (0,-.5) -- (1.4,-.5);
 \draw[ultra thick, dashed, red] (0,.7) -- (1.4,.7);
\draw[ultra thick, red] (0,-.5) -- (0,.7);
\draw[ultra thick, red] (1.4,-.5) -- (1.4,.7);
\draw[ultra thick, blue] (0,.7) -- (.7,.3);
\draw[ultra thick, black] (.7,.3) -- (1.4,.7);
\draw[ultra thick, blue] (0,-.5) -- (.7,-.9);
\draw[ultra thick, black] (.7,-.9) -- (1.4,-.5);
\draw[ultra thick, dashed, blue] (.65,.3) -- (.65,-.9);
 \draw[ultra thick, dashed, black] (.75,.3) -- (.75,-.9);
\draw[ultra thick, dashed, blue] (-.12,.7) -- (-.12,-.5);
 \draw[ultra thick, dashed, black] (1.52,.7) -- (1.52,-.5);
\node[circle, fill=white, draw=black, text=black,  inner sep=1.5pt] at (-.1,.7) {$v$};
\node[circle, fill=white, draw=black, text=black,  inner sep=1.5pt] at (.7,.3) {$u$};
\node[circle, fill=white, draw=black, text=black,  inner sep=1.5pt] at (1.5,.7) {$t$};
\node[circle, fill=white, draw=black, text=black,  inner sep=1.5pt] at (-.1,-.5) {$w$};
\node[circle, fill=white, draw=black, text=black,  inner sep=1.5pt] at (0.7,-.9) {$p$};
\node[circle, fill=white, draw=black, text=black,  inner sep=1.5pt] at (1.5,-.5) {$q$};
\end{tikzpicture}
\caption{}
\label{fig:prism}
\end{subfigure}
\caption[Particle Siagrams for the $6$'th moment]{(a) illustrates the particle diagram for the term $6A_{ptqq}\left (A_{twvu}A_{upwv} \right )$. This is almost identical to Fig. \ref{fig:qtermfig} but with the equivalent of Fig. \ref{fig:qterm_1} and Fig. \ref{fig:qterm_2} juxtaposed on the same diagram as in Fig \ref{fig:combinedqterm}. The additional tail adds a factor ${m\choose k}{{l-m+k}\choose k}$. (b) illustrates the particle diagram of $3A_{putq}A_{qwvt}A_{upwv}$ where all the bonds implied by the expression are illustrated in a single 3D triangular prism. Using the diagram it becomes simpler to identify the single-particle states which must overlap maximally in order to maximise the argument of the whole sum.

\hrulefill}
\label{fig:principle}
\end{figure}
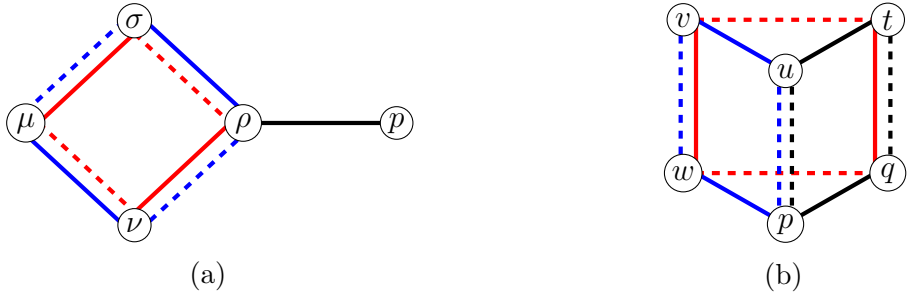

\noindent The argument of this term is $k+2r$ where $ r = min(k, m-k)$ is defined as before, so it will only survive in the limit of $h$ as $l \to \infty$ for $k \le m-k$.

\subsubsection{The term $\mathbf{A_{putq}A_{qwvt}A_{upwv}}$}
The particle diagram for the fourth term $A_{putq}A_{qwvt}A_{upwv}$ of (\ref{eq:s3}) is illustrated in Fig. \ref{fig:prism} as a single three dimensional triangular prism denoting the interrelated conditions on the states that must be satisfied for the expression to be non-zero. The three faces correspond to the three factors. Since these are fermionic states as before, adjacent bonds on the same face cannot share single-particle states. In order to maximise the number of participating single-particle states in the sum, and hence the argument, the four overlaps of $k$ states between $|v\rangle$, $|t\rangle$, $|q\rangle$ and $|w\rangle$ are chosen to be disjoint. This imperative will be shown later using the full method involving loops, but for now it will be momentarily assumed in order to complete the calculation of the result \emph{sans} loops. Next choose the $m-2k$ additional states participating in the bond $v\feyn{f}w$ but not in $v\feyn{h}w$. Note that the many-particle state $|v\rangle$ is given by the combination of the bonds $v\feyn{f}w$ and $v\feyn{h}t$, and equivalently by the combination of $v\feyn{h}w$ and $v\feyn{f}u$. These combinations can coincide only if $v\feyn{f}u$ involves the overlap $v\feyn{h}t$ as well as the aforementioned $m-2k$ states. Hence the $m-2k$ states participate not only in $|v\rangle$ and $|w\rangle$ but also in $|u\rangle$. An analogous argument shows that they also have to be included in $|p\rangle$, and thus in all many-particle states of the ``left'' face in Fig. \ref{fig:prism}. To fully determine $|u\rangle$ and $|p\rangle$ it is still necessary to choose the $k$ states participating in the ``left'' of the two bonds $u\feyn{h}p$, which altogether gives $m-k=m-2k+k$ states in addition to the original four sets of $k$ states. Considering the ``right'' face  in an analogous way one obtains the same choice of $m-k$ states but now broken down differently into a choice of $m-2k$ and a choice of $k$ states. Hence our choice is restricted to selecting four sets of $k$ states and one set of $m-k$ states from the $l$ available states, and then splitting the latter into sets of $m-2k$ and $k$ states in two independent ways. This gives
\allowdisplaybreaks[1]
\begin{figure}
\begin{subfigure}{.33\textwidth}
  \centering
  \includegraphics[width=.8\linewidth]{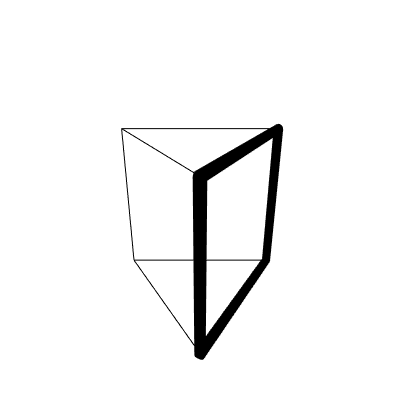}
  \caption{The loop \normalsize$\overrightarrow{utqpu}$}
  \label{fig:L}
\end{subfigure}%
\begin{subfigure}{.33\textwidth}
  \centering
  \includegraphics[width=.8\linewidth]{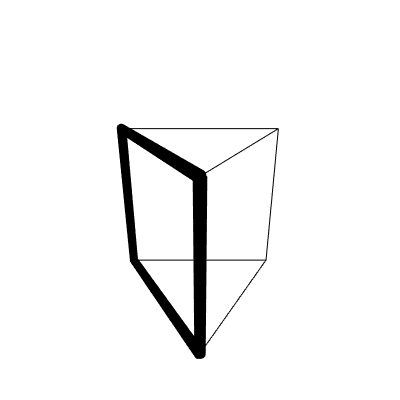}
  \caption{The loop \normalsize$\overrightarrow{uvwpu}$}
  \label{fig:R}
\end{subfigure}%
\begin{subfigure}{.33\textwidth}
  \centering
  \includegraphics[width=.8\linewidth]{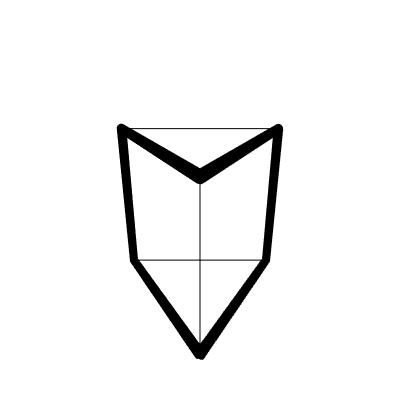}
  \caption{The loop \normalsize$\overrightarrow{vutqpwv}$}
  \label{fig:S}
\end{subfigure}\\
\begin{subfigure}{.33\textwidth}
  \centering
  \includegraphics[width=.8\linewidth]{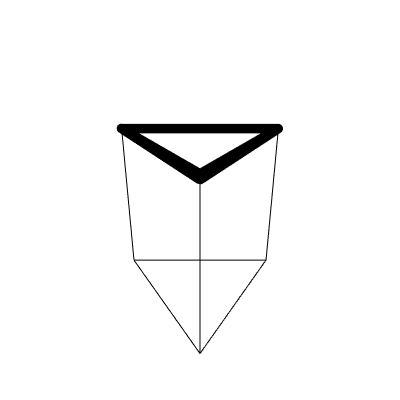}
  \caption{The loop \normalsize$\overrightarrow{vutv}$}
 \label{fig:T}
\end{subfigure}%
\begin{subfigure}{.33\textwidth}
  \centering
  \includegraphics[width=.8\linewidth]{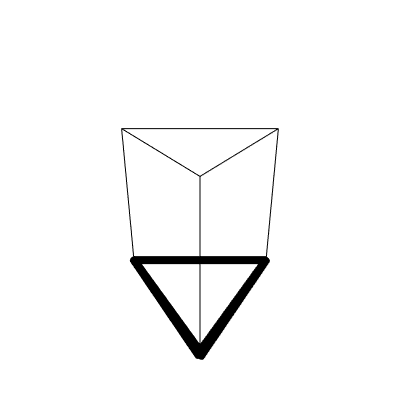}
  \caption{The loop \normalsize$\overrightarrow{qpwq}$}
  \label{fig:B}
\end{subfigure}%
\begin{subfigure}{.33\textwidth}
  \centering
  \includegraphics[width=.8\linewidth]{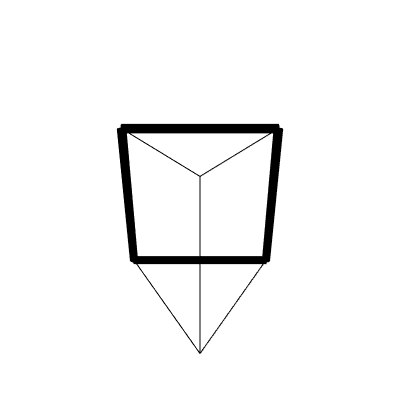}
  \caption{The loop \normalsize$\overrightarrow{vtqwv}$}
 \label{fig:D}
\end{subfigure}\\
\begin{subfigure}{.33\textwidth}
  \centering
  \includegraphics[width=.8\linewidth]{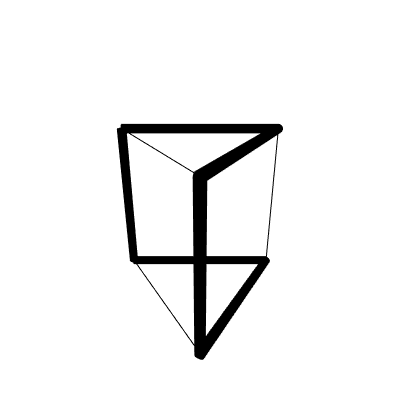}
  \caption{The loop \normalsize$\overrightarrow{qputvwq}$}
  \label{fig:DL}
\end{subfigure}%
\begin{subfigure}{.33\textwidth}
  \centering
  \includegraphics[width=.8\linewidth]{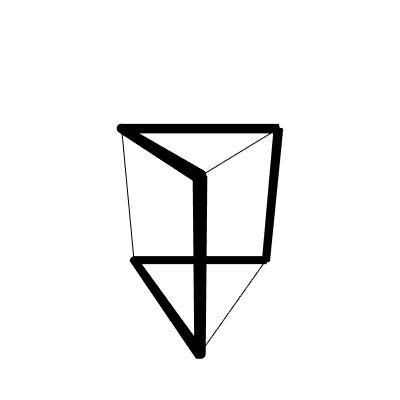}
  \caption{The loop \normalsize$\overrightarrow{wpuvtqw}$}
 \label{fig:DR}
\end{subfigure}%
\begin{subfigure}{.33\textwidth}
  \centering
  \includegraphics[width=.8\linewidth]{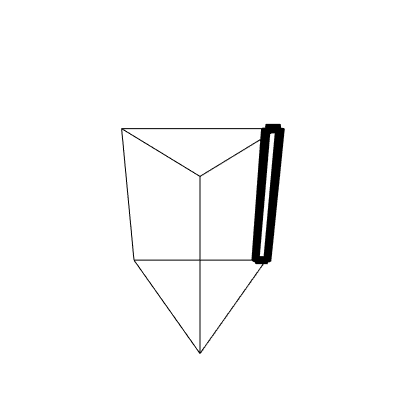}
  \caption{The loop \normalsize$\overrightarrow{tqt}$}
  \label{fig:TBR}
\end{subfigure}\\
\begin{subfigure}{.33\textwidth}
  \centering
  \includegraphics[width=.8\linewidth]{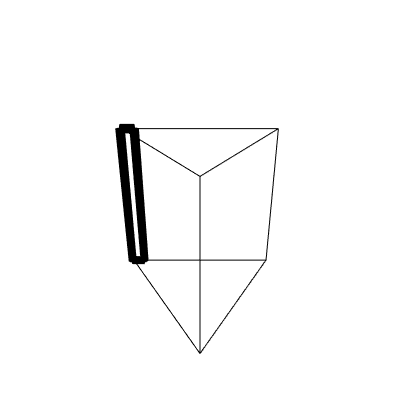}
  \caption{The loop \normalsize$\overrightarrow{vwv}$}
 \label{fig:TBL}
\end{subfigure}%
\begin{subfigure}{.33\textwidth}
  \centering
  \includegraphics[width=.8\linewidth]{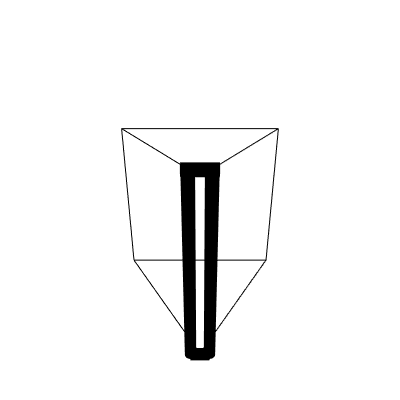}
  \caption{The loop \normalsize$\overrightarrow{pup}$}
  \label{fig:TBC}
\end{subfigure}
\caption[Loops Illustration]{Illustration of all possible loops within the prism-like particle diagram of Fig \ref{fig:prism} corresponding to the term $A_{putq}A_{qwvt}A_{upwv}$.}
\label{fig:paths_prism}
\end{figure}
\begin{equation}\label{eq:s7}A_{putq}A_{qwvt}A_{upwv}\sim{l \choose k\;k\;k\;k\;m-k}{m-k\choose k}^2.\end{equation}
How to calculate this utilising loops? As before the first step is simply to identify all possible loops in the diagram, in this case the prism-like diagram of Fig \ref{fig:prism}. These loops are illustrated in Fig \ref{fig:paths_prism}. Giving each loop a label indicating the number of single-particle state labels contained in the loop gives
\begin{align}
\alpha &=\overrightarrow{utqpu} & \eta &=\overrightarrow{qputvwq}\notag\\
\beta &=\overrightarrow{uvwpu} & \phi  &=\overrightarrow{wpuvtqw}\notag\\
\gamma &=\overrightarrow{vutqpwv} & \varphi &=\overrightarrow{tqt}\notag\\
\delta &=\overrightarrow{vutv} & \pi &=\overrightarrow{vwv}\notag\\
\epsilon &=\overrightarrow{qpwq} & \omega &=\overrightarrow{pup}\notag\\
\kappa &=\overrightarrow{vtqwv} & &
\end{align}
Since the number of labels passing through any bond $\feyn{h}$ must sum to $k$ and the number of labels passing through any solid bond $\feyn{f}$ must sum to $m-k$ it is straightforward to write down the following conservation equations which restrict the values of the loops to
\begin{align}
\alpha+\eta+\omega&=k\label{eq:s8}\\
\alpha+\gamma+\delta+\eta&=m-k\label{eq:s9}\\
\alpha+\gamma+\varphi&=m-k\label{eq:s10}\\
\alpha+\gamma+\epsilon+\eta&=m-k\label{eq:s11}\\
\notag\\
\beta +\phi+\omega&=k\label{eq:s12}\\
\beta+\gamma+\delta+\phi&=m-k\label{eq:s13}\\
\beta+\gamma+\pi&=m-k\label{eq:s14}\\
\beta+\gamma+\epsilon+\phi&=m-k\label{eq:s15}\\
\notag\\
\epsilon+\kappa+\eta+\phi&=k.\label{eq:s16}
\end{align}
The argument of the diagram is simply the sum of all degrees of freedom, which is the sum of the number of labels in each loop
\begin{equation}\label{eq:s17}
arg(A_{putq}A_{qwvt}A_{upwv}) = \alpha+\beta+\gamma+\delta+\epsilon+\kappa+\eta+\phi+\varphi+\pi+\omega.
\end{equation}
Substituting (\ref{eq:s9}) and (\ref{eq:s12}) yields
\begin{equation}\label{eq:s18}
arg(A_{putq}A_{qwvt}A_{upwv}) = m+\epsilon+\kappa+\varphi+\pi
\end{equation}
and finally inserting  (\ref{eq:s16}) gives
\begin{equation}\label{eq:s19}
arg(A_{putq}A_{qwvt}A_{upwv}) = m+k+\varphi+\pi-\eta-\phi
\end{equation}
so that the maximal argument $m+3k$ is achieved when $\varphi=\pi=k$ and $\eta=\phi=0$. Plugging these values back into (\ref{eq:s8} -- \ref{eq:s16}) reveals that $\varphi=\pi=\delta=\epsilon=k$ and $\kappa=\eta=\phi=0$. Additionally
\begin{align}
\alpha+\gamma&=m-2k\notag\\
\beta+\gamma&=m-2k\notag\\
\alpha+\omega&=k\notag\\
\beta + \omega&=k\label{eq:s20}
\end{align}
so that $\alpha, \beta$ and $\gamma$ are functions of $\omega$ and
\begin{align}
A_{putq}A_{qwvt}A_{upwv} &\sim \sum_{\omega}{{l}\choose{k\; k\; k\;k \; \omega\;  \alpha(\omega) \; \beta(\omega) \; \gamma(\omega)}}\notag\\
&= \sum_{\omega}{{l}\choose{k\; k\; k\;k\;  \omega\;  k-\omega\;  k-\omega\;  m-3k+\omega}}\notag\\
&= \sum_{\omega}{{l}\choose{k\; k\; k\;k\;  m-k}}{{m-k}\choose{k}}{{k}\choose{\omega}}{{m-2k}\choose{k-\omega}}\notag\\
&= {l \choose k\;k\;k\;k\;m-k}{m-k\choose k}^2
\label{eq:s21}
\end{align}
where the final step uses Vandermonde's identity $\sum_{\omega} = {{k}\choose{\omega}}{{m-2k}\choose{k-\omega}} = {m-k\choose k}$. This is the same result claimed earlier in (\ref{eq:s7}).

\subsubsection{The term $\mathbf{A_{pvuq}A_{qwvt}A_{tpwu}}$}
For the final term $A_{pvuq}A_{qwvt}A_{tpwu}$ of (\ref{eq:s3}) illustrated in Fig. \ref{fig:diamond} it is useful first to notice that the argument never exceeds $2m$. This can be observed on sight since the states $|w\rangle$ and $|t\rangle$ together define all the single-particle states in $|u\rangle$, $|v\rangle$, $|p\rangle$ and $|q\rangle$. And with an argument never exceeding $2m$ it can be concluded that this term will only contribute to the limit value of $h$ for $3k\le m$. Since $|w\rangle$ and $|t\rangle$ together determine all the single-particle states in the diagram and hence all the $m$-body compound states, there will be maximal degrees of freedom in the choice of the single-particle states in $|w\rangle$ and $|t\rangle$ when $|w\rangle$ and $|t\rangle$ overlap minimally. This also determines when the argument for the entire sum (diagram) is maximal. Single-particle states shared by both $|w\rangle$ and $|t\rangle$ must necessarily be manifested in the bonds which make up the diagram, so for instance if a single-particle state $\alpha$ is in both $|w\rangle$ and $|t\rangle$, then it must have ``travelled upwards'' from the bottom node of the diagram at $|w\rangle$, proceeded ``through'' (read: be a member of) the bonds $w\feyn{f}p$, $w\feyn{f}v$, $w\feyn{h}q$ or $w\feyn{h}u$ and continued towards the uppermost node of the diagram $|t\rangle$ ``through'' bonds $q\feyn{f}t$, $u\feyn{f}t$, $p\feyn{h}t$ or $v\feyn{h}t$. More concretely, taking the bond $w\feyn{f}p$ to represent the set of $m-k$ single-particle states in both $|w\rangle$ and $|p\rangle$ but not in $|u\rangle$, the bond $w\feyn{h}u$ to represent the set of $k$ single-particle states in both  $|w\rangle$ and $|u\rangle$ but not in $|p\rangle$, and so on, it can be seen that for any single-particle state $\alpha$ in both  $|w\rangle$ and $|t\rangle$ it follows both that
\begin{equation}\alpha \in w\feyn{f}p \cup w\feyn{f}v \cup w\feyn{h}q \cup w\feyn{h}u\\
\end{equation}
and
\begin{equation}\alpha \in q\feyn{f}t \cup u\feyn{f}t \cup p\feyn{h}t \cup v\feyn{h}t
\end{equation}
whereas for a given single-particle state $\tilde{\alpha}$ in $|w\rangle$ but not in $|t\rangle$ it follows both that
\begin{equation}\tilde{\alpha} \in w\feyn{f}p \cup w\feyn{f}v \cup w\feyn{h}q \cup w\feyn{h}u\\
\end{equation}
and
\begin{equation}\tilde{\alpha} \notin q\feyn{f}t \cup u\feyn{f}t \cup p\feyn{h}t \cup v\feyn{h}t.
\end{equation}
\begin{figure}[!b]
\hrulefill
\vspace{1cm}

\centering
\begin{tikzpicture}
\draw[ultra thick, black] (0,0) -- (.9,-1);
\draw[ultra thick, black] (1.75,0) -- (2.7,-1);
\draw[ultra thick, dashed, black] (1.75,0) -- (.9,-1);
\draw[ultra thick, dashed, black] (0,0) -- (2.7,-1);
\draw[ultra thick, blue] (.9,-1) -- (1.4,2);
\draw[ultra thick, dashed, blue] (2.7,-1) -- (1.4,2);
\draw[ultra thick, dashed, red] (0,0) -- (1.4,2);
\draw[ultra thick, red] (1.75,0) -- (1.4,2);
\draw[ultra thick, dashed, blue] (.9,-1) -- (1.4,-2.9);
\draw[ultra thick, blue] (2.7,-1) -- (1.4,-2.9);
\draw[ultra thick, red] (0,0) -- (1.4,-2.9);
\draw[ultra thick, dashed, red] (1.75,0) -- (1.4,-2.9);
\node[circle, fill=white, draw=black, text=black,  inner sep=1pt] at (1.4,2) {$t$};
\node[circle, fill=white, draw=black, text=black,  inner sep=1pt] at (0,0) {$p$};
\node[circle, fill=white, draw=black, text=black,  inner sep=1pt] at (1.75,0) {$u$};
\node[circle, fill=white, draw=black, text=black,  inner sep=1pt] at (2.7,-1) {$v$};
\node[circle, fill=white, draw=black, text=black,  inner sep=1pt] at (.9,-1) {$q$};
\node[circle, fill=white, draw=black, text=black,  inner sep=1pt] at (1.4,-2.9) {$w$};
\end{tikzpicture}
\caption[Octohedron particle diagram]{The particle diagram for the term $A_{pvuq}A_{qwvt}A_{tpwu}$ which takes the form of a regular octahedron, or two square pyramids with a shared base determined by the plane on which the sub-diagram for $A_{pvuq}$ is illustrated. The states $|t\rangle$ and $|w\rangle$ together determing the states $|v\rangle$ and $|q\rangle$ through the bonds defined by $A_{qwvt}$ just as they determine the states $|u\rangle$ and $|p\rangle$ through the bonds defined by $A_{tpwu}$. Hence the maximal degrees of freedom in our choice of single-particle states from the set of $l$ possibilities is bounded by $2m$.}
\label{fig:diamond}
\end{figure}
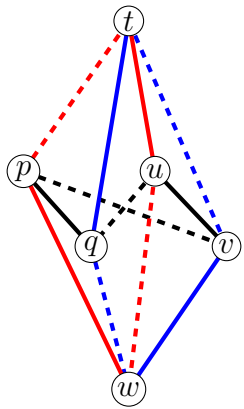
This implies that states such as $\alpha$ in both $|w\rangle$ and $|t\rangle$ can be found in a connected non-repeating sequence of neighbouring nodes in the diagram starting at $|w\rangle$ passing through $|t\rangle$ and looping back to $|w\rangle$. Likewise for a single-particle state $\tilde{\alpha}$ which is in $|w\rangle$ but not in $|t\rangle$ one can trace a non self-crossing loop of bonds in the diagram starting from $|w\rangle$ and returning back to $|w\rangle$ without passing through the node defined by the state $|t\rangle$. In conclusion, since it is imperative to maximise the number of single-particle states that are in $|w\rangle$ but not $|t\rangle$ and vice versa, and since the number of single-particle states which are an element of a loop excluding the node $|t\rangle$ is the complement of the number of loops passing through the states (nodes) $|w\rangle$ and $|t\rangle$, of all the particle diagrams taking the form of Fig. \ref{fig:diamond} those with the largest argument are the ones consisting of the maximum possible number of loops from $|w\rangle$ back to $|w\rangle$ without passing through $|t\rangle$, and from $|t\rangle$ back to $|t\rangle$ without passing through $|w\rangle$. Any diagram with even one less single-particle state than the maximal number partaking in these loops will have an argument less than the diagrams containing the maximal number of loops, so these do not have to be calculated. As before it is prohibited to include a path containing bonds which are mutually exclusive; neighbouring bonds from the same factor of $A$. In Fig. \ref{fig:diamond} for example the bonds $w\feyn{h}q$ and $w\feyn{f}v$ are mutually exclusive as they both appear in the diagram due to the same factor $A_{qwvt}$ in $A_{pvuq}A_{qwvt}A_{tpwu}$. The $k$ single-particle states in $w\feyn{h}q$ are therefore the complement of the $m-k$ single-particle states in $w\feyn{f}v$, the union being the set of single-particle states contained in $|w\rangle$. On the other hand the bonds $w\feyn{h}q$ and $w\feyn{f}p$ appear in the diagram due to separate factors $A_{qwvt}$ and $A_{tpwu}$ respectively so these are allowed to share single-particle states. Hence the loop $\overrightarrow{wquvw}$ is not a valid loop for the diagram of Fig. \ref{fig:diamond} as it contains mutually exclusive bonds, whereas $\overrightarrow{wqpw}$ is a valid loop. To summarise, a single-particle state which is an element of both $|w\rangle$ and $|t\rangle$ must necessarily be an element of every bond in a loop passing through $|w\rangle$ and $|t\rangle$. A single-particle state in $|w\rangle$ and not in $|t\rangle$ must necessarily be an element of every bond in a loop from $|w\rangle$ to $|w\rangle$ not passing through $|t\rangle$ and conversely for a single-particle state in $|t\rangle$ but not in $|w\rangle$. Since the argument of the diagram is maximised by maximising the number of single-particle states contained in every bond of a loop passing through only one of $|w\rangle$ or $|t\rangle$, the only diagrams which will contribute in the asymptotic regime are those where $k$ single-particle states from each of the bonds in the loop
\begin{equation}\label{eq:loop1}p_1 = \overrightarrow{wqpw}
\end{equation}
are equal to the $k$ single-partical states in the bond $w\feyn{h}q$, each of the bonds in the loop
\begin{equation}\label{eq:loop2}p_2 = \overrightarrow{wuvw}
\end{equation}
contain $k$ single-particle states identical to the single-particle states in the bond $w\feyn{h}u$, and so on for all the loops $p_3 = \overrightarrow{wpvw}$ with the bond $p\feyn{h}v$, $p_4 = \overrightarrow{tpqt}$ with the bond $t\feyn{h}p$, $p_5 = \overrightarrow{tuvt}$ with the bond $v\feyn{h}t$ and $p_6 = \overrightarrow{tuqt}$ with the bond $u\feyn{h}q$. Borrowing set notation again one has for the case of (\ref{eq:loop1}) for example, that
\begin{equation} w\feyn{h}q \cap p\feyn{f}q = w\feyn{h}q
\end{equation}
and
\begin{equation} w\feyn{h}q \cap w\feyn{f}p = w\feyn{h}q.
\end{equation}
In this way the $3k$ single-particle states from $|w\rangle$ which participate in the loops $p_1$, $p_2$ and $p_3$ do not overlap with $|t\rangle$ and the $3k$ single-particle states from $|t\rangle$ which participate in the loops $p_4$, $p_5$ and $p_6$ do not overlap with $|w\rangle$. The remaining $m-3k$ single-particle states from $|w\rangle$ and $|t\rangle$ which cannot be included in loops must necessarily be included in a loop through $|w\rangle$ and $|t\rangle$. That is, the number of overlapping single-particle states between the $m$-body states $|w\rangle$ and $|t\rangle$ is $m-3k$. This has profound implications for the value of the sixth moment and we will see a related feature appear in the eighth moment as well. To make the comparison with the corresponding calculation (\ref{eq:arg11}) for the fourth moment more explicit, define $r=min(k,m-k)$ as before and $\tilde{s} = m-3r$. We then have
\begin{equation}
A_{pvuq}A_{qwvt}A_{tpwu} \sim {{l}\choose{\tilde{s}~r~r~r~r~r~r}}
\label{eq:sixth_sr}
\end{equation}
It should be noted that this time a relatively complex particle diagram has been evaluated with loops, albeit not by identifying all loops and maximising the argument as before, but instead by using arguments unique to the particular diagram at hand to determine what the size of each loop must be. It is useful to keep in mind that this is sometimes possible for larger particle diagrams, as it offers a shortcut to a result which would otherwise require a potentially far more laborious calculation. From (\ref{eq:sixth_sr}) it follows that
\begin{equation}
\lim_{N\to\infty} \frac{{\frac{1}{N}}A_{pvuq}A_{qwvt}A_{tpwu}}{\left(\frac{1}{N}\mathrm{tr}(\overline{V_{k}^2})\right)^3} = \frac{{{m-k}\choose{k}}{{m-2k}\choose{k}}}{{m\choose k}^2}.\label{eq:s7.2}
\end{equation}
Taking the quotient for $h$ using the above expressions (\ref{eq:s5}), (\ref{eq:s6}) (\ref{eq:s21}) and (\ref{eq:s7.2})  gives the final result for the sixth moment
\begin{equation}\label{eq:s8} {\lim_{N\to\infty}}h = 5 + \frac{{{m-k}\choose{k}}{{m-2k}\choose{k}}}{{m\choose k}^2} + 6\frac{{{m-k}\choose{k}}}{{m\choose k}} + 3\frac{{{m-k}\choose{k}}^2}{{m\choose k}^2}.\end{equation}
\begin{figure}[h]
\centering
\includegraphics[scale=.7]{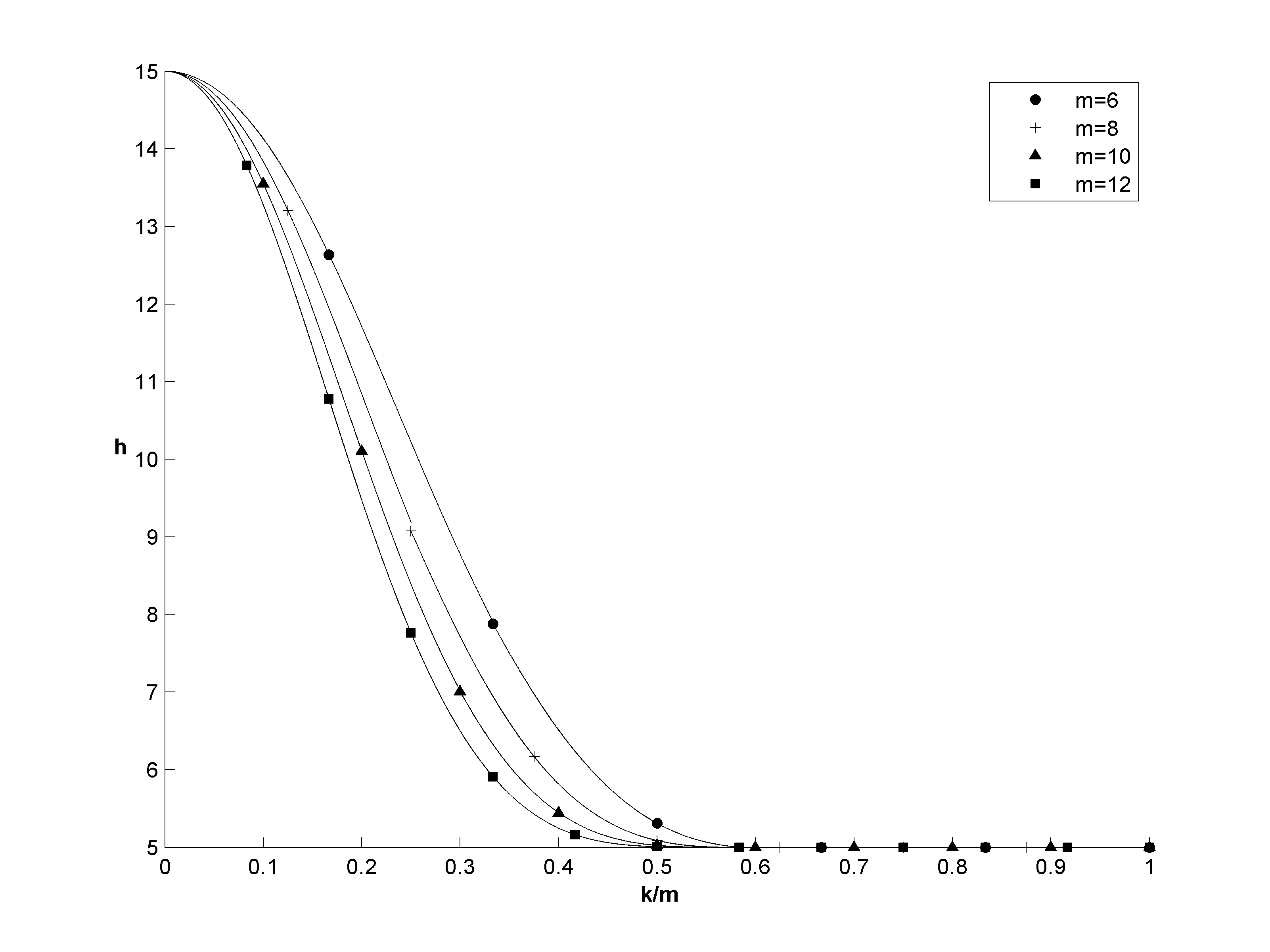}
\caption[The sixth moment $h$ against $k/m$]{$h$ against $k/m$ for $m=6, 8, 10, 12$. Once again we see a domain $k/m > \frac{1}{2}$ for which the sixth moment takes the semi-circular value $h=5$ and a transition thereafter towards a gaussian moment, $h=15$ at $k=0$. Higher values of $m$ give faster convergence to the semi-circular moment.}
\label{fig:sixth_moment}
\end{figure}

\section[The Eighth Moment]{The Eighth Moment}
\noindent The eighth moment will exhibit the same features as the lower moments; a transition starting immediately after $2k=m$ from a semi-circular moment $\tau = 14$, to a gaussian moment $\tau=105$ and a single term of $\mathrm{tr}~\overline{V_k^8}$ taking the form
\begin{equation}{{l}\choose{\tilde{s}~r~r~r~r~r~r~r~r}}\end{equation}
analogous to (\ref{eq:arg11}) and (\ref{eq:sixth_sr}) and with $\tilde{s} := m-\frac{n}{2}k = m-4k$. Of course we must again calculate products of ensemble averaged pairs of the matrix elements of ${H_k}$ for which there are now $\frac{8!}{2^{8/2} (\frac{8}{2})!} = 105$ components in the sum. As may be expected, the method of particle diagrams can be applied analogously to the lower moments though with the diagrams now increasing in size and complexity. Some of these will be identical to particle diagrams which have already been seen in calculations of the lower moments, but with the addition of tails leading to extra factors of ${{l-m+k}\choose k}{m\choose k}$. This feature was seen in the transition from figure {\ref{fig:qtermfig}} to figure {\ref{fig:fish}} for example. For progressively higher powers of ${H_k}$ the particle diagrams become progressively larger and, as for the diagrams illustrated in Fig \ref{fig:prism} and Fig. \ref{fig:diamond} one can think of some of these as graphs in three dimensions, rather than as a flat structure. The larger structure of the subsequent diagrams makes it more difficult to identify just two states which determine all others (as was the case in Fig. \ref{fig:qtermfig} and Fig. \ref{fig:diamond}). Instead we will see diagrams where three or more states determine all others (as was the case in Fig. \ref{fig:prism}). In these cases calculations will rely more on the analytical method of solving the diagrams utilising loops as first introduced in section \ref{sec:TFMWL}. Calculating all loops in a diagram and maximising the argument over these paths is a method which works for all diagrams. However, using the method blindly can lead to unecessary work, as there are frequently shortcuts to determining the answer, as was seen in the case of Fig \ref{fig:diamond}.

\noindent The normalised eighth moment is given by
\begin{equation}\label{eq:s1}\tau = \frac{\frac{1}{N}\mathrm{tr}(\overline{V_{k}^8})}{\left(\frac{1}{N}\mathrm{tr}(\overline{V_{k}^2})\right)^4}.\end{equation}
In this case Wick's theorem gives (dropping the indices $k$)
\begin{align}
\label{a}
\mathrm{tr}(\overline{\Vk^8})&=
2\langle\mathrm{tr}
\contraction[2ex]{}{\Vk}{}{\Vk}
\contraction[2ex]{\Vk\Vk}{\Vk}{}{\Vk}
\contraction[2ex]{\Vk\Vk\Vk\Vk}{\Vk}{}{\Vk}
\contraction[2ex]{\Vk\Vk\Vk\Vk\Vk\Vk}{\Vk}{}{\Vk}
\Vk \Vk \Vk \Vk\Vk \Vk\Vk\Vk\rangle
+
8\langle\mathrm{tr}
\contraction[2ex]{}{\Vk}{}{\Vk}
\contraction[2ex]{\Vk\Vk}{\Vk}{}{\Vk}
\contraction[4ex]{\Vk\Vk\Vk\Vk}{\Vk}{\Vk\Vk}{\Vk}
\contraction[2ex]{\Vk\Vk\Vk\Vk\Vk}{\Vk}{}{\Vk}
\Vk \Vk \Vk \Vk\Vk \Vk\Vk\Vk\rangle
+4\langle\mathrm{tr}
\contraction[2ex]{\Vk\Vk\Vk}{\Vk}{}{\Vk}
\contraction[4ex]{\Vk\Vk}{\Vk}{\Vk\Vk}{\Vk}
\contraction[6ex]{\Vk}{\Vk}{\Vk\Vk\Vk\Vk}{\Vk}
\contraction[8ex]{}{\Vk}{\Vk\Vk\Vk\Vk\Vk\Vk}{\Vk}
\Vk \Vk \Vk \Vk\Vk \Vk\Vk\Vk\rangle
\\
&+8\langle\mathrm{tr}
\contraction[2ex]{}{\Vk}{}{\Vk}
\contraction[2ex]{\Vk\Vk}{\Vk}{}{\Vk}
\contraction[4ex]{\Vk\Vk\Vk\Vk\Vk}{\Vk}{\Vk}{\Vk}
\contraction[2ex]{\Vk\Vk\Vk\Vk}{\Vk}{\Vk}{\Vk}
\Vk \Vk \Vk \Vk\Vk \Vk\Vk\Vk\rangle
+
8\langle\mathrm{tr}
\contraction[4ex]{}{\Vk}{\Vk\Vk}{\Vk}
\contraction[2ex]{\Vk}{\Vk}{}{\Vk}
\contraction[4ex]{\Vk\Vk\Vk\Vk\Vk}{\Vk}{\Vk}{\Vk}
\contraction[2ex]{\Vk\Vk\Vk\Vk}{\Vk}{\Vk}{\Vk}
\Vk \Vk \Vk \Vk\Vk \Vk\Vk\Vk\rangle
+8\langle\mathrm{tr}
\contraction[2ex]{}{\Vk}{}{\Vk}
\contraction[2ex]{\Vk\Vk\Vk}{\Vk}{}{\Vk}
\contraction[6ex]{\Vk\Vk\Vk\Vk\Vk}{\Vk}{\Vk}{\Vk}
\contraction[4ex]{\Vk\Vk}{\Vk}{\Vk\Vk\Vk}{\Vk}
\Vk \Vk \Vk \Vk\Vk \Vk\Vk\Vk\rangle\nonumber\\
&+4\langle\mathrm{tr}
\contraction[2ex]{}{\Vk}{}{\Vk}
\contraction[2ex]{\Vk\Vk\Vk\Vk}{\Vk}{}{\Vk}
\contraction[6ex]{\Vk\Vk\Vk}{\Vk}{\Vk\Vk\Vk}{\Vk}
\contraction[4ex]{\Vk\Vk}{\Vk}{\Vk\Vk\Vk}{\Vk}
\Vk \Vk \Vk \Vk\Vk \Vk\Vk\Vk\rangle\\
&\label{c}+8\langle\mathrm{tr}
\contraction[2ex]{}{\Vk}{}{\Vk}
\contraction[2ex]{\Vk\Vk}{\Vk}{\Vk\Vk}{\Vk}
\contraction[4ex]{\Vk\Vk\Vk}{\Vk}{\Vk\Vk}{\Vk}
\contraction[6ex]{\Vk\Vk\Vk\Vk}{\Vk}{\Vk\Vk}{\Vk}
\Vk \Vk \Vk \Vk\Vk \Vk\Vk\Vk\rangle\\
&\label{d}+8\langle\mathrm{tr}
\contraction[2ex]{}{\Vk}{}{\Vk}
\contraction[2ex]{\Vk\Vk}{\Vk}{\Vk}{\Vk}
\contraction[2ex]{\Vk\Vk\Vk \Vk \Vk}{\Vk}{\Vk}{\Vk}
\contraction[4ex]{\Vk\Vk\Vk}{\Vk}{\Vk \Vk}{\Vk}
\Vk \Vk \Vk \Vk\Vk \Vk\Vk\Vk\rangle
+16\langle\mathrm{tr}
\contraction[2ex]{\Vk}{\Vk}{}{\Vk}
\contraction[4ex]{}{\Vk}{\Vk\Vk\Vk}{\Vk}
\contraction[6ex]{\Vk\Vk\Vk}{\Vk}{\Vk\Vk}{\Vk}
\contraction[4ex]{\Vk\Vk\Vk\Vk\Vk}{\Vk}{\Vk}{\Vk}
\Vk \Vk \Vk \Vk\Vk \Vk\Vk\Vk\rangle\\
&\label{e}+4\langle\mathrm{tr}
\contraction[2ex]{}{\Vk}{\Vk}{\Vk}
\contraction[4ex]{\Vk}{\Vk}{\Vk}{\Vk}
\contraction[2ex]{\Vk \Vk \Vk \Vk}{\Vk}{\Vk}{\Vk}
\contraction[4ex]{\Vk \Vk \Vk \Vk\Vk}{\Vk}{\Vk}{\Vk}
\Vk \Vk \Vk \Vk\Vk \Vk\Vk\Vk\rangle
\\
&+\label{f}
4\langle\mathrm{tr}
\contraction[8ex]{}{\Vk}{\Vk\Vk\Vk}{\Vk}
\contraction[2ex]{\Vk\Vk\Vk}{\Vk}{\Vk}{\Vk}
\contraction[4ex]{\Vk\Vk}{\Vk}{\Vk\Vk\Vk}{\Vk}
\contraction[6ex]{\Vk}{\Vk}{\Vk\Vk\Vk\Vk\Vk}{\Vk}
\Vk \Vk \Vk \Vk\Vk \Vk\Vk\Vk\rangle
+2\langle\mathrm{tr}
\contraction[4ex]{}{\Vk}{\Vk\Vk\Vk\Vk}{\Vk}
\contraction[2ex]{\Vk}{\Vk}{\Vk\Vk}{\Vk}
\contraction[8ex]{\Vk\Vk}{\Vk}{\Vk\Vk\Vk\Vk}{\Vk}
\contraction[6ex]{\Vk\Vk\Vk}{\Vk}{\Vk\Vk}{\Vk}
\Vk \Vk \Vk \Vk\Vk \Vk\Vk\Vk\rangle
+
4\langle\mathrm{tr}\contraction[4ex]{}{\Vk}{\Vk\Vk\Vk\Vk}{\Vk}
\contraction[2ex]{\Vk}{\Vk}{\Vk\Vk}{\Vk}
\contraction[8ex]{\Vk\Vk\Vk}{\Vk}{\Vk\Vk\Vk}{\Vk}
\contraction[6ex]{\Vk\Vk}{\Vk}{\Vk\Vk\Vk}{\Vk}
\Vk \Vk \Vk \Vk\Vk \Vk\Vk\Vk\rangle\nonumber\\
&+
8\langle\mathrm{tr}
\contraction[2ex]{}{\Vk}{\Vk}{\Vk}
\contraction[4ex]{\Vk}{\Vk}{\Vk\Vk}{\Vk}
\contraction[2ex]{\Vk\Vk\Vk}{\Vk}{\Vk\Vk}{\Vk}
\contraction[4ex]{\Vk\Vk\Vk\Vk\Vk}{\Vk}{\Vk}{\Vk}
\Vk \Vk \Vk \Vk\Vk \Vk\Vk\Vk\rangle
+
8\langle\mathrm{tr}
\contraction[4ex]{}{\Vk}{\Vk}{\Vk}
\contraction[2ex]{\Vk}{\Vk}{\Vk\Vk\Vk}{\Vk}
\contraction[4ex]{\Vk\Vk\Vk}{\Vk}{\Vk\Vk}{\Vk}
\contraction[6ex]{\Vk\Vk\Vk\Vk}{\Vk}{\Vk\Vk}{\Vk}
\Vk \Vk \Vk \Vk\Vk \Vk\Vk\Vk\rangle
+\langle\mathrm{tr}
\contraction[2ex]{}{\Vk}{\Vk\Vk\Vk}{\Vk}
\contraction[4ex]{\Vk}{\Vk}{\Vk\Vk\Vk}{\Vk}
\contraction[6ex]{\Vk\Vk}{\Vk}{\Vk\Vk\Vk}{\Vk}
\contraction[8ex]{\Vk\Vk\Vk}{\Vk}{\Vk\Vk\Vk}{\Vk}
\Vk \Vk \Vk \Vk\Vk \Vk\Vk\Vk\rangle
\end{align}
where the prefactors indicate the number of equivalent contributions that can be obtained by cyclic permutation of the trace. The prefactor 16 in the second term of (\ref{d}) also incorporates equivalent contributions obtained by reverting the order of $V$'s. Reassuringly the prefactors sum to $105=(8-1)!!$, the overall number of possible contractions between eight elements. Writing these in terms of (\ref{eq:ee46}) it can be shown that
\begin{align}\label{eq:s1.5}\mathrm{tr} (\overline{V_k^8}) &= 14 A_{ppuu}A_{uuww}A_{wwvv}A_{vvqq} + 28 A_{ttqq}A_{qqpp}A_{puvw}A_{pwvu}\notag\\
&~~~~~ + 24 A_{eeww}A_{wvup}A_{upqt}A_{vtqw} + 4 A_{cewu}A_{uwvt}A_{ttvqp}A_{pceq}\notag\\
&~~~~~ + 2 A_{petq}A_{qtuv}A_{euwc}A_{cwvp} + 8 A_{tpce}A_{evqc}A_{puwq}A_{uwvt}\notag\\
&~~~~~ + 4 A_{tucp}A_{tpcu}A_{cewv}A_{cvwe} + 8 A_{qwvt}A_{uwpt}A_{pvuq}A_{ccpp}\notag\\
&~~~~~ + 4 A_{uwvt}A_{cqvt}A_{pueq}A_{pwec} + 8 A_{uwce}A_{epqc}A_{tqvu}A_{twvp}\notag\\
&~~~~~ + A_{uvqt}A_{twvc}A_{cewp}A_{pueq}.
\end{align}
The particle diagrams for each of the terms for $\tau$ are illustrated in Figures {\ref{fig:s2tail}} -- {\ref{fig:box}} and the process for evaluating them is the same as for the fourth and sixth moments; (1) diagrams which are otherwise the same as diagrams for lower order moments but with the addition of tails simply gain additional factors of ${{l-m+k} \choose k}{m\choose k}$ for each tail. (2)  Some diagrams, although initially appearing distinct, actually ``collapse'' -- taking the same value as a known diagram -- when the condition is imposed on the bonds of the diagram that it's argument be maximal, ie., when we look only at those instances of the diagram which attain the maximal argument of the diagram. This process will become clearer when looking at some actual calculations. (3) In the absence of tails the argument of the particle diagram will be maximised by using the method of particle diagrams. This involves determining all loops in the particle diagram and identifying which of these, by containing more single-particle states, increases the argument of the particle diagram. These  methods do not need to be used in isolation. For example loop counting can be used initially, followed by the observation that a diagram will ``collapse'', just as it is perfectly legitimate to observe a diagram collapse almost fully, except for the addition of a remaining tail.

\subsubsection{A Single Chain}
For the first term $A_{ppuu}A_{uuww}A_{wwvv}A_{vvqq}$ of $\tau$ the particle diagram is simply a chain of bonds
\begin{equation}p~\feyn{f}~u\feyn{f}~w\feyn{f}~v\feyn{f}~q
\end{equation}
so that we first choose $m$ states to determine $|p\rangle$, then add factors of ${m\choose k}{{l-m+k}\choose{k}}$ for each additional tail, giving
\begin{equation}\label{eq:ASC1}A_{ppuu}A_{uuww}A_{wwvv}A_{vvqq}={l\choose m}\left[{m\choose k}{{l-m+k}\choose{k}}\right]^4.
\end{equation}

\subsubsection{Diagrams of Lower Order Terms + Tails}
The next term $A_{ttqq}A_{qqpp}A_{puvw}A_{pwvu}$ illustrated in Fig {\ref{fig:s2tail}} is identical in form to the standard diagram but with the addition of two tails, or alternatively Fig {\ref{fig:fish}} with the addition of a single tail so that
\begin{equation}A_{ttqq}A_{qqpp}A_{puvw}A_{pwvu} \sim {m\choose k}^2{{l-m+k}\choose{k}}^2{{l}\choose{s~r~r~r~r}}.
\end{equation}
Likewise the term $A_{eeww}A_{wvup}A_{upqt}A_{vtqw}$ with particle diagram as in Fig. {\ref{fig:p1tail}} is identical to the particle diagram of Fig. {\ref{fig:prism}} but with the addition of a single tail, giving
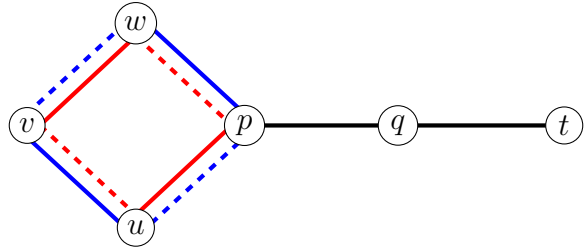
\begin{figure}[h]
\hrulefill

\vspace{1cm}
\centering
\begin{tikzpicture}[scale=1.7]
\draw[ultra thick, black] (1.5,0) -- (2.8,0);
\draw[ultra thick, black] (2.8,0) -- (4.1,0);
\node[circle, fill=white, draw=black, text=black,  inner sep=2pt] at (2.8,0) {$q$};
\node[circle, fill=white, draw=black, text=black,  inner sep=2pt] at (4.1,0) {$t$};
\draw[ultra thick, red] (0,0) -- (0.75,.7);
\draw[ultra thick, dashed, red] (0.75,.7) -- (1.5,0);\draw[ultra thick, dashed, red] (0,0) -- (0.75,-.7);\draw[ultra thick, red] (0.75,-.7) -- (1.5,0);
\draw[ultra thick, dashed, blue] (-0.2,0) -- (0.72,.85);
\draw[ultra thick, blue] (0.78,.85) -- (1.7,0);
\draw[ultra thick, blue] (-.2,0) -- (0.72,-.85);
\draw[ultra thick, dashed, blue] (0.78,-.85) -- (1.7,0);
\node[circle, fill=white, draw=black, text=black, inner sep=2pt] at (-.1,0) {$v$}; 
\node[circle, fill=white, draw=black, text=black,  inner sep=2pt] at (1.6,0) {$p$}; 
\node[circle, fill=white, draw=black, text=black,  inner sep=2pt] at (0.75,-.8) {$u$}; 
\node[circle, fill=white, draw=black, text=black,  inner sep=2pt] at (0.75,.8) {$w$};
\end{tikzpicture}
\caption[Particle Diagram for the term $A_{ttqq}A_{qqpp}A_{puvw}A_{pwvu}$]{Particle diagram for the term $A_{ttqq}A_{qqpp}A_{puvw}A_{pwvu}$ which is identical to the particle diagram of Fig. \ref{fig:fish} with the addition of a tail, which is the same as the standard diagram illustrated in Fig. \ref{fig:combinedqterm} but with two tails.

\hrulefill}
\label{fig:s2tail}
\end{figure}
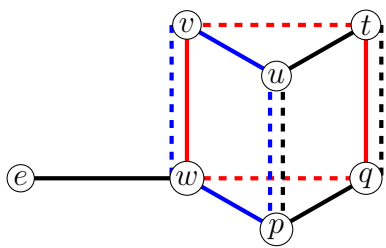
\begin{figure}[h]
\centering
\begin{tikzpicture}[scale=1.7]
\draw[ultra thick, black] (-1.3,-.5) -- (0,-.5);
\node[circle, fill=white, draw=black, text=black,  inner sep=1pt] at (-1.3,-.5) {$e$};
 \draw[ultra thick, dashed, red] (0,-.5) -- (1.4,-.5);
 \draw[ultra thick, dashed, red] (0,.7) -- (1.4,.7);
\draw[ultra thick, red] (0,-.5) -- (0,.7);
\draw[ultra thick, red] (1.4,-.5) -- (1.4,.7);
\draw[ultra thick, blue] (0,.7) -- (.7,.3);
\draw[ultra thick, black] (.7,.3) -- (1.4,.7);
\draw[ultra thick, blue] (0,-.5) -- (.7,-.9);
\draw[ultra thick, black] (.7,-.9) -- (1.4,-.5);
\draw[ultra thick, dashed, blue] (.65,.3) -- (.65,-.9);
 \draw[ultra thick, dashed, black] (.75,.3) -- (.75,-.9);
\draw[ultra thick, dashed, blue] (-.12,.7) -- (-.12,-.5);
 \draw[ultra thick, dashed, black] (1.52,.7) -- (1.52,-.5);
\node[circle, fill=white, draw=black, text=black,  inner sep=1pt] at (0,.7) {$v$};
\node[circle, fill=white, draw=black, text=black,  inner sep=1pt] at (.7,.3) {$u$};
\node[circle, fill=white, draw=black, text=black,  inner sep=1pt] at (1.4,.7) {$t$};
\node[circle, fill=white, draw=black, text=black,  inner sep=1pt] at (0,-.5) {$w$};
\node[circle, fill=white, draw=black, text=black,  inner sep=1pt] at (0.7,-.9) {$p$};
\node[circle, fill=white, draw=black, text=black,  inner sep=1pt] at (1.4,-.5) {$q$};
\end{tikzpicture}
\caption[Particle Diagram for the term $A_{eeww}A_{wvup}A_{upqt}A_{vtqw}$]{The particle diagram for the term $A_{eeww}A_{wvup}A_{upqt}A_{vtqw}$ which is identical to the particle diagram of the term $A_{putq}A_{qwvt}A_{upwv}$ illustrated in Fig \ref{fig:prism} giving this term the same value but with an additional factor of ${m\choose k}{{l-m+k}\choose{k}}$ with the result given by (\ref{eq:prism_tail}).
}
\label{fig:p1tail}
\end{figure}
\begin{align}\label{eq:prism_tail}A_{eeww}&A_{wvup}A_{upqt}A_{vtqw} \sim {m\choose k}{{l-m+k}\choose{k}}\left[{l \choose k\;k\;k\;k\;m-k}{m-k\choose k}^2 \right].
\end{align}
The same applies to $A_{qwvt}A_{uwpt}A_{pvuq}A_{ccpp}$. Since this is equivalent to the particle diagram of Fig. \ref{fig:diamond} but with the addition of a tail we have simply
\begin{equation}A_{qwvt}A_{uwpt}A_{pvuq}A_{ccpp}\sim {m\choose k}{{l-m+k}\choose{k}}{{l}\choose{\tilde{s}~r~r~r~r~r~r}}.
\end{equation}

\subsubsection{The term $\mathbf{A_{tucp}A_{tpcu}A_{cewv}A_{cvwe}}$}
Finally, although we do not have simply a tail extention of a previously calculated diagram for the term $A_{tucp}A_{tpcu}A_{cewv}A_{cvwe}$ illustrated in Fig \ref{fig:standard_squared}, this is a composition of two copies of the standard diagram with each copy sharing the state $|c\rangle$. Choosing the state $|c\rangle$ will give a factor of $l\choose m$, after which the reasoning follows identically for the left and right instances as it does for the standard diagram so that the binomial terms for each are the same as in (\ref{eq:arg11}), giving
\begin{equation}A_{tucp}A_{tpcu}A_{cewv}A_{cvwe}\sim{l \choose m}\left[{{l - m} \choose r}{{l - m - r}\choose r}{m \choose r}{{m - r}\choose r}\right]^2.
\end{equation}
\begin{figure}[t]
\centering
\begin{tikzpicture}[scale=1.7]
\draw[ultra thick, dashed, green] (0,0) -- (0.75,.7);
\draw[ultra thick, green] (0.75,.7) -- (1.5,0);
\draw[ultra thick, green] (0,0) -- (0.75,-.7);
\draw[ultra thick, dashed, green] (0.75,-.7) -- (1.5,0);
\draw[ultra thick, blue] (-0.2,0) -- (0.72,.85);
\draw[ultra thick, dashed, blue] (0.78,.85) -- (1.7,0);
\draw[ultra thick, dashed, blue] (-.2,0) -- (0.72,-.85);
\draw[ultra thick, blue] (0.78,-.85) -- (1.7,0);
\node[circle, fill=white, draw=black, text=black, inner sep=2pt] at (-.1,0) {$t$}; 
\node[circle, fill=white, draw=black, text=black,  inner sep=2pt] at (0.75,-.8) {$p$}; 
\node[circle, fill=white, draw=black, text=black,  inner sep=2pt] at (0.75,.8) {$u$};

\begin{scope}[shift={(1.7,0)}]
\draw[ultra thick, dashed, red] (0,0) -- (0.75,.7);
\draw[ultra thick, red] (0.75,.7) -- (1.5,0);
\draw[ultra thick, red] (0,0) -- (0.75,-.7);
\draw[ultra thick, dashed, red] (0.75,-.7) -- (1.5,0);
\draw[ultra thick, black] (-0.2,0) -- (0.72,.85);
\draw[ultra thick, dashed, black] (0.78,.85) -- (1.7,0);
\draw[ultra thick, dashed, black] (-.2,0) -- (0.72,-.85);
\draw[ultra thick, black] (0.78,-.85) -- (1.7,0);
\node[circle, fill=white, draw=black, text=black, inner sep=4pt] at (-.1,0) {$c$}; 
\node[circle, fill=white, draw=black, text=black,  inner sep=2pt] at (1.6,0) {$w$}; 
\node[circle, fill=white, draw=black, text=black,  inner sep=2pt] at (0.75,-.8) {$v$}; 
\node[circle, fill=white, draw=black, text=black,  inner sep=2pt] at (0.75,.8) {$e$};
\end{scope}
\end{tikzpicture}
\caption[Particle Diagram for the term $A_{tucp}A_{tpcu}A_{cewv}A_{cvwe}$]{The particle diagram for $A_{tucp}A_{tpcu}A_{cewv}A_{cvwe}$, although not an extention of a previously seen diagram with a tail, is simply the product of two standard diagrams with a single shared center node given by the state $|c\rangle$.

\hrulefill
}
\label{fig:standard_squared}
\end{figure}
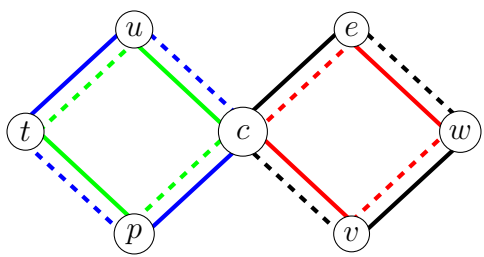
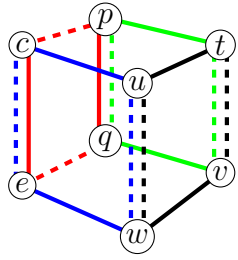
\begin{figure}
\centering
\begin{tikzpicture}[scale=1.7]
\draw[ultra thick, dashed, red] (0,0) -- (0.7,.2);
\draw[ultra thick, red] (0.05,0) -- (0.05,-1.1);
\draw[ultra thick, dashed, blue] (-.05,0) -- (-.05,-1.1);
\draw[ultra thick, dashed, red] (0,-1.1) -- (.7,-.75);
\draw[ultra thick, green] (.7,-.75) -- (1.55,-1);
\draw[ultra thick, dashed, black] (1.6,-1) -- (1.6,0);
\draw[ultra thick, dashed, green] (1.5,-1) -- (1.5,0);
\draw[ultra thick, black] (1.55,0) -- (0.9,-.3);
\draw[ultra thick, blue] (0,-1.1) -- (.9,-1.5);
\draw[ultra thick, black] (.9,-1.5) -- (1.55,-1);
\draw[ultra thick, green] (1.55,0) -- (.7,.2);
\draw[ultra thick, dashed, green] (.7,.2) -- (.7,-.75);
\draw[ultra thick, red] (.6,.2) -- (.6,-.75);
\draw[ultra thick, blue] (0,0) -- (0.9,-.3);
\node[circle, fill=white, draw=black, text=black,  inner sep=1pt] at (0.65,-.75) {$q$};
\draw[ultra thick, dashed, blue] (.85,-.3) -- (.85,-1.5);
\draw[ultra thick, dashed, black] (.95,-.3) -- (.95,-1.5);
\node[circle, fill=white, draw=black, text=black, inner sep=1pt] at (0,0) {$c$}; 
\node[circle, fill=white, draw=black, text=black,  inner sep=1pt] at (0.65,.2) {$p$}; 
\node[circle, fill=white, draw=black, text=black,  inner sep=1pt] at (1.55,0) {$t$};
\node[circle, fill=white, draw=black, text=black, inner sep=1pt] at (.9,-.3) {$u$}; 
\node[circle, fill=white, draw=black, text=black,  inner sep=1pt] at (0,-1.1) {$e$}; 
\node[circle, fill=white, draw=black, text=black, inner sep=1pt] at (1.55,-1) {$v$}; 
\node[circle, fill=white, draw=black, text=black,  inner sep=1pt] at (0.9,-1.5) {$w$}; 
\end{tikzpicture}
\caption[Particle Diagram for the term $A_{cewu}A_{uwvt}A_{ttvqp}A_{pceq}$]{Particle diagram for the term $A_{cewu}A_{uwvt}A_{ttvqp}A_{pceq}$ of $\tau$ which takes the form of a cuboid structure. The contributing sum is calculated by determining all permitted loops in the particle diagram. Then, maximising the argument over all permitted loops reveals the structure exclusively of those loops which maximise the argument of the particle diagram (see Figs \ref{fig:k_top} -- \ref{fig:beta}) }
\label{fig:cuboid}
\end{figure}

\begin{figure}\label{fig:paths}
\begin{subfigure}{.33\textwidth}
  \centering
  \includegraphics[width=.8\linewidth]{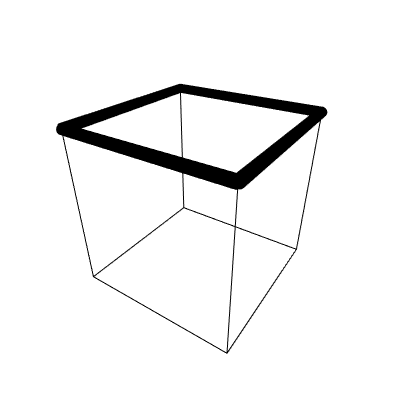}
  \caption{The loop \normalsize$\overrightarrow{cutpc}$}
  \label{fig:k_top}
\end{subfigure}%
\begin{subfigure}{.33\textwidth}
  \centering
  \includegraphics[width=.8\linewidth]{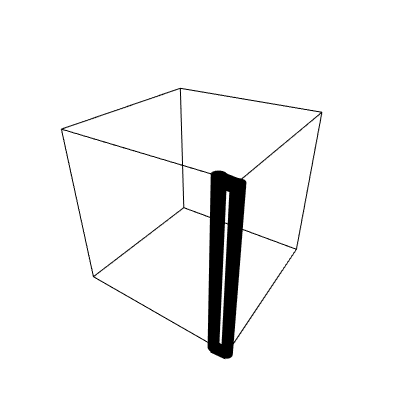}
  \caption{The loop \normalsize$\overrightarrow{uwu}$}
  \label{fig:delta_L}
\end{subfigure}%
\begin{subfigure}{.33\textwidth}
  \centering
  \includegraphics[width=.8\linewidth]{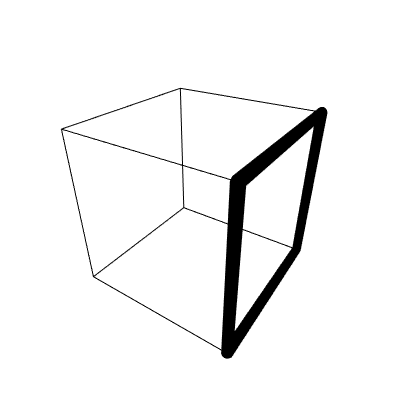}
  \caption{The loop \normalsize$\overrightarrow{utvwu}$}
  \label{fig:alpha_F}
\end{subfigure}\\
\begin{subfigure}{.33\textwidth}
  \centering
  \includegraphics[width=.8\linewidth]{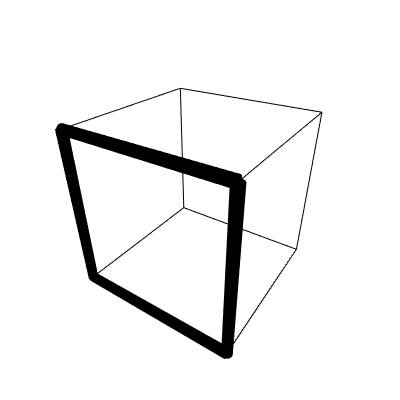}
  \caption{The loop \normalsize$\overrightarrow{wecuw}$}
 \label{fig:alpha_L}
\end{subfigure}%
\begin{subfigure}{.33\textwidth}
  \centering
  \includegraphics[width=.8\linewidth]{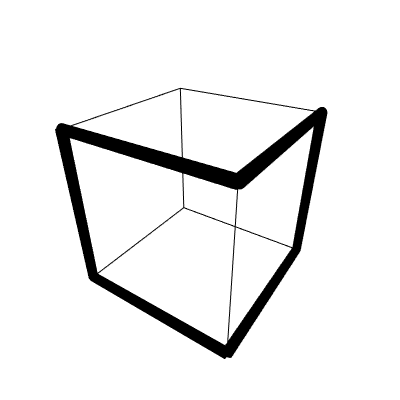}
  \caption{The loop \normalsize$\overrightarrow{cutvwec}$}
  \label{fig:gamma_L}
\end{subfigure}%
\begin{subfigure}{.33\textwidth}
  \centering
  \includegraphics[width=.8\linewidth]{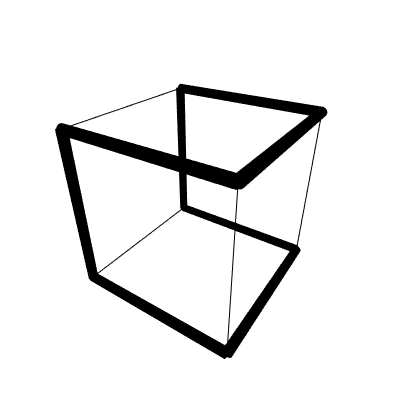}
  \caption{The loop \normalsize$\overrightarrow{cutpqvwec}$}
 \label{fig:beta}
\end{subfigure}
\caption[Loops Illustration]{Illustration of a selection of the loops within the particle diagram for $A_{cewu}A_{uwvt}A_{ttvqp}A_{pceq}$ illustrated in Fig \ref{fig:cuboid}.}
\end{figure}
\subsubsection{The Cuboid Diagram}
The term $A_{cewu}A_{uwvt}A_{ttvqp}A_{pceq}$ of (\ref{eq:s1.5}) has a cuboid particle diagram, Fig. \ref{fig:cuboid}. One begines by identifying all the permitted loops in the diagram and ends by maximising the argument by tuning the number of single-particle states participating in each path. The permitted paths for the cuboid diagram are $\overrightarrow{cutpc}$,  $\overrightarrow{ewvqe}$, $\overrightarrow{cuwec}$, $\overrightarrow{vtpqv}$, $\overrightarrow{utpqvwu}$, $\overrightarrow{cutvwec}$, $\overrightarrow{pcuweqp}$, $\overrightarrow{cptvqec}$, $\overrightarrow{cpqec}$, $\overrightarrow{utvwu}$, $\overrightarrow{cuweqvtpc}$,  $\overrightarrow{cutpqvwec}$, $\overrightarrow{utvweqpcu}$, $\overrightarrow{ptuwvqecp}$, $\overrightarrow{cec}$, $\overrightarrow{uwu}$,  $\overrightarrow{tvt}$ and $\overrightarrow{pqp}$. However, if we check the loops which maximise the argument of the diagram we find the \textit{contributing loops} which, unlike the permitted non-contributing loops, can contain a non-zero number of single-particle states without decreasing the argument of the diagram from its maximal value $m-4k$. The contributing loops for the cuboid diagram are
\[ \begin{array}{lll}
\overrightarrow{ewvqe} \hspace{1cm}& \overrightarrow{cutpc} & \hspace{1cm}\mathrm{(see~Fig.~\ref{fig:k_top})} \label{eq:cutpc}\\
\overrightarrow{cec} \hspace{1cm}& \overrightarrow{cutpqvwec} & \hspace{1cm}\mathrm{(see~Fig.~\ref{fig:beta})}\\
\overrightarrow{pqp} \hspace{1cm}& \overrightarrow{wuw} & \hspace{1cm}\mathrm{(see~Fig.~\ref{fig:delta_L})}\\
\overrightarrow{tvt} \hspace{1cm}& \overrightarrow{utvwu} & \hspace{1cm}\mathrm{(see~Fig.~\ref{fig:alpha_F})}\\
\overrightarrow{vtpqv} \hspace{1cm}& \overrightarrow{cuwec} & \hspace{1cm}\mathrm{(see~Fig.~\ref{fig:alpha_L})} \\
\overrightarrow{utpqvwu} & \overrightarrow{cutvwec} & \hspace{1cm}\mathrm{(see~Fig.~\ref{fig:gamma_L})}
\end{array} \]
Moreover, maximising the argument over the contributing loops in analogy to the process introduced in section \ref{sec:TFMWL} yields the expression
\begin{equation}\label{eq:TCD1}A_{cewu}A_{uwvt}A_{ttvqp}A_{pceq}\sim\sum_{\alpha\beta}{{l} \choose {m~k~k~\alpha~\beta~k-\alpha~k-\beta}}{{m}\choose{k}}{{m-k}\choose{k}}{{s}\choose{\alpha}}{{s}\choose{\beta}}.
\end{equation}
\subsubsection{The term $\mathbf{A_{tpce}A_{evqc}A_{puwq}A_{uwvt}}$}
We next look at the particle diagram of the term $A_{tpce}A_{evqc}A_{puwq}A_{uwvt}$ of (\ref{eq:s1.5}), illustrated in Fig. \ref{fig:standard_collapse_2}. To leading order it is the standard diagram in disguise, with the addition of $2k$ degrees of freedom which require explaining. In fact the overlapping of the bond $t\feyn{f}e$ with $e\feyn{f}c$ and the overlapping of the bond $e\feyn{f}c$ with $c\feyn{f}p$ on the left sub-diagram must necessarily be minimised in order to maximise the argument of the combined diagram. The same applies to the corresponding states in the right sub-diagram of Fig. \ref{fig:standard_collapse_2}. The minimum overlap is of course $m-2k$ as before, giving
\begin{equation}e\feyn{h}v \subset e\feyn{f}t
\end{equation}
\begin{equation}c\feyn{h}q \subset c\feyn{f}p
\end{equation}
\begin{equation}e\feyn{h}c \subset e\feyn{f}c
\end{equation}
The additional coincidence in $m-2k$ states between say $|t\rangle$ and $|p\rangle$ which already share $k$ single-particle states through the bond $t\feyn{h}p$ brings the total overlap between them to $m-k$. This gives an additional factor ${{{m-k}\choose{k}}}^2$, one factor each for the left and right sub-diagrams which together form Fig. \ref{fig:standard_collapse_2}. One also gains the double factor ${{{l-2k-s}\choose{k}}}^2$ from the choice of single-particles contained in $e\feyn{h}c \cap e\feyn{f}c$ and correspondingly $u\feyn{h}w \cap u\feyn{f}w$. In this way we find
\begin{equation}\label{eq:COLL1}A_{tpce}A_{evqc}A_{puwq}A_{uwvt} \sim {l \choose {s~k~k~k~k}}{{m-k}\choose k}^2{{l-2k-s}\choose{k}}^2.
\end{equation}
This is an instance of a particle diagram ``collapsing'' partially to a diagram seen before in the calculation of a lower moment, in this case the standard diagram of Fig \ref{fig:combinedqterm} which has been a presence in calculations from the fourth moment onwards.
\FloatBarrier
\begin{figure}[t]
\centering
\begin{tikzpicture}[scale=1.7]
\draw[ultra thick, black] (0,0) -- (1,0);
\draw[ultra thick, black] (0,-1) -- (1,-1);
\draw[ultra thick, dashed, black] (0,0) -- (0,-1);
\draw[ultra thick, dashed, black] (.95,0) -- (.95,-1);
\draw[ultra thick, dashed, green] (1,0) -- (2,0);
\draw[ultra thick, dashed, green] (1,-1) -- (2,-1);
\draw[ultra thick, green] (1.05,0) -- (1.05,-1);
\draw[ultra thick, green] (2,0) -- (2,-1);
\node[circle, fill=white, draw=black, text=black, inner sep=1pt] at (0,0) {$t$}; 
\node[circle, fill=white, draw=black, text=black,  inner sep=1pt] at (1,0) {$e$}; 
\node[circle, fill=white, draw=black, text=black,  inner sep=1pt] at (2,0) {$v$};
\node[circle, fill=white, draw=black, text=black, inner sep=1pt] at (0,-1) {$p$}; 
\node[circle, fill=white, draw=black, text=black,  inner sep=1pt] at (1,-1) {$c$}; 
\node[circle, fill=white, draw=black, text=black, inner sep=1pt] at (2,-1) {$q$};
\begin{scope}[shift={(3,0)}]
 \draw[ultra thick, dashed, blue] (0,0) -- (1,0);
\draw[ultra thick, dashed, blue] (0,-1) -- (1,-1);
\draw[ultra thick, blue] (0,0) -- (0,-1);
\draw[ultra thick, blue] (.95,0) -- (.95,-1);
\draw[ultra thick, red] (1,0) -- (2,0);
\draw[ultra thick, red] (1,-1) -- (2,-1);
\draw[ultra thick, dashed, red] (1.05,0) -- (1.05,-1);
\draw[ultra thick, dashed, red] (2,0) -- (2,-1);
\node[circle, fill=white, draw=black, text=black, inner sep=1pt] at (0,0) {$p$}; 
\node[circle, fill=white, draw=black, text=black,  inner sep=1pt] at (1,0) {$u$}; 
\node[circle, fill=white, draw=black, text=black,  inner sep=1pt] at (2,0) {$t$};
\node[circle, fill=white, draw=black, text=black, inner sep=1pt] at (0,-1) {$q$}; 
\node[circle, fill=white, draw=black, text=black,  inner sep=1pt] at (1,-1) {$w$}; 
\node[circle, fill=white, draw=black, text=black, inner sep=1pt] at (2,-1) {$v$};
\end{scope}
\end{tikzpicture}
\caption[Particle diagram of the term $A_{tpce}A_{evqc}A_{puwq}A_{uwvt}$]{The $A_{tpce}A_{evqc}A_{puwq}A_{uwvt}$  which to leading order is the same as the standard diagram, but with an additional $k$ degrees of freedom. This is analogous but not exactly the same as the case where we have a known diagram with an extra tail. For $k\le m-k$ the overlap of the bonds linking the states $|e\rangle$, $|c\rangle$ and $|u\rangle$, $|w\rangle$ respectively implies that the states $|t\rangle$ and $|p\rangle$ overlap by an additional $s=m-2k$ states, bringing their total overlap to $m-k$ and likewise for the states  $|t\rangle$ and  $|v\rangle$.}
\label{fig:standard_collapse_2}
~ ~\\
\centering
\begin{tikzpicture}[scale=1.7]
\draw[ultra thick, black] (0,0) -- (1,0);
\draw[ultra thick, black] (0,-1) -- (1,-1);
\draw[ultra thick, dashed, black] (0,0) -- (0,-1);
\draw[ultra thick, dashed, black] (.95,0) -- (.95,-1);
\draw[ultra thick, green] (1,0) -- (2,0);
\draw[ultra thick, green] (1,-1) -- (2,-1);
\draw[ultra thick, dashed, green] (1.05,0) -- (1.05,-1);
\draw[ultra thick, dashed, green] (2,0) -- (2,-1);
\node[circle, fill=white, draw=black, text=black, inner sep=1pt] at (0,0) {$p$}; 
\node[circle, fill=white, draw=black, text=black,  inner sep=1pt] at (1,0) {$q$}; 
\node[circle, fill=white, draw=black, text=black,  inner sep=1pt] at (2,0) {$v$};
\node[circle, fill=white, draw=black, text=black, inner sep=1pt] at (0,-1) {$e$}; 
\node[circle, fill=white, draw=black, text=black,  inner sep=1pt] at (1,-1) {$t$}; 
\node[circle, fill=white, draw=black, text=black, inner sep=1pt] at (2,-1) {$u$};
\begin{scope}[shift={(3,0)}]
 \draw[ultra thick, blue] (0,0) -- (1,0);
\draw[ultra thick, blue] (0,-1) -- (1,-1);
\draw[ultra thick, dashed, blue] (0,0) -- (0,-1);
\draw[ultra thick, dashed, blue] (.95,0) -- (.95,-1);
\draw[ultra thick, red] (1,0) -- (2,0);
\draw[ultra thick, red] (1,-1) -- (2,-1);
\draw[ultra thick, dashed, red] (1.05,0) -- (1.05,-1);
\draw[ultra thick, dashed, red] (2,0) -- (2,-1);
\node[circle, fill=white, draw=black, text=black, inner sep=1pt] at (0,0) {$e$}; 
\node[circle, fill=white, draw=black, text=black,  inner sep=1pt] at (1,0) {$c$}; 
\node[circle, fill=white, draw=black, text=black,  inner sep=1pt] at (2,0) {$p$};
\node[circle, fill=white, draw=black, text=black, inner sep=1pt] at (0,-1) {$u$}; 
\node[circle, fill=white, draw=black, text=black,  inner sep=1pt] at (1,-1) {$w$}; 
\node[circle, fill=white, draw=black, text=black, inner sep=1pt] at (2,-1) {$v$};
\end{scope}
\end{tikzpicture}
\caption[Particle diagram of $A_{petq}A_{qtuv}A_{euwc}A_{cwvp}$]{Illustrated is the particle diagram of $A_{petq}A_{qtuv}A_{euwc}A_{cwvp}$ which has some similarities to the standard diagram. For the purpose of evaluation however, it is not similar enough. To calculate the leading order summation for the terms satisfying this diagram one needs to use the method of particle diagrams, availing of loop summation.

\hrulefill}
\label{fig:standard_collapse_1}
\end{figure}
\subsubsection{The term $\mathbf{A_{petq}A_{qtuv}A_{euwc}A_{cwvp}}$}
The calculation for the term $A_{petq}A_{qtuv}A_{euwc}A_{cwvp}$ of (\ref{eq:s1.5}) with diagram shown in Fig. \ref{fig:standard_collapse_1} will be calculated using the method of particle diagrams by utilising loop summation. Explicitly, the number of single-particle states contained in each loop is optimised in order to maximise the argument. The contributing loops are those which do not necessarily have to contain zero elements in order to maximise the argument given by the sum $\mathrm{arg} = \sum_i n_i$ where $n_i$ is the number of elements in loop $i$. The loops on the particle diagram of the term $A_{petq}A_{qtuv}A_{euwc}A_{cwvp}$ are
\allowdisplaybreaks[1]
\[ \begin{array}{ll}
\alpha ~=~\overrightarrow{pvq} & \xi ~=~\overrightarrow{qvpeut} \\
\beta ~=~\overrightarrow{pec} & \theta ~=~\overrightarrow{qpvuet}\\
\gamma ~=~\overrightarrow{eut} & \lambda ~=~\overrightarrow{cpeuvw}\\
\delta ~=~\overrightarrow{uvw} & \nu ~=~\overrightarrow{cepvuw} \\
~ & ~ \\
\epsilon ~=~\overrightarrow{qtuv} & \pi ~=~\overrightarrow{pqvwutec} \\
\eta ~=~\overrightarrow{qtep} & \sigma ~=~\overrightarrow{qt} \\
\omega ~=~\overrightarrow{cwvp} & \tau ~=~\overrightarrow{cw} \\
\mu ~=~\overrightarrow{cwue} & \rho ~=~\overrightarrow{pvue} \\
\end{array}\]
Because the number of single-particle states contained in all paths which pass through a bond $\feyn{h}$ must sum to $k$ and similarly the number of single-particle states contained in all paths which pass through a bond $\feyn{f}$ must sum to $m-k$ one can immediately read off the following conservation equations
\allowdisplaybreaks[1]
\begin{align}
n_{\alpha}+n_{\xi}+n_{\theta}+n_{\nu}+n_{\rho}&=k\notag\\
n_{\beta}+n_{\xi}+n_{\lambda}+n_{\nu}+n_{\rho}&=k\notag\\
n_{\gamma}+n_{\xi}+n_{\theta}+n_{\lambda}+n_{\rho}&=k\notag\\
n_{\delta}+n_{\theta}+n_{\lambda}+n_{\nu}+n_{\rho}&=k\\
\notag\\
n_{\epsilon}+ n_{\xi}+ n_{\sigma}&=k\notag\\
n_{\eta}+n_{\theta}+n_{\sigma}&=k\notag\\
n_{\omega}+n_{\lambda}+n_{\tau}&=k\notag\\
n_{\mu}+n_{\nu}+n_{\tau}&=k\\
\notag\\
n_{\epsilon}+n_{\xi}+n_{\omega}+n_{\pi}+n_{\alpha}&=m-k\notag\\
n_{\epsilon}+n_{\xi}+n_{\mu}+n_{\pi}+n_{\gamma}&=m-k\notag\\
n_{\epsilon}+n_{\omega}+n_{\lambda}+n_{\pi}+n_{\delta}&=m-k\notag\\
n_{\epsilon}+n_{\mu}+n_{\nu}+n_{\pi}+n_{\delta}&=m-k\\
\notag\\
n_{\eta}+n_{\theta}+n_{\omega}+n_{\pi}+n_{\alpha}&=m-k\notag\\
n_{\eta}+n_{\omega}+n_{\lambda}+n_{\pi}+n_{\beta}&=m-k\notag\\
n_{\eta}+n_{\mu}+n_{\nu}+n_{\pi}+n_{\beta}&=m-k\notag\\
n_{\eta}+n_{\theta}+n_{\mu}+n_{\pi}+n_{\gamma}&=m-k
\end{align}
For a particular loop to contribute the argument must additionally satisfy
\begin{align}
\mathrm{arg} &= \sum_i n_i = n_{\alpha} +n_{\beta} +n_{\gamma}+n_{\delta} +n_{\epsilon} +n_{\eta }+n_{\omega}+n_{\mu} +n_{\xi} +n_{\theta} +n_{\lambda} +n_{\nu}+n_{\pi} +n_{\rho} +n_{\sigma} +n_{\tau}\notag\\
& = m+4k
\end{align}
since $m+4k$ is the maximal argument that this diagram can attain and therefore the $n_i$ satisfying this equality are the only terms of interested in the limit $l\to \infty$. Imposing this contraint on the conservation equations yields $n_{\alpha} ~=~n_{\beta} ~=~n_{\gamma} ~=~n_{\delta} ~=~k$ and $n_{\xi} ~=~n_{\theta} ~=~n_{\lambda} ~=~n_{\nu} ~=~n_{\rho} ~=~0$. Moreover
\[\begin{array}{ll}
n_{\epsilon} = n_{\eta}, & \hspace{1cm}n_{\pi} = n_{\pi}\\
n_{\omega} = n_{\mu}, & \hspace{1cm}n_{\sigma}  = n_{\sigma}\\
~~ & \hspace{1cm}n_{\tau} =n_{\tau}.
\end{array}\]
Inserting these back into the conservation equations gives
\begin{align}
&n_{\epsilon} = k-n_{\sigma} \\
&n_{\omega} = k-n_{\tau}\\
&n_{\epsilon} +n_{\omega}+n_{\pi}=m-2k
\end{align}
so that in the limit $l\to\infty$ the surviving term reads
\begin{align}\label{eq:hahn_term} &A_{petq}A_{qtuv}A_{euwc}A_{cwvp} \sim \sum{{l}\choose{n_{\alpha}~n_{\beta}~n_{\gamma}~n_{\delta}~n_{\sigma}~n_{\tau}~n_{\epsilon}~n_{\eta}~n_{\omega}~n_{\mu}~n_{\pi}}}\notag\\
&= \sum_{n_{\alpha}n_{\beta}}{l \choose {k~k~k~k~n_{\sigma}~n_{\tau}~k-n_{\sigma}~k-n_{\sigma}~k-n_{\tau}~k-n_{\tau}~{m-4k+n_{\sigma}+n_{\tau}}}}.
\end{align}
Dividing (\ref{eq:hahn_term}) by the normalisation term $\left({\frac{1}{N}}\mathrm{tr}(\overline{V_{k}^2})\right)^4$ and summing out $\tau$ one extraordinarily attains the following \textit{Hahn polynomial}
\begin{equation}\label{eq:HAHN}
\lim_{N\to\infty} \frac {{\frac{1}{N}}{A_{petq}A_{qtuv}A_{euwc}A_{cwvp}}} { \left({\frac{1}{N}} \mathrm{tr}(\overline{V_{k}^2})\right)^4 }
 = \frac{{{m-k}\choose{k}}}{{{m\choose {k}}}^3} {\sum_{n_{\sigma}} {{m-k-n_{\sigma}}\choose{k}} {{m-2k}\choose{n_{\sigma}}} {{k}\choose{n_{\sigma}}}}.
\end{equation}
This is unusual because no previous term has explicitly required a Hahn polynomial in order to be expressed properly. Additionally, this is an interesting development because it hints that it may be possible to express all previous terms as Hahn polynomials and perhaps even find a set of tools with which to calculate the moments using a ``Hahn Method'' as apposed to using the method of particle diagrams which this thesis sets out to present and explain.

\subsection{Hahn Polynomials}
The author's proof of the following lemma will briefly be highlighted to express (\ref{eq:hahn_term}) as a quotient of binomials in $m$ and $k$. The proof uses the formalism of the class of \textit{Hahn polynomials} for which definitions can be found in the compendium \cite{koekoek}.

\vspace{.2cm}
\noindent\textit{Lemma.}
\vspace{.2cm}
\begin{equation}{\frac{{{m-k}\choose{k}}}{{{{m}\choose{k}}^3}}}{\sum_{\alpha} {{m-k-\alpha}\choose{k}}{{k}\choose{\alpha}}{{m-2k}\choose{\alpha}}} = {\frac{{{{m-k}\choose{k}}^2}}{{{{m}\choose{k}}^3}}} \sum_p \frac{{{{{k}\choose {p}}^2}{{m-2k}\choose {k-p}}}}{{{m-k}\choose{p}}}\end{equation}\\
\vspace{.2cm}
\noindent\textit{Proof:}\\ 
\vspace{.2cm}
Writing the l.h.s as an hypergeometric function
\begin{equation}{\sum {{m-k-\alpha}\choose{k}}{{k}\choose{\alpha}}{{m-2k}\choose{\alpha}}} = {{m-k}\choose{k}} {~}_3 F_2\left( \begin{array}{cc}
-k, 2k-m, 2k-m\\
1, k-m \end{array} ;1\right)  \end{equation}
and recalling the definition of a Hahn polynomial
\begin{equation}Q_n (x; \alpha, \beta, N) := {~}_3 F_2\left( \begin{array}{cc}
-n, n+\alpha+\beta+1,-x\\
\alpha+1, -N \end{array} ;1\right) \end{equation}
we have $n=k$, $\alpha=0$, $\beta = k-m-1$ and $x=m-2k$. To express this as a series it should then be noted that
\begin{equation}
{~}_1 F_1\left( \begin{array}{cc}
-x\\
\alpha+1\end{array} ;-t\right)  {~}_1 F_1\left( \begin{array}{cc}
x-N\\
\beta+1\end{array} ;t\right) = \sum{\frac{(-N)_n}{(\beta+1)_n n!}} Q_n (x; \alpha, \beta, N) t^n.
 \end{equation}
(see \cite{koekoek} chapter 9). Furthermore, by definition
\begin{equation}{~}_1 F_1\left( \begin{array}{cc}
a\\
b\end{array} ;t\right) := \sum_{n=0}^{\infty} \frac{(a)_n}{(b)_n}  \frac{t^n}{n!}
 \end{equation}
where
\begin{equation}
(a)_n := \frac{\Gamma(a+n)}{\Gamma(a)} = (a+1)(a+2)\ldots(a+n)
\end{equation}
and the symmetric extention of this is
\begin{equation}
(-x)_n = (-1)^n (x-n+1)_n.
\end{equation}
Using the definition it then follows straightforwardly that
\begin{equation}{~}_1 F_1\left( \begin{array}{cc}
2k-m\\
1\end{array} ;-t\right) = \sum_{n=0}^{\infty} \frac{\Gamma(m-2k+1)}{\Gamma(m-2k+1-n)n!} \frac{t^n}{n!}=\sum_{n=0}^{\infty} {{m-2k}\choose{n}} \frac{t^n}{n!}
 \end{equation}
and similarly
\begin{equation}{~}_1 F_1\left( \begin{array}{cc}
{-k}\\
{-(m-k)}\end{array} ;t\right) = \sum_{n=0}^{\infty}\frac{\Gamma(k+1)\Gamma(m-k-n+1)}{\Gamma(k-n+1)\Gamma(m-k+1)}  \frac{t^n}{n!}=\sum_{n=0}^{\infty} {\frac{{{{k}\choose {n}}}}{{{m-k}\choose{n}}}}{\frac{t^n}{n!}}
 \end{equation}
so that the complete series can be written as
\begin{equation}{~}_1 F_1\left( \begin{array}{cc}
2k-m\\
1\end{array} ;-t\right) {~}_1 F_1\left( \begin{array}{cc}
{-k}\\
{-(m-k)}\end{array} ;t\right) = \sum_{n,p=0}^{\infty} \frac{{{m-2k}\choose n}{k\choose p}}{{{m-k}\choose p}}\frac{{t^{n+p}}}{{n!p!}}.\end{equation}
Finally, comparing the coefficients of $t^k$ gives
 \begin{equation}Q_k(x; \alpha, \beta, N) = \sum_{p=0}^{k} \frac{{{{{k}\choose {p}}^2}{{m-2k}\choose {k-p}}}}{{{m-k}\choose{p}}} \end{equation}
which completes the proof.\hfill$\square$

\subsection{Remaining Diagrams}
There are just three remaining terms left in the calculation of (\ref{eq:s1.5}) for which we have already fully covered the techniques required; \emph{(1)} comparison to diagrams representing lower order terms,  \emph{(2)} the ``collapse'' of diagrams to others of familiar form such as in the case of Fig \ref{fig:standard_collapse_2} which collapses to take the form of the standard diagram with a tail when we just look at those components of the diagram which attain the maximal argument, and finally \emph{(3)} the method of particle diagrams in which we maximise the argument ( $=$ degrees of freedom) of the diagram by identifying all \emph{permitted loops} in the diagram and from these finding all \emph{contributing loops}. The path summation method can be used in every case, even those where (1) and (2) do not apply.
\subsubsection{The term $\mathbf{A_{uwvt}A_{cqvt}A_{pueq}A_{pwec}}$}
Identifying paths and maximising the argument for $A_{uwvt}A_{cqvt}A_{pueq}A_{pwec}$ of (\ref{eq:s1.5}) with corresponding diagram Fig. \ref{fig:diamond_collapse_1} in the same manner as described previously we yields the following paths
\[
\begin{array}{ll}
\alpha ~=~\overrightarrow{utvwe} & \hspace{.5cm}\sigma ~=~\overrightarrow{ewvtupcq} \\
\beta ~=~\overrightarrow{tcpqv} & \hspace{.5cm}\tau ~=~\overrightarrow{pqvtceuw} \\
\gamma ~=~\overrightarrow{uew} & \hspace{.5cm}\phi ~=~\overrightarrow{pctvqewu}\\
\delta ~=~\overrightarrow{cpq} & \hspace{.5cm}\theta ~=~\overrightarrow{ectvq} \\
\epsilon ~=~\overrightarrow{utce} & \hspace{.5cm}\lambda ~=~\overrightarrow{putvw} \\
\eta ~=~\overrightarrow{wvqe} & \hspace{.5cm}\nu ~=~\overrightarrow{ecq} \\
\omega ~=~\overrightarrow{qvwp} & \hspace{.5cm}\pi ~=~\overrightarrow{puw} \\
\mu ~=~\overrightarrow{ctup} & \hspace{.5cm}\psi ~=~\overrightarrow{euwpqc}\\
\xi ~=~\overrightarrow{eutcpqvw} & \hspace{.5cm}\kappa ~=~\overrightarrow{pcqewu}\\
\rho ~=~\overrightarrow{eutvwpqc} & \hspace{.5cm}\chi ~=~\overrightarrow{tv}
\end{array}
\]
and the following loop conservation equations
\allowdisplaybreaks[1]
\begin{align}
n_{\alpha}+n_{\lambda}+n_{\sigma}+n_{\rho}+n_{\chi}&=k\notag\\
n_{\beta}+n_{\theta}+n_{\tau}+n_{\phi}+n_{\chi}&=k\\
\notag\\
n_{\epsilon}+n_{\theta}+n_{\nu}+n_{\rho}+n_{\tau}+n_{\psi}&=k\notag\\
n_{\eta}+n_{\theta}+n_{\nu}+n_{\sigma}+n_{\phi}+n_{\kappa}&=k\notag\\
n_{\omega}+n_{\lambda}+n_{\pi}+n_{\rho}+n_{\tau}+n_{\psi}&=k\notag\\
n_{\mu}+n_{\lambda}+n_{\pi}+n_{\sigma}+n_{\phi}+n_{\kappa}&=k\notag\\
n_{\gamma}+n_{\pi}+n_{\tau}+n_{\phi}+n_{\psi}+n_{\kappa}&=k\notag\\
n_{\delta}+n_{\nu}+n_{\rho}+n_{\sigma}+n_{\psi}+n_{\kappa}&=k\\
\notag\\
n_{\alpha}+n_{\gamma}+n_{\epsilon}+n_{\xi}+n_{\rho}+n_{\tau}+n_{\psi}&=m-k\notag\\
n_{\alpha}+n_{\gamma}+n_{\eta}+n_{\xi}+n_{\sigma}+n_{\phi}+n_{\kappa}&=m-k\notag\\
n_{\alpha}+n_{\epsilon}+n_{\mu}+n_{\xi}+n_{\lambda}+n_{\rho}+n_{\sigma}&=m-k\notag\\
n_{\alpha}+n_{\eta}+n_{\omega}+n_{\xi}+n_{\lambda}+n_{\rho}+n_{\sigma}&=m-k\\
\notag\\
n_{\beta}+n_{\epsilon}+n_{\mu}+n_{\xi}+n_{\theta}+n_{\tau}+n_{\phi}&=m-k\notag\\
n_{\beta}+n_{\delta}+n_{\mu}+n_{\xi}+n_{\sigma}+n_{\phi}+n_{\kappa}&=m-k\notag\\
n_{\beta}+n_{\delta}+n_{\omega}+n_{\xi}+n_{\rho}+n_{\tau}+n_{\psi}&=m-k\notag\\
n_{\beta}+n_{\eta}+n_{\omega}+n_{\xi}+n_{\theta}+n_{\tau}+n_{\phi}&=m-k.
\end{align}
Solving for
\begin{align}
\mathrm{arg} &= \sum_i n_i = n_{\alpha}  +n_{\beta} +n_{\gamma}+n_{\delta}+n_{\epsilon}+n_{\eta}+n_{\omega}+n_{\mu} +n_{\xi} +n_{\theta}+n_{\lambda} +n_{\nu}\notag\\&\hspace{2cm}+n_{\pi} +n_{\rho} +n_{\sigma}+n_{\tau} +n_{\phi} +n_{\psi} +n_{\kappa} +n_{\chi} = m+4k
\end{align}
gives the following restrictions on the number of single-particles in each of these loops
\begin{align}
n_{\theta} ~=~n_{\lambda }~=~n_{\nu }~=~n_{\pi }~=~n_{\rho }~=~n_{\sigma }~=~n_{\tau }~=~n_{\phi }~=~n_{\psi }~=~n_{\kappa }~=~0\\
n_{\gamma} ~=~n_{\delta }~=~n_{\epsilon }~=~n_{\eta }~=~n_{\omega }~=~n_{\mu }~=~k
\end{align}
with $n_{\alpha }~=~n_{\beta }$, as well as the following identities
\begin{align}
n_{\chi} &= k-n_{\alpha}\\
n_{\xi} &= m-3k-n_{\alpha}.
\end{align}
Hence
\begin{figure}[t]
\centering
\begin{tikzpicture}[scale=1.7]
\draw[ultra thick, black] (0,0) -- (1,0);
\draw[ultra thick, black] (0,-1) -- (1,-1);
\draw[ultra thick, dashed, black] (0,0) -- (0,-1);
\draw[ultra thick, dashed, black] (.95,0) -- (.95,-1);
\draw[ultra thick, green] (1,0) -- (2,0);
\draw[ultra thick, green] (1,-1) -- (2,-1);
\draw[ultra thick, dashed, green] (1.05,0) -- (1.05,-1);
\draw[ultra thick, dashed, green] (2,0) -- (2,-1);
\node[circle, fill=white, draw=black, text=black, inner sep=1pt] at (0,0) {$u$}; 
\node[circle, fill=white, draw=black, text=black,  inner sep=1pt] at (1,0) {$t$}; 
\node[circle, fill=white, draw=black, text=black,  inner sep=1pt] at (2,0) {$c$};
\node[circle, fill=white, draw=black, text=black, inner sep=1pt] at (0,-1) {$w$}; 
\node[circle, fill=white, draw=black, text=black,  inner sep=1pt] at (1,-1) {$v$}; 
\node[circle, fill=white, draw=black, text=black, inner sep=1pt] at (2,-1) {$q$};
\begin{scope}[shift={(3,-.5)}]
\draw[ultra thick, red] (0,0) -- (0.75,.7);\draw[ultra thick, dashed, red] (0.75,.7) -- (1.5,0);\draw[ultra thick, dashed, red] (0,0) -- (0.75,-.7);\draw[ultra thick, red] (0.75,-.7) -- (1.5,0);\node[circle, fill=white, draw=black, text=black, inner sep=1pt] at (0,0) {$p$}; \node[circle, fill=white, draw=black, text=black,  inner sep=1pt] at (1.5,0) {$e$}; \node[circle, fill=white, draw=black, text=black,  inner sep=1pt] at (0.75,-.7) {$u$}; \node[circle, fill=white, draw=black, text=black,  inner sep=1pt] at (0.75,.7) {$q$};
\begin{scope}[shift={(2,0)}]
\draw[ultra thick, dashed, blue] (0,0) -- (0.75,.7);\draw[ultra thick, blue] (0.75,.7) -- (1.5,0);\draw[ultra thick, blue] (0,0) -- (0.75,-.7);\draw[ultra thick, dashed, blue] (0.75,-.7) -- (1.5,0);\node[circle, fill=white, draw=black, text=black, inner sep=1pt] at (0,0) {$p$}; \node[circle, fill=white, draw=black, text=black, inner sep=1pt] at (1.5,0) {$e$}; \node[circle, fill=white, draw=black, text=black, inner sep=1pt] at (0.75,-.7) {$c$}; \node[circle, fill=white, draw=black, text=black, inner sep=1pt] at (0.75,.7) {$w$};\end{scope}
\end{scope}
\end{tikzpicture}
\caption[Particle Diagram for the term $A_{uwvt}A_{cqvt}A_{pueq}A_{pwec}$]{Illustration of the particle diagram for the term $A_{uwvt}A_{cqvt}A_{pueq}A_{pwec}$.}
\label{fig:diamond_collapse_1}
~ ~\\
\centering
\begin{tikzpicture}[scale=1.7]
\draw[ultra thick, black] (0,0) -- (1,0);
\draw[ultra thick, black] (0,-1) -- (1,-1);
\draw[ultra thick, dashed, black] (0,0) -- (0,-1);
\draw[ultra thick, dashed, black] (.95,0) -- (.95,-1);
\draw[ultra thick, dashed, green] (1,0) -- (2,0);
\draw[ultra thick, dashed, green] (1,-1) -- (2,-1);
\draw[ultra thick, green] (1.05,0) -- (1.05,-1);
\draw[ultra thick, green] (2,0) -- (2,-1);
\node[circle, fill=white, draw=black, text=black, inner sep=1pt] at (0,0) {$w$}; 
\node[circle, fill=white, draw=black, text=black,  inner sep=1pt] at (1,0) {$c$}; 
\node[circle, fill=white, draw=black, text=black,  inner sep=1pt] at (2,0) {$q$};
\node[circle, fill=white, draw=black, text=black, inner sep=1pt] at (0,-1) {$u$}; 
\node[circle, fill=white, draw=black, text=black,  inner sep=1pt] at (1,-1) {$e$}; 
\node[circle, fill=white, draw=black, text=black, inner sep=1pt] at (2,-1) {$p$};
\begin{scope}[shift={(3,-.5)}]
\draw[ultra thick, red] (0,0) -- (0.75,.7);\draw[ultra thick, dashed, red] (0.75,.7) -- (1.5,0);\draw[ultra thick, dashed, red] (0,0) -- (0.75,-.7);\draw[ultra thick, red] (0.75,-.7) -- (1.5,0);\node[circle, fill=white, draw=black, text=black, inner sep=1pt] at (0,0) {$t$}; \node[circle, fill=white, draw=black, text=black,  inner sep=1pt] at (1.5,0) {$v$}; \node[circle, fill=white, draw=black, text=black,  inner sep=1pt] at (0.75,-.7) {$q$}; \node[circle, fill=white, draw=black, text=black,  inner sep=1pt] at (0.75,.7) {$u$};
\begin{scope}[shift={(2,0)}]
\draw[ultra thick, dashed, blue] (0,0) -- (0.75,.7);\draw[ultra thick, blue] (0.75,.7) -- (1.5,0);\draw[ultra thick, blue] (0,0) -- (0.75,-.7);\draw[ultra thick, dashed, blue] (0.75,-.7) -- (1.5,0);\node[circle, fill=white, draw=black,  text=black, inner sep=1pt] at (0,0) {$t$}; \node[circle, fill=white, draw=black, text=black, inner sep=1pt] at (1.5,0) {$v$}; \node[circle, fill=white, draw=black, text=black, inner sep=1pt] at (0.75,-.7) {$p$}; \node[circle, fill=white, draw=black, text=black, inner sep=1pt] at (0.75,.7) {$w$};\end{scope}
\end{scope}
\end{tikzpicture}
\caption[The particle diagram for $A_{uwce}A_{epqc}A_{tqvu}A_{twvp}$]{The particle diagram for $A_{uwce}A_{epqc}A_{tqvu}A_{twvp}$. By acknowledging the overlaps between the bonds of the left sub-diagram we can conclude that the complete diagram is identical to Fig \ref{fig:diamond} which we have already calculated for the sixth moment, with an additional factor ${{m-k}\choose{k}}{{l-m-k}\choose k}$ due to the degrees of freedom between $|c\rangle$ and $|e\rangle$.

\hrulefill}
\label{fig:diamond_collapse_2}
\end{figure}

\begin{figure}\label{fig:paths_blah}
\begin{subfigure}{.33\textwidth}
\begin{tikzpicture}[node distance=1cm, xscale=.75]
\node (w) [draw, circle, inner sep=1pt, fill=blue!20, font=\bfseries] {\footnotesize w};
\node (c) [draw, circle, inner sep=1pt, fill=blue!20, font=\bfseries, right of=w] {\footnotesize c};
\node (q) [draw, circle, inner sep=1pt, fill=blue!20, font=\bfseries, right of=c] {\footnotesize q};
\node (t) [draw, circle, inner sep=1pt, fill=blue!20, font=\bfseries, below left of =w] {\footnotesize t};
\node (v) [draw, circle, inner sep=1pt, fill=blue!20, font=\bfseries, below right of =q] {\footnotesize v};
\node (u) [draw, circle, inner sep=1pt, fill=blue!20, font=\bfseries, below right of=t] {\footnotesize u};
\node (p) [draw, circle, inner sep=1pt, fill=blue!20, font=\bfseries, below left of=v] {\footnotesize p};
\node (e) [draw, circle, inner sep=1pt, fill=blue!20, font=\bfseries, left of=p] {\footnotesize e};
\draw (t) to [out=90, in=130] (q);
\draw (u) to [out=-50, in=-90] (v);
\draw (q) -- (c) -- (e) -- (p) -- (v);
\draw (t) -- (w) -- (u);
\end{tikzpicture}
  \caption{The loop \normalsize$\overrightarrow{wtqcepvuw}$}
  \label{fig:pb_1}
\end{subfigure}%
\begin{subfigure}{.33\textwidth}
  \begin{tikzpicture}[node distance=1cm, xscale=.75, yscale=.75]
\node (w) [draw, circle, inner sep=1pt, fill=apricot, font=\bfseries] {\footnotesize w};
\node (c) [draw, circle, inner sep=1pt, fill=apricot, font=\bfseries, right of=w] {\footnotesize c};
\node (q) [draw, circle, inner sep=1pt, fill=apricot, font=\bfseries, right of=c] {\footnotesize q};
\node (t) [draw, circle, inner sep=1pt, fill=apricot, font=\bfseries, below left of =w] {\footnotesize t};
\node (v) [draw, circle, inner sep=1pt, fill=apricot, font=\bfseries, below right of =q] {\footnotesize v};
\node (u) [draw, circle, inner sep=1pt, fill=apricot, font=\bfseries, below right of=t] {\footnotesize u};
\node (p) [draw, circle, inner sep=1pt, fill=apricot, font=\bfseries, below left of=v] {\footnotesize p};
\node (e) [draw, circle, inner sep=1pt, fill=apricot, font=\bfseries, left of=p] {\footnotesize e};
\draw (t) to [out=-90, in=-130] (p);
\draw (w) to [out=50, in=90] (v);
\draw (w) -- (c) -- (e) -- (u) -- (t);
\draw (p) -- (q) -- (v);
\end{tikzpicture}
  \caption{The loop \normalsize$\overrightarrow{wvqptuecw}$}
  \label{fig:pb2}
\end{subfigure}%
\begin{subfigure}{.33\textwidth}
 \begin{tikzpicture}[node distance=1cm, xscale=.75, yscale=.75]
\node (w) [draw, circle, inner sep=1pt, fill=blue!20, font=\bfseries] {\footnotesize w};
\node (c) [draw, circle, inner sep=1pt, fill=blue!20, font=\bfseries, right of=w] {\footnotesize c};
\node (q) [draw, circle, inner sep=1pt, fill=blue!20, font=\bfseries, right of=c] {\footnotesize q};
\node (t) [draw, circle, inner sep=1pt, fill=blue!20, font=\bfseries, below left of =w] {\footnotesize t};
\node (v) [draw, circle, inner sep=1pt, fill=blue!20, font=\bfseries, below right of =q] {\footnotesize v};
\node (u) [draw, circle, inner sep=1pt, fill=blue!20, font=\bfseries, below right of=t] {\footnotesize u};
\node (p) [draw, circle, inner sep=1pt, fill=blue!20, font=\bfseries, below left of=v] {\footnotesize p};
\node (e) [draw, circle, inner sep=1pt, fill=blue!20, font=\bfseries, left of=p] {\footnotesize e};
\draw (t) to [out=-90, in=-130] (p);
\draw (w) to [out=50, in=90] (v);
\draw (p) -- (e) -- (c) -- (q) -- (v);
\draw (w) -- (u) -- (t);
\end{tikzpicture}
  \caption{The loop \normalsize$\overrightarrow{wvqceptuw}$}
  \label{fig:pb3}
\end{subfigure}\\
\begin{subfigure}{.33\textwidth}
\begin{tikzpicture}[node distance=1cm, xscale=.75, yscale=.75]
\node (w) [draw, circle, inner sep=1pt, fill=apricot, font=\bfseries] {\footnotesize w};
\node (c) [draw, circle, inner sep=1pt, fill=apricot, font=\bfseries, right of=w] {\footnotesize c};
\node (q) [draw, circle, inner sep=1pt, fill=apricot, font=\bfseries, right of=c] {\footnotesize q};
\node (t) [draw, circle, inner sep=1pt, fill=apricot, font=\bfseries, below left of =w] {\footnotesize t};
\node (v) [draw, circle, inner sep=1pt, fill=apricot, font=\bfseries, below right of =q] {\footnotesize v};
\node (u) [draw, circle, inner sep=1pt, fill=apricot, font=\bfseries, below right of=t] {\footnotesize u};
\node (p) [draw, circle, inner sep=1pt, fill=apricot, font=\bfseries, below left of=v] {\footnotesize p};
\node (e) [draw, circle, inner sep=1pt, fill=apricot, font=\bfseries, left of=p] {\footnotesize e};
\draw (t) to [out=90, in=130] (q);
\draw (u) to [out=-50, in=-90] (v);
\draw (t) -- (w) -- (c) -- (e) -- (u);
\draw (v) -- (p) -- (q);
\end{tikzpicture}
  \caption{The loop \normalsize$\overrightarrow{tqpvuecwt}$}
  \label{fig:pb_1}
\end{subfigure}%
\begin{subfigure}{.33\textwidth}
  \begin{tikzpicture}[node distance=1cm, xscale=.75, yscale=.75]
\node (w) [draw, circle, inner sep=1pt, fill=blue!20, font=\bfseries] {\footnotesize w};
\node (c) [draw, circle, inner sep=1pt, fill=blue!20, font=\bfseries, right of=w] {\footnotesize c};
\node (q) [draw, circle, inner sep=1pt, fill=blue!20, font=\bfseries, right of=c] {\footnotesize q};
\node (t) [draw, circle, inner sep=1pt, fill=blue!20, font=\bfseries, below left of =w] {\footnotesize t};
\node (v) [draw, circle, inner sep=1pt, fill=blue!20, font=\bfseries, below right of =q] {\footnotesize v};
\node (u) [draw, circle, inner sep=1pt, fill=blue!20, font=\bfseries, below right of=t] {\footnotesize u};
\node (p) [draw, circle, inner sep=1pt, fill=blue!20, font=\bfseries, below left of=v] {\footnotesize p};
\node (e) [draw, circle, inner sep=1pt, fill=blue!20, font=\bfseries, left of=p] {\footnotesize e};
\draw (t) to [out=90, in=130] (q);
\draw (u) [white] to [out=-50, in=-90] (v); 
\draw (t) -- (w) -- (c) -- (q);
\end{tikzpicture}
  \caption{The loop \normalsize$\overrightarrow{tqcwt}$}
  \label{fig:pb2}
\end{subfigure}%
\begin{subfigure}{.33\textwidth}
 \begin{tikzpicture}[node distance=1cm, xscale=.75, yscale=.75]
\node (w) [draw, circle, inner sep=1pt, fill=apricot, font=\bfseries] {\footnotesize w};
\node (c) [draw, circle, inner sep=1pt, fill=apricot, font=\bfseries, right of=w] {\footnotesize c};
\node (q) [draw, circle, inner sep=1pt, fill=apricot, font=\bfseries, right of=c] {\footnotesize q};
\node (t) [draw, circle, inner sep=1pt, fill=apricot, font=\bfseries, below left of =w] {\footnotesize t};
\node (v) [draw, circle, inner sep=1pt, fill=apricot, font=\bfseries, below right of =q] {\footnotesize v};
\node (u) [draw, circle, inner sep=1pt, fill=apricot, font=\bfseries, below right of=t] {\footnotesize u};
\node (p) [draw, circle, inner sep=1pt, fill=apricot, font=\bfseries, below left of=v] {\footnotesize p};
\node (e) [draw, circle, inner sep=1pt, fill=apricot, font=\bfseries, left of=p] {\footnotesize e};
\draw (w) to [out=50, in=90] (v);
\draw (u) [white] to [out=-50, in=-90] (v); 
\draw (w) -- (c) -- (q) -- (v);
\end{tikzpicture}
  \caption{The loop \normalsize$\overrightarrow{wvqcw}$}
  \label{fig:pb3}
\end{subfigure}\\
\begin{subfigure}{.33\textwidth}
\begin{tikzpicture}[node distance=1cm, xscale=.75, yscale=.75]
\node (w) [draw, circle, inner sep=1pt, fill=blue!20, font=\bfseries] {\footnotesize w};
\node (c) [draw, circle, inner sep=1pt, fill=blue!20, font=\bfseries, right of=w] {\footnotesize c};
\node (q) [draw, circle, inner sep=1pt, fill=blue!20, font=\bfseries, right of=c] {\footnotesize q};
\node (t) [draw, circle, inner sep=1pt, fill=blue!20, font=\bfseries, below left of =w] {\footnotesize t};
\node (v) [draw, circle, inner sep=1pt, fill=blue!20, font=\bfseries, below right of =q] {\footnotesize v};
\node (u) [draw, circle, inner sep=1pt, fill=blue!20, font=\bfseries, below right of=t] {\footnotesize u};
\node (p) [draw, circle, inner sep=1pt, fill=blue!20, font=\bfseries, below left of=v] {\footnotesize p};
\node (e) [draw, circle, inner sep=1pt, fill=blue!20, font=\bfseries, left of=p] {\footnotesize e};
\draw (t) [white] to [out=90, in=130] (q); 
\draw (t) to [out=-90, in=-130] (p);
\draw (t) -- (u) -- (e) -- (p);
\end{tikzpicture}
  \caption{The loop \normalsize$\overrightarrow{tpeut}$}
  \label{fig:pb_1}
\end{subfigure}%
\begin{subfigure}{.33\textwidth}
  \begin{tikzpicture}[node distance=1cm, xscale=.75, yscale=.75]
\node (w) [draw, circle, inner sep=1pt, fill=apricot, font=\bfseries] {\footnotesize w};
\node (c) [draw, circle, inner sep=1pt, fill=apricot, font=\bfseries, right of=w] {\footnotesize c};
\node (q) [draw, circle, inner sep=1pt, fill=apricot, font=\bfseries, right of=c] {\footnotesize q};
\node (t) [draw, circle, inner sep=1pt, fill=apricot, font=\bfseries, below left of =w] {\footnotesize t};
\node (v) [draw, circle, inner sep=1pt, fill=apricot, font=\bfseries, below right of =q] {\footnotesize v};
\node (u) [draw, circle, inner sep=1pt, fill=apricot, font=\bfseries, below right of=t] {\footnotesize u};
\node (p) [draw, circle, inner sep=1pt, fill=apricot, font=\bfseries, below left of=v] {\footnotesize p};
\node (e) [draw, circle, inner sep=1pt, fill=apricot, font=\bfseries, left of=p] {\footnotesize e};
\draw (t) [white] to [out=90, in=130] (q); 
\draw (u) to [out=-50, in=-90] (v);
\draw (u) -- (e) -- (p) -- (v);
\end{tikzpicture}
  \caption{The loop \normalsize$\overrightarrow{uvpeu}$}
  \label{fig:pb2}
\end{subfigure}%
\begin{subfigure}{.33\textwidth}
 \begin{tikzpicture}[node distance=1cm, xscale=.75, yscale=.75]
\node (w) [draw, circle, inner sep=1pt, fill=blue!20, font=\bfseries] {\footnotesize w};
\node (c) [draw, circle, inner sep=1pt, fill=blue!20, font=\bfseries, right of=w] {\footnotesize c};
\node (q) [draw, circle, inner sep=1pt, fill=blue!20, font=\bfseries, right of=c] {\footnotesize q};
\node (t) [draw, circle, inner sep=1pt, fill=blue!20, font=\bfseries, below left of =w] {\footnotesize t};
\node (v) [draw, circle, inner sep=1pt, fill=blue!20, font=\bfseries, below right of =q] {\footnotesize v};
\node (u) [draw, circle, inner sep=1pt, fill=blue!20, font=\bfseries, below right of=t] {\footnotesize u};
\node (p) [draw, circle, inner sep=1pt, fill=blue!20, font=\bfseries, below left of=v] {\footnotesize p};
\node (e) [draw, circle, inner sep=1pt, fill=blue!20, font=\bfseries, left of=p] {\footnotesize e};
\draw (t) [white] to [out=90, in=130] (q); 
\draw (u) [white] to [out=-50, in=-90] (v); 
\draw (t) -- (w) -- (c) -- (q) -- (v) -- (p) -- (e) -- (u) -- (t);
\end{tikzpicture}
  \caption{The loop \normalsize$\overrightarrow{twcqvpeut}$}
  \label{fig:pb3}
\end{subfigure}\\
\begin{subfigure}{.33\textwidth}
\begin{tikzpicture}[node distance=1cm, xscale=.75, yscale=.75]
\node (w) [draw, circle, inner sep=1pt, fill=apricot, font=\bfseries] {\footnotesize w};
\node (c) [draw, circle, inner sep=1pt, fill=apricot, font=\bfseries, right of=w] {\footnotesize c};
\node (q) [draw, circle, inner sep=1pt, fill=apricot, font=\bfseries, right of=c] {\footnotesize q};
\node (t) [draw, circle, inner sep=1pt, fill=apricot, font=\bfseries, below left of =w] {\footnotesize t};
\node (v) [draw, circle, inner sep=1pt, fill=apricot, font=\bfseries, below right of =q] {\footnotesize v};
\node (u) [draw, circle, inner sep=1pt, fill=apricot, font=\bfseries, below right of=t] {\footnotesize u};
\node (p) [draw, circle, inner sep=1pt, fill=apricot, font=\bfseries, below left of=v] {\footnotesize p};
\node (e) [draw, circle, inner sep=1pt, fill=apricot, font=\bfseries, left of=p] {\footnotesize e};
\draw (t) [white] to [out=90, in=130] (q); 
\draw (u) [white] to [out=-50, in=-90] (v); 
\draw (t) -- (w) -- (u) -- (t);
\end{tikzpicture}
  \caption{The loop \normalsize$\overrightarrow{twut}$}
  \label{fig:pb_1}
\end{subfigure}%
\begin{subfigure}{.33\textwidth}
  \begin{tikzpicture}[node distance=1cm, xscale=.75, yscale=.75]
\node (w) [draw, circle, inner sep=1pt, fill=blue!20, font=\bfseries] {\footnotesize w};
\node (c) [draw, circle, inner sep=1pt, fill=blue!20, font=\bfseries, right of=w] {\footnotesize c};
\node (q) [draw, circle, inner sep=1pt, fill=blue!20, font=\bfseries, right of=c] {\footnotesize q};
\node (t) [draw, circle, inner sep=1pt, fill=blue!20, font=\bfseries, below left of =w] {\footnotesize t};
\node (v) [draw, circle, inner sep=1pt, fill=blue!20, font=\bfseries, below right of =q] {\footnotesize v};
\node (u) [draw, circle, inner sep=1pt, fill=blue!20, font=\bfseries, below right of=t] {\footnotesize u};
\node (p) [draw, circle, inner sep=1pt, fill=blue!20, font=\bfseries, below left of=v] {\footnotesize p};
\node (e) [draw, circle, inner sep=1pt, fill=blue!20, font=\bfseries, left of=p] {\footnotesize e};
\draw (t) [white] to [out=90, in=130] (q); 
\draw (u) [white] to [out=-50, in=-90] (v); 
\draw (q) -- (p) -- (v) -- (q);
\end{tikzpicture}
  \caption{The loop \normalsize$\overrightarrow{qvpq}$}
  \label{fig:pb2}
\end{subfigure}%
\begin{subfigure}{.33\textwidth}
 \begin{tikzpicture}[node distance=1cm, xscale=.75, yscale=.75]
\node (w) [draw, circle, inner sep=1pt, fill=apricot, font=\bfseries] {\footnotesize w};
\node (c) [draw, circle, inner sep=1pt, fill=apricot, font=\bfseries, right of=w] {\footnotesize c};
\node (q) [draw, circle, inner sep=1pt, fill=apricot, font=\bfseries, right of=c] {\footnotesize q};
\node (t) [draw, circle, inner sep=1pt, fill=apricot, font=\bfseries, below left of =w] {\footnotesize t};
\node (v) [draw, circle, inner sep=1pt, fill=apricot, font=\bfseries, below right of =q] {\footnotesize v};
\node (u) [draw, circle, inner sep=1pt, fill=apricot, font=\bfseries, below right of=t] {\footnotesize u};
\node (p) [draw, circle, inner sep=1pt, fill=apricot, font=\bfseries, below left of=v] {\footnotesize p};
\node (e) [draw, circle, inner sep=1pt, fill=apricot, font=\bfseries, left of=p] {\footnotesize e};
\draw (t) [white] to [out=90, in=130] (q); 
\draw (u) [white] to [out=-50, in=-90] (v); 
\draw (t) -- (w) -- (c) -- (e) -- (u) -- (t);
\end{tikzpicture}
  \caption{The loop \normalsize$\overrightarrow{twceut}$}
  \label{fig:pb3}
\end{subfigure}\\
\end{figure}
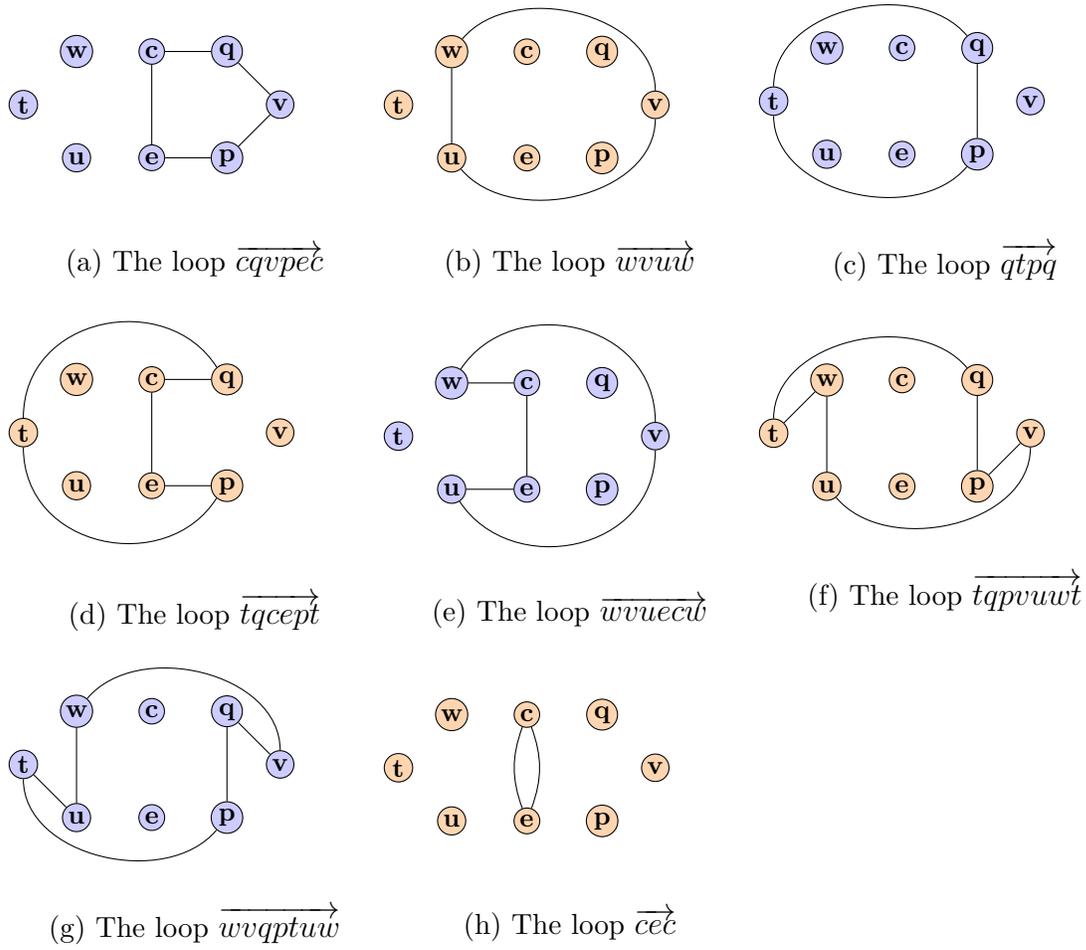
\begin{figure}
\ContinuedFloat
\begin{subfigure}{.33\textwidth}
\begin{tikzpicture}[node distance=1cm, xscale=.75, yscale=.75]
\node (w) [draw, circle, inner sep=1pt, fill=blue!20, font=\bfseries] {\footnotesize w};
\node (c) [draw, circle, inner sep=1pt, fill=blue!20, font=\bfseries, right of=w] {\footnotesize c};
\node (q) [draw, circle, inner sep=1pt, fill=blue!20, font=\bfseries, right of=c] {\footnotesize q};
\node (t) [draw, circle, inner sep=1pt, fill=blue!20, font=\bfseries, below left of =w] {\footnotesize t};
\node (v) [draw, circle, inner sep=1pt, fill=blue!20, font=\bfseries, below right of =q] {\footnotesize v};
\node (u) [draw, circle, inner sep=1pt, fill=blue!20, font=\bfseries, below right of=t] {\footnotesize u};
\node (p) [draw, circle, inner sep=1pt, fill=blue!20, font=\bfseries, below left of=v] {\footnotesize p};
\node (e) [draw, circle, inner sep=1pt, fill=blue!20, font=\bfseries, left of=p] {\footnotesize e};
\draw (t) [white] to [out=90, in=130] (q); 
\draw (u) [white] to [out=-50, in=-90] (v); 
\draw (c) -- (q) -- (v) -- (p) -- (e) -- (c);
\end{tikzpicture}
  \caption{The loop \normalsize$\overrightarrow{cqvpec}$}
  \label{fig:pb_1}
\end{subfigure}%
\begin{subfigure}{.33\textwidth}
  \begin{tikzpicture}[node distance=1cm, xscale=.75, yscale=.75]
\node (w) [draw, circle, inner sep=1pt, fill=apricot, font=\bfseries] {\footnotesize w};
\node (c) [draw, circle, inner sep=1pt, fill=apricot, font=\bfseries, right of=w] {\footnotesize c};
\node (q) [draw, circle, inner sep=1pt, fill=apricot, font=\bfseries, right of=c] {\footnotesize q};
\node (t) [draw, circle, inner sep=1pt, fill=apricot, font=\bfseries, below left of =w] {\footnotesize t};
\node (v) [draw, circle, inner sep=1pt, fill=apricot, font=\bfseries, below right of =q] {\footnotesize v};
\node (u) [draw, circle, inner sep=1pt, fill=apricot, font=\bfseries, below right of=t] {\footnotesize u};
\node (p) [draw, circle, inner sep=1pt, fill=apricot, font=\bfseries, below left of=v] {\footnotesize p};
\node (e) [draw, circle, inner sep=1pt, fill=apricot, font=\bfseries, left of=p] {\footnotesize e};
\draw (w) to [out=50, in=90] (v);
\draw (u) to [out=-50, in=-90] (v);
\draw (w) -- (u);
\end{tikzpicture}
  \caption{The loop \normalsize$\overrightarrow{wvuw}$}
  \label{fig:pb2}
\end{subfigure}%
\begin{subfigure}{.33\textwidth}
 \begin{tikzpicture}[node distance=1cm, xscale=.75, yscale=.75]
\node (w) [draw, circle, inner sep=1pt, fill=blue!20, font=\bfseries] {\footnotesize w};
\node (c) [draw, circle, inner sep=1pt, fill=blue!20, font=\bfseries, right of=w] {\footnotesize c};
\node (q) [draw, circle, inner sep=1pt, fill=blue!20, font=\bfseries, right of=c] {\footnotesize q};
\node (t) [draw, circle, inner sep=1pt, fill=blue!20, font=\bfseries, below left of =w] {\footnotesize t};
\node (v) [draw, circle, inner sep=1pt, fill=blue!20, font=\bfseries, below right of =q] {\footnotesize v};
\node (u) [draw, circle, inner sep=1pt, fill=blue!20, font=\bfseries, below right of=t] {\footnotesize u};
\node (p) [draw, circle, inner sep=1pt, fill=blue!20, font=\bfseries, below left of=v] {\footnotesize p};
\node (e) [draw, circle, inner sep=1pt, fill=blue!20, font=\bfseries, left of=p] {\footnotesize e};
\draw (t) to [out=90, in=130] (q);
\draw (t) to [out=-90, in=-130] (p);
\draw (q) -- (p);
\end{tikzpicture}
  \caption{The loop \normalsize$\overrightarrow{qtpq}$}
  \label{fig:pb3}
\end{subfigure}\\
\begin{subfigure}{.33\textwidth}
\begin{tikzpicture}[node distance=1cm, xscale=.75]
\node (w) [draw, circle, inner sep=1pt, fill=apricot, font=\bfseries] {\footnotesize w};
\node (c) [draw, circle, inner sep=1pt, fill=apricot, font=\bfseries, right of=w] {\footnotesize c};
\node (q) [draw, circle, inner sep=1pt, fill=apricot, font=\bfseries, right of=c] {\footnotesize q};
\node (t) [draw, circle, inner sep=1pt, fill=apricot, font=\bfseries, below left of =w] {\footnotesize t};
\node (v) [draw, circle, inner sep=1pt, fill=apricot, font=\bfseries, below right of =q] {\footnotesize v};
\node (u) [draw, circle, inner sep=1pt, fill=apricot, font=\bfseries, below right of=t] {\footnotesize u};
\node (p) [draw, circle, inner sep=1pt, fill=apricot, font=\bfseries, below left of=v] {\footnotesize p};
\node (e) [draw, circle, inner sep=1pt, fill=apricot, font=\bfseries, left of=p] {\footnotesize e};
\draw (t) to [out=90, in=130] (q);
\draw (t) to [out=-90, in=-130] (p);
\draw (q) -- (c) -- (e) -- (p);
\end{tikzpicture}
  \caption{The loop \normalsize$\overrightarrow{tqcept}$}
  \label{fig:pb_1}
\end{subfigure}%
\begin{subfigure}{.33\textwidth}
  \begin{tikzpicture}[node distance=1cm, xscale=.75]
\node (w) [draw, circle, inner sep=1pt, fill=blue!20, font=\bfseries] {\footnotesize w};
\node (c) [draw, circle, inner sep=1pt, fill=blue!20, font=\bfseries, right of=w] {\footnotesize c};
\node (q) [draw, circle, inner sep=1pt, fill=blue!20, font=\bfseries, right of=c] {\footnotesize q};
\node (t) [draw, circle, inner sep=1pt, fill=blue!20, font=\bfseries, below left of =w] {\footnotesize t};
\node (v) [draw, circle, inner sep=1pt, fill=blue!20, font=\bfseries, below right of =q] {\footnotesize v};
\node (u) [draw, circle, inner sep=1pt, fill=blue!20, font=\bfseries, below right of=t] {\footnotesize u};
\node (p) [draw, circle, inner sep=1pt, fill=blue!20, font=\bfseries, below left of=v] {\footnotesize p};
\node (e) [draw, circle, inner sep=1pt, fill=blue!20, font=\bfseries, left of=p] {\footnotesize e};
\draw (u) to [out=-50, in=-90] (v);
\draw (w) to [out=50, in=90] (v);
\draw (w) -- (c) -- (e) -- (u);
\end{tikzpicture}
  \caption{The loop \normalsize$\overrightarrow{wvuecw}$}
  \label{fig:pb2}
\end{subfigure}%
\begin{subfigure}{.33\textwidth}
 \begin{tikzpicture}[node distance=1cm, xscale=.75, yscale=.75]
\node (w) [draw, circle, inner sep=1pt, fill=apricot, font=\bfseries] {\footnotesize w};
\node (c) [draw, circle, inner sep=1pt, fill=apricot, font=\bfseries, right of=w] {\footnotesize c};
\node (q) [draw, circle, inner sep=1pt, fill=apricot, font=\bfseries, right of=c] {\footnotesize q};
\node (t) [draw, circle, inner sep=1pt, fill=apricot, font=\bfseries, below left of =w] {\footnotesize t};
\node (v) [draw, circle, inner sep=1pt, fill=apricot, font=\bfseries, below right of =q] {\footnotesize v};
\node (u) [draw, circle, inner sep=1pt, fill=apricot, font=\bfseries, below right of=t] {\footnotesize u};
\node (p) [draw, circle, inner sep=1pt, fill=apricot, font=\bfseries, below left of=v] {\footnotesize p};
\node (e) [draw, circle, inner sep=1pt, fill=apricot, font=\bfseries, left of=p] {\footnotesize e};
\draw (t) to [out=90, in=130] (q);
\draw (u) to [out=-50, in=-90] (v);
\draw (q) -- (p) -- (v);
\draw (t) -- (w) -- (u);
\end{tikzpicture}
  \caption{The loop \normalsize$\overrightarrow{tqpvuwt}$}
  \label{fig:pb3}
\end{subfigure}\\
\begin{subfigure}{.33\textwidth}
\begin{tikzpicture}[node distance=1cm, xscale=.75, yscale=.75]
\node (w) [draw, circle, inner sep=1pt, fill=blue!20, font=\bfseries] {\footnotesize w};
\node (c) [draw, circle, inner sep=1pt, fill=blue!20, font=\bfseries, right of=w] {\footnotesize c};
\node (q) [draw, circle, inner sep=1pt, fill=blue!20, font=\bfseries, right of=c] {\footnotesize q};
\node (t) [draw, circle, inner sep=1pt, fill=blue!20, font=\bfseries, below left of =w] {\footnotesize t};
\node (v) [draw, circle, inner sep=1pt, fill=blue!20, font=\bfseries, below right of =q] {\footnotesize v};
\node (u) [draw, circle, inner sep=1pt, fill=blue!20, font=\bfseries, below right of=t] {\footnotesize u};
\node (p) [draw, circle, inner sep=1pt, fill=blue!20, font=\bfseries, below left of=v] {\footnotesize p};
\node (e) [draw, circle, inner sep=1pt, fill=blue!20, font=\bfseries, left of=p] {\footnotesize e};
\draw (t) to [out=-90, in=-130] (p);
\draw (w) to [out=50, in=90] (v);
\draw (p) -- (q) -- (v);
\draw (t) -- (u) -- (w);
\end{tikzpicture}
  \caption{The loop \normalsize$\overrightarrow{wvqptuw}$}
  \label{fig:pb_1}
\end{subfigure}%
\begin{subfigure}{.33\textwidth}
  \begin{tikzpicture}[node distance=1cm, xscale=.75, yscale=.75]
\node (w) [draw, circle, inner sep=1pt, fill=apricot, font=\bfseries] {\footnotesize w};
\node (c) [draw, circle, inner sep=1pt, fill=apricot, font=\bfseries, right of=w] {\footnotesize c};
\node (q) [draw, circle, inner sep=1pt, fill=apricot, font=\bfseries, right of=c] {\footnotesize q};
\node (t) [draw, circle, inner sep=1pt, fill=apricot, font=\bfseries, below left of =w] {\footnotesize t};
\node (v) [draw, circle, inner sep=1pt, fill=apricot, font=\bfseries, below right of =q] {\footnotesize v};
\node (u) [draw, circle, inner sep=1pt, fill=apricot, font=\bfseries, below right of=t] {\footnotesize u};
\node (p) [draw, circle, inner sep=1pt, fill=apricot, font=\bfseries, below left of=v] {\footnotesize p};
\node (e) [draw, circle, inner sep=1pt, fill=apricot, font=\bfseries, left of=p] {\footnotesize e};
\draw (t) [white] to [out=90, in=130] (q); 
\draw (u) [white] to [out=-50, in=-90] (v); 
\draw (c) to [out=-70, in=70] (e) ;
\draw (e) to [out=110, in=-110] (c);
\end{tikzpicture}
  \caption{The loop \normalsize$\overrightarrow{cec}$}
  \label{fig:pb2}
\end{subfigure}%
\caption[Loops Illustration]{Illustration of all the loops in the particle diagram for the term $A_{uwce}A_{epqc}A_{tqvu}A_{twvp}$ represented in Fig \ref{fig:diamond_collapse_2}.}
\end{figure}
\begin{align}\label{eq:penpen}A_{uwvt}A_{cqvt}A_{pueq}A_{pwec}& \sim \sum_{n_{\alpha}}{l \choose {k~k~k~k~k~k~n_{\alpha}~n_{\beta}~n_{\chi}~n_{\xi}}}\notag\\
&=\sum_{n_{\alpha}}{l \choose {k~k~k~k~k~k~n_{\alpha}~n_{\alpha}~k-n_{\alpha}~m-3k-n_{\alpha}}}
\end{align}
giving
\begin{align}\lim_{N\to\infty} \frac{{\frac{1}{N}}A_{uwvt}A_{cqvt}A_{pueq}A_{pwec}}{\left({\frac{1}{N}} \mathrm{tr}(\overline{V_{k}^2})\right)^4 } = \frac{{{m-k}\choose{k}}{{m-2k}\choose{k}}^2}{{m\choose k}^3}.
\end{align}
\subsubsection{The term $\mathbf{A_{uwce}A_{epqc}A_{tqvu}A_{twvp}}$}
The penultimate term $A_{uwce}A_{epqc}A_{tqvu}A_{twvp}$ of (\ref{eq:s1.5}) yields the particle diagram of Fig. \ref{fig:diamond_collapse_2} which has the particle loops
\[
\begin{array}{ll}
\alpha ~=~\overrightarrow{wtqcepvu} & \hspace{.5cm}\lambda ~=~\overrightarrow{qvp} \\
\beta ~=~\overrightarrow{wceutpqv} & \hspace{.5cm}\rho ~=~\overrightarrow{wvu} \\
\gamma ~=~\overrightarrow{wutpecqv} & \hspace{.5cm} \sigma ~=~\overrightarrow{tqp} \\
\delta ~=~\overrightarrow{wceuvpqt} & \hspace{.5cm}\nu ~=~\overrightarrow{wtuec} \\
\epsilon ~=~\overrightarrow{wtqc}  & \hspace{.5cm}\pi ~=~\overrightarrow{cqvpe} \\
\eta ~=~\overrightarrow{wvqc}  & \hspace{.5cm}\tau ~=~\overrightarrow{tqcep}\\
\omega ~=~\overrightarrow{tuep} & \hspace{.5cm} \phi ~=~\overrightarrow{wceuv}\\
\mu ~=~\overrightarrow{uepv}  & \hspace{.5cm}\psi ~=~\overrightarrow{wtqpvu}\\
\xi ~=~\overrightarrow{twcqvpeu} & \hspace{.5cm}\kappa ~=~\overrightarrow{wvqptu}\\
\theta ~=~\overrightarrow{wtu} & \hspace{.5cm}\chi ~=~\overrightarrow{ce}\\
\end{array}
\]
and the following particle conservation equations
\allowdisplaybreaks[1]
\begin{align}
n_{\alpha}+n_{\delta}+n_{\epsilon }+n_{\xi }+n_{\theta}+ n_{\nu}+n_{\psi}&=k\\
n_{\alpha}+n_{\delta}+n_{\mu}+n_{\xi}+n_{\lambda}+n_{\pi}+n_{\psi}&=k\\
n_{\alpha}+n_{\gamma}+n_{\epsilon}+n_{\eta}+n_{\xi}+n_{\pi}+n_{\tau}&=k\\
n_{\alpha}+n_{\gamma}+n_{\omega}+n_{\mu}+n_{\xi}+n_{\pi}+n_{\tau}&=k\\
n_{\alpha}+n_{\delta}+n_{\epsilon}+n_{\tau}+n_{\psi}+n_{\sigma}&=k\\
n_{\alpha}+n_{\delta}+n_{\mu}+n_{\rho}+n_{\phi}+n_{\psi}&=k\\
n_{\alpha}+n_{\gamma}+n_{\theta}+n_{\rho}+n_{\psi}+n_{\kappa}&=k\\
n_{\alpha}+n_{\gamma}+n_{\pi}+n_{\tau}+n_{\chi}&=k\\
\notag\\
n_{\beta}+n_{\gamma}+n_{\omega}+n_{\xi}+n_{\theta}+n_{\nu}+n_{\kappa}&=m-k\\
n_{\beta}+n_{\gamma}+n_{\eta}+n_{\xi}+n_{\lambda}+n_{\pi}+n_{\kappa}&=m-k\\
n_{\beta}+n_{\delta}+n_{\epsilon}+n_{\eta}+n_{\xi}+n_{\nu}+n_{\phi}&=m-k\\
n_{\beta}+n_{\delta}+n_{\omega}+n_{\mu}+n_{\xi}+n_{\nu}+n_{\phi}&=m-k\\
\notag\\
n_{\beta}+n_{\gamma}+n_{\eta}+n_{\rho}+n_{\phi}+n_{\kappa}&=m-k\\
n_{\beta}+n_{\gamma}+n_{\omega}+n_{\sigma}+n_{\tau}+n_{\kappa}&=m-k\\
n_{\beta}+n_{\delta}+n_{\lambda}+n_{\sigma}+n_{\psi}+n_{\kappa}&=m-k\\
n_{\beta}+n_{\delta}+n_{\nu}+n_{\phi}+n_{\chi}&=m-k.
\end{align}
Solving for
\begin{align}
\mathrm{arg} &= \sum_i n_i \notag\\
&=n_{\alpha } +n_{\beta }+n_{\gamma}+n_{\delta }+n_{\epsilon}+n_{\eta}+n_{\omega}+n_{\mu }+n_{\xi }+n_{\theta}+n_{\lambda }+n_{\nu}\notag\\
&\hspace{2cm}+n_{\pi }+n_{\rho }+n_{\sigma}+n_{\tau }+n_{\phi }+n_{\psi }+n_{\kappa }+n_{\chi }\notag\\
&= m+4k
\end{align}
gives the following identities
\begin{align}
n_{\beta}+n_{\kappa}&=m-3k\\
n_{\theta} + n_{\nu} &= k\\
n_{\theta }+ n_{\kappa} &= n_{\phi}\\
n_{\rho} +n_{\kappa} &= n_{\nu}
\end{align}
as well as the restrictions $n_{\alpha} ~=~n_{\gamma} ~=~n_{\delta} ~=~n_{\epsilon} ~=~n_{\mu }~=~n_{\xi} ~=~n_{\pi} ~=~n_{\tau} ~=~n_{\psi} ~=~0$ and $n_{\eta} ~=~n_{\omega} ~=~n_{\lambda} ~=~n_{\sigma} ~=~n_{\chi }~=~k$ on the number of single-particles in each of these loops. In addition one finds
\begin{align}
n_{\beta} &~=~m-3k-n_{\kappa}\notag\\
n_{\theta }&~=~n_{\phi}-n_{\kappa} = k-n_{\rho}-n_{\kappa}\notag\\
n_{\nu} &~=~n_{\rho}+n_{\kappa}\notag\\
n_{\phi} &~=~k-n_{\rho}\\
\notag\\
n_{\rho} &~=~n_{\rho} \notag\\
n_{\kappa }&~=~n_{\kappa}
\end{align}
so that the final expression takes the value
\begin{equation}\label{eq:pen}A_{uwce}A_{epqc}A_{tqvu}A_{twvp} \sim \sum_{\kappa\rho}{l \choose {k~k~k~k~k~\kappa~m-3k-\kappa~\rho ~\rho+\kappa~k-\rho-\kappa}}
\end{equation}
which then gives
\begin{equation}\lim_{N\to\infty} \frac{{\frac{1}{N}}A_{uwce}A_{epqc}A_{tqvu}A_{twvp}}{\left({\frac{1}{N}} \mathrm{tr}(\overline{V_{k}^2})\right)^4 } = \frac{{{m-k}\choose{k}}^2{{m-2k}\choose{k}}}{{m\choose k}^3}
\end{equation}
\subsubsection{The final term $\mathbf{A_{uvqt}A_{twvc}A_{cewp}A_{pueq}}$}
For the evaluation of the final term $A_{uvqt}A_{twvc}A_{cewp}A_{pueq}$ of (\ref{eq:s1.5}) one can use the same method as outlined in the evaluation of $A_{pvuq}A_{qwvt}A_{tpwu}$ (Fig. \ref{fig:diamond}). That is, note that the states $|u\rangle$ and $|v\rangle$ together determine all other states in the diagram, Fig \ref{fig:box}. This means that to maximise the argument of the diagram we minimise the overlap between $|u\rangle$ and $|v\rangle$. As before this is done by identifying as many loops as possible within the diagram from $|u\rangle$ to $|u\rangle$ without passing through $|v\rangle$ and from $|v\rangle$ to $|v\rangle$ without passing through $|u\rangle$. The single-particle states which cannot be incorporated into these loops must necessarily be in the set that is shared by both $|u\rangle$ and $|v\rangle$. And whereas for the case of Fig \ref{fig:diamond} each loop is composed of three bonds, for this analogous term of the eighth moment each loop consists of $\frac{n}{2}=4$ bonds. The loops passing through the state $|u\rangle$ without passing through $|v\rangle$ are
\begin{align}p_1&=\overrightarrow{utcpu}\\
p_2&=\overrightarrow{utweu}\\
p_3&=\overrightarrow{uewvu}\\
p_4&=\overrightarrow{utcpweu}
\end{align}
and similarly the loops passing through the state $|u\rangle$ without passing through $|v\rangle$ are
\begin{align}p_5&=\overrightarrow{qpcvq}\\
p_6&=\overrightarrow{qewvq}\\
p_7&=\overrightarrow{qpctq}\\
p_8&=\overrightarrow{qvwecpq}.
\end{align}
\begin{figure}[t]
\centering
\begin{tikzpicture}[scale=1.7]
\draw[ultra thick, dashed, blue] (0,0) -- (0.7,.2);
\draw[ultra thick, red] (0,0) -- (0,-1.1);
\draw[ultra thick, black] (0,-1.1) -- (.7,-.8);
\draw[ultra thick, dashed, black] (.7,-.8) -- (1.55,-1);
\draw[ultra thick, red] (1.55,-1) -- (1.55,0);
\draw[ultra thick, dashed, blue] (1.55,0) -- (0.9,-.3);
\draw[ultra thick, dashed, black] (0,-1.1) -- (.9,-1.5);
\draw[ultra thick, black] (.9,-1.5) -- (1.55,-1);
\draw[ultra thick, blue] (1.55,0) -- (.7,.2);
\draw[ultra thick, green] (.7,.2) -- (.7,-.8);
\node[circle, fill=white, draw=black, text=black,  inner sep=1pt] at (0.7,-.8) {$e$};
\draw[ultra thick, dashed, blue] (.9,-.3) -- (.9,-1.5);
\draw[ultra thick, green] (.9,-.3) -- (.9,-1.5);
\draw[ultra thick, blue] (0,0) -- (0.9,-.3);
\node[circle, fill=white, draw=black, text=black, inner sep=1pt] at (0,0) {$t$}; 
\node[circle, fill=white, draw=black, text=black,  inner sep=1pt] at (0.7,.2) {$w$}; 
\node[circle, fill=white, draw=black, text=black,  inner sep=1pt] at (1.55,0) {$v$};
\node[circle, fill=white, draw=black, text=black, inner sep=1pt] at (.9,-.3) {$c$}; 
\node[circle, fill=white, draw=black, text=black,  inner sep=1pt] at (0,-1.1) {$u$}; 
\node[circle, fill=white, draw=black, text=black, inner sep=1pt] at (1.55,-1) {$q$}; 
\node[circle, fill=white, draw=black, text=black,  inner sep=1pt] at (0.9,-1.5) {$p$}; 
\begin{scope}[shift={(2.5,-.5)}]
\draw[ultra thick, green] (0,0) -- (0.75,.7);\draw[ultra thick, dashed, green] (0.75,.7) -- (1.5,0);\draw[ultra thick, dashed, green] (0,0) -- (0.75,-.7);\draw[ultra thick, green] (0.75,-.7) -- (1.5,0);\node[circle, fill=white, draw=black, text=black, inner sep=1pt] at (0,0) {$e$}; \node[circle, fill=white, draw=black, text=black,  inner sep=1pt] at (1.5,0) {$p$}; \node[circle, fill=white, draw=black, text=black,  inner sep=1pt] at (0.75,-.7) {$c$}; \node[circle, fill=white, draw=black, text=black,  inner sep=1pt] at (0.75,.7) {$w$};
\begin{scope}[shift={(2,0)}]
\draw[ultra thick, dashed, red] (0,0) -- (0.75,.7);\draw[ultra thick, red] (0.75,.7) -- (1.5,0);\draw[ultra thick, red] (0,0) -- (0.75,-.7);\draw[ultra thick, dashed, red] (0.75,-.7) -- (1.5,0);\node[circle, fill=white, draw=black, text=black, inner sep=1pt] at (0,0) {$q$}; \node[circle, fill=white, draw=black, text=black, inner sep=1pt] at (1.5,0) {$u$}; \node[circle, fill=white, draw=black, text=black, inner sep=1pt] at (0.75,-.7) {$v$}; \node[circle, fill=white, draw=black,  text=black, inner sep=1pt] at (0.75,.7) {$t$};\end{scope}
\end{scope}
\end{tikzpicture}
\caption[The particle diagram for $A_{uvqt}A_{twvc}A_{cewp}A_{pueq}$]{The particle diagram for $A_{uvqt}A_{twvc}A_{cewp}A_{pueq}$ which takes the form $l\choose{\hat{s}~k~k~k~k~k~k~k~k}$ with $\hat{s}=m-4k$.

\hrulefill}
\label{fig:box}
\end{figure}
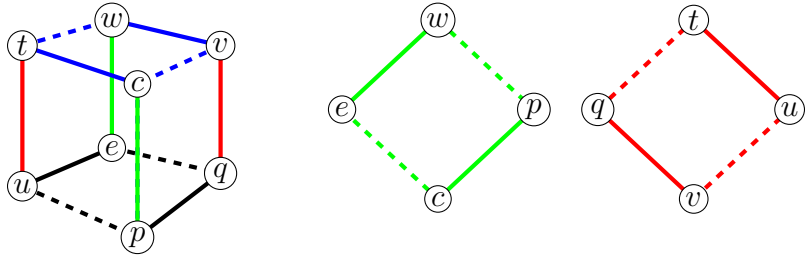
The single-particle states which are not an element of each of the bonds in a loop are necessarily an element of all the bonds in a loop through $|u\rangle$ and $|v\rangle$ and there are $\hat{s}:=m-4k$ of these. The final expression then reads
\begin{equation}\label{eq:ultimate}A_{uvqt}A_{twvc}A_{cewp}A_{pueq} \sim {l\choose{\hat{s}~k~k~k~k~k~k~k~k}}.
\end{equation}
Bringing all terms (\ref{eq:ASC1} -- \ref{eq:TCD1}), (\ref{eq:COLL1}), (\ref{eq:HAHN}), (\ref{eq:penpen}), (\ref{eq:pen}) and (\ref{eq:ultimate}) together gives
\begin{align}\label{eq:s11} &{lim_{N\to\infty}}\tau = 14 + \frac{{{m-k}\choose{k}}{{m-2k}\choose{k}}{{m-3k}\choose{k}}}{{m\choose k}^3} + 4\frac{{{m-k}\choose{k}}{{m-2k}\choose{k}}^2}{{m\choose k}^3} + 8\frac{{{m-k}\choose{k}}{{m-2k}\choose{k}}}{{m\choose k}^2}\notag\\
&~~~~~~~ + 8\frac{{{m-k}\choose{k}}^2{{m-2k}\choose{k}}}{{m\choose k}^3} + 8\frac{{{m-k}\choose{k}}^3}{{m\choose k}^3} + 4\frac{{{m-k}\choose{k}}^2}{{m\choose k}^2} + 28\frac{{{m-k}\choose{k}}}{{m\choose k}} + 24\frac{{{m-k}\choose{k}}^2}{{m\choose k}^2} \notag\\
&~~~+ 4\frac{{{m-k}\choose{k}}^3}{{m\choose k}^3} + 2 \frac{{{m-k-\alpha}\choose{k}}}{{{m\choose {k}}}^3} {\sum_{\alpha} {{m-k-\alpha}\choose{k}} {{m-2k}\choose{\alpha}} {{k}\choose{\alpha}}}.\end{align}
Illustrated in Fig \ref{fig:eighth_moment} the eighth moment takes the same form as the fourth and sixth moments, transitioning from a gaussian value $(2n-1)!!=105$ at $k=0$ to a semi-circular moment for all $k > \frac {m}{2}$. Unlike a the lower moments the eighth moment contains a Hahn term. Fortunately the method of particle diagrams can be used to yield all such terms, and as we have seen the procedure is the same, independent of how complex a particular diagram is. These results and the method of particle diagrams will be discussed in greater detail in the next chapter.
\begin{figure}[H]
\centering
\includegraphics[scale=.7]{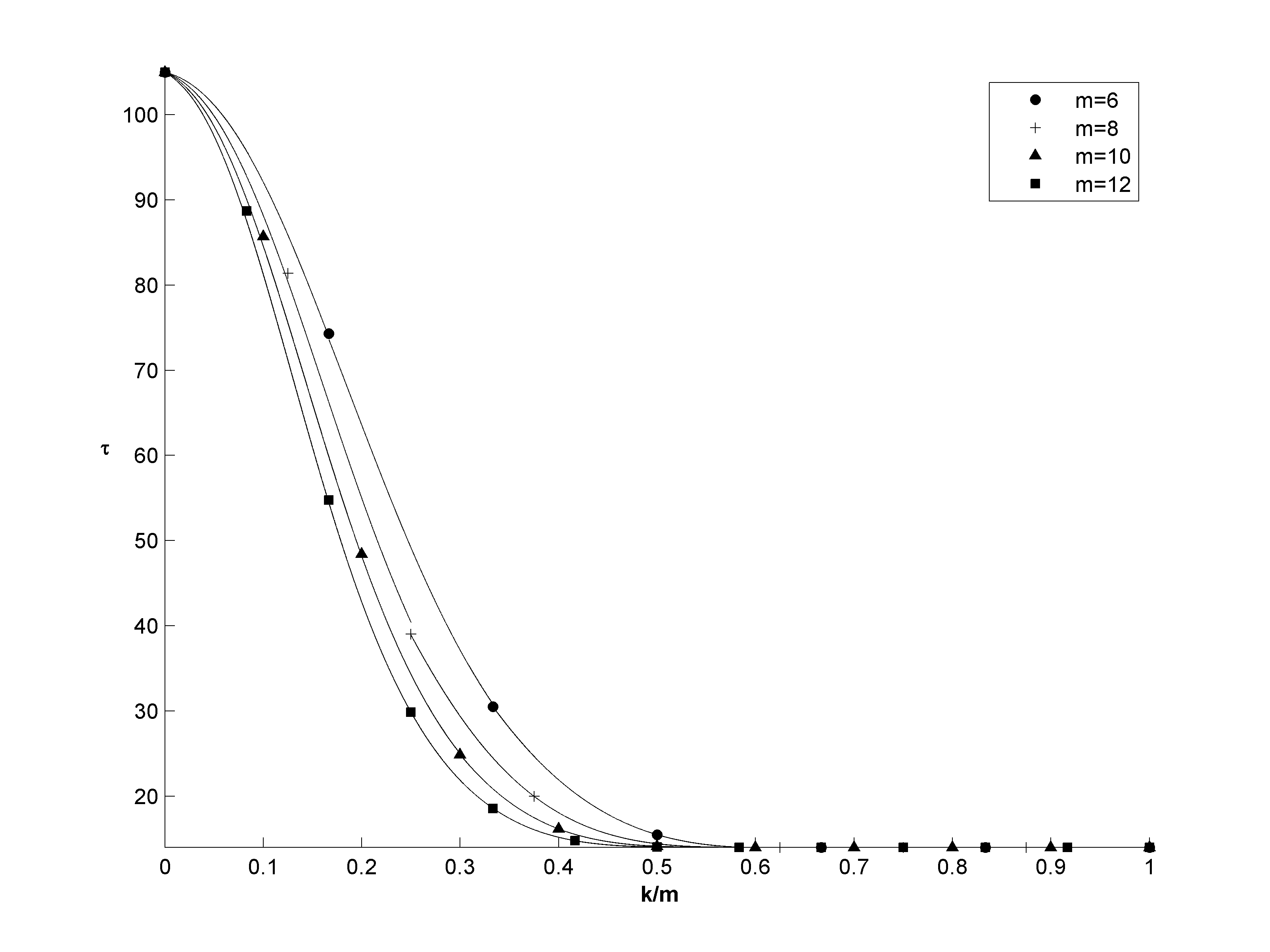}
\caption[The eighth moment $\tau$ against $k/m$]{The eighth moment $\tau$ of the level density against $k/m$ for $m=6, 8, 10, 12$ exhibiting the same properties as the fourth and sixth; a semi-circular domain, $\tau =14$ for $k/m > \frac{1}{2}$ converging to a gaussian value, $\tau=105$ for $k=0$. Higher values of $m$ give faster convergence to the semi-circular moment.}
\label{fig:eighth_moment}
\end{figure}
\begin{center}.....\end{center}

\chapter[Conclusions]{Conclusions}
\section{Moments}\label{sec:mmnts}
This thesis has focused on the problem of characterising the statistics of unified RMT, more commonly known as embedded many-body random matrix theory. As noted in chapter \ref{ch:intro} this is a relatively young area of research, with few practitioners and a modest body of literature. The approach established here has involved the method of particle diagrams, a mathematical toolkit invented by the author, with which we have investigated the calculation of moments of the level density $\rho(E)$ of the random potential derived in chapter \ref{ch:RMTs} given by
\begin{equation}\label{eq:CC1}
{H}_k = \sum_{\bj \bi}v_{\bj\bi} a_{\bj}^{\dag} a_{\bi}.
\end{equation}
These potentials are characterised by the parameter set $\{k, m ,l, S\}$ where $k$ is the order of the potential, $m$ is the number of particles in the experiment, $l$ is the number of single-particle energy levels and $S$ represents any additional symmetry conditions placed on the potential $H_k$. Together these determine the state space of the potential, and thereby the state space of unified random matrix theory. This is a superspace of canonical random matrix theory. Canonical RMT places a single random variable into each matrix cell and coincides with the point $\{m, m , \infty, S\}$, with $S$ usually enforcing orthogonal ($\beta=1$), unitary $\beta=2$ or symplectic ($\beta=4$) symmetry on $H_k$. Here we have made calculations of the fourth, sixth and eighth moments for all $k\le m$ for the embedded GUE (see section \ref{sec:egue}). It was noted in (\ref{eq:smrv3}) that the moments of a Gaussian are given by
\begin{equation}\label{eq:LC1}
\overline{v^{2n}} = \sqrt{\frac{1}{2\pi}}\int v^{2n} e^{- {v^2}/{2}}dv = (2n-1)!!
\end{equation}
and the moments of a semi-circle, those characterising Wigner's semi-circle law, are given by
\begin{equation}\label{eq:LC3}
\frac{2}{\pi r^2} \int_{-r}^{r} x^{2n} \sqrt{r^2 - x^2}~dx = C_n
\end{equation}
where the Catalan numbers (\ref{eq:dw8}) are
\begin{equation}\label{eq:LC2}
C_n = {\frac{1}{n+1}} {{2n}\choose{n}}.
\end{equation}
Either through direct calculation or by using the method of particle diagrams we have seen in section \ref{sec:egue} that the fourth moment of the level density is
\begin{align}\label{eq:LC4}\lim_{N\to\infty}\kappa = 2 + \lim_{N\to\infty}\frac{\frac{1}{N}A_{\mu\nu\rho\sigma}A_{\sigma\mu\nu\rho}}{\left [ \frac{1}{N}\sum A_{\mu\mu\rho\rho} \right ] ^2}
= 2 + \frac{{{m - k} \choose k}}{{m \choose k}}\end{align}
which is illustrated in Fig \ref{fig:fourth_moment}. The figure reveals a transition from a Gaussian fourth moment $(2n-1)!!=3$ at $k=0$ to a semi-circular moment $C_n=2$ for all $k > \frac{m}{2}$. This includes $k=m$ which we know already from Wigner's semi-circle law should yield $\kappa=2$. Hence there is a portion of the domain in the unified theory $(\frac{m}{2}, m]$ in which $k$ varies (and therefore the action of the potential $H_k$ changes) but the fourth moment \emph{remains fixed} to a semi-circular value.

It is shown in chapter \ref{ch:THM} that the sixth moment is given by the combinatorial sum
\begin{equation}\label{eq:LC5} {\lim_{N\to\infty}}h = 5 + \frac{{{m-k}\choose{k}}{{m-2k}\choose{k}}}{{m\choose k}^2} + 6\frac{{{m-k}\choose{k}}}{{m\choose k}} + 3\frac{{{m-k}\choose{k}}^2}{{m\choose k}^2}.\end{equation}
This was illustrated previously in Fig \ref{fig:sixth_moment}. Again we can observe a gradual transition from a Gaussian valued moment $(2n-1)!! = 15$ at $k=0$ to a semi-circular moment $C_n = 5$ not just for the canonical case $k=m$, but for all $k > \frac{m}{2}$.

The eighth moment, also derived in chapter \ref{ch:THM}, was shown to be equal to the rambling expression
\begin{align}\label{eq:LC6} &{lim_{N\to\infty}}\tau = 14 + \frac{{{m-k}\choose{k}}{{m-2k}\choose{k}}{{m-3k}\choose{k}}}{{m\choose k}^3} + 4\frac{{{m-k}\choose{k}}{{m-2k}\choose{k}}^2}{{m\choose k}^3} + 8\frac{{{m-k}\choose{k}}{{m-2k}\choose{k}}}{{m\choose k}^2}\notag\\
&~~~~~~~ + 8\frac{{{m-k}\choose{k}}^2{{m-2k}\choose{k}}}{{m\choose k}^3} + 8\frac{{{m-k}\choose{k}}^3}{{m\choose k}^3} + 4\frac{{{m-k}\choose{k}}^2}{{m\choose k}^2} + 28\frac{{{m-k}\choose{k}}}{{m\choose k}} + 24\frac{{{m-k}\choose{k}}^2}{{m\choose k}^2} \notag\\
&~~~+ 4\frac{{{m-k}\choose{k}}^3}{{m\choose k}^3} + 2 \frac{{{m-k}\choose{k}}}{{{m\choose {k}}}^3} {\sum_{\alpha} {{m-k-\alpha}\choose{k}} {{m-2k}\choose{\alpha}} {{k}\choose{\alpha}}}\end{align}
which has been illustrated in Fig \ref{fig:eighth_moment}. From the figure we can immediately note the way in which the value of the moment transitions from a gaussian value $(2n-1)!!=105$ on the left-most edge of the phase space $k=0$ to a semi-circular moment $C_n = 14$ for all values $\frac{m}{2} < k \le m$ where despite the potential $H_k$ being fundamentally different, the eighth moment remains constant. 

\section{Internal Structure of the Moments}\label{sec:ISM}
The results shown above suggest that the unified RMT state space for the eGUE consists of two distinct domains in the order $k$. Firstly, each of the moments reveal a \emph{canonical domain} defined by $\frac{m}{2} < k \le m$ (see Fig \ref{fig:all_moments}). In this region of the phase space the action of the potential is changing, but the moments of the level density remain fixed to semi-circular values. Hence, we may expect that in this region Wigner's semi-circle law applies for all values of $k$.

Secondly, again with reference to Fig \ref{fig:all_moments}, there is a \emph{critical domain} defined by $0\le k \le \frac{m}{2}$. Here the value of $k$ is crucial for determining the value of the moments, which vary for all $k$ within this region of the state space.

\begin{figure}[!ht]
\centering
\includegraphics[width=\linewidth]{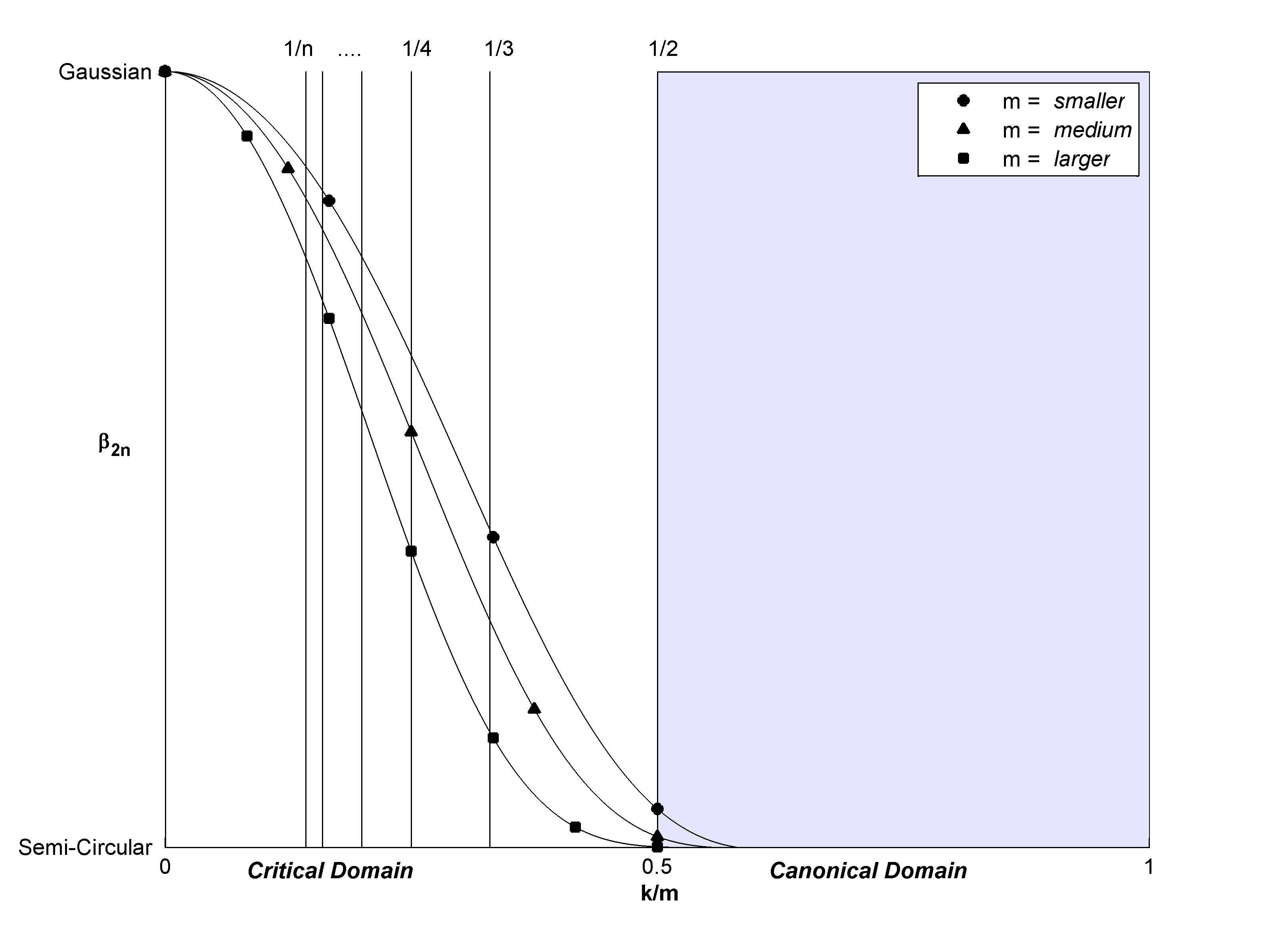}
\caption[General form of the moments]{Illustration of the general form of all the moments, showing a transition from a Gaussian value to a semi-circular value for all $k$ in the canonical domain, and some additional structure in the critical domain at the discrete points $k= \frac{m}{2}, \frac{m}{3}, \ldots, \frac{m}{n}$.

\hrulefill}
\label{fig:all_moments}
\end{figure}
Finally, the critical domain for each of the moments contains an internal structure. For the $2n$'th moment this additional detail is defined by the discrete points $\frac{m}{2}, \frac{m}{3}, \ldots, \frac{m}{n}$ at which the combinatorial expressions (\ref{eq:LC4}--\ref{eq:LC6}) contain terms which become zero for all values of $k$ thereafter. For example the quotient
\begin{equation}\label{eq:LC7}
 \frac{{{m - k} \choose k}}{{m \choose k}}
\end{equation}
giving the contribution of the standard diagram to the fourth moment (\ref{eq:LC4}) will be zero for all $k > m-k$ while the quotient
\begin{equation}\label{eq:LC8}
\frac{{{m-k}\choose{k}}{{m-2k}\choose{k}}}{{m\choose k}^2}
\end{equation}
from  (\ref{eq:LC5}) will be zero for all $k > m-2k$. The gradual removal of these terms at the discrete values of $k= \frac{m}{2}, \frac{m}{3}, \ldots, \frac{m}{n}$ is what produces the decline in the moments from gaussian values to semi-circular values and explaining the \emph{physical} ramifications of this (if any) is an open question which would make for some very interesting reading. Related to this, but offering perhaps less detail, is a physical explanation for why the random potential, being different for each value of $k$, should nonetheless have the same level density for all $k$ in the canonical domain $k>\frac{m}{2}$.

A fundamental component of the internal structure of the critical domain is the existence, in each of the moments, of a single diagram which takes the value
\begin{equation}\label{eq:LC9}
{l}\choose{s \underbrace{\;k \;k \ldots \;k}_{2n~times}\;}
\end{equation}
where for the $2n$'th moment $s := m - nk$ and the sequence of $k$'s repeats $2n$ times so that the sum is $s + \sum_{2n} k = m + nk$. This is the argument attained by any diagram contributing to the limit value of the $2n$'th moment as $l\to\infty$. For the fourth moment the diagram which takes the form of (\ref{eq:LC9}) is the standard diagram, Fig \ref{fig:combinedqterm}, while for the sixth and eighth moments it is Figures \ref{fig:diamond} and \ref{fig:box} respectively. The fact that these diagrams take the form of (\ref{eq:LC9}) is due to the existence in each case of just two $m$-body states which together determine all other single-particle labels present in the diagram.

\section{The Dilute Limit $k \ll m \ll l$}\label{sec:EVS}
Another result of the analysis is the conclusion that in the limit case $k/m \to 0$ the moments are those of a Gaussian distribution. This can be seen in equations (\ref{eq:LC4}--\ref{eq:LC6}) by taking the limit $m \to \infty$ and applying Stirling's approximation to the resulting binomials, yielding the moments for the special case $k \ll m \ll l$ which is also known as the \emph{dilute limit}. In this limit each binomial expression contributes just a factor $1$, so that the moments become equal to the sum of the coefficients. Since the coefficients sum to the number of unique pairwise partitions of a set of $2n$ objects the result is always $(2n-1)!!$, the Gaussian moments. The fact that each combinatorial term gives a contribution of unity can also be seen by noting that the number of combinatorial expressions in $m$ in the numerator is always equal to the number of combinatorial terms in $m$ in the denominator. For example the component
\begin{equation}
\frac{{{m-k}\choose{k}}{{m-2k}\choose{k}}{{m-3k}\choose{k}}}{{m\choose k}^3}
\end{equation}
of the eighth moment (\ref{eq:LC6}) has three binomial terms ${{{m-k}\choose{k}}{{m-2k}\choose{k}}{{m-3k}\choose{k}}}$ in $m$ forming the numerator and three binomial terms ${{m\choose k}^3}={m\choose k}{m\choose k}{m\choose k}$ in $m$ forming the denominator. Likewise, the Hahn expression
\begin{equation}
2 \frac{{{m-k-\alpha}\choose{k}}}{{{m\choose {k}}}^3} {\sum_{\alpha} {{m-k-\alpha}\choose{k}} {{m-2k}\choose{\alpha}} {{k}\choose{\alpha}}}
\end{equation}
has the three factors ${{m-k}\choose{k}} {{m-k-\alpha}\choose{k}} {{m-2k}\choose{\alpha}}$ in $m$ forming the numerator and the three factors ${{m\choose {k}}}^3$ in the denominator.

\section{Bosonic states}\label{bosons}
Using the method of particle diagrams it is possible to make a final remark regarding the statistics of the hermitian eGUE potential (\ref{eq:ee27}) given a bosonic state space. From section \ref{sec:second.quantization} it is known that bosonic many-particle states may contain multiple copies of the same single-particle states. Retaining the same potential (\ref{eq:ee27}), but now assuming a bosonic state space, it is possible to use what has been learned already for the eGUE embedded in a fermionic state space to make far-reaching conclusions about the bosonic case. Namely, it is straightforward to see that in the limit $l\to\infty$ the results for all moments agree with fermionic systems. Intuitively, this happens because in this limit contributions arise only for those choices of many-particle states which maximise the number of participating single-particle states. This means that in the bosonic case multiple occupancy of the same single-particle states is penalised, such that for the terms which survive in the limit $l\to\infty$ there is no difference from the fermionic case.

For a formal derivation consider bosonic $m$-particle states containing repeats of $z\le m$ unique single-particle states. There are ${l \choose z}$ ways to select the participating single-particle states, and ${m-1\choose z-1}$ ways to select their multiplicities to obtain altogether $z$ particles. As stated before in section \ref{sec:osym} the number of many-particle states is thus
\begin{align}\label{eq:bosons}N = \sum_{z=1}^{m}{l\choose z}{{m-1}\choose{z-1}} = {{l+m-1}\choose{m}}\end{align}
where the overall sum as well as the summand $z=m$ have the argument $m$ whereas all other summands have lower arguments. As a consequence the asymptotic form of $N$ coincides with the fermionic case where $N={l\choose m}$.

The same logic applies when evaluating the particle diagrams, where the dashed bonds $\feyn{h}$ contain only unique sets of $k$ single-particle state labels whereas in contrast to the fermionic case the solid bonds $\feyn{f}$ contain $m-k$ single-particle state labels with repeats now permitted. In the fermionic case, the contributions to the particle diagrams with maximal argument $m+nk$ can always be factorized into a term ${l\choose {m+nk}}$ counting the number of ways in which the $m+nk$ participating states can be chosen, and further $l$-independent factors counting the number of ways in which these $m+nk$ states can be distributed among different sets while obeying the conditions implied by the diagram. In the bosonic case we instead have to  select $1\leq z\leq m+nk$ states with multiplicities summing to $m+nk$. The contribution of each diagram turns into a sum over $z$, and for each $z$ a single $l$-dependent term ${l\choose z}$ is obtained, together with further $l$-independent finite factors counting the number of ways to distribute these states among sets containing $m$ particles. The latter choices have to be compatible both with the restrictions implied by the diagram and the requirement on the multiplicities. However, the only summand attaining the maximal argument will be the last, for which $z=m+nk$. For this summand all multiplicities are 1 meaning that we only have to consider bosonic states composed entirely of distinct single-particle states. Since these are mathematically equivalent to fermionic states it can be concluded without any further calculations that \textit{all bosonic and fermionic moments are equal in the limit} $l\to\infty$.

\section{The Method of Particle Diagrams}
The focus of this thesis has been the study of quantum many-body potentials taking the second quantised form of (\ref{eq:CC1}). This involved the study of a unified phase space of random matrix theories composed of the parameters $k, m, l$ which are the order of interaction, number of particles and number of available states respectively. These, in addition to the symmetry constraints on $H_k$, determine each point in the phase space. The canonical form of random matrix theory corresponds to the single point $\{m,m,\infty\}$. In their seminal paper Benet, Rupp and Weidenm\"uller were able to calculate the fourth moment of the level density of the eGUE in the domain $\{0\le k \le m, m, \infty\}$\cite{weid}. In so doing they also illustrated the difficulty involved in studying the more general embedded RMT state space, suggesting that new methods would be needed beyond those which are specialised for the canonical $k=m$ case. The author's response to this was the creation of the method of particle diagrams which has been illustrated in the previous chapters to calculate the fourth, sixth and eighth moments of the embedded GUE. This is a modest success in itself, since the sixth and eighth moments were not known before in the critical domain $k \le \frac{m}{2}$. Moreover the method is unmatched in it's simplicity, involving no superalgebra and no complex mathematics beyond basic combinatorics. The basic ingredients are as follows.
\begin{enumerate}
\item \emph{Arguments} (order of magnitude).
The first ingredient of the method of particle diagrams is the quantity called the \emph{argument} which is the power in $l$ of a binomial expression and can be shown using Stirling's formula $lim_{n\to\infty}n! = \sqrt{2\pi n}{\left (\frac{n}{e}\right )}^n$ to be given by
\begin{equation}\label{eq:LC10}\arg\left [\prod_n{{l - a_n} \choose b_n}^{i_n}\right ] = \sum_n i_n b_n. \end{equation}
Arguments are a simplifying feature of the method of particle diagrams, working to seperate those terms which will not contribute to the limit value of the moments as $l\to\infty$ but simultaneously leaving enough mathematical structure behind so that the limiting result can be calculated. Arguments also provide the constraints neccessary to optimise loops.

\item \emph{Particle Diagrams} (A graph structure to represent factors of $A_{\mu\nu\rho\sigma}$).
The particle diagrams, although themselves not essential to the calculations, aid us in visualising ``what is going on'' as we proceed to calculate the arguments and identify loops in the diagram for each term of the higher moments. Particle diagrams form the bridge between arguments and loops. They also provide a way of identifying visually, as opposed to analytically, all the loops in a given term. Particle diagrams found in calculations for lower moments re-appear in calculations for higher moments so that calculating successive moments makes later work easier. It also means that the standard diagram, Fig \ref{fig:combinedqtermreview}, is found in calculations for all the higher moments, even though it's maiden appearance is in the calculation of the fourth moment.
\begin{figure}[t]
\centering
\begin{tikzpicture}[scale=1.5]
\draw[ultra thick, red] (-.2,0) -- (0.75,.9);
\draw[ultra thick, dashed, red] (0.75,.9) -- (1.7,0);
\draw[ultra thick, dashed, red] (-.2,0) -- (0.75,-.9);
\draw[ultra thick, red] (0.75,-.9) -- (1.7,0);
\draw[ultra thick, dashed, blue] (0,0) -- (0.75,.7);
\draw[ultra thick, blue] (0.75,.7) -- (1.5,0);
\draw[ultra thick, blue] (0,0) -- (0.75,-.7);
\draw[ultra thick, dashed, blue] (0.75,-.7) -- (1.5,0);
\node[circle, fill=white, draw=black, text=black, inner sep=1.2pt] at (-.05,0) {$\mu$}; 
\node[circle, fill=white, draw=black, text=black,  inner sep=1.2pt] at (1.55,0) {$\rho$}; 
\node[circle, fill=white, draw=black, text=black,  inner sep=1.2pt] at (0.75,-.78) {$\nu$};
 \node[circle, fill=white, draw=black, text=black,  inner sep=1.2pt] at (0.75,.78) {$\sigma$};
\end{tikzpicture}
\caption[The Standard Diagram Review]{The particle diagram of the term $A_{\mu\nu\rho\sigma}A_{\sigma\mu\nu\rho}$ is the standard diagram. Since diagrams from lower order moments appear again in calculations of higher moments the standard diagram appears in calculations of all the moments. The standard diagram is also an example of a diagram whose value takes the form of (\ref{eq:LC9}) since the states $|\mu\rangle$ and $|\rho\rangle$ together determine the single-particle states contained in $|\sigma\rangle$ and $|\nu\rangle$.

\hrulefill}
\label{fig:combinedqtermreview}
\end{figure}
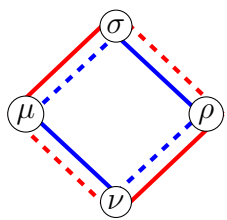

\item \emph{Loops} (Intersections between states). Loops represent intersections between states. These are the single-particle state labels shared between the many-body states (which are sets of $m$ state labels). Loops also represent the sets which must be summed over in order to calculate the trace which yields the moments
\begin{equation}\label{eq:LC11}
\beta_{2n} = \frac{\frac{1}{N}\mathrm{tr}({\overline{H^{2n}_k}})}{\left(\frac{1}{N}\mathrm{tr}({\overline{H^2_k}})\right)^n}.
\end{equation}
Finally and most importantly, loops determine the argument of a diagram, so that the sum of loops in a diagram must be maximised.
\end{enumerate}
Together these three ingredients form the method of particle diagrams. Although the method has been used here to calculate moments for the eGUE it can be used to calculate statistics for other symmetry classes as well. The most interesting of these is the embedded gaussian symplectic ensemble (eGSE) about which virtually nothing is currently known. The author has also calculated the fourth and sixth moments for the case where the potential satisfies the eGOE symmetry of (\ref{eq:os5}) and moreover is embedded in a fermionic state space. In this instance it is seen that, conditional on taking the limit $l\to \infty$, these moments are identical for the eGUE and eGOE.

To end, it should be noted that although particle diagrams are used here to calculate moments for all $0\le k \le m$, the results coincide with Wigner's semi-circle law as expected for the canonical case $k=m$ as well. In fact, for $k=m$ the dashed bonds $\feyn{h}$ must necessarily contain $k=m$ elements, whereas the solid bonds $\feyn{f}$ contain $m-k=0$ elements. This in turn results in diagrams which are equivalent to those shown in section \ref{sec:polygons} where cycles are represented as partitions of regular polygons, or equivalently Dyck paths.

\section{Related Literature and Wider Context}
\subsection{Many-Body Potentials Revisited}

The unification of random matrix theory offers the possibility that the landscape of statistical quantities of random matrices can be studied \emph{en masse}, with a single equation for each quantity. Each statistical quantity found in canonical random matrix theory would then be just a special case ($k=m$) of an equation determined by the parameters $k, m$ and $l$ as well as the symmetry conditions placed on the matrix elements. There is another paradigm, however, which is larger still than the embedded state space and which amalgamates and connects a vast array of physical and mathematical models. This is the paradigm of \emph{many-body potentials}. The set of many-body potentials includes random matrix theory, both its canonical and unified form (\ref{eq:CC1}), as well as countless other models. These models do not share the same hamiltonian but are unified by the similarities in their mathematical structure. Together they explain a diverse array of disparate phenomena from the study of quantum spin hypergraphs to the application of neural networks implemented algorithmically by computers (essentially ``artificial brains'') which underly most modern online services from Google, Amazon, Netflix and countless other technology-driven enterprises.
The fundamental form underlying all many-body potentials is given by the equation
 \begin{equation}\label{eq:MBPS}
      H = \sum_{\bm{k}_1 \bm{k}_2} v(\bm{k}_1, \bm{k}_2)~A^{\dag}_{\bm{k}_1}B_{\bm{k}_2}
      \end{equation}
where $|\bm{k}_1| = |\bm{k}_2| = k$ are $k$-tuples. Here $A_{\bm{k}_1}$ and $B_{\bm{k}_2}$ are \emph{any} $k$-tuple operators. They can be complex and multidimensional so, for example, they can be matrices. The coefficients $v(\bm{k}_1, \bm{k}_2)$ are numbers, possibly complex and if they are random variables one refers to (\ref{eq:MBPS}) as a random many-body problem. The defining feature of many-body potentials is the fact that the operators $A_{\bm{k}_1}$ and $B_{\bm{k}_2}$ are $k$-tuple operators.
\begin{figure}[t]
\centering
\begin{tikzpicture}
\node[circle, fill=blue!10,  inner sep=87pt] at (7,7) {}; 
\node[circle, fill=red!20,  inner sep=14pt] at (5,4.8) {}; 

\node[text=blue,  inner sep=0pt] at (7,10.5) {\bf {\it m}-tuples}; 
\node[text=red,  inner sep=0pt] at (5,4.8) {\bf {\it k}-tuples}; 

\node[draw, ultra thick, shape=circle, fill=black, inner sep=0pt, minimum size=.15cm](1) at (9,4.4) {};
\node[draw, ultra thick, shape=circle, fill=black, inner sep=0pt, minimum size=.15cm](1) at (3.5,5.9) {};
\node[draw, ultra thick, shape=circle, fill=black, inner sep=0pt, minimum size=.15cm](1) at (3.2,6) {};
\node[draw, ultra thick, shape=circle, fill=black, inner sep=0pt, minimum size=.15cm](1) at (3.9,7.1) {};
\node[draw, ultra thick, shape=circle, fill=black, inner sep=0pt, minimum size=.15cm](1) at (3,8.3) {};
\node[draw, ultra thick, shape=circle, fill=black, inner sep=0pt, minimum size=.15cm](1) at (3.7,9) {};

\node[draw, ultra thick, shape=circle, fill=black, inner sep=0pt, minimum size=.15cm](1) at (4.7,4.4) {};
\node[draw, ultra thick, shape=circle, fill=black, inner sep=0pt, minimum size=.15cm](1) at (4.9,5.1) {};
\node[draw, ultra thick, shape=circle, fill=black, inner sep=0pt, minimum size=.15cm](1) at (4,6.9) {};
\node[draw, ultra thick, shape=circle, fill=black, inner sep=0pt, minimum size=.15cm](1) at (4.3,7.7) {};
\node[draw, ultra thick, shape=circle, fill=black, inner sep=0pt, minimum size=.15cm](1) at (4.8,8.1) {};
\node[draw, ultra thick, shape=circle, fill=black, inner sep=0pt, minimum size=.15cm](1) at (4.2,9.3) {};
\node[draw, ultra thick, shape=circle, fill=black, inner sep=0pt, minimum size=.15cm](1) at (5,4.5) {};
\node[draw, ultra thick, shape=circle, fill=black, inner sep=0pt, minimum size=.15cm](1) at (5.9,5.2) {};
\node[draw, ultra thick, shape=circle, fill=black, inner sep=0pt, minimum size=.15cm](1) at (5,6.1) {};
\node[draw, ultra thick, shape=circle, fill=black, inner sep=0pt, minimum size=.15cm](1) at (5.7,7.3) {};
\node[draw, ultra thick, shape=circle, fill=black, inner sep=0pt, minimum size=.15cm](1) at (5.3,8.8) {};
\node[draw, ultra thick, shape=circle, fill=black, inner sep=0pt, minimum size=.15cm](1) at (5.5,9.5) {};
\node[draw, ultra thick, shape=circle, fill=black, inner sep=0pt, minimum size=.15cm](1) at (6,4.9) {};
\node[draw, ultra thick, shape=circle, fill=black, inner sep=0pt, minimum size=.15cm](1) at (6.6,5) {};
\node[draw, ultra thick, shape=circle, fill=black, inner sep=0pt, minimum size=.15cm](1) at (6.9,6.4) {};
\node[draw, ultra thick, shape=circle, fill=black, inner sep=0pt, minimum size=.15cm](1) at (6.1,7.5) {};
\node[draw, ultra thick, shape=circle, fill=black, inner sep=0pt, minimum size=.15cm](1) at (6.3,8.9) {};
\node[draw, ultra thick, shape=circle, fill=black, inner sep=0pt, minimum size=.15cm](1) at (6.2,9.6) {};

\node[draw, ultra thick, shape=circle, fill=black, inner sep=0pt, minimum size=.15cm](1) at (7.4,4.7) {};
\node[draw, ultra thick, shape=circle, fill=black, inner sep=0pt, minimum size=.15cm](1) at (7.9,5) {};
\node[draw, ultra thick, shape=circle, fill=black, inner sep=0pt, minimum size=.15cm](1) at (7.1,6.1) {};
\node[draw, ultra thick, shape=circle, fill=black, inner sep=0pt, minimum size=.15cm](1) at (7.8,7.2) {};
\node[draw, ultra thick, shape=circle, fill=black, inner sep=0pt, minimum size=.15cm](1) at (7.6,8.5) {};
\node[draw, ultra thick, shape=circle, fill=black, inner sep=0pt, minimum size=.15cm](1) at (7.3,9.6) {};
\node[draw, ultra thick, shape=circle, fill=black, inner sep=0pt, minimum size=.15cm](1) at (8.9,4.9) {};
\node[draw, ultra thick, shape=circle, fill=black, inner sep=0pt, minimum size=.15cm](1) at (8,5.6) {};
\node[draw, ultra thick, shape=circle, fill=black, inner sep=0pt, minimum size=.15cm](1) at (8.6,6.3) {};
\node[draw, ultra thick, shape=circle, fill=black, inner sep=0pt, minimum size=.15cm](1) at (8.7,7.8) {};
\node[draw, ultra thick, shape=circle, fill=black, inner sep=0pt, minimum size=.15cm](1) at (8.3,8.1) {};
\node[draw, ultra thick, shape=circle, fill=black, inner sep=0pt, minimum size=.15cm](1) at (8.1,9.2) {};
\node[draw, ultra thick, shape=circle, fill=black, inner sep=0pt, minimum size=.15cm](1) at (9.9,4.9) {};
\node[draw, ultra thick, shape=circle, fill=black, inner sep=0pt, minimum size=.15cm](1) at (9.3,5) {};
\node[draw, ultra thick, shape=circle, fill=black, inner sep=0pt, minimum size=.15cm](1) at (9,6.1) {};
\node[draw, ultra thick, shape=circle, fill=black, inner sep=0pt, minimum size=.15cm](1) at (9.5,7.4) {};
\node[draw, ultra thick, shape=circle, fill=black, inner sep=0pt, minimum size=.15cm](1) at (9.3,8.8) {};
\node[draw, ultra thick, shape=circle, fill=black, inner sep=0pt, minimum size=.15cm](1) at (9.8,9) {};

\node[draw, ultra thick, shape=circle, fill=black, inner sep=0pt, minimum size=.15cm](1) at (7,4) {};
\node[draw, ultra thick, shape=circle, fill=black, inner sep=0pt, minimum size=.15cm](1) at (10.5,5.9) {};
\node[draw, ultra thick, shape=circle, fill=black, inner sep=0pt, minimum size=.15cm](1) at (10.7,6) {};
\node[draw, ultra thick, shape=circle, fill=black, inner sep=0pt, minimum size=.15cm](1) at (10.1,7.3) {};
\node[draw, ultra thick, shape=circle, fill=black, inner sep=0pt, minimum size=.15cm](1) at (10.3,8.5) {};
\node[draw, ultra thick, shape=circle, fill=black, inner sep=0pt, minimum size=.15cm](1) at (9,9.5) {};

\end{tikzpicture}
\caption[Many Body Potential]{Illustration of a many-body problem, given by (\ref{eq:MBPS}). The hamiltonian is a sum over $k$-tuples in an $m$-tuple with $k\le m$. The $m$-tuples usually represent quantum many-body states, and the $k$-tuple operators $A^{\dag}_{\bm{k}_1}B_{\bm{k}_2}$ with coefficient $v(\bm{k}_1, \bm{k}_2)$ give the interaction terms between $k$-tuples of single-particle states within the compound $m$-body state.

\hrulefill}
\label{fig:many_body_dots}
\end{figure}
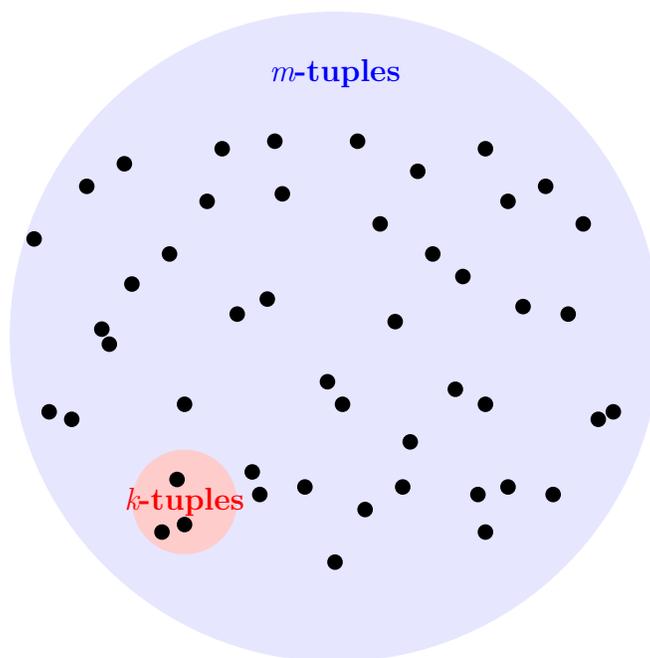

\subsection{Other Unified RMT Results}
The embedded form of RMT (\ref{eq:CC1}) introduced by Mon and French\cite{mon} coincides with the general form (\ref{eq:MBPS}) when the $k$-tuple operators
\begin{align}A_{\bm{k}_1}=a_{\bj}^{\dag}\notag\\
B_{\bm{k}_2}=a_{\bi}
\end{align}
and $v(\bm{k}_1, \bm{k}_2) = v_{\bj\bi}$ where $v_{\bj\bi}$ is a Gaussian random variable and as before $\bj = (j_1, j_2, \ldots, j_k)$. These ensembles have been studied before by \cite{mon, wong, weid, kota2, nakata, srednicki} among others. For the limit case $k/m \to \infty$ it was found by \cite{mon} that the distribution is a Gaussian. This is corroborated by the results as discussed in section \ref{sec:EVS} above. Additionally, before the method of particle diagrams was used to calculate the same results Benet, Rupp and Weidenm\"uller (BRW) had already shown using their eigenvector expansion method that the fourth moment of the eGUE is given by (\ref{eq:LC4}). They also showed using a mixture of supersymmetry and eigenvector expansions that the moments are semi-circular in the canonical domain $k > \frac{m}{2}$\cite{weid}. The method of particle diagrams, which is the latest addition to the family of embedded many-body methods, reproduces the fourth moment of the eGUE as shown by BRW and adds the sixth and eighth moments for all $0 \le k \le m$. It also reveals the internal structure of the moments in the critical domain as discussed in section \ref{sec:ISM}.

\subsection{Quantum Spin Graphs}
An area of current research which yields fascinating similarities to the embedded many-body state space is the study of quantum spin graphs. These are many-body potentials taking the form
 \begin{equation}
      H_n = \sum_{(j_1,\cdots,j_k)\in\Gamma_n}^n \sum_{a_1,\cdots,a_k=1}^3 \alpha_{a_1,\cdots,a_k(j_1,\cdots,j_k)}\sigma_{j_1}^{(a_1)}\cdots\sigma_{j_k}^{(a_k)}
\label{eq:qsg}
      \end{equation}
 Comparing this with (\ref{eq:MBPS}) shows that the $k$-tuple operator is
\begin{align}A_{\bm{k}_1}B_{\bm{k}_2}=\sigma_{j_1}^{(a_1)}\cdots\alpha_{j_k}^{(a_k)}
\end{align}
and the coefficients are
\begin{equation}
v(\bm{k}_1, \bm{k}_2) = \alpha_{a_1,\cdots,a_k(j_1,\cdots,j_k)}.
\end{equation}
In this context $k$ is the number of vertices in every hyperedge of the graph $\Gamma$. Viewing (\ref{eq:qsg}) in the context of the general form (\ref{eq:MBPS}) shows us that this is a generic many-body hamiltonian with Pauli spin tuples acting as the $k$-tuple operators. It is also known as a quantum spin glass model. If $k=2$ the potential coincides with a quantum spin chain model. These hamiltonians, particularly the form with $k=2$, are the focus of a growing body of research which includes \cite{schroeder, hess, gubin, huw}. The work of \cite{schroeder} is particularly interesting in the context of this thesis, since it was proved there that the level density of a random hypergraph (\ref{eq:qsg}) takes the form of a Gaussian distribution for $k\ll\sqrt{m}$, a semi-circle for $k \gg \sqrt{m}$ and a compactly supported density function for values of $k$ and $m$ satisfying $lim_{m\to\infty} \frac{k}{\sqrt{m}}=\lambda > 0$. This shows that the statistical features of quantum hypergraphs are strikingly similar to the results shown here for the unified RMT state space. Both show a transition from a Gaussian regime to a semi-circular one. While the point of transition in the case of the eGUE is $k=\frac{m}{2}$, the results of \cite{schroeder} suggest it is at $k\sim\sqrt{m}$ for the quantum spin hypergraph potential. Finally, whereas for the eGUE one takes $l\to\infty$, for spin graphs there are only two available states, up and down, so that $l=2$. Hence, although the embedded RMT potential (\ref{eq:CC1}) and the quantum spin graph potential (\ref{eq:qsg}) are clearly different, they also share many statistical features and both models take the form of a generic many-body potential (\ref{eq:MBPS}). Exploring their statistical similarities further could yield some remarkable insights into many-body potentials.

\subsection{Neural Networks}
Many-body potentials of the form (\ref{eq:MBPS}) are also present in research into artificial intelligence, notably machine learning neural network models which attempt to simulate some basic functionality of a generic brain. These machine learning networks are used by the vast majority of modern technology companies such as Google (search), Amazon (online shopping), Netflix (movie streaming) and IBM (most famously in it's Watson machine which is now the world champion of the popular quiz show Jeopardy). Neural network models taking the form of a many-body potential (\ref{eq:MBPS}) are referred to variously as higher-order Boltzmann machines\cite{sejnowski}, higher-order neural networks\cite{giles} or simply spin-glass models\cite{amit}. The general form of the quantity analogous to the potential in neural networks is given by
\begin{equation}\label{eq:nnm}
H=\frac{1}{k}\sum_{i_1 i_2\ldots i_k \in \Gamma} v_{i_1 i_2\ldots i_k} s_{i_1} s_{i_2}\ldots s_{i_k}
\end{equation}
where $\Gamma$ is the graph of the neural network, the $v_{\bj}$ are weights and the $s_i$ take binary values\cite{sejnowski}. Models used in most applications are optimised in order to make decisions, and so are deterministic. However, by optimising the paramaters of (\ref{eq:nnm}) over a state space of inputs and outputs, $s_i$, the optimisation phase of these models involves the artificial brain exploring a subset of the randomised phase space of the model.

\begin{center}.....\end{center}

\bibliographystyle{alpha}
\bibliography{thesis_bib}
\end{document}